\documentclass[twoside,12pt]{report}

\usepackage{latexsym}
\usepackage{graphicx}
\usepackage{fancyhdr}

\begin{document}   

% Definizione comandi speciali

% Feynman slash 
\newbox\SlashedBox  
\def\fs#1{\setbox\SlashedBox=\hbox{#1} 
\hbox to 0pt{\hbox to 1\wd\SlashedBox{\hfil/\hfil}\hss}{#1}} 
\def\hboxtosizeof#1#2{\setbox\SlashedBox=\hbox{#1} 
\hbox to 1\wd\SlashedBox{#2}} 
\def\littleFraction#1#2{\hbox{$#1\over#2$}} 
\def\ms#1{\setbox\SlashedBox=\hbox{$#1$}
\hbox to 0pt{\hbox to 1\wd\SlashedBox{\hfil/\hfil}\hss}#1}
\def\partialslash{\mathslashed{\partial}}
%%% 
\newcommand{\dd}{\raisebox{10pt}
           {\tiny$\scriptscriptstyle\longleftrightarrow$}\hspace{-11.7pt}}
\newcommand{\db}{\raisebox{10pt}
           {\tiny$\scriptscriptstyle\longleftarrow$}\hspace{-11pt}}
\newcommand{\dbo}{\raisebox{11.5pt}
           {\tiny$\scriptscriptstyle\longleftarrow$}\hspace{-11pt}}           
%%%
\newcommand{\Dsm}{\,{\raisebox{1pt}{$/$} \hspace{-8.7pt} D}}
\newcommand{\Ds}{\,{\raisebox{1pt}{$/$} \hspace{-15.7pt} D}}
\newcommand{\desm}{\,{\raisebox{1pt}{$/$} \hspace{-7pt} \partial}}
\newcommand{\des}{\,{\raisebox{1pt}{$/$} \hspace{-14.5pt} $\partial$}}
\newcommand{\napp}{\raisebox{0.5pt}{/} \hspace{-10.5pt} \in}
%%% 
\def\noblackbox{\overfullrule=0pt}
%%%
\def\I {1 \hspace{-1.1mm} {\rm I}}
\def\R {{\rm I}\! {\rm R}}
\def\C {\rule{0.6pt}{8pt} \hspace{-1.35mm} {\rm C}}
\def\sC {\mbox{\rule{0.3pt}{4pt}} \hspace{-0.75mm} \mbox{C}} 
\def\Z {\mbox{\sf{Z}} \hspace{-1.6mm} \mbox{\sf{Z}} \hspace{0.4mm}}
%%%
\newcommand{\tr}{{\rm tr}}
\newcommand{\ie}{{\em i.e.~}}
\newcommand{\eg}{{\em e.g.~}}
%%%
\newcommand{\RR}{$R\otimes R$ }
\newcommand{\NSNS}{$NS\otimes NS$ }
\newcommand{\RNS}{$R\otimes NS$ }
\newcommand{\NSR}{$NS\otimes R$ }
\newcommand{\alpr}{{\alpha^{\prime}}}
\newcommand{\gym}{g_{_{{\rm YM}}}}
\newcommand{\tym}{\theta_{_{{\rm YM}}}}
\newcommand{\be}{\begin{equation}}
\newcommand{\ee}{\end{equation}}
\newcommand{\bea}{\begin{eqnarray}}
\newcommand{\eea}{\end{eqnarray}}
\newcommand{\beq}{\begin{eqnarray}}
\newcommand{\eeq}{\end{eqnarray}}
\newcommand{\ba}{\begin{eqnarray}}
\newcommand{\ea}{\end{eqnarray}}
\def\a{\alpha}
\def\b{\beta}
\def\g{\gamma}
\def\G{\Gamma}
\def\d{\delta}
\def\e{\epsilon}
\def\z{\zeta}
\def\h{\eta}
\def\L{\Lambda}
\def\th{\theta}
\def\k{\kappa}
\def\l{\lambda}
\def\m{\mu}
\def\n{\nu}
\def\x{\xi}
\def\X{\Xi}
\def\p{\pi}
\def\P{\Pi}
\def\r{\rho}
\def\s{\sigma}
\def\S{\Sigma}
\def\t{\tau}
\def\f{\phi}
\def\F{\Phi}
\def\c{\chi}
\def\w{\omega}
\def\W{\Omega}
\def\de{\partial}

%%%

%%%

% FRONT PAGE AND ABSTRACT

\thispagestyle{empty}

\vspace*{-2cm}
% \begin{flushright}
% ROM2F/2002/04
% \end{flushright}
\hbox{\hskip 12cm NIKHEF/2002-002 \hfil}
\hbox{\hskip 12cm ROM2F/2002/04  \hfil}
\hbox{\hskip 12cm hep-th/0203157 \hfil}

\vspace{1cm}

\begin{center}

{\LARGE {\bf Low-energy structure of six-dimensional open-string vacua}}

\vspace{1cm}
{\large Fabio Riccioni} \\ 
\vspace{0.6cm} 
{\large {\it NIKHEF}}\\
{\large {\it Kruislaan 409}}\\
{\large {\it 1098 SJ \ Amsterdam, \ The Netherlands}}\\
{\large {\it and}}\\
{\large {\it Dipartimento di Fisica, \ Universit{\`a} di Roma \  
``Tor Vergata''}} \\  
{\large {\it I.N.F.N.\ -\ Sezione di Roma \ ``Tor Vergata''}} \\ 
{\large {\it Via della Ricerca  Scientifica, 1}} \\ 
{\large {\it 00133 \ Roma, \ Italy}} 
\end{center}

\vspace*{1cm}
{\small 
\begin{center} {\bf Abstract} \end{center}
\noindent
This dissertation reviews some properties of the low-energy effective 
actions for six dimensional open-string models. 
The first chapter is a pedagogical introduction about supergravity theories.
In the second chapter closed strings are analyzed, with particular 
emphasis on type IIB, whose orientifold projection, 
in order to build type-I models, is the subject of the third chapter. 
Original results are reported in chapters 4 and 5.
In chapter 4 we describe the complete coupling of (1,0) six-dimensional 
supergravity to tensor, vector and hypermultiplets. The generalized 
Green-Schwarz mechanism implies that the resulting theory embodies 
factorized gauge and supersymmetry anomalies, to be disposed of by 
fermion loops. Consequently, the low-energy theory is determined by 
the Wess-Zumino consistency conditions, rather than by the requirement 
of supersymmetry, and this procedure does not fix a quartic coupling 
for the gauginos.
In chapter 5 we describe the low-energy effective actions 
for type-I models with brane supersymmetry breaking, resulting form
the simultaneous presence of supersymmetric bulks, with one or 
more gravitinos, and non-supersymmetric combinations of BPS branes.
The consistency of the resulting gravitino couplings implies that local  
supersymmetry is non-linearly realized on some branes. We analyze in detail 
the ten-dimensional $USp(32)$ model and the six-dimensional (1,0) models. 
}

\newpage

\baselineskip=18pt                 %

% TITLE

\thispagestyle{empty}

\vspace*{-1.5truecm}

\hspace*{-0.6cm}\makebox[\textwidth]{
	\includegraphics[height=1.45truecm]{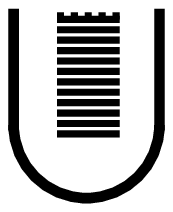}
	\raisebox{0.68cm}{
	\begin{minipage}[h]{13truecm}
        \hspace*{0.2cm}{\LARGE UNIVERSIT\`A DEGLI STUDI DI ROMA 
        \hspace*{3.76cm}``TOR \rule{0pt}{24pt}VERGATA''}
    \end{minipage}}}

\vspace{0.8cm}

\begin{center}
FACOLT\`A DI SCIENZE MATEMATICHE, FISICHE E NATURALI \\
Dipartimento di Fisica \\
\end{center}

\vspace{2cm}

\begin{center}
{\LARGE {\bf Low-energy structure of six-dimensional
open-string vacua}}
\end{center}
   
\vspace{1.5cm}
                  
\begin{center}
{\large Tesi di dottorato di ricerca in Fisica} \\
\rule{0pt}{22pt} presentata da \rule{0pt}{22pt} \\ 
\rule{0pt}{22pt}{\large {\em Fabio Riccioni}} \rule{0pt}{22pt}
\end{center}

\vspace{1.5cm}                  
\noindent
Relatore \\
\rule{0pt}{20pt}{\large Prof. {\em Augusto Sagnotti} \\ 
\rule{0pt}{20pt}} \\ 
Coordinatore del dottorato \rule{0pt}{30pt} \\ 
\rule{0pt}{20pt}{\large Prof. {\em Piergiorgio Picozza}} \\ 

\vspace{1cm}
\begin{center}
\underline{Ciclo XIV} \\
\vspace{0.5cm}
Anno Accademico 2000-2001
\end{center}

% \newpage ~                  %  per lasciare una pagina bianca
% \thispagestyle{empty}       %  

%%%

% % Definizione dei parametri di stampa
% 
% \baselineskip=19pt                 %
% \textheight=22truecm               %   30 righe per pagina
% \textwidth=15.2truecm              %
% 
% \headheight=15pt                   %
% \headsep=1.2cm                     %

%%% 

\pagestyle{fancy}
\renewcommand{\headrulewidth}{0.4pt}
\addtolength{\headwidth}{1cm}
\fancyhead{}
\fancyfoot{}

%%%
 
% PREFACE

\pagenumbering{roman}

\setcounter{page}{1}         %  per avere la corretta numerazione
                             %  quando si include il titolo

\renewcommand{\headrulewidth}{0.4pt}

% INDEX

\fancyhead[RO,LE]{\thepage}
\fancyhead[RE,LO]{{\footnotesize {\rm Table of contents}}}

\tableofcontents

\newpage
\fancyhead{}
\fancyfoot{}
\renewcommand{\headrulewidth}{0pt}

\addcontentsline{toc}{chapter}{Acknowledgments}
% \fancyfoot[C]{\thepage}
\thispagestyle{plain}

{\Large \textit{\textbf{Acknowledgments}}}

\vspace*{0.5cm}
\noindent
{\em 
This dissertation is based on research done at the Physics Department 
of the {\rm Universit\`a di Roma ``Tor Vergata''}, during the period 
November 1998 - October 2001, under the supervision of Prof. Augusto Sagnotti,
that I would like to thank for his suggestions and encouragement. 
I am also grateful to Dr. Gianfranco Pradisi for the enjoyable and 
fruitful collaboration during the last part of my Ph.D., on the subject 
reported in the last chapter. During my studies, I attended the ``Semester on 
Supergravity, Superstrings and M-Theory'' at the Centre Emile Borel,
Insitut Henri Poincar\'e, Paris (september-december 2000), 
and the Les Houches 2001 Summer School 
``Unity from duality: gravity, gauge theory and strings''. 
I would like to thank the organizers of these two schools for having given
me the opportunity to be there and the partecipants for the very 
stimulating and enjoyable atmosphere. Finally, 
I am grateful to the whole theoretical physics group at the Department 
of Physics of the  {\rm Universit\`a di Roma ``Tor Vergata''} for its hospitality.
}

\vspace*{0.5cm}
\begin{center}\rule{3cm}{0.4pt}\end{center}

% CHAPTERS

\fancyhead{}
\fancyfoot{}
\renewcommand{\headrulewidth}{0.4pt}

\newpage ~                  %  per lasciare una pagina bianca
\thispagestyle{empty}       %  tra l'indice e il primo capitolo

% \newpage ~                  %  
% \thispagestyle{empty}       %  

\chapter*{Introduction}
\addcontentsline{toc}{chapter}{Introduction}
\pagenumbering{arabic}
\setcounter{page}{1}
% \newpage
\vspace*{2cm}
\fancyhead[RO,LE]{\thepage}
\fancyhead[RE,LO]{{\footnotesize {\rm Introduction}}} 

\noindent
String theory is an extremely powerful tool for the quantization of  
gravity and the unification of fundamental 
interactions. In the spectrum of excitations of closed strings, a
massless spin 2 field is always present, and its interactions 
are described at low-energies by the Einstein-Hilbert action. Typically, requiring
that in the low-energy limit gravitational interactions be regulated by Newton's constant fixes 
the string scale to be of the order of the Planck scale. This explains why particles behave 
as pointlike objects at low energies, and
if one restricts the world-sheet action of the string to be 
supersymmetric, consistency selects the space-time dimension to be equal
to 10. Moreover, in ten dimensions there are five superstring theories, \ie
five theories that have a supersymmetric spectrum. In order to obtain 
models that are phenomenologically interesting, one has then to
suitably compactify them, so that the resulting vacua be four-dimensional.
Although these theories
have different perturbative spectra, at the non-perturbative level they are 
related by dualities, and the picture that emerges is that the
(unique and unknown) complete theory behind them is described by
different string theories in different regimes. 

While four of these five theories contain only closed strings at the perturbative level, 
type-I string theory contains open strings as well. As we will see 
throughout this thesis, this peculiarity of type-I strings will give rise to several interesting
physical phenomena, that in closed string theories correspond to non-perturbative effects. More 
precisely, the hyper-surfaces on which the open strings end (D-branes) are dynamical objects whose 
excitations are open-string modes, and  they correspond to non-perturbative states of the closed string 
theories. 
It is then interesting to study compactifications of type-I models, and in this respect 
the analysis of type-I vacua with minimal supersymmetry in six-dimensions 
turns out to be particularly rich, and will be the main topic of this dissertation.

Before I started my PhD, during the period of my INFN Pre-Doctoral Fellowship, I analyzed
minimal six-dimensional supergravity in collaboration with S. Ferrara and A. Sagnotti \cite{frs}, 
concentrating in particular on the couplings of supergravity to tensor multiplets and to 
non-abelian vector multiplets. 
The result we ended up with is that this theory is not completely determined by supersymmetry, 
since consistency conditions leave a quartic gaugino coupling undetermined. Subsequently, in
\cite{rs1,rs2,rs3}, we also came to a better understanding of the properties of these models. This
dissertation is partly a continuation of these results: during the first part of my PhD I
completed the low-energy effective action of six-dimensional supergravity coupled to 
tensor multiplets and abelian vector multiplets \cite{fr2}, showing that in the abelian case 
additional couplings can be added, and then in \cite{fr3} I obtained the complete action
of supergravity coupled to vector, tensor and hypermultiplets. These results are the 
subject of Sections (4.4) and (4.5) of this thesis.

Another peculiar feature of type-I strings is that one can consider brane-world scenarios, in which our 
universe is confined on some coincident branes, where the standard model lives, while gravity invades 
the whole 
bulk. In this scenarios, one can naturally lower the string scale to the order 
of the supersymmetry scale, since 
hierarchy is generated by compactification, and thus the Planck scale is obtained dynamically.
Moreover, one can consistently obtain ``brane supersymmery breaking'' models, 
in which supersymmetry is realized in the bulk,
at least to lowest order, while it is broken on some branes.
During the last part of my PhD, I have analyzed in collaboration 
with G. Pradisi the low-energy effective action 
corresponding  to type-I ``brane supersymmetry breaking'' models \cite{pr}. 
These results are collected in Chapter 5.

The thesis is organized as follows. Chapter 1 is a 
pedagogical introduction about supergravity theories, centered on topics  
that will become useful for the following, and 
in particular for Chapter 4.
In Chapter 2 I analyze closed strings, with particular emphasis on type IIB, while  
its orientifold projection, 
in order to build type-I models, will be the subject of Chapter 3. Chapter 4 is devoted to a 
detailed description of minimal supergravity in six dimensions.
In Chapter 5 I derive to lowest order in the Fermi fields the effective action for type-I brane supersymmetry breaking
models. Finally, the last chapter is devoted to the conclusions, 
while the Appendix contains some conventions 
and useful identities.

\chapter{Generalities about supergravity theories}
\label{cap1}
% \pagenumbering{arabic}
% \setcounter{page}{1}
% \newpage
\vspace*{2cm}
\fancyhead[RO,LE]{\thepage}
\fancyhead[RE]{{\footnotesize {\rm Chapter 1.}~~{\it Generalities about supergravity theories}}} 

\noindent
Supersymmetry is a space-time symmetry that combines bosonic and fermionic
fields. The simplest example of a model invariant under supersymmetry 
transformations is the four-dimensional Wess-Zumino model \cite{wesszumino}, consisting of
a complex scalar and a Weyl fermion. The lagrangian of the model,
$$
{\cal L}= \de_\m {\phi}^* \de^\m \phi + i\bar{\psi} \g^\m \de_\m \psi \quad ,
$$
is invariant under the supersymmetry transformations
$$
\d \phi = \bar{\e } \psi \quad , \qquad
\d \psi = - i \g^\m \e \de_\m \phi \quad ,
$$
where $\e$ is a constant Weyl spinor with opposite chirality with respect to $\psi$. 
There are two important properties that this model shares with all other supersymmetric theories:
the first is the fact that the commutator of two supersymmetry transformations on the 
bosonic field generates a translation, 
$$
[\d_1 , \d_2 ] \phi = - 2 i (\bar{\e}_2 \g^\m \e_1 ) \de_\m \phi 
\quad ,
$$ 
the second is the fact that on the spinor, the same algebra is realized only on-shell. 
The first result is a general property of the supersymmetry algebra, and 
corresponds to the fact that the anticommutator of two supersymmetry generators gives the momentum, 
the generator of translations.
The second result is due to the fact that the matching of bosonic and fermionic degrees of freedom
actually only holds on-shell. In order to have the same matching also off-shell, we should add suitable 
auxiliary fields, and then, in the complete set of fields, the supersymmetry algebra will close exactly. 
We will not consider
the (partly unsolved) problem of finding the off-shell representations of supersymmetry
throughout this thesis, and in fact in 
Chapter 4 we will use the property that the supersymmetry algebra 
closes only on shell in order to determine the field equations for the fermionic fields of six-dimensional
supergravity.  

An additional step is to try to make this symmetry local. 
There is a field that can naturally be regarded as the gauge field of supersymmetry,  
the Rarita-Schwinger field $\psi_\m$, whose linearized equation
$$
i \g^{\m\n\r} \de_\n \psi_\r =0
$$
is invariant under
$$
\d \psi_\m = \de_\m \e \quad .
$$
From the supersymmetry algebra we learn also that if the parameter 
of supersymmetry is local (\ie space-time dependent), then the algebra generates a translation 
with a local parameter, a general coordinate transformation. If a field theory is invariant 
under general coordinate transformations, it must contain general relativity. So the important lesson
that we learn is that a theory invariant under local supersymmetry, containing the field $\psi_\m$, 
must contain gravity as well. The field $\psi_\m$ is then the supersymmetric partner of the 
graviton, and for this reason it is called gravitino. 

In this chapter we want to give a general introduction to supergravity theories. We begin in Section 1 
with a brief discussion about torsion in general relativity, and in Section 2
we describe the ${\cal N}=1$ four-dimensional model, that contains only
a graviton and a gravitino, but already reveals some of the subtleties common to other 
supergravity models. In Section 3 we consider D=11 supergravity and finally in Section 4
we briefly describe the various supergravity theories that arise in ten dimensions.

\section{Torsion in General Relativity}
\label{torsion}
\fancyhead[LO]{{\footnotesize 1.1~~{\it Torsion in general relativity}}}

In this section we report some known results on general relativity \cite{wald} that 
are essential for the following, in particular for Chapter 4.
First of all, we fix the notations that we will follow throughout the thesis. The metric has signature
$(+,-,...,-)$, 
the covariant derivative for a vector has the form
\be
\nabla_\m V_\n =\de_\m V_\n -\G^\a{}_{\m\n}V_\a \quad,
\ee
where $\G$ is the Christoffel connection, and this covariant derivative can be rewritten
as
\be
D_\m V^m = \de_\m V^m +\w_{\m}{}^m{}_n V^n \quad ,
\ee
where $\w$ is the spin connection, in terms of 
\be
V^m = e_\m{}^m V^\m
\ee
whose indices are on the locally tangent space. The covariant derivative for a spinor is then
\be
D_\m \chi = \de_\m \chi +\frac{1}{4} \w_{\m mn} \g^{mn} \chi \quad ,
\ee
and these two covariant derivatives are consistent because the vierbein is
covariantly constant:
\be
\nabla_\m e_\n{}^m = \de_\m e_\n{}^m - \G^\r{}_{\m\n} e_\r{}^m + \w_{\m}{}^m{}_n e_\n{}^n =0 \quad .
\ee
A priori, the Christoffel connection has no determined symmetry with respect to its lower indices.
Nevertheless, it is important to observe that if the connection is chosen to be symmetric, then
it is completely determined by the condition that the metric be covariantly constant. Denoting with
$\{ \}$ this connection, the result is
\be
\{^\r{}_{\m\n} \}= \frac{1}{2} g^{\r\s} (\de_\m g_{\s\n} +\de_\n g_{\s\m} - \de_\s g_{\m\n} ) \quad .
\ee
Correspondingly, the spin connection in this case is also completely determined in terms of the vielbein, 
and 
\be
e_\n{}^m e_\r{}^n \w^0_{\m mn} = \frac{1}{2} [ e_{\r p} (\de_\m e_\n{}^p - \de_\n e_\m{}^p )
- e_{\m p} (\de_\m e_\r{}^p - \de_\r e_\n{}^p )+ e_{\n p} (\de_\r e_\m{}^p - \de_\m e_\r{}^p )]\quad .
\ee 
In general, one can consider a Christoffel connection with an antisymmetric part, that is called torsion,
\be
\G^\r_{\m\n} = \G_0^\r{}_{\m\n} + S^\r{}_{\m\n}\quad ,
\ee
where $\G_0$ is symmetric and $S$ is antisymmetric under the interchange of the lower indices. 
It is important to observe that this different choice of connection simply corresponds to the addition
of covariant tensors to the minimal choice $\{ \}$. In fact, it can be shown that the torsion 
is a covariant tensor, and the symmetric part of the connection becomes
\be
\G_0^\r{}_{\m\n} = \{^\r{}_{\m\n} \}+ S_\m{}^\r{}_\n + S_\n{}^\r{}_\m \quad .
\ee
In the presence of torsion, the spin connection is also modified by the addition of covariant terms:
\be
\w_{\m mn} = \w^0_{\m mn} + S_{m \m n} + S_{\m mn} + S_{nm \m} \quad ,\label{spinconn}
\ee
while the condition that the metric and the vierbein be covariantly constant
is independent of the torsion.

In our notations, the Riemann tensor has the form
\be
R^\a{}_{\m\n\r} (\G )= \de_\n \G^\a{}_{\r\m} -\de_\r \G^\a{}_{\n\m} + \G^\b{}_{\r\m} \G^\a{}_{\n\b}
- \G^\b{}_{\n\m} \G^\a{}_{\r\b}
\ee
in terms of the Christoffel conection, or
\be
R_{\m\n}{}^{mn} (\w )= \de_\m \w_\n{}^{mn} - \de_\n \w_\m{}^{mn}+\w_{\m}{}^{mp} \w_{\n p}{}^n
- \w_{\n}{}^{mp} \w_{\m p}{}^n
\ee
in terms of the spin connetion, and the fact that the metric is covariantly constant implies that these 
two curvature tensors are equivalent, so that 
\be
e_\m{}^n R_{\s\t mn} (\w ) = R^\a{}_{\m\s\t} (\G ) e_{\a m} \quad .
\ee
Consequently, the Ricci tensor is
\be
R_{\m\n} (\G ) = e_\m{}^n e^{\t m} R_{\t\n mn} (\w )
\ee
and the Ricci scalar is
\be
R (\G ) = e^\s{}_m e^{\t}{}_n R_{\s\t}{}^{mn} (\w )\quad .
\ee
This curvature tensor is the object that naturally appears in the commutator of two covariant derivatives, 
while the presence of torsion corresponds to the addition in the commutators of a term containing 
the covatiant derivative of the vector, so that 
\be
[ \nabla_\r , \nabla_\n ] V_\m = R^\a{}_{\m\n\r} (\G )V_\a - 2 S^\a{}_{\r\n} \nabla_\a V_\m \quad .
\ee

One of the consequences of torsion is the fact that the Ricci tensor is no longer 
symmetric, and precisely
\be
R_{\m\n} -R_{\n\m} = 2 \nabla_\a S^\a{}_{\n\m}-2 \nabla_\n S^\a{}_{\a\m} +2 \nabla_\m S^\a{}_{\a\n}
- 4 S^\b{}_{\n\m} S^\a{}_{\a\b} \quad ,
\ee
while it can also be shown that the relation
\be
\G^\a{}_{\m\a} = \de_\m \ln \sqrt{-g}
\ee
is still true in the presence of torsion.

There are three different lagrangian formulations of general relativity. In the second 
order formulation  one assumes
from the start that the torsion vanishes, and considers
the theory as a function of the metric $g_{\m\n}$ only.  The 
connections are determined in terms of the metric by the requirement that the torsion be zero. 
In the first order (Palatini) formulation, the metric and the connection are assumed to be independent.
It is also possible to consider a pure connection formulation, in which the metric is obtained as a 
composite field. 
We now show that for pure gravity, the first order and the second order formulations coincide.
Fixing Newton's constant appropriately,  
the Einstein-Hilbert action takes the form
\be
S= -\frac{1}{4} \int d^D x \sqrt{-g} R \quad .
\ee
The variation of the Riemann tensor is
\be
\d R_{\m\n} = \nabla_\a \d \G^\a{}_{\n\m} - \nabla_\n \d \G^\a{}_{\a\m} + 2S^\b{}_{\a\n} \d \G^\a{}_{\b\m}
\quad ,
\ee
where the first two terms do not contribute to the field equation, and 
as a consequence, one sees that the second order formulation, where  the absence
of torsion is required as a condition, gives the same equation as the first order formulation, where the
absence of torsion comes from the equation of the connection. 
The field equation is then completely determined by the variation of the metric, and is 
\be
R_{\m\n}-\frac{1}{2} g_{\m\n} R = 0 \quad .
\ee
The same result is obtained if one expresses the Riemann tensor in terms of the spin connection, 
and indeed 
varying the action
\be
S= -\frac{1}{4} \int d^D x e e^\m{}_m e^\n{}_n R_{\m\n}^{mn} (\w )
\ee
with respect to the spin connection $\w$ gives 
\be
\d S = \frac{3}{2} \int d^D x e e^\m{}_{[ m} e^\n{}_n e^\r{}_{p ]} D_\m e_\r{}^p \d \w_\n{}^{mn}\quad ,
\ee
where the covariant derivative $D$ contains only the spin connection. The field equation for $\w$
then becomes $D_{[\m} e_{\n ]}^m = 0$, that implies the absence of torsion.
Now suppose that matter is present, and consider the simple case of a single spinor, whose lagrangian is
\be
{\cal L} = \frac{i}{2} \bar{\chi} \g^\m D_\m \chi \quad ,
\ee
where
\be
D_\m \chi =\de_\m \chi +\frac{1}{4} \w_{\m mn}\g^{mn} \chi \quad .
\ee
The equation for $\w$ now implies that the spin connection is 
\be
\w_{\m mn}= \w^0_{\m mn} -\frac{i}{4} \bar{\chi}\g_{\m mn} \chi \quad ,
\ee
and from eq. (\ref{spinconn}) this corresponds to the presence of the torsion
\be
S_{\m\n\r} =\frac{i}{4} \bar{\chi}\g_{\m \n\r} \chi \quad .
\ee

In supergravity theories it is necessary  
to consider the spin connection rather than the Christoffel connection in
the lagrangian, since these theories contain spinors. This is also natural, 
since these theories typically contain $p$-forms, antisymmetric 
tensors with $p$ space-time indices, that generalize vectors in higher dimensions. The field strength for a
$p$-form $A$ is $F = d A$, that in components means
\be
F_{\m_1 ... \m_p } = p \de_{[ \m_1 } A_{\m_2 ... \m_p ]} \quad ,
\ee
and the fact that the indices are completely antisymmetrized implies that this object is covariant without 
the addition of any connection. The gauge invariance of $F$ under
\be
\d A_p = d \L_{p-1} 
\ee
is preserved only if no torsion term is present. Moreover, the Rarita-Schwinger action is covariant in terms
of the covariant derivative $D$, with only the spin connection, and thus in general Christoffel and spin connection enter very differently the relevant couplings.  

The simplest supergravity theory is ${\cal N}  =1$ 
supergravity in four dimensions \cite{ffvn}, that contains only  the graviton and a gravitino. 
The invariance of this 
model under supersymmetry was originally shown in the second order formalism in \cite{ffvn} and then 
in the first order formalism in \cite{deszum}.  
One could actually make things simpler, and combine the advantages of both formalisms, considering 
the spin connection as an independent field, but always imposing that it satisfy
its field equations (see Ref. \cite{vannieu}). This formalism, known as the 1.5 formalism,
is the most natural way to formulate supergravity theories, and is the one we will use in the next 
sections and in Chapter 4.
In the next section we will explicitly 
see how all this works for minimal supergravity in four dimensions, 
while in Section 3 we will consider 11-dimensional supergravity.

\section{Minimal supergravity in four dimensions}
\label{susyreps}
\fancyhead[LO]{{\footnotesize 1.2~~{\it Minimal supergravity in four dimensions}}}

The ${\cal N}=1$ four-dimensional supergravity multiplet contains just the graviton and a Majorana 
gravitino. In this section we show how the lagrangian for this model is constructed, 
showing  that the lagrangian
\be
{\cal L} = -\frac{1}{4}e R -\frac{ie}{2}\bar{\psi}_\m \g^{\m\n\r} D_\n \psi_\r \label{lagd=4}
\ee
is invariant under the supersymmetry transformations
\bea
\d e_\m{}^m &=& - i \bar{\e} \g^m \psi_\m \quad ,\nonumber \\ 
\d \psi_\m &=& D_\m \e \quad .\label{susyd=4}
\eea
First of all, we prove that supersymmetry holds to lowest (\ie quadratic) order in the fermions.  
This means that we have to consider the equation for the metric without gravitino terms, and the equation
for the gravitino without cubic terms in the gravitino itself. At this level the spin 
connection does not contain torsion, that as we saw in the previous section is quadratic 
in the spinors. To lowest order, the variation of the lagrangian is
\be
\d {\cal L} = -\frac{i}{2}e (\bar{\e} \g_\m \psi_\n ) [R^{\m\n} -\frac{1}{2}g^{\m\n} R ]
- i e( D_\m \bar{\e} \g^{\m\n\r} D_\n \psi_\r ) \quad ,
\ee
where the covariant derivative contains the torsion-free spin connection.
The second term, integrated by parts, gives
\be
\frac{i}{8} e \bar{\e}\g^{\m\n\r}\g^{mn}\psi_\r R_{\m\n mn}\quad , \label{gamma5riem}
\ee
and using the properties of the Clifford algebra and the cyclic identity 
for the Riemann tensor one can show that it cancels with the variation of 
the Einstein-Hilbert action, so that 
supersymmetry holds to lowest order in the 
Fermi fields.
We emphasize that this result applies in any dimension, although 
the actual coefficient in front of the 
Rarita-Schwinger action depends on the reality properties of the spinor.

We now want to show that supersymmetry holds to all orders in the fermi fields, with a suitable 
definition of $\w$. In eq. (\ref{susyd=4}),
the variation of the vielbein does not receive any further correction, while one could add terms 
proportional to $\psi^2 \e$ to the variation of $\psi$. This is a general feature of supersymmetric
models: as we will see also in other cases, the transformations of the bosonic fields are quadratic in 
the fermions, and are competely determined at the lowest order, while the transformation of the 
gravitino contains a derivative term and terms cubic in the fermions. For
this reason the lagrangian can contain two-derivative terms, one-derivative terms contracted with
fermionic bilinears and four-fermi terms. Any higher order term in the fermions would 
correspond by supersymmetry to higher derivative terms.

The spin connection that satisfies its field equation is in this case 
\be
\w_{\m mn}= \w^0_{\m mn } -\frac{i}{2} e^\n{}_m e^\r{}_n [(\bar{\psi}_\m \g_\n \psi_\r )+
(\bar{\psi}_\n \g_\r \psi_\m ) +(\bar{\psi}_\n \g_\m \psi_\r ) ] \quad ,\label{spinconn4}
\ee
and consequently the coupling of gravity with the gravitino in four dimensions induces the torsion
\be
S_{\m\n\r} = -\frac{i}{2} (\bar{\psi}_\n \g_\m \psi_\r ) \quad .
\ee
As we will see in the next section, these two relations will be modified 
by a bilinear in the gravitino with five $\g$-matrices in dimensions greater than 4. Observe that 
the connection defined in eq. (\ref{spinconn4}) is supercovariant, since its supersymmetry variation does
not contain terms proportional to the derivative of $\e$.
Only for minimal supergravity in four dimensions the supercovariant 
spin connection is the one that satisfies its field equation, while 
in more complicated models additional fermions contribute to the torsion and to the connection
as well. 
The concept of supercovariance is very useful in order to construct supergravity theories, and  we 
will also use it in Chapter 4, when we will determine minimal six-dimensional supergravity theories
to all orders in the Fermi fields. Working in the 1.5 formalism, we do not consider the variation of 
the connection (\ref{spinconn4}). 

Because of the Fierz identity
\be
\g_m \psi_{[ \m} (\bar{\psi}_{\n} \g^m \psi_{\r ]} )= 0 \quad ,
\ee
that holds in four dimensions as a consequence of $\g_m \g^{np} \g^m = 0$, 
one can show that the gravitino field equation
resulting from the variation of the complete lagrangian in the 1.5 formalism,
\be
-i \g^{\m\n\r} D_\n \psi_\r = 0 \quad ,
\ee
is supercovariant, 
with the connection (\ref{spinconn4}).
Using similar relations one can then show that the lagrangian of eq. (\ref{lagd=4}) is invariant 
under the variations 
(\ref{susyd=4}), with the spin connection defined in eq. (\ref{spinconn4}). 
In this model supercovariance completely determines the supersymmetry
transformation and the field equation for the gravitino, and one can show using the Fierz relations 
collected in the Appendix that the complete variation of the action vanishes. 

We now show that the closure of the supersymmetry algebra gives exactly the same 
gravitino equation, from which one can recover 
the complete lagrangian (\ref{lagd=4}). Given a supergravity
model, the commutator of two supersymmetry transformations on any field in the multiplet gives all 
the possible local transformations for that field (supersymmetry, general coordinate, 
local Lorentz and gauge transformations). In addition, for fermionic fields, the algebra closes on shell. 
For instance, the commutator of two supersymmetry transformations acting on the vielbein is 
\be
[\d_{\e_1} , \d_{\e_2} ] e_\m{}^m =  \d_{gct}e_\m{}^m +\d_{lL}e_\m{}^m + \d_{susy}e_\m{}^m \quad ,
\ee
where the parameters of general coordinate, local Lorentz and supersymmetry transformations are
\be
\xi_\m = i (\bar{\e}_2 \g_\m \e_1 ) \quad , \qquad \W^{mn} = -\xi^\n \w_\n{}^{mn} \quad ,\qquad
\zeta = \xi^\n \psi_\n \quad .\label{param4}
\ee
Performing the same commutator on $\psi$, one can extract from it all the local symmetries, with the
same parameters as in eq. (\ref{param4}). The terms that are left must be zero on-shell, and thus from 
them one can read the field equation for $\psi$. From the field equation for the gravitino, one can
determine the lagrangian completely. 
\footnote{Observe that the connection must not be kept fixed when computing the commutator, 
since it is fixed in the lagrangian just because its variation generates the field equation.}
The end result is
\bea
[\d_{\e_1} , \d_{\e_2} ] e_\m{}^a &=&  \d_{gct} \psi +\d_{lL} \psi + \d_{susy} \psi \nonumber\\
&+& 3i \xi_\n \g^\n [ (eq.\psi)_\m -\frac{1}{3} \g_\m (\g -{\rm trace} ) ]
-2 i \xi^\n \g_\m [(eq.\psi)_\n -\frac{1}{2} \g_\n (\g -{\rm trace} ) ]\nonumber \\
&-& \frac{1}{4} (\bar{\e}_1 \g^{\n\r} \e_2 ) \g^\r \g_\m 
[(eq.\psi)_\n -\frac{1}{2} \g_\n (\g -{\rm trace} ) ]\quad ,
\eea
where with $\g -{\rm trace}$ we mean the gravitino equation contracted with a $\g$ matrix, \ie
\be
- 2i \g^{\n\r} D_\n \psi_\r =0 \quad .
\ee

\section{Eleven-dimensional supergravity}
\label{superspace1}
\fancyhead[LO]{{\footnotesize 1.3~~{\it Eleven-dimensional supergravity}}}

The number of components of the supersymmetry charge for minimal supersymmetry in four dimensions is 
4. The chiral multiplet that we described in the introductory section, 
the gravity multiplet and the vector multiplet, 
containing a vector and a Weyl spinor, are all the multiplets with minimal supersymmetry in four dimensions.
For all these multiplets, the number of bosonic and fermionic on-shell degrees of freedom is 2. 
In four dimensions one can also consider theories with extended supersymmetry. For instance the
${\cal N} =2$ gravity multiplet contains the graviton, two gravitinos and an abelian vector (the 
graviphoton). The on-shell matching of Bose and Fermi degrees of freedom is straightforward, and  
the number of supercharges is in this case 8. Continuing this way, one can show that the maximal number 
of supercharges compatible with representations of ordinary (\ie spin $\leq$2) fields is 32. Moreover, 
in this case one has a single representation of the supersymmetry algebra that contains fields
with spin $\leq$2, that is the ${\cal N} = 8$ gravity multiplet \cite{cj}. The corresponding lagrangian can be 
obtained by dimensional reduction from eleven-dimensional supergravity. Indeed, analyzing the properties 
of the spinors in various dimensions (see the Appendix), one can show that in $D=11$ the only possibility 
is to have 32 supercharges \cite{nahm}, and the corresponding multiplet is the gravity multiplet, containing the 
graviton (44 on-shell degrees of freedom), an antisymmetric 3-form (84 on-shell degrees of freedom) and
a Majorana gravitino (128  on-shell degrees of freedom). Dimensional reduction of this theory gives maximal 
supergravity in any dimension. As we will see in the next section and in Chapter 4, one 
can also have theories with 16 supercharges starting from ten dimensions, and theories with 8 supercharges
starting from six dimensions \cite{salamsez}. $D=11$ is the maximal dimension for 
which one can realize supersymmetry
in terms of an ordinary supergravity theory.

Returning to the eleven-dimensional supergravity \cite{cjs}, to lowest order, the lagrangian 
\bea
{\cal L} &=& -\frac{e}{4} R -\frac{ie}{2} ( \bar{\psi}_\m \g^{\m\n\r} D_\n \psi_\r )-\frac{e}{48}
F_{\m\n\r\s} F^{\m\n\r\s} \nonumber\\
&+& \frac{e}{96} F_{\r\s\d\t} (\bar{\psi} \g^{\m\n\r\s\d\t} \psi_\n ) +\frac{e}{8}
F_{\m\n\r\s} (\bar{\psi}^\m \g^{\n\r} \psi^\s ) \nonumber \\
&+& \frac{1}{144 \cdot 72} \e^{\a_1 ...\a_4 \b_1 ...\b_4
\m\n\r } F_{\a_1 ...\a_4 } F_{\b_1 ... \b_4 } A_{\m\n\r} \label{lag11lo}
\eea
is invariant under the supersymmetry transformations
\bea
& & \d e_\m{}^m = -i (\bar{\e} \g^m \psi_\m ) \nonumber \\
& & \d \psi_\m = D_\m \e +\frac{i}{144} F^{\n\r\s\t} \g_{\m\n\r\s\t} \e -\frac{i}{18} F_{\m\n\r\s}
\g^{\n\r\s} \e \nonumber\\
& & \d A_{\m\n\r} =\frac{3}{2} ( \bar{\e} \g_{[\m\n } \psi_{\r ]} )\quad ,
\eea 
where the field strength
\be
F_{\m_1 \m_2 \m_3 \m_4} = 4 \de_{[ \m_1} A_{\m_2 \m_3 \m_4 ]}
\ee
is invariant under the gauge transformation $\d A_{\m\n\r} = 3 \de_{[\m} \L_{\n\r ]}$.
In order to prove the invariance of the lagrangian, observe that 
the first term in the variation of the gravitino 
produces eq. (\ref{gamma5riem}), while this term cancels as in $D=4$ against 
the variation of the Einstein-Hilbert
action; the additional term containing five antisymmetrized $\g$ matrices, absent in $D=4$, vanishes
because of the cyclic identity
\be
R^\a{}_{[ \m\n\r ]}= 0 \quad .
\ee
The cancellation of the terms containg $F$ and a derivative or $F^2$ can be proved using
the relation
\be
\g^{\m_1 ... \m_n}= \frac{i (-1)^{[n/2]}}{e(11-n)!}\e^{\m_1 ...\m_n \n_1
...\n_{11-n}}\g_{\n_1 ...\n_{11-n}} \quad ,\label{gammamatrices11}
\ee 
derived from the similar ten-dimensional relation (\ref{gammamatrices10}) obtained in the Appendix. 
Observe the presence of the Wess-Zumino term $A \wedge F \wedge F$ in the last line of eq. (\ref{lag11}).
Similar terms typically  appear also in lower-dimensional supergravity theories, and 
in the next chapters we will see their implications, in particular for anomaly cancellations.

We now want to prove supersymmetry to all orders in the Fermi fields, working in the 1.5 order 
formalism.
The spin connection that satisfies its field equation is
\bea
\w_{\m mn} &=& \w^0_{\m mn } -\frac{i}{2} e^\n{}_m e^\r{}_n [(\bar{\psi}_\m \g_\n \psi_\r )+
(\bar{\psi}_\n \g_\r \psi_\m ) +(\bar{\psi}_\n \g_\m \psi_\r ) ] \nonumber \\
&-& \frac{i}{4} (\bar{\psi}^\n \g_{\m mn \n \r} \psi^\r ) 
\quad ,\label{spinconn11}
\eea
and differs from the supercovariant spin connection, that is given as
in $D=4$ by
\be
\hat{\w}_{\m mn}= \w^0_{\m mn } -\frac{i}{2} e^\n{}_m e^\r{}_n [(\bar{\psi}_\m \g_\n \psi_\r )+
(\bar{\psi}_\n \g_\r \psi_\m ) +(\bar{\psi}_\n \g_\m \psi_\r ) ] \quad . \label{scsc11}
\ee
Similarly, one can define the supercovariant 4-form field strength
\be
\hat{F}_{\m\n\r\s} = F_{\m\n\r\s} -3 (\bar{\psi}_{[ \m} \g_{\n\r} \psi_{\s ]} )\quad ,
\ee
and as in four dimensions, one can determine the lagrangian that gives a supercovariant 
gravitino field equation. The result is 
\bea
{\cal L} &=& -\frac{e}{4} R -\frac{ie}{2} ( \bar{\psi}_\m \g^{\m\n\r} D_\n (\frac{\w +\hat{\w}}{2})
\psi_\r )-\frac{e}{48}
F_{\m\n\r\s} F^{\m\n\r\s} \nonumber\\
&+& \frac{e}{192}( F+ \hat{F})_{\r\s\d\t} (\bar{\psi} \g^{\m\n\r\s\d\t} \psi_\n ) +\frac{e}{16}
(F+ \hat{F})_{\m\n\r\s} (\bar{\psi}^\m \g^{\n\r} \psi^\s ) \nonumber \\
&+& \frac{1}{144 \cdot 72} \e^{\a_1 ...\a_4 \b_1 ...\b_4
\m\n\r } F_{\a_1 ...\a_4 } F_{\b_1 ... \b_4 } A_{\m\n\r} \label{lag11}\quad . 
\eea
One can show that the resulting gravitino equation is
\be
-ie \g^{\m\n\r} D_\n (\hat{\w}) \psi_\r +\frac{e}{48} \hat{F}_{\r\s\d\t}\g^{\m\n\r\s\d\t}\psi_\n
+\frac{e}{4} \hat{F}^{\m\n\r\s} \g_{\n\r} \psi_\s =0 \label{supercov11}
\ee
using the Fierz identity
\bea
& & \frac{1}{8}\g^{\m\n\a\b\g\d} \psi_\n (\bar{\psi}_\a \g_{\b\g} \psi_\d )-\frac{1}{8}
\g_{\b\g} \psi_\n (\bar{\psi}_\a \g^{\m\n\a\b\g\d} \psi_\d ) \nonumber \\
& &  -\frac{1}{4} \g^{\m\n\a\b\g}
\psi_\n (\bar{\psi}_\a \g_\b \psi_\g ) +\frac{1}{4} \g_\b (\bar{\psi}_\a \g^{\m\n\a\b\g} \psi_\g
\nonumber \\
& & +\g^{\a\b\g} \psi_\b (\bar{\psi}_\a \g^\m \psi_\g ) -2 \g^{\m\a\b}\psi_\b (\bar{\psi}_\a \g^\g \psi_\g )
+ \g^{\m\a\b}\psi_\g (\bar{\psi}_\a \g^\g \psi_\b ) =0 \quad .
\eea
Using similar Fierz identities one can then show that the lagrangian (\ref{lag11}) 
is invariant under the supersymmetry transformations
\bea
& & \d e_\m{}^m = -i (\bar{\e} \g^m \psi_\m ) \nonumber \\
& & \d \psi_\m = D_\m (\hat{\w }) \e +\frac{i}{144} \hat{F}^{\n\r\s\t} \g_{\m\n\r\s\t} \e -\frac{i}{18} 
\hat{F}_{\m\n\r\s}
\g^{\n\r\s} \e \nonumber\\
& & \d A_{\m\n\r} =\frac{3}{2} ( \bar{\e} \g_{[\m\n } \psi_{\r ]} )\quad .\label{susy11}
\eea 
Observe again that both the field equation 
and the supersymmetry transformation of the gravitino are supercovariant.

As we showed in the four dimensional case, one can arrive at the same result imposing the 
closure of the supersymmetry algebra. Indeed, the commutator of two supersymmetry 
transformations gives local Lorentz, general coordinate and
supersymmetry transformations on the vielbein, while in the case of the 3-form gives an additional
gauge transformation and in the case of the gravitino additional terms proportional to  
its field equation. The parameters of general coordinate, 
local Lorentz, supersymmetry and 3-form gauge transformations are
\bea
& & \xi_\m = i (\bar{\e}_2 \g_\m \e_1 ) \quad , \nonumber \\
& & \W^{mn} = -\xi^\n \w_\n{}^{mn} +\frac{1}{72} \hat{F}_{\m\n\r\s} (\bar{\e}_2 
\g^{\m\n\r\s mn} \e_1 ) +\frac{1}{3} \hat{F}^{\m\n mn} (\bar{\e}_2 \g_{\m\n} \e_1 )\quad ,\nonumber \\
& & \zeta = \xi^\n \psi_\n \quad , \nonumber \\
& & \L_{\m\n} =\frac{1}{2} (\bar{\e}_2 \g_{\m\n} \e_1 ) + \xi^\r A_{\m\n\r} \quad .\label{param11}
\eea
Performing the same commutator on $\psi$, and extracting from it all the local symmetries, with the
same parameters as in eq. (\ref{param11}), one can read the field equation for $\psi$, that results to be 
the supercovariant equation (\ref{supercov11}).
Once the field equation for the gravitino is known, integrating it one determines 
the lagrangian completely. 

By dimensional analysis, one can reinsert Newton's constant in eqs. (\ref{lag11}) and (\ref{susy11}),
obtaining
\bea
{\cal L} &=& -\frac{e}{4 \kappa^2} R -\frac{ie}{2} ( \bar{\psi}_\m \g^{\m\n\r} D_\n (\frac{\w +\hat{\w}}{2})
\psi_\r )-\frac{e}{48}
F_{\m\n\r\s} F^{\m\n\r\s} \nonumber\\
&+& \frac{e \kappa}{192}( F+ \hat{F})_{\r\s\d\t} (\bar{\psi} \g^{\m\n\r\s\d\t} \psi_\n ) +\frac{e}{16}
(F+ \hat{F})_{\m\n\r\s} (\bar{\psi}^\m \g^{\n\r} \psi^\s ) \nonumber \\
&+& \frac{\kappa}{144 \cdot 72} \e^{\a_1 ...\a_4 \b_1 ...\b_4
\m\n\r } F_{\a_1 ...\a_4 } F_{\b_1 ... \b_4 } A_{\m\n\r} \label{lag11b} 
\eea
and
\bea
& & \d e_\m{}^m = -i \kappa (\bar{\e} \g^m \psi_\m ) \nonumber \\
& & \d \psi_\m = \frac{1}{\kappa}
D_\m \e +\frac{i}{144} \hat{F}^{\n\r\s\t} \g_{\m\n\r\s\t} \e -\frac{i}{18} 
\hat{F}_{\m\n\r\s}
\g^{\n\r\s} \e \nonumber\\
& & \d A_{\m\n\r} =\frac{3}{2} ( \bar{\e} \g_{[\m\n } \psi_{\r ]} )\quad .\label{susy11b}
\eea 

\section{Ten-dimensional supergravities}
\label{susymod}
\fancyhead[LO]{{\footnotesize 1.4~~{\it Ten-dimensional supergravities}}}

In this section we shortly describe supergravity theories in ten dimensions. There are two supergravity
theories with 32 supercharges, the ${\cal N} = 2a$ and ${\cal N}= 2b$ supergravities, and one theory 
with 16 supercharges, the ${\cal N}= 1$ supergravity, that can be coupled to a Yang-Mills vector 
multiplet.

\subsection{${\cal N}=2a$ supergravity}

Dimensional reduction of eleven-dimensional supergravity gives ${\cal N}=2a$
(or ${\cal N}=(1,1)$)
supergravity in ten dimensions (the notation $(1,1)$ means that 
the supersymmetry charge is a non-chiral Majorana spinor,
corresponding to 16 left-handed  and 16 right-handed supercharges). It is straightforward to derive the field content of
the multiplet, given by the graviton, a 3-form, a 2-form, a vector and a scalar in the bosonic sector and
by a Majorana gravitino and a Majorana spinor (both non-chiral) in the fermionic sector. One can verify 
that the number of bosonic and fermionic on-shell degrees of freedom coincide.

In order to compactify the lagrangian (\ref{lag11b}) on a circle of radius $r$ 
(in eleven-dimensional units), we make the ansatz 
\be
\hat{e}_M{}^A = \left( 
\begin{array}{cc} \Phi^\a e_\m^a  & \Phi A_\m \\ 0 &  \Phi \end{array} \right) \quad ,\label{vb11}
\ee
for the vielbein, 
that corresponds to the ansatz
\be
ds^2 = \Phi^{2 \a }g_{\m\n} dx^\m d x^\n - ( \Phi A_\m d x^\m + \Phi d x_{10} )^2 
\ee
for the metric. We are then interested in the massless sector of the resulting ten-dimensional theory. 
Denoting with $\Delta$ the compact dimension, the original eleven-dimensional general coordinate 
transformation for $A_{MNP}$ becomes for $A_{\m\n\r}$ 
a ten-dimensional general coordinate transformation plus an 
additional gauge transformation with respect to the gauge field $A_\m$ defined in eq. (\ref{vb11}).
More precisely, defining 
\bea
& & F_{\m\n\r\s} = 4 \de_{[\m } A_{\n\r\s ]} \nonumber \\
& & {\cal F}_{\m\n\r} = 3 \de_{[\m } {\cal A}_{\n\r ]} \quad , 
\eea
where ${\cal A}_{\m\n}= A_{\m\n\Delta}$, one obtains that the 4-form 
\be
H_{\m\n\r\s} = F_{\m\n\r\s} + 4 A_{[\m } {\cal F}_{\n\r\s ]}
\ee
is gauge invariant. 
With these definitions, and denoting with $\kappa_{11}$ the eleven-dimensional Newton constant,
it can be verified that the dimensional reduction of the bosonic sector 
of eleven-dimensional supergravity gives the ten-dimensional lagrangian
\bea
{\cal L}_{2a} &=& -\frac{e}{4 \kappa^2} \Phi^{8\a +1} R -\frac{9\a (8\a +2) e}{4 \kappa^2}
\Phi^{8\a -1} \de_\m \Phi \de^\m \Phi -\frac{e}{16 \kappa^2} \Phi^{6\a +3 }F_{\m\n} F^{\m\n} 
\nonumber \\
&-& \frac{(2 \p r) e}{48}\Phi^{2\a +1} H_{\m\n\r\s} H^{\m\n\r\s} 
+ \frac{(2 \p r) e}{12} \Phi^{4\a -1} {\cal F}_{\m\n\r} {\cal F}^{\m\n\r}\nonumber \\
&+& \frac{(2 \p r)^{\frac{3}{2}} \kappa}{8 \cdot 144}\e^{\a_1 ... \a_4 \b_1 ...\b_4 \m\n} 
F_{\a_1 ... \a_4} F_{\b_1 ... \b_4} {\cal A}_{\m\n} \quad ,
\eea
where $\kappa^2 = \kappa^2_{11} / 2 \p r$ defines the ten-dimensional Newton's constant.
 
The fermionic terms and the supersymmetry tranformations can be derived directly 
by dimensional reduction, and will not be considered here. Rather, we want to emphasize that
different choices of $\a$, corresponding to Weyl rescalings, give the same theory in 
different frames.  
The choice $\a =-1/8$ corresponds to the {\it Einstein frame}, in which  
the Einstein-Hilbert term 
becomes no more dependent on the scalar, and the action assumes the form
(setting for simplicity $\kappa = 2\p r =1$)
\bea
{\cal L} &=& -\frac{e}{4}  R +\frac{e}{2}
\de_\m \phi \de^\m \phi -\frac{e}{16 } e^{3\phi}F_{\m\n} F^{\m\n} 
\nonumber \\
&-& \frac{e}{48} e^\phi  H_{\m\n\r\s} H^{\m\n\r\s} 
+ \frac{e}{12} e^{-2\phi} {\cal F}_{\m\n\r} {\cal F}^{\m\n\r}\nonumber \\
&+& \frac{1}{8 \cdot 144}\e^{\a_1 ... \a_4 \b_1 ...\b_4 \m\n} 
F_{\a_1 ... \a_4} F_{\b_1 ... \b_4} {\cal A}_{\m\n} \quad ,
\eea
where $\Phi =e^{\frac{4}{3}\phi}$. We will see in the next chapter that it is interesting 
to consider the same lagrangian in the {\it string frame}, in which the dependence on the 
scalar vanishes in both the kinetic terms for the 1-form and for the 3-form. This corresponds 
to putting $\a = -1/2$, and the resulting lagrangian is
\bea
{\cal L} &=& e e^{- 2\phi} [-\frac{1}{4}  R -
\de_\m \phi \de^\m \phi + \frac{1}{12} {\cal F}_{\m\n\r} {\cal F}^{\m\n\r}]\nonumber\\
&-& \frac{e}{16 } F_{\m\n} F^{\m\n} 
- \frac{e}{48}  H_{\m\n\r\s} H^{\m\n\r\s} 
\nonumber \\
&+& \frac{1}{8 \cdot 144}\e^{\a_1 ... \a_4 \b_1 ...\b_4 \m\n} 
F_{\a_1 ... \a_4} F_{\b_1 ... \b_4} {\cal A}_{\m\n} \quad ,
\eea
where $\Phi = e^{\frac{2}{3}\phi}$.

\subsection{${\cal N}=2b$ supergravity}

In this subsection we describe the lowest order terms of 
${\cal N}=2b$ (or ${\cal N}=(2,0)$) supergravity \cite{schwarz,howewest}. This theory 
can not be obtained by reduction from a higher-dimensional lagrangian, and  
contains the graviton, two scalars, two 2-forms and a self-dual 4-form in the bosonic sector, together 
with a complex left-handed gravitino and a complex right-handed spinor in the fermionic sector. 

In \cite{schwarz} the field equations for this model were derived to lowest order in the Fermi fields
requiring the closure of the supersymmetry algebra. We will see that 
all these equations, with the exception of the 
self-duality condition for the field strength of the 4-form, 
\be
F^{\m_1 ... \m_5 } = \frac{1}{5 !e} \e^{\m_1 ...\m_5 \n_1 ... \n_5 }F_{\n_1 ...\n_5}
\quad , \label{sd5}
\ee
can be derived from a lagrangian, imposing eq. (\ref{sd5}) only after varying. More recently, a lagrangian 
formulation for self dual forms has been developed by Pasti, Sorokin and Tonin \cite{pst}, 
and then applied in \cite{pst2b}
to the ten-dimensional ${\cal N}=2b$ supergravity. This PST method corresponds 
to the introduction of an  additional scalar auxiliary field, and 
the self-duality condition results from the gauge fixing (that can not be imposed directly on the action)
of additional (PST) local symmetries.
It will be
used in Chapters 4 and 5 in order to derive the action for six-dimensional supergravity theories, while 
here we only assume that we have already fixed the PST gauge so that eq. (\ref{sd5}) holds on-shell.

Let us now summarize the field content of the theory. The two scalars parametrize the coset $SU(1,1)/U(1)$, that 
can be described in terms of the $SU(1,1)$ matrix 
\be
U = ( \ V_-^\a \ \ V_+^\a \ ) \quad ,
\ee
satisfying the constraint
\be
V_-^\a V_+^\b - V_+^\a V_-^\b = \e^{\a\b} \quad ,
\ee
with $(V^\a_- )^* =\e_{\a\b} V^\b_+$, 
where $\a=1,2$ is an $SU(1,1)$ index and $+$ and $-$ denote the $U(1)$ charge. 
From the left-invariant 1-form
\be
U^{-1} \de_\m U = \left( \begin{array}{cc} -i Q_\m & P_\m \\ P_\m^* & iQ_\m \end{array} \right) 
\ee
one reads the $U(1)$-covariant quantity
\be
P_\m = \e_{\a\b}  V_+^\b  \de_\m V_+^\a \quad , 
\ee
that has charge 2, and the $U(1)$ connection 
\be
Q_\m = i \e_{\a\b}  V_+^\b  \de_\m V_-^\a 
\quad .
\ee
The 2-forms are collected in an $SU(1,1)$ doublet $A^\a_{\m\n}$ satisfying the constraint
\be
(A^\a_{\m\n} )^* = \e_{\a\b} A^\b_{\m\n} \quad ,
\ee
while the 4-form is invariant under $SU(1,1)$, and varies as
\be
\d A_{\m\n\r\s} = -\frac{i}{4} \e_{\a\b} \L^{\a}_{[\m} F^\b_{\n\r\s ]}  
\ee 
under 2-form gauge transformations, 
where $\d A^\a_{\m\n} = 2 \de_{[ \m} \L^\a_{\n ]}$ and
$F^\a_{\m\n\r} = 3 \de_{[\m} A^\a_{\n\r ]}$, so that the proper gauge-invariant 5-form field-strength
is
\be
F_{\m\n\r\s\t} =5 \de_{[\m} A_{\n\r\s\t ]}+ \frac{5i}{8} \e_{\a\b} A^\a_{[\m\n} F^\b_{\r\s\t ]} \quad .
\ee
This 5-form satisfies the self-duality condition (\ref{sd5}). 
It is convenient to define the complex 3-form 
\be
G_{\m\n\r} = - \e_{\a\b} V^\a_+ F^\b_{\m\n\r} \quad ,
\ee
that is an $SU(1,1)$ singlet with $U(1)$ charge 1, and  
finally the gravitino has $U(1)$ charge $1/2$, while the spinor has $U(1)$ charge $3/2$ \cite{schwarz}.

The lagrangian 
\bea
{\cal L} &=& -\frac{e}{4}R +\frac{e}{2} P^*_\m P^\m +\frac{e}{48}G^*_{\m\n\r} G^{\m\n\r}
\nonumber\\
&+& \frac{e}{5!} F_{\m_1... \m_5 }F^{\m_1 ...\m_5 } +\frac{i}{4 \cdot 12^3 } 
\e_{\a\b} \e^{\m_1 ...\m_{10}} A_{\m_1 ...\m_4 } F^\a_{\m_5 ...\m_7 }  F^\b_{\m_8 ...\m_{10} }
\nonumber \\
&-& ie (\bar{\psi}_\m \g^{\m\n\r}D_\n \psi_\r )+ ie ( \bar{\chi} \g^\m D_\m \chi )\nonumber \\
&-& \frac{e}{12} F_{\m_1 ...\m_5 }(\bar{\psi}^{\m_1} \g^{\m_2 \m_3 \m_4} \psi^{\m_5 } )
+\frac{e}{2 \cdot 5!}   F_{\m_1 ...\m_5 } (\bar{\psi}_\m \g^{\m\n \m_1 ...\m_5} \psi_\n )\nonumber \\
&+& \biggl\{ \frac{ie}{2 \sqrt{2}} G_{\m\n\r} (\bar{\psi}^\m \g^\n \psi^\r_C ) -\frac{ie}{12 \sqrt{2}}
G_{\m\n\r} (\bar{\psi}_\s \g^{\m\n\r\s\d} \psi_{\d C} )\nonumber \\
&-& \frac{e }{\sqrt{2}} P_\m (\bar{\psi}_\n \g^\m \g^\n \chi_C ) +\frac{e}{2 \cdot 5!} G_{\m\n\r}^*
(\bar{\psi}_\s \g^{\m\n\r} \g^\s \chi ) + {\rm h. \ c. } \biggr\} \nonumber \\
&-& \frac{e}{5!} F_{\m_1... \m_5 } (\bar{\chi} \g^{\m_1... \m_5 } \chi ) 
\eea
is invariant under the supersymmetry transformations
\bea 
& & \d e_\m{}^a = -i (\bar{e} \g^a \psi_\m ) + {\rm h. \ c. }\quad ,\nonumber \\
& & \d V^\a_+ =\sqrt{2} V^\a_- (\bar{\e}_C \chi ) \quad , \nonumber \\
& & \d V^\a_- =\sqrt{2} V^\a_+ (\bar{\e} \chi_C  ) \quad , \nonumber \\
& & \d A^\a_{\m\n} = -\frac{1}{2} V^\a_- (\bar{\e} \g_{\m\n} \chi ) - i\sqrt{2}
V^\a_- ( \bar{\e}_C \g_{[\m} \psi_{\n ]} ) +
{\rm h. \ c. }\quad ,\nonumber \\
& & \d A_{\m\n\r\s} =4 (\bar{\e} \g_{[ \m\n\r} \psi_{\s ]} + {\rm h. \ c. }
-\frac{3i}{8} \e_{\a\b} A^\a_{[\m\n} \d A^\b_{\r \s ]} \quad , \nonumber \\
& & \d \psi_\m = D_\m \e -\frac{i}{48} F_{\m\m_1 ...\m_4 } \g^{\m_1 ...\m_4 } \e \nonumber \\
& & \quad \quad +\frac{1}{24 \sqrt{2}} G^{\n\r\s} \g_{\m\n\r\s} \e_C -\frac{3}{8\sqrt{2}} 
G_{\m\n\r} \g^{\n\r} \e_C \quad , \nonumber \\
& & \d \chi = -\frac{i}{\sqrt{2}} P_\m \g^\m \e_C -\frac{i}{12} G_{\m\n\r} \g^{\m\n\r} \e \quad ,
\eea 
provided one imposes the self-duality 
condition of eq. (\ref{sd5}) after varying.

It is interesting to study in more detail the kinetic term for the scalar fields. The complex variable
\be
z=\frac{V_+^1}{V^2_+} 
\ee
is invariant under local $U(1)$ transformations, and so it is a good coordinate for the scalar 
manifold. Under the $SU(1,1)$ transformation
\be
 \left( \begin{array}{c} V^1_+ \\ V^2_+ \end{array} \right) 
\rightarrow \left( \begin{array}{cc} \a & \b \\ \bar{\b} & \bar{\a} \end{array} \right)  
\left( \begin{array}{c} V^1_+ \\ V^2_+ \end{array} \right) \quad ,\label{su11transf}
\ee
that is an isometry of the scalar manifold, $z$ transforms as 
\be
z \rightarrow \frac{\a z +\b }{\bar{\b} z +\bar{\a}} \quad .
\ee
The variable $z$ parametrizes the unit disc, $\vert z \vert < 1$, and the kinetic 
term assumes the form
\be
{\cal L}_{scalar} =\frac{e}{2} \frac{\de_\m z \de^\m \bar{z}}{(1 - z\bar{z})^2} \quad .
\ee
The further change of variables
\be
z = \frac{w-i}{w+i}
\ee
maps the disc in the complex upper-half plane, ${\rm Im} w > 0$, and 
in terms of $w$ the transformations (\ref{su11transf}) become
\be
w \rightarrow \frac{a w+ b}{cw +d } \quad ,\label{sl2r}
\ee
where 
\be
\left( \begin{array}{cc} a & b \\ c & d \end{array} \right) \in SL(2,R) \quad ,
\ee
while the scalar lagrangian takes the form
\be
{\cal L}_{scalar} =\frac{e}{8} \frac{\de_\m w \de^\m \bar{w}}{({\rm Im} w)^2} \quad .
\ee
We now want to see how these redefinitions modify the form of the bosonic part of the 
action. First of all, we define $F_{\m\n\r}$ such that 
\be
G_{\m\n\r} = - V^1_+ F^*_{\m\n\r} + V^2_+ F_{\m\n\r} \quad ,
\ee  
and 
\be
F_{\m\n\r} = ({\cal F} + i {\cal G})_{\m\n\r} \quad ,
\ee
with ${\cal F}$ and ${\cal G}$ real. 
We also define
\be
{\cal F}^\prime = {\cal F} + {\rm Re} w {\cal G} \quad .
\ee
Writing $w = \rho + i e^\phi $, the bosonic part of the action becomes
\bea
{\cal L} &=& -\frac{e}{4} R + \frac{e}{8} e^{-2 \phi} (\de_\m \rho )^2 +\frac{e}{8} (\de_\m \phi )^2 
+ \frac{e}{48} e^{-\phi} ({\cal F}^\prime_{\m\n\r} )^2 \nonumber \\
&+& \frac{e}{48} e^\phi ({\cal G}_{\m\n\r} )^2 +\frac{e}{5!} (F_{\m_1 ...\m_5 })^2
+ \frac{i}{6 \cdot 24^2 } \e^{\m_1 ...\m_{10} } A_{\m_1 ...\m_4 } F_{\m_5 ...\m_7 }
F^*_{\m_8 ...\m_{10} } \quad .
\eea
If we now perform the Weyl rescaling $g_{\m\n} \rightarrow e^{\phi /2 } g_{\m\n}$, 
we end up with $2b$ supergravity in the string frame, 
\bea
{\cal L} &=& e^{-2 \phi} [-\frac{e}{4} R  
+\frac{e}{8} (\de_\m \phi )^2 +\frac{e}{48} e^\phi ({\cal G}_{\m\n\r} )^2 ] \nonumber \\
&+& \frac{e}{8} e^{-2 \phi} (\de_\m \rho )^2 
+\frac{e}{48} e^{-\phi} ({\cal F}^\prime_{\m\n\r} )^2 
+  \frac{e}{5!} (F_{\m_1 ...\m_5 })^2 \nonumber \\
&+& \frac{i}{6 \cdot 24^2 } \e^{\m_1 ...\m_{10} } A_{\m_1 ...\m_4 } F_{\m_5 ...\m_7 }
F^*_{\m_8 ...\m_{10} } \quad .
\eea
Observe that in this frame the $SL(2,R) \simeq SU(1,1)$ symmetry is no longer manifest. We will see in the next 
chapter the implications of this in the context of string theory.

\subsection{${\cal N}=1$ supergravity}

The representations of the supersymmetry algebra with 16 supercharges in ten dimensions are 
the gravity multiplet, containing the graviton, 
a 2-form, a left-handed Majorana gravitino and a right-handed Majorana spinor, 
and the Yang-Mills multiplet, containing a gauge vector and a left-handed Majorana gaugino.

The lagrangian for ${\cal N}=1$ supergravity coupled to vector multiplets in terms of these fields
is  \cite{bdrdwvn,chapmant}
\bea
e^{-1} {\cal{L}} = 
&-& \frac{1}{4} R +\frac{1}{2}\de_\m \phi \de^\m \phi +\frac{1}{6}
e^{-2\phi}H_{\m\n\r} H^{\m\n\r} -\frac{1}{2}e^{-\phi}\tr (F_{\m\n}F^{\m\n} )
\nonumber\\
&-& \frac{i}{2} (\bar{\psi}_\m \g^{\m\n\r} D_\n \psi_\r )+\frac{i}{2}(\bar{\chi}
\g^\m D_\m \chi )+\frac{1}{\sqrt{2}}(\bar{\psi}_\m \g^\n \g^\m \chi )\de_\n \phi 
\nonumber\\
&-& \frac{i}{12\sqrt{2}} e^{-\phi} H_{\m\n\r} (\bar{\psi}_\s \g^{\s\d\m\n\r}\psi_\d
)+\frac{i}{2\sqrt{2}}e^{-\phi} H_{\m\n\r} (\bar{\psi}^\m \g^\n \psi^\r)\nonumber \\
&+& \frac{1}{12}
e^{-\phi}H_{\m\n\r} (\bar{\psi}_\s \g^{\m\n\r} \g^\s \chi) +i \tr (\bar{\l}\g^\m
D_\m \l )\nonumber \\
&-& \frac{1}{2} e^{-\frac{1}{2}\phi } \tr [F^{\m\n} (\bar{\l} \g_{\m\n} \chi)]
+\frac{i}{\sqrt{2}}e^{-\frac{1}{2}\phi } \tr [F^{\m\n} (\bar{\l} \g_\r \g_{\m\n}
\psi^\r )]\nonumber\\
&-& \frac{i}{6\sqrt{2}}e^{-\phi }H^{\m\n\r} \tr (\bar{\l} \g_{\m\n\r} \l ) \ , 
\label{lag1}
\eea
up to quartic terms in the fermions.
The 3-form 
$H_{\m\n\r}$ includes a Chern-Simons coupling, so that 
\be
H_{\m\n\r}= 3 \de_{[\m} B_{\n\r ]} +\sqrt{2} \w_{\m\n\r} \quad ,\label{3form}
\ee
where $\w_{\m\n\r}$ is the Chern-Simons 3-form defined as 
\be
\w = A dA -\frac{2i}{3} A^3 \quad ,
\ee
while the supersymmetry transformations are
\bea
& & \d e_\m{}^m = -i (\bar{\e} \g^m \psi_\m )\quad , \nonumber\\
& & \delta \phi =-\frac{1}{\sqrt{2}} (\bar{\e} \chi )\quad ,\nonumber\\
& & \d B_{\m\n}= -\frac{i}{\sqrt{2}}e^\phi (\bar{\e}\g_{[\m}\psi_{\n]} )-\frac{1}{4}
e^\phi (\bar{\e}\g_{\m\n} \chi ) +2\sqrt{2} \tr (A_{[\m } \d A_{\n ]}) \quad , \nonumber \\
& & \d A_\m =-\frac{i}{\sqrt{2}} e^{\frac{1}{2}\phi}(\bar{\e} \g_\m \l ) \quad ,
\nonumber\\
& & \d \psi_\m =D_\m \e +\frac{1}{24\sqrt{2}}e^{-\phi}H^{\n\r\s}\g_{\m\n\r\s} \e
-\frac{3}{8\sqrt{2}}e^{-\phi} H_{\m\n\r} \g^{\n\r}\e \quad , \nonumber\\
& & \d \chi =-\frac{i}{\sqrt{2}}\de_\m \phi \g^\m \e-\frac{i}{12}e^{-\phi}
H^{\m\n\r} \g_{\m\n\r} \e \quad , \nonumber \\ 
& & \d \l =-\frac{1}{2\sqrt{2}}  e^{-\frac{1}{2}\phi}F^{\m\n}\g_{\m\n}\e \quad , 
\label{susy1}
\eea
and gauge invariance of $H$ requires that under vector gauge
transformations $B$ transform as 
\be
\d B = -\sqrt{2} \tr (\L dA ) \ .
\ee
This lagrangian was intitially written in \cite{chamseddine} in terms of the 6-form dual
to the 2-form. We will return to this point in Chapter 5. 

In concluding this chapter, we observe that  one can perform a Weyl rescaling, in order 
to map eq.  (\ref{lag1}) to a lagrangian in a different frame. In particular, 
the bosonic part of the action is 
\be
{\cal L} = e^{-2 \phi} [ -\frac{e}{4}R - e (\de_\m \phi )^2 +\frac{e}{12} H_{\m\n\r}^2 
-\frac{e}{2} F_{\m\n}^2 ] 
\ee
in the {\it heterotic-string frame}, 
and 
\be
{\cal L} = e^{-2 \phi} [ -\frac{e}{4}R - e (\de_\m \phi )^2 ] +\frac{e}{12} H_{\m\n\r}^2 
-\frac{e}{2} e^{-\phi} F_{\m\n}^2  
\ee
in the {\it type I-string frame}. These two lagrangans are mapped one in the other by the relations
\bea
& & g_{H, \m\n} = e^{-\phi_I} g_{I, \m\n} \quad ,\nonumber \\
& & B_{H,\m\n} = B_{I,\m\n}  \quad , \nonumber \\
& & A_{H,\m} = A_{I,\m } \quad , \nonumber \\
& & \phi_H = -\phi_I \quad .\label{dualhetso32}
\eea
We will see that these and analogous relations in supergravity theories correspond 
to dualities between different string theories. In particular, the last relation in eq. (\ref{dualhetso32})
is a manifestation of a strong-weak coupling duality between the $SO(32)$ heterotic string and 
the type-I string in ten dimension.

% \newpage ~                  %  per lasciare una pagina bianca
% \thispagestyle{empty}       %  

\chapter{Closed strings}
\label{cap2}
% \newpage
\vspace*{2cm}
\fancyhead[RO,LE]{\thepage}
\fancyhead[RE]{{\footnotesize {\rm Chapter 2.}~~{\it Closed strings}}} 
\fancyhead[LO]{}

\noindent 
This chapter is a brief introduction to oriented closed strings. Most of the analysis is devoted to
the derivation of one-loop vacuum amplitudes, and the results obtained 
will be applied in the next chapter to the derivation of type-I models. 
As we will see, the consistency of the models is guaranteed requiring modular invariance for the one-loop 
vacuum amplitudes, and this naturally implies anomaly cancellation in the low-energy effective action.
We will also comment on the non-perturbative states of these models, that will turn out to have 
a role in perturbative type-I vacua. The picture one ends up with, after 
the analysis of this and the next chapter,
is that perturbative type-I models embody some peculiar features 
in their dynamics, that typically are revealed in the closed-string setting only at the 
non-perturbative level.

The content of this chapter is the following. In Section 1 we shortly 
describe how to build the spectrum of closed strings, 
while Section 2 is devoted to the partition function of various superstring theories. 
The rules for writing partition functions are then applied to type IIB compactified on
the $T^4 /Z_2$ orbifold in Section 3. We will use these results in the next chapter when we 
discuss orientifolds of type IIB. In Section 4 we discuss gauge and gravitational 
anomalies in field theory, showing that two-dimensional consistency (\ie modular invariance) of 
closed superstring theories always gives rise to low-energy effective
actions that are anomaly-free. Finally, in Section 5 we introduce the concept of D-branes and 
describe how the various supersymmetric string theories can be connected 
at the non-perturbative level by dualities.

\section{The spectrum of closed oriented superstrings}
\label{spectrum}
\fancyhead[LO]{{\footnotesize 2.1~~{\it The spectrum of closed oriented superstrings}}}

Before considering superstrings, we present an introduction 
to the bosonic string \cite{gsw,polchinski}.
The action for a bosonic string in flat space-time is 
\be
S= -\frac{1}{4 \p \a^\prime} \int d^2 \xi \sqrt{-g} g^{\a\b} \de_\a X^\m \de_\b X^\n
\eta_{\m\n} \quad ,
\ee
where $X^\m (\xi )$ are the coordinates of the string in D space-time dimensions and  
$\xi^\a$ $(\a=1,2 )$ parametrize the world-sheet, whose metric is $g_{\a\b}$, and   
$\a^\prime$ is a dimensionful constant, related to the string tension by 
$T=\frac{1}{2\p \a^\prime}$.
The field equation for the metric implies that the energy momentum tensor
\be
T_{\a\b} = \de_\a X^\m \de_\b X_\m -\frac{1}{2} g_{\a\b} \de^\g X^\m \de_\g X_\m 
\ee
vanishes, while the field equation for $X^\m$ is 
\be
\de_\a (\sqrt{- g} g^{\a\b} \de_\b X^\m ) =0 \quad .\label{eqX}
\ee
In two dimensions, one can use reparametrization invariance to prove that 
every metric is conformally equivalent to the flat metric. With $\xi^\a =(\t, \s )$,  
eq. (\ref{eqX}) reduces to the standard wave equation 
\be
\left( \frac{\de^2}{\de \t^2}- \frac{\de^2}{\de \s^2 } \right) X^\m =0 \quad ,
\ee
and for a closed oriented string, with the periodicity condition $X^\m (\t, \s )= 
X^\m (\t, \s +2\p )$, this equation has the solution
\be
X^\m = x^\m +\a^\prime p^\m \t +i \sqrt{\frac{\a^\prime}{2}} \sum_{n \neq 0} \left[ \frac{\a^\m_n}{n}
e^{-in(\t -\s )} +  \frac{\tilde{\a}^\m_n}{n} e^{-in(\t +\s )} \right]\quad .
\ee
The quantization condition for $X$ results in the commutation relations
\be
[\a^\m_m , \a^\n_n ] = n \d_{m+n,0} \eta^{\m\n} \quad , \qquad [\tilde{\a}^\m_m , 
\tilde{\a}^\n_n ] = n \d_{m+n,0} \eta^{\m\n} \quad 
\ee
for the expansion modes, that thus 
behave like creation and annihilation operators. 

In fact, the original invariance under Weyl 
rescalings and reparametrizations leaves a residual gauge symmetry that
corresponds to arbitrary (anti)analytic reparametrizations, and 
can be used to eliminate 
the oscillators in the $+$ direction, 
where $X^\pm = (X^0 \pm X^{D-1})/2 $. In this {\it light cone gauge}, 
the Virasoro operators, \ie the modes of the energy-momentum 
tensor, are written in terms of the transverse 
oscillators as
\bea
& & L_m =\frac{1}{2} : \sum_n \a^i_{m-n} \a^i_n : \quad ,\nonumber \\
& & \bar{L}_m =\frac{1}{2} : \sum_n \tilde{\a}^i_{m-n} \tilde{\a}^i_n : \quad .
\eea
The $L_m$ and $\bar{L}_m$ are mutually commuting, and they satisfy the Virasoro algebra
with central charge $c=D-2$:
\be
[ L_m , L_n ] = (m-n) L_{m+n} +\frac{D-2}{12} m (m^2 -1 ) \delta_{m+n, 0} \quad .
\ee 
The zero modes $L_0$ and $\bar{L}_0$ define the mass-shell condition for physical states,
\be
M^2 =\frac{2}{\a^\prime } \left( L_0 +\bar{L}_0 -\frac{D-2}{24} \right) \quad ,
\ee
where the shift originates from the normal ordering of $L_0$ and $\bar{L}_0$, 
together with the level-matching condition $L_0 = \bar{L}_0$. This relation determines 
the dimension of space-time: 
a Lorentz invariant spectrum is obtained with $D=26$, with a massless first excited level, 
$\a^i_{-1} \tilde{\a}^j_{-1} | 0 \tilde{0} \rangle$, corresponding to a metric 
fluctuation $h_{\m\n}$, an antisymmetric
tensor $B_{\m\n}$ and a scalar $\phi$, called the dilaton, whose vacuum expectation 
value weights the perturbative expansion. The ground state of this model is a tachyon.

We can now analyze the supersymmetric version of the string action,
\be
S= -\frac{1}{4 \p \a^\prime} \int d^2 \xi \left(  \de^\a X^\m \de_\a X^\n
\eta_{\m\n} +i \bar{\psi}^\m \g^\a \de_\a \psi^\n \eta_{\m\n} \right) \quad ,
\ee
where the $\psi$'s are Majorana spinors whose mode expansion is  
\be
\psi^\m = \frac{1}{\sqrt{2}}\sum_{r} \left[ \l^\m_r
e^{-ir(\t -\s )} +  \tilde{\l}^\m_r e^{-ir(\t +\s )} \right]\quad ,\label{modespin}
\ee
and where the oscillators $\l$ satisfy the anticommutation relations
\be
\{ \l^\m_r , \l^\n_s \} =-\eta^{\m\n} \delta_{r+s} \quad .
\ee
Actually, the periodicity of the currents of space-time symmetries 
is guaranteed if $\psi$ is periodic or antiperiodic, \ie 
if $r$ is integer or half-integer. 
The Ramond (R) sector corresponds to integer $r$, and the anticommutation relations 
for the zero mode $\l_0$ give rise to the Clifford algebra, resulting in a fermionic vacuum.
The Neveu-Schwarz (NS) sector, on the other hand, corresponds to half-integer $r$, so that there are no zero modes, 
and the vacuum is a scalar. The dimension of space-time is determined to be $D=10$ requiring Lorentz invariance
of the spectrum, while
the super-Virasoro modes assume the form
\be
L_m =\frac{1}{2} : \sum_n \a^i_{m-n} \a^i_n : + \frac{1}{2} : \sum_r (r-m ) \l^i_{m-r} \l^i_r :
+ \d_{m,0} \Delta \quad ,
\ee
where the normal-ordering shift is  $-\frac{1}{2}$ in the NS sector and vanishes in the R sector.

In the next section we will determine the spectrum of superstring theories from the analysis of 
their partition functions. Consistency at the one loop level imposes a suitable projection of the states,
so that the spectrum one ends up with is supersymmetric and free of tachyons. This projection selects 
two different ten-dimensional theories of oriented closed strings, 
type IIA and type IIB, whose massless sectors correspond to the field contents of 
$2a$ and $2b$ supergravities, respectively. There is another consistent 
model that can be built, the {\it heterotic string}, a closed oriented string whose left movers
are bosonic and right movers are supersymmetric. The spectrum of the theory has ${\cal N}=1$ supersymmetry, 
and at the massless level it contains the supergravity multiplet plus vector multiplets, while 
consistency imposes
that the gauge group be $SO(32)$ or $E_8 \times E_8$. Actually, 
we will see that consistent non-supersymmetric models can be
generated as well, but the corresponding spectra contain tachyons. 

The perturbative spectra of these theories consist of a finite number 
of massless particles, as well as an infinite tower
of massive excitations, with masses proportional to the string tension. 
Since this is the only scale in the theory, and 
the massless degrees of freedom always contain a graviton, this string scale 
must be naturally of the order of the Planck scale. 
The effective theory for the massless modes results from integrating all the 
massive ones, and since the expansion in derivatives
is suppressed by powers of $E/M$, where $E$ is the energy scale we are considering 
and $M$ is the string scale, we
expect that at low energy only two-derivative terms are relevant. 

There are two ways to compute the effective action for the massless modes. 
The first consists in computing the S-matrix 
elements in string theory, and then extracting the low-energy limit. The second consists in studying 
string propagation in a curved background, and determining 
the equations for the background fields imposing conformal invariance of the 
world-sheet action. In theories with 32 or 16 supercharges, 
the low-energy effective action is also completely constrained by supersymmetry.
The end result is that the low-energy effective actions of type IIA and type IIB superstrings are respectively $2a$ and $2b$
supergravity in the string frame, while the low-energy effective action for the heterotic string is ${\cal N}=1$ supergravity 
coupled to $SO(32)$ or $E_8 \times E_8$ vector multiplets in the heterotic string frame (see Section (1.4)). 
In the case of type I, that we will analyze in the next chapter, the low-energy effective action is 
${\cal N}=1$ supergravity 
coupled to $SO(32)$ vector multiplets in the type I string frame.
The reason for the dilaton dependence of the massless string fields in the various models is that 
the coupling constant regulating the loop expansion in string theory is 
the exponential of the vacuum expectation value of the dilaton.

\section{Partition function for closed oriented strings}
\label{partfunc}
\fancyhead[LO]{{\footnotesize 2.2~~{\it Partition funcion for closed oriented strings}}}

The one-loop partition function measures the vacuum energy $\Gamma$ of a given theory. 
For the simple case of a free scalar field of mass $M$ in $D$ dimensions, it is given by
\be
e^{-\Gamma} = \int [D \phi ] e^{-S_E} \sim {\det}^{-\frac{1}{2}} (- \Delta + M^2 )
\ee
where $S_E$ is the euclidean free-field action.
The $M$ dependence of $\Gamma$ can be extracted from the identity
\be
{\log}\ {\det} A = -\int_\e^{\infty} \frac{dt}{t} \tr ( e^{-tA})
\ee
where $\e$ is an ultraviolet cutoff and $t$ is a Schwinger parameter. The result is 
\be
\Gamma = \frac{V}{2} \int_\e^{\infty} \frac{dt}{t} e^{-t M^2} \int \frac{d^D p}{(2 \p )^D} 
e^{-t p^2} \quad ,
\ee
where $V$ denotes the volume of space-time, and
performing the gaussian integral yields the final expression
\be
\Gamma =  \frac{V}{2 (4 \p )^{D/2}}  \int_\e^{\infty} \frac{dt}{t^{D/2 +1}} e^{-t M^2}
\quad .\label{partft}
\ee
Generalizing this result in order to include generic Bose and Fermi fields, one 
obtains
\be
\Gamma_{tot} =  \frac{V}{2 (4 \p )^{D/2} } \int_\e^{\infty} {\rm Str}\frac{dt}{t^{D/2 +1}} e^{-t M^2}
\quad ,\label{partft2}
\ee
where Str counts the multiplicities of the fields, with a minus sign in the case of fermions,
on account of the Grassmann nature of the fermionic path integral. 
From eq. (\ref{partft2}) it follows that $\Gamma$ vanishes identically for supersymmetric models.
Nevertheless, also in this case one can read from its integrand 
the masses and multiplicities of Bose and Fermi
fields independently. 

We now want to derive the vacuum amplitude for oriented closed strings. 
As a starting point, consider the closed bosonic string in $D=26$, whose spectrum is encoded in
\be
M^2 = \frac{2}{\a^\prime} ( L_0 + \bar{L}_0 -2 ) \quad ,
\ee
together with the level-matching condition 
\be
L_0 = \bar{L}_0 \quad .
\ee
Applying eq. (\ref{partft2}) in this case gives 
\be
\Gamma = \frac{V}{2 (2 \p )^{13}} \int_{-\infty}^{\infty} ds  \int_\e^{\infty} \frac{dt}{t^{14}}
\tr ( e^{-\frac{2}{\a^\prime} ( L_0 + \bar{L}_0 -2 ) t} e^{2\p i (L_0 - \bar{L}_0 ) s} ) \quad .
\label{bos1}
\ee
Defining the ``complex'' Schwinger parameter 
\be
\t = \t_1 +i \t_2 = s + \frac{it}{\a^\prime \p } \quad ,
\ee
and denoting
\be
q = e^{2\p i \t } 
\quad ,
\ee
eq. (\ref{bos1}) then takes the form
\be
\Gamma = \frac{V}{2 (4 \p^2 \a^\prime )^{13}} \int_{-\infty}^{\infty} d\t_1  
\int_\e^{\infty} \frac{d\t_2}{\t_2^{14}}
\tr \ q^{L_0 -1 } \bar{q}^{\bar{L}_0 -1 } \quad .
\ee
In order to compute this vacuum ampitude, we should recall that $L_0$ and 
$\bar{L}_0$ are effectively number operators 
for two infinite sets of harmonic oscillators, and in terms of 
conventionally normalized creation and annihilation operators, 
for each transverse space-time dimension we have
\be
L_0 = \sum_k k a_k^\dagger a_k \quad .
\ee
Thus, for any mass-level $k$, we have
\be
\tr \ q^{ k a_k^\dagger a_k } = 1+q^k +q^{2k}+... = \frac{1}{1-q^k} \quad ,
\ee
and putting all the contributions together, in the light-cone gauge, one gets
\be
\Gamma = \int_{C_+} \frac{d^2 \t}{\t_2^2} \frac{1}{\t_2^{12}} 
\frac{1}{\vert \eta (\t ) \vert^{48}} \quad , \label{vacampl}
\ee
where the integral is performed over the complex upper-half plane $C_+$ 
and $\eta$ is the Dedekind function
\be
\eta (\t ) = q^{\frac{1}{24}} \prod_{n=1}^\infty (1 - q^n ) \quad .
\ee
However, a more careful analysis of the amplitude of eq. (\ref{vacampl}) shows that it diverges,
since the integrand is invariant under the {\it modular} transformations
\be
\t \rightarrow \frac{a \t +b}{c\t +d } \quad ,\label{mod}
\ee
with $ad-bc =1$, and $(a,b,c,d) \in {Z}$. These transformations connect different
points in the upper-half plane, and in order to get a finite integral we have 
to mod it out. We now want to explain the origin of this invariance.

\begin{figure}[!h]
\centering
\includegraphics[width=6truecm]{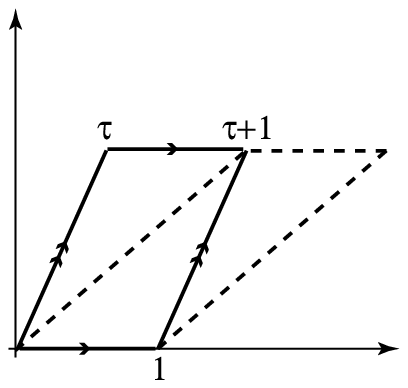}
\caption{Torus}
\label{torus}
\end{figure}

\begin{figure}[!h]
\centering
\includegraphics[width=8truecm]{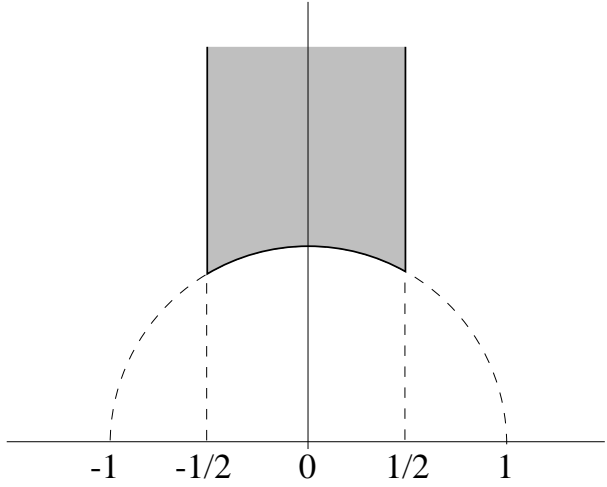}
\caption{Fundamental region}
\label{fundamental}
\end{figure}

The one-loop diagram for a closed oriented string is a torus, whose points are in 
correspondence with the points in a cell of the periodic lattice of fig. (\ref{torus}). 
The parameter $\t$ in the figure is the complex modulus (Teichmuller parameter) 
of the torus, and periodicity of the lattice
means that the modulus is invariant under the $PSL(2, { Z}) = SL(2, {Z})/{Z}_2$ 
modular group, whose action on $\t$ is exactly like in eq. (\ref{mod}). 
This group is generated by the two transformations 
\be
T: \t \rightarrow \t +1 \quad , \qquad S: \t \rightarrow -\frac{1}{\t} \quad ,
\ee
that satisfy the relation 
\be
S^2 = (ST )^3 \quad .
\ee
As a result, all different values of $\t$ can be mapped by a modular transformation to the values 
within a fundamental region, for instance the region 
\be
{\cal F} = \{ -1/2 < \t_1 \leq 1/2 , \vert \t \vert \geq 1 \}
\ee
of fig. (\ref{fundamental}). 

Coming back to our amplitude, using the transformation properties of the $\eta$ function under $S$ and 
$T$,
\be
T: \eta (\t +1 ) = e^{\frac{i\p}{12}} \eta (\t ) \quad , \qquad S: \eta (-1 /\t ) = \sqrt{-i \t_2} 
\eta (\t ) \quad ,
\ee
one can prove that the integrand of eq. (\ref{vacampl}) is modular invariant. The Schwinger parameter $\t$
is the complex modulus of the torus, and the torus amplitude is obtained restricting 
the integration region to a fundamental region, for instance  ${\cal F}$:
\be
{\cal T} = \int_{\cal F} \frac{d^2 \t}{\t_2^2} \frac{1}{\t_2^{12}} 
\frac{1}{\vert \eta (\t ) \vert^{48}} \quad . \label{torusamplbos}
\ee
From the amplitude, one could read the spectrum just expanding the integrand of eq. (\ref{torusamplbos})
in powers of $q$ and $\bar{q}$, and taking only the terms containing $q \bar{q}$ because of the level 
matching condition. The factor $\t_2^{-12}$ comes from the integral over the momentum, so we have just to expand
\be
\frac{1}{\vert \eta (\t ) \vert^{48}} \simeq \frac{1}{q \bar{q}} [ 1 +24^2 (q \bar{q} ) +... ]
\quad .
\ee
The first term corresponds to the tachyon, while the second corresponds to the graviton, the 2-form and
the scalar that form the massless spectrum of the bosonic string.

Our next task is to write the torus partition function for IIA and IIB superstrings in ten dimensions. 
As we have seen in the previous section, the Virasoro generators
in the light cone gauge have the form
\be
L_m =\frac{1}{2} : \sum_n \a^i_{m-n} \a^i_n : + \frac{1}{2} : \sum_r (r-m ) \l^i_{m-r} \l^i_r :
+ \d_{m,0} \Delta \quad ,
\ee
where $r$ is half-odd integer in the NS sector and integer in the R sector. The normal-ordering shift is determined 
by the following rule: each fermionic coordinate contributes $-\frac{1}{24}$ in the NS sector and $\frac{1}{12}$ 
in the R sector while, as for the bosonic string, each bosonic coordinate contributes $-\frac{1}{12}$. As a result, 
the total shift in ten dimensions is $-\frac{1}{2}$ in the NS sector and vanishes in the R sector.

Let us first consider the NS sector. In this case the (antiperiodic) transverse fermions $\l^i$ do not 
have zero modes, and so the vacuum is a scalar. The vacuum amplitude receives contributions from
these fermionic oscillators, and since
\be
\tr (q^{\sum_r r \l^\dagger_r \l_r}) = \prod_r \tr (q^{r a^\dagger_r a_r} ) = \prod_r (1+ q^r )^8 \quad ,
\ee
including the contribution of the bosonic oscillators, in this sector we have 
\be
\tr (q^{L_0}) = \frac{\prod_{m=1}^\infty (1 + q^{m-1/2})^8}{q^{1/2}\prod_{m=1}^\infty (1- q^m )^8 } \quad .\label{trns}
\ee

In the R sector, the zero modes of the $\l^i$ imply that the vacuum carries 
a 16-dimensional representation of the $SO(8)$ 
Clifford algebra, and is thus a space-time spinor, like all its excitations. Consequently, 
the analogue of eq. (\ref{trns}) is 
\be
\tr (q^{L_0}) =16 \frac{\prod_{m=1}^\infty (1 + q^{m})^8}{\prod_{m=1}^\infty (1- q^m )^8 } 
\quad .\label{trr}
\ee
The factor $q^{1/2}$ is now absent in the denominator since, as we have seen, 
the R sector has $\Delta =0$, and 
so it starts with massless modes. 

In order to build modular-invariant quantities from these expressions, 
one has to suitably project the spectrum.
The simplest possibility is then to project out all states created by even numbers of fermionic oscillators. 
This prescription, originally proposed by Gliozzi, Scherk and Olive (GSO) \cite{gso}, 
has the virtue of removing the tachyon, giving rise to models that 
are supersymmetric in the target space. The corresponding GSO-projected NS sector is described by
\be
\tr \left( \frac{1-(-1)^F}{2} q^{L_0}\right) =\frac{1}{2}
\frac{\prod_{m=1}^\infty (1 + q^{m-1/2})^8 -  \prod_{m=1}^\infty (1 - q^{m-1/2})^8 }{q^{1/2}\prod_{m=1}^\infty (1- q^m )^8 } \quad .
\ee
In the R sector, the GSO-projection consists in taking space-time fermions of a given
chirality at each mass-level. Precisely, the term $q^{L_0}$ is projected by the operator
\be
\frac{1}{2} (1 \pm \g_{11} (-1)^{ : \sum_r \l^i_{-r} \l^i_r :} ) \quad .
\ee
As we will see, one can consider the 
case in which the left and right R vacua have the same chirality (type IIB) 
or opposite chirality (type IIA). 
Since $\g_{11}$ is traceless, the second term does not contribute to the vacuum energy. This simply means 
that the trace has numerically the same value if the vacuum is a left or a right spinor.

We now want to prove that the properly projected partition function is modular invariant. 
With this in mind, let us introduce the Jacobi theta functions 
\be
\theta [^\a_\b](z|\t )=\sum_{n\in Z}q^{{1\over 2}\left( n +\a \right)^2}
e^{2\p i\left( n +\a\right) \left( z+\b \right)} \quad ,
\label{t1}
\ee
that transform under $S$ and $T$ as 
\bea
& & \theta [^\a_\b](z|\t +1 )= e^{-i \p \a (\a-1)} \th[^{\quad \a}_{\b+\a -1/2}](z|\t ) \quad ,\nonumber \\
& & \theta [^\a_\b](\frac{z}{\t}| -\frac{1}{\t} )= e^{2 i \p \a \b + i \p z^2/ \theta}
(-i \t )^{1/2}  \theta [^{\ \b}_{-\a}](z|\t ) \quad .\label{thetamod}
\eea
Since the fermions $\l^i$ are (anti)periodic, we are interested in theta functions with vanishing argument
$z$ and with $\a$ and $\b$ equal to $0$ and $\frac{1}{2}$. Denoting
\be
\theta_1 = \theta [^{1/2}_{1/2}] \quad , \qquad \theta_2 = \theta [^{1/2}_{\ 0}] \quad\qquad 
\theta_3 = \theta [^0_0 ]\quad ,\qquad \theta_4 = \theta [^{\ 0}_{1/2}] \quad ,
\ee
one can then define the $so(2n)$ characters
\bea
& & O_{2n} =\frac{ \theta_3^n +\theta_4^n }{2 \eta^n} \quad ,\nonumber \\
& & V_{2n}=\frac{ \theta_3^n -\theta_4^n }{2 \eta^n} \quad ,\nonumber \\
& & S_{2n}=\frac{ \theta_2^n +i^{-n}\theta_1^n }{2 \eta^n} \quad ,\nonumber \\
& & C_{2n}=\frac{ \theta_2^n - i^{-n}\theta_1^n }{2 \eta^n} \quad ,\label{so2ncharacters}
\eea
whose transformation properties under $T$ and $S$ transformations are summarized by
\be
T_{2n} = e^{-i n \p /12} {\rm diag} (1,-1, e^{i n \p /4}, e^{i n \p /4} ) \label{tmatrix}
\ee
and
\be
S_{2n} = \frac{1}{2}\pmatrix{1 & 1 & 1 & 1  \cr 1 & 1 & -1 & -1 \cr 
1 & -1 & i^{-n} & -i^{-n} \cr 1 & -1 & -i^{-n} & i^{-n}} 
\quad .\label{smatrix}
\ee
One can also prove that the theta  functions with argument $z=0$ have the product expansions
\bea
& & \frac{\theta_2^4 (0 | \t ) }{\eta^{12}(\t ) }= 16 
\frac{ \prod_{m=1}^\infty (1 +q^m )^8}{\prod_{m=1}^\infty 
(1 - q^m )^8} \quad , \nonumber \\
& & \frac{\theta_3^4 (0 | \t ) }{\eta^{12}(\t ) }= \frac{ \prod_{m=1}^\infty (1 +q^{m -1/2})^8}{ q^{1/2}\prod_{m=1}^\infty 
(1 - q^m )^8} \quad , \nonumber \\
& & \frac{\theta_4^4 (0 | \t ) }{\eta^{12}(\t ) }= 
\frac{ \prod_{m=1}^\infty (1 - q^{m -1/2})^8}{ q^{1/2}\prod_{m=1}^\infty 
(1 - q^m )^8} \quad ,
\eea
while $\theta_1$ vanishes when evaluated at $z=0$.

Collecting all these results, one obtains that the GSO-projected torus partition functions
\bea
& & {\cal T}_{IIA} = \frac{V}{2 (4 \p^2 \a^\prime )^5 } \int_{\cal F} \frac{d^2 \t}{\t_2^6 } 
\frac{ (\bar{V}_8 - \bar{S}_8 )( V_8 -C_8 ) }{| \eta (\t )|^{16}} \quad ,\nonumber \\
& & {\cal T}_{IIB} = \frac{V}{2 (4 \p^2 \a^\prime )^5 } \int_{\cal F} \frac{d^2 \t}{\t_2^6 } 
\frac{ | \bar{V}_8 - \bar{S}_8 |^2}{| \eta (\t )|^{16}} \quad ,
\eea
are modular-invariant. 
Leaving the modular integration and the bosonic 
contribution implicit, one can write
\bea
& & {\cal T}_{IIA} = (\bar{V}_8 - \bar{S}_8 )( V_8 -C_8 ) \quad , \nonumber \\
& & {\cal T}_{IIB} = | \bar{V}_8 - \bar{S}_8 |^2 \quad .\label{torus2a2b}
\eea
As anticipated, the IIA amplitude is obtained  by opposite chiral projections in left and right 
R sectors, while the IIB amplitude is obtained by the same chiral projection in the two sectors. 
As in the bosonic case, we can read the spectrum expanding the argument of the integral in power of $q$.
The $O_8$ character starts at the lowest mass level with the tachyon and, in group theoretical language, 
corresponds to the conjugacy class of the singlet in the weight lattice. $V_8$ starts with the massless vector
and corresponds to the conjucacy class of the vector in the weight lattice. 
Finally, $S_8$ and $C_8$ start at the lowest 
mass level with massless left and right-handed spinors, respectively.  
It is then straightforward to see from eqs. (\ref{torus2a2b}) that IIA and IIB have no tachyon, while 
their massless spectra exactly coincide with the field content 
of $2a$ and $2b$ ten-dimensional supergravities.
Numerically, both these amplitudes vanish because of supersymmetry, as can be verified using the 
famous Jacobi's {\it aequatio identica satis abstrusa},
\be
\theta_3^4 - \theta_4^4 - \theta_2^4 =0 \quad .
\ee

Actually, there are other two modular invariant partition functions one can write in ten dimensions,
namely \cite{0a0b}
\bea
& & {\cal T}_{0A} = | O_8 |^2 +| V_8 |^2 + \bar{S}_8 C_8 +  \bar{C}_8 S_8  \quad ,\nonumber \\
& & {\cal T}_{0B} = | O_8 |^2 +| V_8 |^2 + |S _8 |^2  + | C_8 |^2 \quad .
\eea
Both these theories are tachyonic and non-supersymmetric 
(the spectrum is in fact purely bosonic). We will not
analyze these theories, but here we only mention that a suitable orientifold projection of the 0B model
gives rise to a spectrum without tachyons in the closed sector. 
Moreover, the tachyon in the open
sector can be projected out by a proper choice of the gauge group \cite{augusto0b}. 
The details about how to build
orientifold models are the subject of the next chapter.
We conclude by writing also the partition functions corresponding to the heterotic 
strings in ten dimensions \cite{het}, 
\be
{\cal T}_{SO(32)} =  (\bar{V}_8 - \bar{S}_8 ) (O_{32} + S_{32} )
\ee
for the $SO(32)$ case and 
\be
{\cal T}_{E_8 \times E_8} = (\bar{V}_8 - \bar{S}_8 ) (O_{16} + S_{16} )(O_{16} + S_{16} )
\ee
for the $E_8 \times E_8$ case. A suitable projection of the latter model gives a 
tachyon-free string theory without
space-time supersymmetry, whose low-energy limit is a ten-dimensional anomaly-free 
chiral $O(16) \times O(16)$ 
gauge theory coupled to gravity \cite{nonsusyhet}.  
 
\section{IIB superstring on $T^4 / Z_2$}
\label{orbi}
\fancyhead[LO]{{\footnotesize 2.3~~{\it IIB supersting on $T^4 /Z_2$}}}

Type IIB string theory compactified on $K3$ gives rise to a  six-dimensional model with 
$(2,0)$ supersymmetry, whose 
low-energy effective action corresponds to $(2,0)$ supergravity coupled to 21 tensor multiplets. 
The $(2,0)$ gravity multiplet contains the metric, two left-handed gravitinos and five self-dual 
2-forms, while the tensor multiplet contains an antiself-dual 2-form, two right-handed spinors and five 
scalars. The couplings of supergravity to tensor multiplets were described in \cite{romans} to lowest 
order in the Fermi fields, and then completed in \cite{fr1} to all orders in the Fermi fields.
In this section we descibe IIB string theory compactified on a four-dimensional orbifold, 
a singular limit of $K3$, in which all the (infinite) curvature is localized on the fixed points.
In particular, we will consider the orbifold $T^4 /Z_2$, where $Z_2$ changes the sign of all the coordinates of 
$T^4$. This orbifold has 16 fixed points. We derive the torus partition function for this model, and in the 
next chapter we will use these results to derive the partition function for some six-dimensional IIB orientifolds.

In order to describe the orbifold compactification, we first 
derive the partition function for IIB on $T^4$. 
When a closed string is compactified on a circle, the spectrum includes, in addition to the usual Kaluza-Klein 
momentum modes, an infinity of topologically distinct sectors, associated 
to closed strings wrapped $n$ times around the circle. The mode expansion for the coordinate along the string 
then reads
\be
X= x + \a^\prime \frac{m}{R} +n R \s +({\rm oscillators}) \quad ,
\ee
where $m$ is the KK momentum and $n$ is the winding number.
In terms of left and right oscillators the same expansion becomes
\be
X_{L,R} = \frac{1}{2}x +\frac{\a^\prime}{2}p_{L,R}(\t \pm \s ) +({\rm oscillators})_{L,R}\quad ,
\ee
where 
\be
p_{L,R} = \frac{m}{R} \pm \frac{nR}{\a^\prime} \quad .
\ee
Thus, if a non-compact coordinate is replaced with a compact one, the continuous integration
over internal momenta is replaced by a lattice sum, so that
\be
\frac{1}{\t_2 \eta (q) \eta (\bar{q}) } \rightarrow \sum_{m,n\in Z} \frac{q^{\a^\prime p_L^2 /4}
\bar{q}^{\a^\prime p_R^2 /4}}{ \eta (q) \eta (\bar{q})} \quad .
\ee
In order to generalize this result to $T^4$, we shall consider for simplicity 
the case in which the torus is a product of circles of equal radii and the internal NS $B$ field vanishes.
In the partition function, the continuous integration over the internal momenta 
is now replaced by
\be
\frac{1}{\t_2^{2} |\eta (q) |^8 } \rightarrow \sum_{m,n\in Z^4} \frac{q^{\a^\prime p_L^2 /4}
\bar{q}^{\a^\prime p_R^2 /4}}{ |\eta (q)|^8 } \quad ,
\ee
where $p_{L,R}$ is a 4-vector with components
\be
p^a_{L,R} = \frac{m^a}{R} \pm \frac{n^a R}{\a^\prime} \quad . 
\ee 
Finally, we need to analyze the effect of the compactification on the fermionic oscillators.
From eq. (\ref{so2ncharacters}) it turns out that the decomposition of the $SO(8)$ characters 
in terms of $SO(4) \times SO(4)$ is 
\bea
& & V_8 = V_4 O_4 + O_4 V_4 \quad ,\nonumber \\
& & S_8 = S_4 S_4 + C_4 C_4 \quad ,\nonumber \\
& & O_8 = O_4 O_4 + V_4 V_4 \quad , \nonumber \\
& & C_8 = S_4 C_4 + C_4 S_4  \quad ,
\eea
where the second $SO(4)$ is associated to the four compact dimensions.
Collecting all the results, the torus partition function for IIB compactified on a flat
$T^4$ is (again neglecting the measure and the non-compact bosonic oscillators)
\be
{\cal T}_{T^4} = | V_4 O_4 + O_4 V_4 - S_4 S_4 - C_4 C_4 |^2 
\sum_{m,n\in Z^4} \frac{q^{\a^\prime p_L^2 /4}
\bar{q}^{\a^\prime p_R^2 /4}}{ |\eta (q)|^8 } \quad .\label{torust4}
\ee
Applying the rules we derived in the ten dimensional case, it is straightforward to determine 
the massless spectrum: it consists of the metric, 16 vectors, 5 antisymmetric tensors and 25 scalars 
in the bosonic sector, and 2 Dirac gravitinos and 10 Dirac spinors in the fermionic sector. This is the field 
content of maximal six-dimensional supergravity, whose scalars parametrize 
the coset manifold $SO(5,5)/SO(5) \times SO(5)$.

Starting from this result, we want to project out of (\ref{torust4}) the states that are not invariant 
under the orbifold projection. It is convenient to define the supersymmetric combination of characters
\bea
& & Q_o = V_4 O_4 - C_4 C_4 \quad , \qquad Q_v = O_4 V_4 - S_4 S_4 \quad , \nonumber \\
& & Q_s = O_4 C_4 - S_4 O_4 \quad , \qquad Q_c = V_4 S_4  -C_4 V_4 \quad .
\eea
Since the internal $V_4$ and one of the two spinorial characters, \eg $C_4$, change sign under the
orbifold projection, the $Q$'s are eigenvectors of the $Z_2$ generator. Moreover, at the massless level,
$|Q_o |^2$ contains the $(2,0)$ gravity multiplet plus one tensor multiplet,  
$|Q_v |^2$ contains four tensor multiplets and $|Q_s |^2$ contains one tensor multiplet.
The result of the projection is then
\be
\frac{1}{2} \left[
 | Q_o + Q_v |^2 \sum_{m,n\in Z^4} \frac{q^{\a^\prime p_L^2 /4}
\bar{q}^{\a^\prime p_R^2 /4}}{ |\eta |^8 } + | Q_o - Q_v |^2 \left| \frac{2 \eta}{\theta_2} \right|^4 \right]
\quad .
\ee
A straightforward analysis, using eqs. (\ref{tmatrix}) and (\ref{smatrix}), reveals that this amplitude
is not modular invariant. In other words, we are missing some states. More precisely, what we have done is 
just to take the {\it untwisted} states, the subset of the original 
closed string states that are invariant under the orbifold
projetion. But other states can be added, namely the ones that correspond to a string closing up to an 
orbifold transformation. These {\it twisted} states are exactly needed to restore modular 
invariance \cite{orbifold}, 
and the modular invariant torus amplitude is thus
\bea 
& & \frac{1}{2} \biggl\{
 | Q_o + Q_v |^2 \sum_{m,n\in Z^4} \frac{q^{\a^\prime p_L^2 /4}
\bar{q}^{\a^\prime p_R^2 /4}}{ |\eta |^8 } + | Q_o - Q_v |^2 \left| \frac{2 \eta}{\theta_2} \right|^4
\nonumber \\ 
& & +16 \left[ | Q_s + Q_c |^2 \left| \frac{\eta}{\theta_4} \right|^4 
+ | Q_s - Q_c |^2 \left| \frac{\eta}{\theta_3} \right|^4 
\right] \biggr\}
\quad .\label{torusorb2b}
\eea
Observe that the twisted sector has a multiplicity equal to the number of fixed points. 
Using the standard technique, we can then read the massless sector of this amplitude. 
The relevant terms are 
\be
| Q_o |^2 + | Q_v |^2 + 16| Q_s |^2 \quad ,
\ee
and so the massless spectrum corresponds to supergravity with 5 tensor multiplets from the 
untwisted sector and 16 additional tensor multiplets from the twisted sector.
As in general for $K3$ compactifications of type IIB, we thus obtain that the low-energy effective
action of type IIB on $T^4 /Z_2$ is $(2,0)$ supergravity coupled to 21 tensor multiplets.

\section{Anomaly cancellation}
\label{anom}
\fancyhead[LO]{{\footnotesize 2.4~~{\it Anomaly cancellation}}}

An anomaly is a breakdown of a classical symmetry by quantum corrections, originated by diagrams
that do not admit a regulator compatible with simultaneous conservation of all the attached currents.
Here we are interested only in anomalies associated to local symmetries. 
Given a classically gauge invariant theory, consider the effective action $\Gamma (A_\m )$
defined by
\be
e^{-\Gamma} = \int D\psi D \bar{\psi} e^{-S} \quad ,\label{effact}
\ee
obtained integrating over the matter fields $(\psi , \bar{\psi})$ in the theory.
Under an infinitesimal gauge transformation $\delta A_\m = D_\m \L$ one obtains
\be
\delta \Gamma = - {\rm tr} \int d^D x \L D_\m \frac{\delta \Gamma}{\delta A_\m} \quad ,\label{deltagamma}
\ee
and from eq. (\ref{effact}) we thus read that
\be
\frac{\delta \Gamma}{\delta A_\m } = < J^\m > \quad .
\ee
Therefore eq. (\ref{deltagamma}) becomes 
\be
{\cal A}_\L = \delta \Gamma = - {\rm tr} \int d^D x \L D_\m < J^\m > \quad ,
\ee
and a lack of conservation of the expectation value of the current implies that the effective
action is not invariant under infinitesimal gauge transformations.
 
\begin{figure}[!h]
\centering
\includegraphics[width=6truecm]{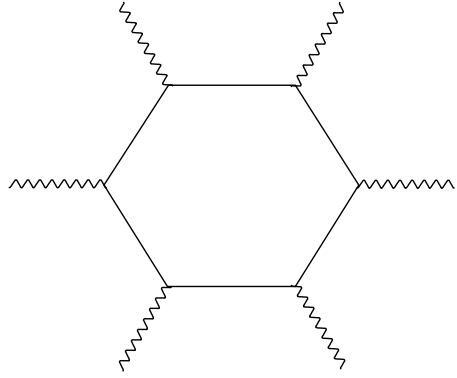}
\caption{Hexagon diagram responsible for ten-dimensional anomalies.}
\label{hexagon}
\end{figure}

Anomalous diagrams can occur only when chiral fermions (or (anti)self-dual bosons) are circulating in the loop,
otherwise one can always construct a gauge invariant mass term that can be used as a Pauli-Villars regulator.
As a consequence, we expect to find purely gravitational anomalies (anomalies corresponding to amplitudes 
whose external fields are only gravitons) only in $2k+2$ dimensions 
(with signature $(1, 2k+1)$), since in these dimensions the 
charge conjugated of a chiral spinor is a spinor of the same chirality (see Appendix (A.1) for details).  
In general, we expect to find anomalies only if the dimension of space-time is even.
Denoting with $D=2n$ the dimension of space-time, the first parity violating amplitude that is potentially anomalous
is a loop with $n+1$ external legs.
In $D=10$ this corresponds to the hexagon diagram in fig. (\ref{hexagon}).

We now resume the techniques for 
deriving anomalies in various dimensions (see Refs. \cite{anomaly} for reviews).
The index of an operator ${\cal O}$ is defined as the difference between the dimension of the kernel of 
${\cal O}$ and that of its adjoint. It can be proved that the anomaly produced by a chiral fermion 
is in correspondence with the index 
of the $(2n+2)$-dimensional Weyl operator $D_+ = \g^\m D_\m P_+$, where $P_+$ is the chirality projector. 
The Atiyah-Singer index theorem \cite{atiyahsinger} relates the index of $D_+$ on a manifold $M$ to topological invariants
of the bundle:
\be
{\rm ind} \ iD_+ =\int_{M} [\hat{A} (M) ch(F)]_{vol} \quad ,
\ee
where $\hat{A}(M)$ is the Dirac genus (or A-roof genus) of $M$ and $ch(F)$ is the Chern character. 
They are defined by
\bea
& & \hat{A} (M) =\prod_{a} \frac{x_a /2}{{\rm sinh} (x_a /2)} \quad , \nonumber \\
& & ch( F) = {\rm tr} e^{\frac{i}{2\p}F} \quad ,
\eea
where $x_a$ are the skew-eigenvalues of the curvature on $M$.  
Analogously, the index of the Rarita-Schwinger operator is given by
\be
{\rm ind} \ iD_{3/2} =\int_{M} [\hat{A} (M) ( ch (R) -1 )ch(F)]_{vol} \quad ,
\ee
while the index for self-dual antisymmetric tensors is given by
\be
{\rm ind} \  iD_{S} =\frac{1}{4}\int_{M} [L(M)]_{vol} \quad ,
\ee
where $L(M)$ is the Hirzebruch polynomial, defined by
\be
L(M) = 2^n \prod_a \frac{ x_a /2}{{\rm tanh}(x_a /2 )} \quad .
\ee
we will need in the following the expansion of $\hat{A}$ and $L$ with respect to the traces of the 
curvature:
\bea
\hat{A}(M) &=& 1 + \frac{1}{(4 \p )^2} \frac{1}{12}\tr R^2 +\frac{1}{(4 \p )^4} \left[
\frac{1}{288} (\tr R^2 )^2 +\frac{1}{360} \tr R^4 \right] \nonumber \\
&+& \frac{1}{(4 \p )^6} \left[ \frac{1}{128 \cdot 81} (\tr R^2 )^3 +\frac{1}{16\cdot 270}
\tr R^2 \tr R^4 +\frac{1}{90 \cdot 63} \tr R^6 \right] +...\nonumber \\
L(M) &=& 1 - \frac{1}{(2 \p )^2} \frac{1}{6}\tr R^2 +\frac{1}{(2 \p )^4} \left[
\frac{1}{72} (\tr R^2 )^2 -\frac{7}{180} \tr R^4 \right] \nonumber \\
&+& \frac{1}{(2 \p )^6} \left[ -\frac{1}{1296} (\tr R^2 )^3 +\frac{7}{1080}
\tr R^2 \tr R^4 -\frac{31}{2835} \tr R^6 \right] +...\quad .
\eea

The anomaly polynomials
\bea
& & I_{1/2}(R,F) =[\hat{A}(M) ch(F) ]_{D+2} \quad ,\nonumber\\
& & I_{3/2} (R) = [\hat{A}(M) (ch(R)-1) ]_{D+2} \quad ,\nonumber\\
& & I_{S} (R) = -\frac{1}{8} [L(M)]_{D+2} \label{anompolyn}
\eea
are closed (D+2)-forms, and locally determine a (D+1)-form whose gauge transformation is exact:
\bea
& & I_{D+2} = d I_{D+1} \quad ,\nonumber \\
& & \delta_\L I_{D+1} =d I_{D}^\L \quad .
\eea
The anomaly is finally determined as the D-dimensional integral of the form $I_{D}^\L$.

In ten dimensions, the anomaly originates from a 12-form, and an analysis of the previous results 
shows that the anomaly polynomial exactly cancels for a theory containing a complex left-handed 
gravitino, a complex right-handed spinor and a self-dual 4-form. This is exactly the chiral content of 
$2b$ supergravity, so that the effective action of 
type-IIB string theory turns out to be anomaly-free \cite{agw}.

We now consider $(2,0)$ 
supersymmetric models in six dimensions. In this case the chiral content corresponds to
two left-handed gravitinos and five self-dual 2-forms from the gravity multiplet, as well as 
$n$ antiself-dual 2-forms and $2n$ right-handed spinors from $n$ tensor multiplets. 
The cancellation of the anomaly polynomial 8-form implies $n=21$ \cite{agw}, that as we have seen in the previous
section is exactly what one obtains compactifying type IIB on $K3$.

\begin{figure}[!h]
\centering
\includegraphics[width=6truecm]{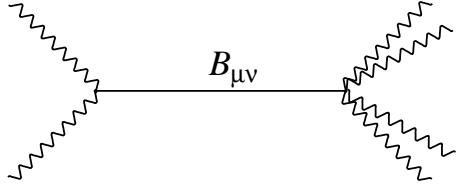}
\caption{Anomalous diagram responsible for the Green-Schwarz mechanism}
\label{bfield}
\end{figure}

Now we consider ten-dimensional $(1,0)$ supergravity coupled to Yang-Mills vector multiplets. 
In this case the chiral content of the theory is a left-handed Majorana gravitino and a right-handed
Majorana spinor from the gravity multiplet, and a left-handed Majorana gaugino in the adjoint of the 
gauge group from the vector multiplet. 
The 12-form anomaly polynomial is proportional to
\bea
& & -\frac{496 -n}{128 \cdot 2835}\tr R^6 -\frac{224+n}{256\cdot 1080}\tr R^2 \tr R^4 -\frac{64 -n}{512 \cdot 1296}
(\tr R^2 )^3 \nonumber \\
& & -\frac{1}{720} {\rm Tr} F^6 +\frac{1}{24\cdot 48} {\rm Tr} F^4 \tr R^2 -\frac{1}{256\cdot 45} {\rm Tr} 
F^2 \tr R^4 -\frac{1}{192} {\rm Tr} F^2 (\tr R^2 )^2 ,
\eea
where ${\rm Tr}$ is the trace in the adjoint representation and $n$ is the dimension of the gauge group.
Imposing $n=496$ cancels the {\it irreducible } term $\tr R^6$, and we are left with the residual 
anomaly polynomial
\bea
& & \frac{1}{8} \tr R^2 \tr R^4 +\frac{1}{32} (\tr R^2 )^3 -\frac{1}{15} {\rm Tr} F^6 \nonumber \\
& & +\frac{1}{24} {\rm Tr} F^4 \tr R^2 -\frac{1}{240} {\rm tr} F^2 \tr R^4 -\frac{1}{192} 
{\rm Tr} F^2 ( \tr R^2 )^2 \quad .
\eea
The result of \cite{gs} is that for the gauge groups $SO(32)$ and $E_8 \times E_8$ this residual anomaly 
polynomial factorizes into the product of a 4-form and an 8-form. In the $SO(32)$ case the result is 
\be
(\tr R^2 -\tr F^2 )( \tr F^4 -\frac{1}{8} \tr R^2 \tr F^2 +\frac{1}{8} \tr R^4 +\frac{1}{32} (\tr R^2 )^2 )\quad .
\ee
Adding to the low-energy effective action the term 
\be
B \wedge ( \tr F^4 -\frac{1}{8} \tr R^2 \tr F^2 +\frac{1}{8} \tr R^4 +\frac{1}{32} (\tr R^2 )^2 )\quad ,
\ee
and demanding that under gauge and Lorentz transformations $B$ transform as\footnote{The first term on 
the right-hand side of eq. 
(\ref{deltabgsmech}) is already determined by supersymmetry, as was shown in the first
chapter.} 
\be
\d B = \w_{2,YM} - \w_{2,L} \quad ,\label{deltabgsmech}
\ee
where $\w_2$ is determined from the Chern-Simons 3-form by $d \w_2 =\delta \w_{CS}$, the total anomaly exactly cancels.
This is the celebrated {\it Green-Schwarz mechanism} \cite{gs}, that guarantees consistency of the low-energy effective action
for both heterotic and type I in ten dimensions. 
It corresponds to the cancellation of part of the anomaly 
resulting from the hexagon diagram of fig. (\ref{hexagon}) against the anomalous 
contribution coming from the tree-level amplitude of fig. (\ref{bfield}).  

We finally consider ${\cal N} =(1,0)$ supersymmetry in six dimensions, 
whose multiplets are the gravity multiplet, 
containing the metric, a self-dual 2-form and a left-handed gravitino, the tensor multiplet, containing
an antiself-dual 2-form, a scalar and a right-handed spinor, the vector multiplet, containing 
a vector and a left-handed gaugino, and the hypermultiplet, containing four scalars and a right-handed spinor (we 
will analyze in detail the couplings of these multiplets in Chapter 4). From eq. (\ref{anompolyn}) we can 
read the 8-form anomaly polynomial corresponding to supergravity coupled to $n_T$ tensor multiplets, 
$n_V$ vector multiplets and $n_H$ hypermultiplets, and 
the cancellation of the coefficient of the irreducible term $\tr R^4$ 
implies the condition \cite{rdsss}
\be
273 -29 n_T + n_V - n_H =0 \quad . 
\ee
In the next chapter we will analyze the reducible part of the anomaly polynomial both for 
the heterotic string 
compactified on $K3$ and for six-dimensional type-IIB orientifolds. 

\section{D-branes and dualities}
\label{Dbranes}
\fancyhead[LO]{{\footnotesize 2.5~~{\it D-branes and dualities}}}

All the theories of oriented closed strings 
we have analyzed so far have in the massless spectrum a 2-form coming from the NS sector. 
Oriented closed strings, in fact, are precisely the sources for this field, 
as electrically charged particles are
sources for vector potentials. One can then ask which are the objects that are charged with respect to
the R-R forms present in type IIA and type IIB. The nature of these objects is non-perturbative, since 
there are no states  belonging to the perturbative 
spectrum of string theories that are charged under R-R forms, and they result to be defined as hyper-planes 
on which open strings can end \cite{dbranes}. More precisely, the excitations 
of these p-dimensional extended 
objects are open strings that satisfy Dirichlet boundary conditions 
in the 9-p directions orthogonal to them,
and for these reason these objects are called D-branes \cite{dbranes}. 
As a consequence type IIA and type IIB, that at a perturbative
level are theories of oriented closed strings only, admit 
non-perturbative open-string excitations. Moreover, analyzing the 
spectrum of the two theories, one sees that type IIA contains D-branes with an even number of spatial directions, while 
type IIB contains D-branes with an odd number of spatial directions. 
The non-perturbative origin of these branes is 
manifest if one consider that in the string frame they have tensions inversely proportional 
to the string coupling constant, and 
so they disappear from the spectrum at weak coupling. Finally, their 
supersymmetric nature corresponds to the 
fact that these objects correspond to BPS solutions 
of the low-energy supergravities, while their supersymmetric
world-sheet action contains the Born-Infeld action and Wess-Zumino terms of the type 
\be
\int_{p+1} (C^{(p+1)} + F \wedge C^{(p-1)}+... )\quad ,\label{wesszumbrane}
\ee
that naturally couple the brane to the R-R potentials. 
In the low-energy limit, these objects decouple from gravity, and the resulting action is supersymmetric 
Yang-Mills theory. In the case of $N$ coincident D-branes, the resulting gauge group is $U(N)$ (we will see in the next 
chapter that if the open strings are not oriented, the gauge group can be orthogonal or symplectic \cite{pc,schwarz82,ms}).

In analyzing IIB supergravity, we showed that the scalar manifold has an $SL(2,R)$ isometry, 
that is a symmetry of the low-energy
theory. In the full non-perturbative string theory the discrete subgroup $SL(2,Z)$ 
survives \cite{ht}, and the theory manifests 
an S-duality, acting on the scalars (see eq. (\ref{sl2r})) as 
\be
w \rightarrow \frac{a w+ b}{cw +d } \quad ,
\ee
with $w = \rho + i e^\phi $, where $\phi$ is the dilaton and $\rho$ the R-R scalar. 
This discrete symmetry is non-perturbative,
as can be seen from the fact that it inverts the coupling, and this is why it is not manifest 
if we express the low-energy action in the string frame (see Section (1.4)). For instance, 
this symmetry justifies
the presence of a D1-brane in the non-perturbative spectrum of type IIB: in the dual description, 
this D1-string 
becomes fundamental. Another remark concerning type IIB has to be made: 
in the non-perturbative spectrum, a
space-time filling D9-brane is also present. 
This object has no dynamics, but his role will appear to be relevant in 
the next chapter. 

Having shown that type IIB is self-dual, 
so that the theory has the same form at strong and at weak coupling,
we want to see what happens to type IIA at strong coupling. 
In Section (1.4) we have shown that dimensional reduction of eleven-dimensional supergravity gives rise to 
IIA supergravity. In particular, in the string frame the relation between the compactification radius and the 
string coupling constant is
\be
R = g_A^{2/3} \quad ,
\ee
and this relation 
shows that, for generic values of the dilaton, the type-IIA string manifests a perturbatively hidden 
eleventh dimension \cite{witten1}.
The strong coupling limit develops
this extra dimension, resulting in an eleven-dimensional theory, called {\it M-theory} \cite{witten1,townsend}. 
The low-energy action of M-theory is
eleven-dimensional supergravity, and the presence of a 3-form potential suggests that the complete 
theory describes an M2-brane and a dual M5-brane. 
Even if this theory is not known, some higher derivative couplings 
can be deduced by supersymmetry and by anomaly considerations. 
Consider as an example the Wess-Zumino term $A \wedge F \wedge F$
of eleven-dimensional supergravity. This term is gauge invariant in 
the absence of M5-branes, but if an M5-brane, 
a magnetic source for the 4-form field strength, is present, 
the Bianchi identity $dF=0$ must be replaced 
by $dF= \delta^{5}(x -x^i )$, where $x$'s are the directions orthogonal 
to the M5-brane, and the Wess-Zumino term is no more gauge invariant. 
The resulting anomaly produced  
in the bulk gravity action exactly cancels the anomalous contribution 
from the Wess-Zumino term (analogous to eq. 
(\ref{wesszumbrane})) in the M5-brane action. This {\it anomaly inflow} 
mechanism \cite{aninflow}, in which the anomaly in the bulk
generated by a source is canceled by the anomaly on the brane \cite{aninflow2}, 
also applies to higher derivative gravitational
anomalies, and can be used to determine higher derivative couplings in the 
low-energy supergravity action. 
The Kaluza-Klein modes of the compactification, that with respect to the 
eleven-dimensional metric have masses
proportional to $1/ R$, in the string metric have masses of the order of 
$e^{-1/3 \phi} / R = 1/ g_A$ \cite{witten1}, and are D0-branes 
in the ten-dimensional theory. An M2-brane wrapped around the circle 
gives a string in ten dimensions, and it can
be shown that its tension in the string frame does not depend on the 
dilaton, as pertains to a fundamental object, while 
an unwrapped M2-brane is a D2-brane, whose tension again scales like $1/g_A$ in the string frame. 
Finally, M-theory compactified on $S^1/Z_2$ corresponds to non perturbative 
$E_8 \times E_8$ heterotic string, where
again the ten-dimensional dilaton is related to the length of the interval \cite{hw}. 
We will return to this identification in the next chapter.

This chain of dualities relating various theories in different regimes 
becomes richer and richer in lower dimensions.  
In this respect, all known (supersymmetric) string theories
represent different charts of a moduli space of an unknown more 
fundamental theory, while the duality relations are transition functions between different charts. 
We will not describe all these duality relations in this
Thesis, see for instance \cite{dualities,bachasdbranes} for reviews.

% \newpage ~                  %  per lasciare una pagina bianca
% \thispagestyle{empty}       %  

\chapter{Open strings}
\label{cap3}
% \newpage
\vspace*{2cm}
\fancyhead[RO,LE]{\thepage}
\fancyhead[RE]{{\footnotesize {\rm Chapter 3.}~~{\it Open strings}}} 
\fancyhead[LO]{}

\noindent
In this chapter we want to describe how to construct type-I models. The basic
idea is that these models are generalized orbifolds ({\it orientifolds}) of
type IIB superstrings \cite{cargese,bs}. More precisely, if we denote with $\W$ the 
operation that exchanges left and right oscillators, the basic relation is 
$$
{\rm Type-I} = ({\rm Type-IIB})/\W
\quad .
$$
In the last chapter we have seen how to write the one-loop partition function for closed
oriented superstring theories. The $\W$ operation is implemented on the torus partition function 
projecting out of the spectrum the states that are not invariant under left-right exchange. This 
corresponds to substituting to the torus amplitude, that is a closed oriented string loop, 
the halved sum of the torus amplitude and the Klein bottle amplitude, in which a closed string flips its 
orientation before closing the loop. 
We have seen in the previous chapter that in order to restore modular invariance in orbifold compactifications
of closed oriented strings, one has to add to the untwisted sector (corresponding to the closed spectrum projected by the
orbifold operation) the twisted sector, corresponding to the spectrum of strings that
close only after the orbifold operation. 
In the construction of type-I models, one encounters a similar problem: the unoriented closed spectrum is typically
inconsistent, and the inconsistency manifests itself by the appearence of tadpoles, that correspond to gauge and
gravitational anomalies in the low-energy effective action. In order to cure this problem, one has to add a suitable 
open sector. 

In Section 1, we will derive the one-loop partition function for the ten-dimensional $SO(32)$ type-I superstring. 
In Section 2 we will see that a different orientifold projection of type IIB in ten dimensions results in a 
consistent non-supersymmetric model which is free of tachyons, and whose gauge group is $USp(32)$. Actually, the 
closed spectrum of this model is supersymmetric, and supersymmetry is only broken in the open sector. 
We will then describe in Section 3 how to obtain six-dimensional type-I models as $T^4 /Z_2$ orientifolds of type IIB. 
Also in six dimensions one can construct consistent non-tachyonic models in which supersymmetry is broken in the 
open sector. Section 4 is devoted to a discussion of anomaly cancellation in type-I models, and finally in
Section 5 we will discuss of 
general features of six-dimensional models that will be relevant for the following
chapters.

\section{The ten-dimensional $SO(32)$ type-I superstring}
\label{so32}
\fancyhead[LO]{{\footnotesize 3.1~~{\it Ten-dimensional $SO(32)$ type-I superstring}}}

\begin{figure}[!h]
\centering
\includegraphics[width=12truecm]{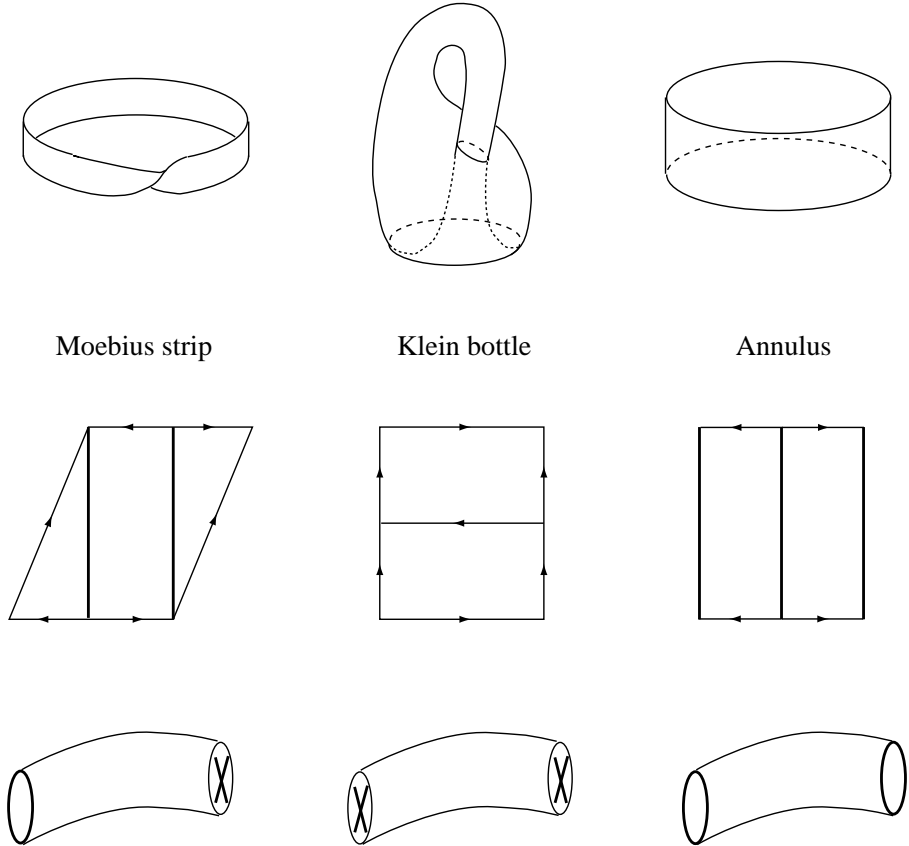}
\caption{M\"obius strip, Klein bottle and annulus}
\label{kma}
\end{figure}

In this section we describe how to obtain the $SO(32)$ type-I model as an orientifold projection
of type IIB. Since type IIB is invariant under $\W$, the operation that
exchanges left and right moving oscillators, we can project out all states
that are not invariant. This projection is realized at the level of 
the partition function by 
\be
{\cal T} \rightarrow \frac{1}{2} ( {\cal T} + {\cal K} ) \quad ,
\ee
where ${\cal K}$ is the {\it Klein bottle} amplitude, correspinding to a closed string 
that flips its orientation before closing the loop (see fig (\ref{kma})).
The Klein bottle is obtained from the torus by the anti-conformal involution
\be
z \rightarrow 1 - \bar{z} + i \t_2 \quad ,
\ee
compatible with the periodicity of the lattice only if $\t$ is purely imaginary. 
The final result is that the Klein bottle amplitude (we omit the overall constants)
\be
{ \cal K} =\frac{1}{2} \int_0^\infty \frac{d \t_2}{\t_2^6 } \frac{(V_8 - S_8 ) (2i \t_2 )}{\eta^8 (2i \t_2 )} \label{ka}
\ee
depends naturally on $2i \t_2$, the modulus of the doubly covering torus of fig. (\ref{kma}).
From the torus and Klein bottle amplitudes we can read the spectrum: at the massless level, 
in the NS-NS sector we are projecting out 
the 2-form, while in the RR sector we are projecting out the self-dual 4-form and the scalar. 
We are thus left with
the metric, the dilaton and the RR 2-form. In the fermionic NS-R and R-NS sector, the Klein bottle does not introduce 
new states, so the projection just halves the number of spinors, and the end result is
the field content of ${\cal N} =1$
ten dimensional supergravity, discussed in Section (1.4). 

While the torus amplitude has no ultraviolet 
divergence, since modular invariance removes the 
dangerous region from the integral, the Klein bottle is not modular invariant, and it manifests a 
divergence for small values of $\t_2$ that we want to analyze. 
Performing an $S$ 
transformation $i \t_2 \rightarrow 1/\t_2$, where $i \t_2$ is the complex parameter of the doubly covering torus,
the one loop amplitude is mapped to a 
tree level amplitude, in which a closed string bounces between two crosscaps\footnote{A 
crosscap can be considered as a boundary with opposite points identified.}. 
The end result is that eq. (\ref{ka}) becomes
the transverse-channel amplitude
\be
\tilde{ \cal K} =\frac{2^5}{2} \int_0^\infty d l \frac{( V_8 - S_8 ) ( il)}{\eta^8 ( il )} \quad , 
\label{katransv}
\ee
while the ultraviolet 
divergence in the direct channel becomes an infrared divergence in the transverse channel, generated 
by massless closed-string exchanges between the two crosscaps. 
From a target space point of view, this can be 
interpreted as closed-string tree-level scattering between two orientifold planes, extended objects that invade 
all the nine spatial dimensions, changing the orientation of the strings they interact with. In this sense, if the 
massless particle that bounces between the two O9-planes is the graviton or the dilaton, the transverse amplitude 
is proportional to the square of the tension of the O9-plane, 
while if the particle is the RR 10-form (a 9-plane can be charged with 
respect to a 10-form) this amplitude is proportional to the square of its charge. 
The overall tension and charge determine the NS and RR tadpoles, whose values can be read from (\ref{katransv}). 
In particular, the presence of a RR tadpole 
makes the model inconsistent, since it corresponds to an overall charge that fills the whole space.

In order to cure this divergence, one adds D9-branes, \ie the open sector.  
The one-loop annulus amplitude is
\be
{ \cal A} =\frac{N^2}{2} \int_0^\infty \frac{d \t_2}{\t_2^6 } \frac{( V_8 - S_8 ) (i \t_2 /2)}{\eta^8 (i \t_2 /2)} \quad ,
\label{annulus}
\ee
where the multiplicity $N$ of the Chan-Paton charge is associated to the number of D9-branes, and 
$i \t_2 /2$ is the modulus of the doubly covering torus of fig. (\ref{kma}).
Performing the $\W$ projection, we then obtain the M\"obius strip amplitude, corresponding to an open string
that flips its orientation before closing the loop. The resulting amplitute,
\be
{ \cal M} = \frac{\e N}{2} \int_0^\infty \frac{d \t_2}{\t_2^6 } \frac{( \hat{V}_8 - \hat{S}_8 ) (i \t_2 
/2 +1/2)}{\hat{\eta}^8 (i \t_2 /2 +1/2)} \quad ,
\label{mobius}
\ee
is written in terms of the ``hatted'' characters\footnote{Given the character $\chi (q) =
q^{h -c /24} \sum_k d_k q^k $, one introduces a real hatted character $\hat{\chi} ( i \t_2 +1/2 )= 
q^{h -c /24} \sum_k (-1 )^k d_k q^k $, where $q = e^{-2\p \t_2}$, that differs from
${\chi} ( i \t_2 +1/2 )$ by the phase $e^{-i \p (h -c /24)}$.}, and $\e= \pm 1$. 
The argument 
$i \t_2 /2+ 1/2$ is again the modulus of the doubly covering torus, as can be deduced from
 fig. (\ref{kma}).

\begin{figure}[!h]
\centering
\includegraphics[width=12truecm]{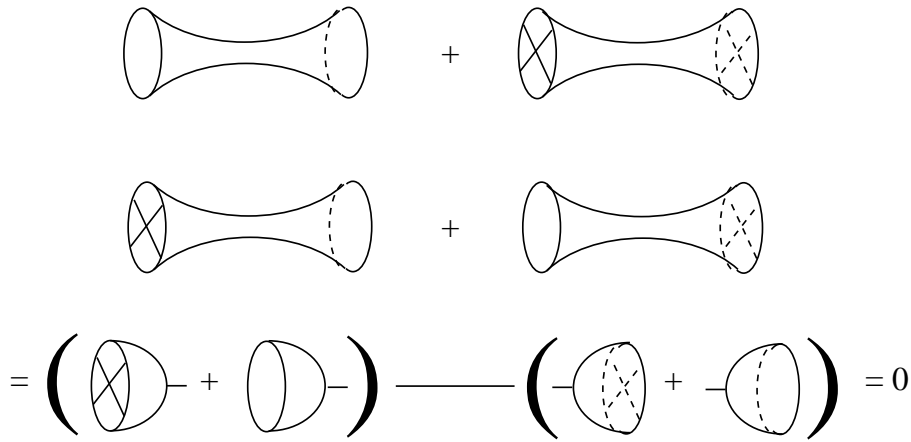}
\caption{Tadpole cancellation}
\label{tadpoles}
\end{figure}

We now want to write the same amplitudes in the transverse channel. 
In the case of the M\"obius amplitude, it is important to observe that since the parameter
of the doubly covering torus has a real part, the modular transformation connecting direct 
and transverse M\"obius amplitudes is no longer an $S$ transfromation, but the transformation
\be
P: \frac{1}{2} +i \frac{\t_2}{2} \rightarrow \frac{1}{2} +i \frac{1}{2 \t_2} \quad .
\ee
The operator that realizes this transformation on the hatted characters is then
\be
P = T^{1/2} S T^2 S T^{1/2} \quad ,
\ee
and the end result is 
\bea
& & \tilde{ \cal A} =\frac{2^{-5} N^2}{2} \int_0^\infty d l \frac{( V_8 - S_8 ) ( il)}{\eta^8 ( il )} \quad ,\nonumber \\
& & \tilde{ \cal M} =2\frac{\e N}{2} \int_0^\infty d l \frac{( \hat{V}_8 - \hat{S}_8 ) ( il +1/2) }{\hat{\eta}^8 ( il +1/2)}\quad .
\eea
We interpret these transverse amplitudes as tree-level amplitudes for closed strings bouncing between 
two boundaries or between a boundary and a crosscap. The request that the orientifold plane and the D-brane 
have opposite charge corresponds then to $\e =-1$, while
the consistency of the model is finally guaranteed imposing the tadpole cancellation 
condition (fig. (\ref{tadpoles})) \cite{tadpoles}
\be
2^5 + 2^{-5} N^2 -2N = 2^{-5} (N -32 )^2 = 0 \quad ,
\ee
that selects $N=32$.   
This condition actually cancels both the RR and the NS tadpoles, so that the resulting configuration has zero charge and zero
tension. This corresponds to the fact that the orientifold plane involved, $O_-$, 
has negative tension and negative charge. 
From the amplitudes (\ref{annulus}) and (\ref{mobius}) in the direct channel we can finally 
read the spectrum: at 
the massless level, we have $N (N-1)/2$ vectors and $N (N-1)/2$ spinors, corresponding to the ten-dimensional  
${\cal N}=1$ vector multiplet of gauge group $SO(32)$. From the results of the previous chapter we can then conclude that
tadpole cancellation reflects the absence of anomalies in the low-energy effective action \cite{tadpoles}. 

\section{Ten-dimensional $USp(32)$ type-I string}
\label{usp32}
\fancyhead[LO]{{\footnotesize 3.2~~{\it Ten-dimensional $USp(32)$ type-I string}}}

The $SO(32)$ model can be modified adding brane-antibrane pairs. Denoting with $N_+$ and $N_-$ the number 
of D9 and D$\bar{9}$ branes, we obtain the transverse amplitudes
\bea
& & \tilde{ \cal A} =\frac{2^{-5}}{2} [ (N_+ + N_- )^2  V_8 - (N_+ -N_- )^2  S_8 ] \quad ,\nonumber \\
& & \tilde{ \cal M} = - \frac{2}{2} [(N_+ + N_- ) \hat{V}_8 - (N_+ -N_- ) \hat{S}_8 ] \quad .\label{so32plusbab}
\eea
RR tadpole cancellation requires
\be
N_+ =32 +N_- \quad ,
\ee
so that this model corresponds to adding $N_-$ brane-antibrane pairs to the stable 
configuration of 32 branes. 
From the direct channel amplitudes
\bea
& & { \cal A} =\frac{1}{2} [ (N_+^2 + N_-^2 )(  V_8 -S_8 ) + N_+ N_- (O_8 -C_8 )] \quad ,\nonumber \\
& & { \cal M} = - \frac{1}{2} [(N_+ + N_- ) \hat{V}_8 - (N_+ -N_- ) \hat{S}_8 ] 
\eea
we derive the open spectrum, that corresponds to the gauge group $SO(32 +N_- )\times SO(N_- )$, where the first gauge factor is supported on the 
branes and the second on the anti-branes. The spinors in the 9-9 sector are in the adjoint representation
of $SO(32+N_- )$,
while the spinors in the $\bar{9}$-$\bar{9}$ sector are in the symmetric (reducible) representation
of $SO(N_- )$. The presence of the tachyon reflects the instability of the vacuum \cite{sen}.

One can actually consider a different orientifold projection, induced by 
$O_+$ planes, with positive tension and positive charge. This is obtained changing the sign of the transverse
M\"obius amplitude in eq. (\ref{so32plusbab}), while RR tadpole cancellation implies in this case
\be
N_- =32 +N_+ \quad .
\ee
From the amplitudes in the direct channel, we then read the massless spectrum: the gauge group 
is $USp(32 +N_+ ) \times USp(N_+ )$, while the spinors in the 9-9 sector are in the adjoint representation
of $USp(N_+ )$
and the spinors in the $\bar{9}$-$\bar{9}$ sector are in the antisymmetric (reducible) representation
of $USp(32+ N_+ )$ \cite{sugimoto}. Observe that this model is consistent only for an even number of brane-antibrane pairs, 
since only in this case a non-degenerate symplectic form exists.
For the particular value $N_+ =0$, the spectrum does not contain a tachyon, and so the model could be considered
stable from this one-loop analysis, even if the NS tadpole, corresponding to a dilaton potential, 
changes the vacuum. The resulting open spectrum is non-supersymmetric. We will analyze the low-energy
action for this model in Chapter 5.
Finally, it is important to observe that all the models mentioned in this section are anomaly-free \cite{sugimoto}, 
as can be checked analyzing their chiral field content.

\section{Six-dimensional type-I models}
\label{6dim}
\fancyhead[LO]{{\footnotesize 3.3~~{\it Six-dimensional type-I models}}}

In Section (2.3) we have derived the partition function for type-IIB theory
compactified on ${T}^4 / {Z}_2$. Here we want to analyze the corresponding
type-I models. 

Six-dimensional orientifold projections result in general in the introduction of $O5$-planes,
in addition to the $O9$-planes present in the ten-dimensional models. In supersymmetric 
models, both the tension and the charge of the $O9_-$ and $O5_-$ planes are canceled by a suitable 
configuration of $D9$ and $D5$ branes. 
In $Z_2$ orbifold compactifications, and in general for 
orbifolds with order-two group generators, one has the option 
of antisymmetrizing some twisted sectors or, equivalently, of inverting
tensions and charges for the corresponding $O5$-planes.  This requires
that $D9$ branes be accompanied by suitable stacks of ${\bar{D}}5$ branes,
with a resulting brane supersymmetry breaking \cite{bsb}.
Concentrating on the torus partition function of eq. (\ref{torusorb2b}), describing 
IIB compactified on $T^4 /Z_2$,
two choices for
the unoriented projection compatible with the crosscap 
constraint of \cite{pss} are described by
\be
{\cal K} = \frac{1}{4} \biggl\{ \ ( \ Q_o + Q_v \ ) \ ( P + W ) 
+ 2 \epsilon \times 16 \ ( \ Q_s +Q_c \ ) \ 
{\biggl( \ \frac{\eta}{\theta_4}\biggr)}^2 \ \biggr\} \ , 
\label{klein6}
\ee
where $P$ ($W$) indicates the momentum (winding) lattice sum and
$\epsilon = \pm 1$.  At the massless level, $\epsilon = 1$
gives $N=(1,0)$ 
supergravity with $1$ tensor multiplet and $20$ hypermultiplets, 
while $\epsilon = - 1$ gives
$N=(1,0)$ supergravity with $17$ tensor multiplets and $4$
hypermultiplets. These closed spectra are both supersymmetric, but
the latter projection introduces $O9_+$ and $O5_-$ planes, and
this leads to an open sector with brane supersymmetry breaking.
This is clearly spelled by the massless contributions to the transverse
Klein-bottle amplitude, that can be read from
\be
\tilde{\cal K}_0 = \frac{2^5}{4} \biggl\{ \  Q_o  \ \biggl( \ \sqrt{v}  +
\epsilon\, 
\frac{1}{\sqrt{v}} \ \biggr)^2 +  Q_v  \ \biggl( \ \sqrt{v}  -
\epsilon \,
\frac{1}{\sqrt{v}} \ \biggr)^2 \ \biggr\} \ , \label{transvklein6}
\ee 
where $v = \sqrt{\det g /(\a^\prime )^4 }$ is the internal volume.
The reflection coefficients
are interchanged in the two cases: $\e=1$ corresponds to the introduction of $O9_-$ and $O5_-$ planes,
both with negative tension and negative charge, while $\e=-1$ corresponds to the introduction 
of $O9_-$ and $O5_+$ planes, where the $O5_+$ planes have positive tension and positive charge.  
We now want to describe the open sector that gives rise to RR tadpole cancellation.

For $\e=1$, the simplest choice corresponds to placing all branes at a single fixed point. 
The corresponding annulus amplitude is
\bea
{\cal A} &=& \frac{1}{4} \biggl\{ (Q_o +Q_v ) \left( N^2 P +D^2 W \right) + ( R_N^2 + R_D^2 ) \left(
\frac{2 \eta }{\theta_2} \right)^2 \nonumber \\
&+& 2ND (Q_s +Q_c ) \left( \frac{ \eta }{\theta_4} \right)^2
+ 2R_N R_D  (Q_s -Q_c ) \left( \frac{ \eta }{\theta_3} \right)^2 \biggr\} \quad , \label{annulus6}
\eea
where $N$ and $D$ count the multiplicities of the string ends with Neumann and Dirichlet boundary conditions, and
$R_N$ and $R_D$ define the orbifold action on the Chan-Paton charges. In the present example, these are associated to the $D9$ 
and $D5$ branes that have to be present to cancel the RR tadpoles. 
The massless contributions to the transverse
annulus amplitude can be read from
\bea
\tilde{\cal A}_0 &=& \frac{2^{-5}}{4}  \biggl\{ Q_o  \left( N \sqrt{v} + \frac{D}{\sqrt{v}}   \right)^2
+ Q_v  \left( N \sqrt{v} - \frac{D}{\sqrt{v}}   \right)^2 \nonumber\\
&+& Q_s [15 R_N^2 +(R_N -4 R_D )^2 ]+ Q_c [15 R_N^2 +(R_N + 4 R_D )^2 ] \biggr\} \quad .\label{transvallulus6}
\eea
Analyzing this expression, one can deduce that, for the untwisted contributions, 
the CP multiplicities $N$ and $D$
determine the overall numbers of $D9$ and $D5$ branes. The structure of these terms matches 
precisely the corresponding 
contributions of $O9$ and $O5$ planes in eq. (\ref{transvklein6}) with $\e=1$. The additional terms
are associated to the exchange of twisted closed-string modes, and 
encode the geometry of the branes sitting at the fixed
points. In this case, where all the $D5$ branes are at the same fixed point, these tadpole terms account precisely for the 15 fixed 
points seen only bt the space-filling $D9$ branes, as well as for the additional single fixed point where also $D5$ branes
are present. From eqs. (\ref{transvklein6}) and (\ref{transvallulus6}) one can deduce 
the massless part of the transverse 
M\"obius amplitude,
\bea
\tilde{\cal M}_0 &=& - \frac{2}{4}  \biggl\{ \hat{Q}_o \left(  \sqrt{v}  +
\epsilon  
\frac{1}{\sqrt{v}} \ \right)  \left( N \sqrt{v} + \frac{D}{\sqrt{v}}   \right) \nonumber \\
&+& \hat{Q}_v  
\left(  \sqrt{v}  - \epsilon  
\frac{1}{\sqrt{v}} \ \right)
\left( N \sqrt{v} - \frac{D}{\sqrt{v}}   \right) \biggr\} \quad ,
\eea
and tadpole cancellation requires $N =D =32$ and $R_N = R_D =0$. In order to read the open spectrum, we have finally
to derive the M\"obius amplitude in the direct channel. This is obtained performing a $P$ modular transformation on the 
complete transverse M\"obius amplitude, and the final result is 
\be
\tilde{\cal M} = - \frac{1}{4}  \biggl\{ (\hat{Q}_o + \hat{Q}_v )\left( N \ P +D \ W \right)  
- (N+D) (\hat{Q}_o - \hat{Q}_v )\left( \frac{2 \hat{\eta}}{\hat{\theta}}  \right) \biggr\} \quad .\label{mobius6}
\ee
The proper parametrization of the Chan-Paton multiplicities,
\bea
& & N= n+\bar{n} \quad , \qquad D= d+\bar{d} \quad ,\nonumber \\
& & R_N = i ( n-\bar{n} ) \quad , \qquad  R_D =i ( d- \bar{d} ) \quad ,
\eea 
with $n=d=16$, identifies the gauge group $U(16) \times U(16 )$ \cite{bs,gp}, 
where one gauge factor lives on the $D9$ branes
and the other on the $D5$ branes. 
Together with the vector multiplets associated 
to this gauge group, the massless open spectrum contains charged 
hypermultiplets in the $({\bf 120} +\bar{\bf 120}, {\bf 1})$ and $({\bf 1},{\bf 120} +\bar{\bf 120})$ 
coming from $DD$ and $NN$ strings, and charged hypermultiplets 
in the $({\bf 16}, \bar{\bf 16})$ coming from $ND$ strings. 
As in the ten-dimensional case, the RR tadpole cancellation implies, 
as a consequence of supersymmetry, that the NS tadpole cancels as well. 
A more general case, where the $D5$ branes are
distributed in the 16 fixed points, can be analyzed following the same technique. 
One can also consider a 
situation in which pairs of image $D5$ branes are moved away from the fixed points \cite{gp}. 
While the $D5$ branes sitting 
at the fixed points lead to unitary gauge groups whose rank is determined by their 
total number, the remaining 
$D5$ branes lead to symplectic gauge groups whose rank is determined by the number of displaced pairs.
Other models can be obtained turning on an internal quantized $B_{ab}$, or by 
non-geometric orbifolds: these models typically
have several tensor multiplets in the closed spectrum 
\cite{bs,severaltensors,iu}, a characteristic that makes these orbifold constructions 
quite peculiar with respect to heterotic models, where the spectrum can only contain 
a single tensor multiplet
at the perturbative level.

Coming back to the Klein-bottle amplitude of eq. (\ref{klein6}), for $\epsilon=-1$ all
R-R tadpoles can be canceled by 32 $D9$ branes and 32
${\bar{D}}5$ branes, since the latter indeed 
revert all the 9-5 R-R contributions to the transverse-channel 
annulus amplitude.  The resulting transverse annulus amplitude is  
\ba
\tilde{\cal A} &=&  \frac{2^{-5}}{4}  \ \biggl\{ \ 
( \ V_4 O_4 + O_4 V_4 - S_4 S_4 - C_4 C_4 \ ) \biggl( \ N^2 v
W  +
\frac{D^2 P}{v} \ \biggr) \nonumber \\
&+& 2 N D \ ( \ V_4 O_4 - O_4 V_4 - S_4 S_4 + C_4 C_4 \ ) \ {\biggl(\frac{2
\eta}{\theta_2}\ \biggr)}^2  \nonumber \\ 
&+&  16 \ ( \ O_4 C_4 + V_4 S_4 - S_4 O_4 - C_4 V_4 \ ) \biggl(
R_N^2 + R_D^2 \ \biggr){\biggl(\frac{
\eta}{\theta_4}\biggr)}^2 \nonumber \\
&+& 8 R_N R_D \ ( \ V_4 S_4 - O_4 C_4 - S_4 O_4 + C_4 V_4 \ ) \ {\biggl(\frac{
\eta}{\theta_3}\ \biggr)}^2 \ \biggr\} \ ,
\label{atra}
\ea
and from $\tilde{\cal K}$ and $\tilde{\cal A}$, by standard methods, it is
straightforward to obtain the open spectra, encoded in the 
direct-channel amplitudes 
\ba 
{\cal A} &=& \frac{1}{4} \ \biggl\{ \ ( \ V_4 O_4 + O_4 V_4 - S_4 S_4 - C_4 C_4 \ ) 
\ ( \ N^2 P  + D^2 W  \ ) \nonumber \\  
&+&  2 N D \ ( \ O_4S_4 + V_4 C_4 - C_4 O_4 - S_4 V_4 \ ) \ 
{\biggl(\frac{\eta}{\theta_4} \ \biggr)}^2 \nonumber \\ 
&+& ( \ R_N^2 + R_D^2 \ ) \  (\ V_4 O_4 - O_4 V_4 + S_4 S_4 - C_4 C_4 \ ) \ {\biggl( \ \frac{2
\eta}{\theta_2} \ \biggr) \ }^2 \nonumber \\
&+& 2 R_N R_D \ ( \ V_4 C_4 - O_4 S_4 + S_4 V_4 - C_4 O_4 \ ) \ 
{\biggl( \ \frac{\eta}{\theta_3} \ \biggr)}^2  \ \biggr\} 
\label{adir}
\ea
and
\ba
{\cal M} = &- & \frac{1}{4} \ \biggl\{ \  N P \ ( \ \hat{O}_4 \hat{V}_4 + \hat{V}_4 \hat{O}_4 - 
\hat{S}_4 \hat{S}_4 - \hat{C}_4 \hat{C}_4 \ ) \nonumber \\
&-&  D W  \ ( \ \hat{O}_4 \hat{V}_4 + \hat{V}_4 
\hat{O}_4 + \hat{S}_4 \hat{S}_4 + \hat{C}_4 \hat{C}_4 \ ) \nonumber \\ 
&-& N \ ( \ \hat{O}_4 \hat{V}_4 \!-\! \hat{V}_4 \hat{O}_4 \!-\! \hat{S}_4 
\hat{S}_4 \!+\! \hat{C}_4 \hat{C}_4 \ ) \ \left(
\ {2{\hat{\eta}}\over{\hat{\theta}}_2} \ \right)^2  \nonumber \\
&+& D \ ( \ \hat{O}_4 
\hat{V}_4 - \hat{V}_4 \hat{O}_4 + \hat{S}_4 \hat{S}_4
 -  \hat{C}_4 \hat{C}_4 \ ) \ \left( \ 
{2{\hat{\eta}}\over{\hat{\theta}}_2} \ \right)^2  \ \biggr\} \ .
\label{mobdir} 
\ea
The R-R tadpole cancellation conditions require
\ba
N \ &=& \ D \ = \ 32 \quad , \nonumber \\
R_N \ &=& \ R_D \ = \ 0 \quad, 
\label{tad}
\ea
and allow a parametrization in terms of real Chan-Paton multiplicities
of the form
$N=n_1+ n_2$, $D=d_1+ d_2$, $R_N=n_1- n_2$ and $R_D=d_1- d_2$,
with $n_1=n_2=d_1=d_2=16$.  
The massless spectrum can be extracted from
\ba  
{\cal A}_0 + {\cal M}_0 &=& \big[ \ \frac{n_1 (n_1-1)}{2} \ + \   
\frac{n_2(n_2-1)}{2} \ + \  \frac{d_1(d_1+1)}{2} \ + \ \frac{d_2(d_2+1)}{2} \  
\big] \ V_4 O_4  \nonumber \\
 &-& \big[ \ \frac{n_1(n_1-1)}{2} \ + \  
\frac{n_2(n_2-1)}{2} \ + \ \frac{d_1(d_1-1)}{2} \ + \ \frac{d_2(d_2-1)}{2} \ 
\big] \ C_4 C_4  \nonumber \\
&+&  ( \ n_1 d_2 + n_2 d_1 \ ) \ O_4 S_4 - ( \ n_1 d_1 + n_2 d_2 \ ) \ C_4 O_4  \nonumber \\
&+& ( \ n_1 n_2 + d_1 d_2 \ ) \ ( \ O_4 V_4 - S_4 S_4 \ ) \quad ,
\label{spect}
\ea 
and the gauge group is thus $[ SO(16) \times SO(16) ]_{9} \times  [ USp(16)
\times USp(16) ]_{\bar{5}}$.  Supersymmetry is realized in the 9-9 sector, 
where the 
vector multiplets of the two $SO(16)$ are accompanied by a hypermultiplet
in the  ${\bf (16,16,1,1)}$, while it is broken on
the ${\bar{D}}5$ branes, where the gauge vectors of the two $USp(16)$ are in the
adjoint representation while the left-handed Weyl fermions are again in
reducible antisymmetric representations, now the ${\bf (1,1,120,1)}$ and
the ${\bf (1,1,1,120)}$. 
In addition, there are
four scalars and two right-handed Weyl fermions in the 
${\bf (1,1,16,16)}$, as well as two scalars in the  ${\bf (16,1,1,16)}$, 
two scalars in the  ${\bf (1,16,16,1)}$ and symplectic Majorana-Weyl 
fermions in the  ${\bf (16,1,16,1)}$ and  ${\bf (1,16,1,16)}$.
This model is free of gauge and gravitational 
anomalies and provides an example of type-I vacuum with a stable non-BPS
configuration of BPS branes.  As in the ten-dimensional $USp(32)$ model,
the breaking of supersymmetry yields a tree-level potential
for the NS-NS moduli related to the uncanceled tadpoles, the dilaton and the
internal volume in this case, that reflects the positive tension
resulting from the $O5_-$ planes and the anti-branes.  

Models with brane supersymmetry breaking exhibit in their spectra
a gauge singlet on the non supersymmetric branes, with the right 
quantum numbers to be a goldstino. As we shall see in the 
last chapter, these goldstinos play 
the role of Volkov-Akulov fields that allow consistent couplings
of the gravitinos to the non supersymmetric matter.  Supersymmetry is thus
linearly realized in the bulk and on some branes, while it is non-linearly 
realized on other (anti)branes. 

\section{Anomaly analysis of six-dimensional models}
\label{anom}
\fancyhead[LO]{{\footnotesize 3.4~~{\it Anomaly analysis of six-dimensional models}}}

In the previous chapter we have analyzed the conditions of anomaly cancellations 
for models of oriented closed strings.
We have seen that in the case of type IIB, both the ten-dimensional model and the chiral 
six-dimensional model, 
obtained by $K3$ compactification, are anomaly-free. Moreover, the consistency of 
the heterotic $SO(32)$ and $E_8 \times E_8$ models is guaranteed, at the level of the low-energy 
effective action, by the presence of the Wess-Zumino term $B \wedge F^4$, whose anomaly exactly 
cancels the one produced by fermion loops. Basically, modular invariance is the guiding
principle for writing consistent models for oriented closed strings.
In this chapter we have seen that for open and unoriented closed strings modular invariance is not a 
property of all amplitudes, and consistency at the one loop level is granted in this case
by R-R tadpole cancellation. As we will see in this section, at the level of the low-energy 
action this corresponds to the cancellation of gauge and gravitational anomalies.

\begin{figure}[!h]
	\centering
    \begin{minipage}[t]{0.4\linewidth}
    \centering
    \includegraphics[width=4cm]{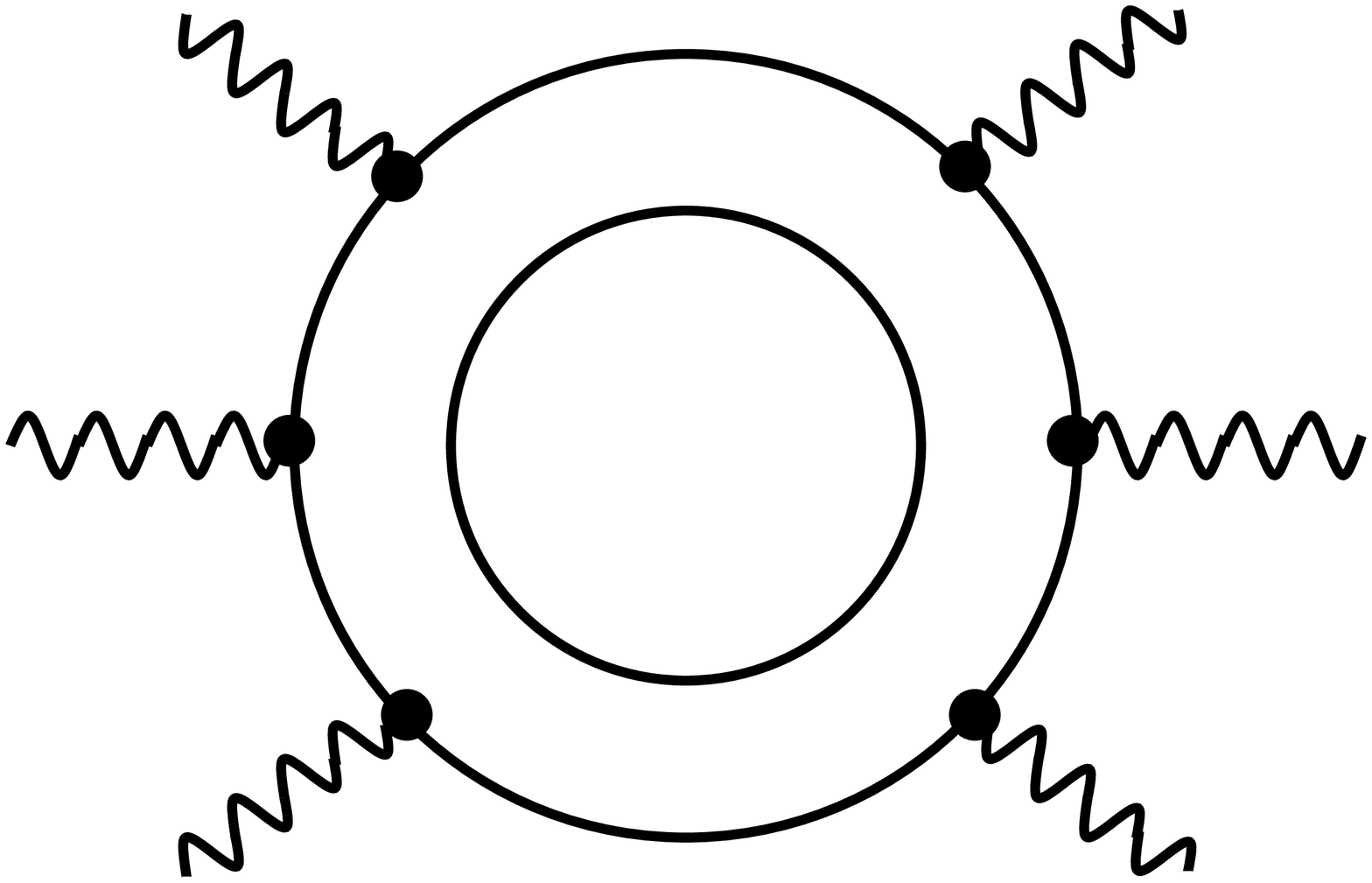}
    \caption{Planar amplitude}
    \label{planar}
    \end{minipage}
	\begin{minipage}[t]{0.1\linewidth} ~
	\end{minipage}
    \begin{minipage}[t]{0.4\linewidth}
	\centering
    \includegraphics[width=4cm]{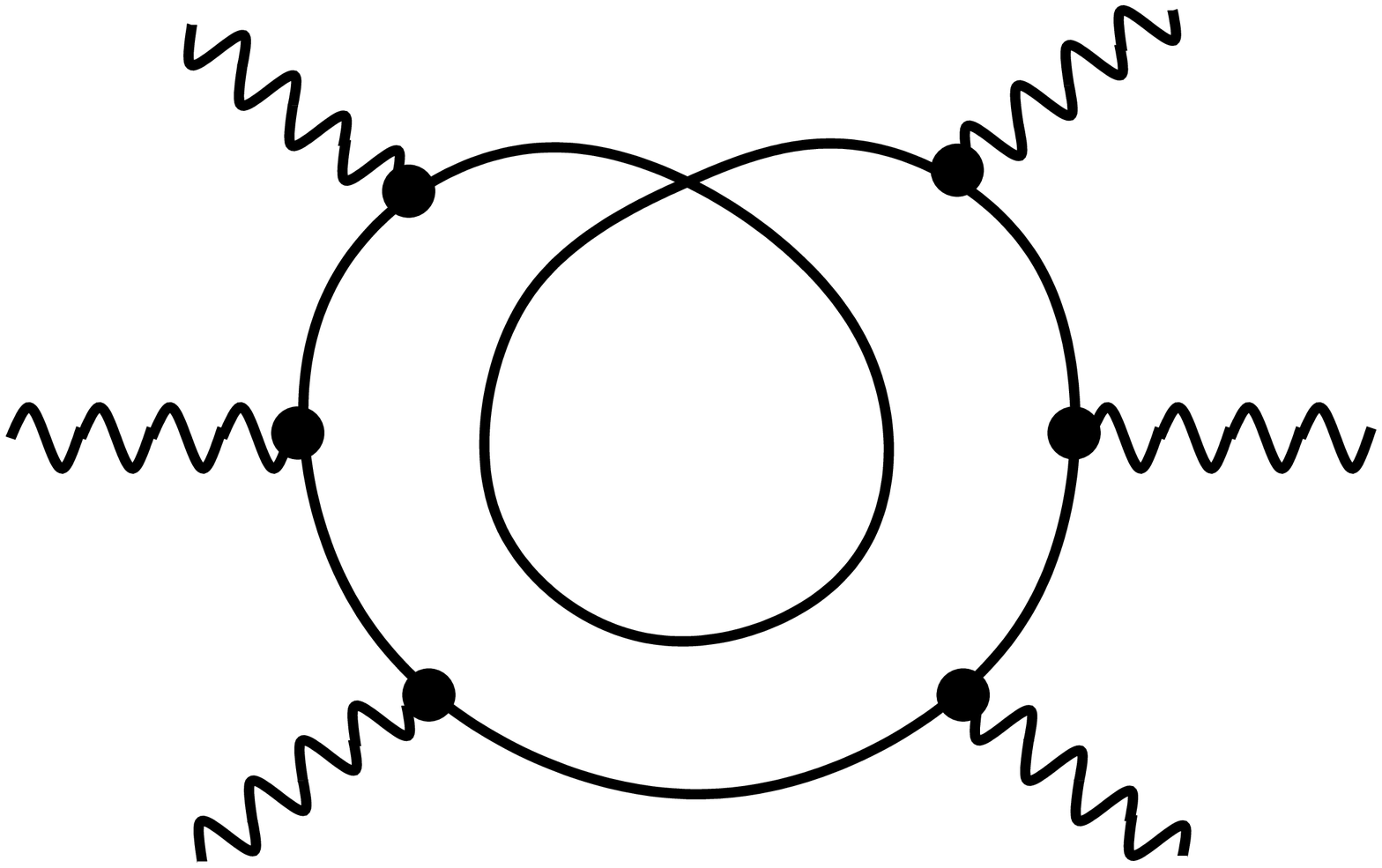}
    \caption{Non-orientable amplitude}
    \label{nonori}
    \end{minipage}
\end{figure}

\begin{figure}[!h]
\centering
\includegraphics[width=4truecm]{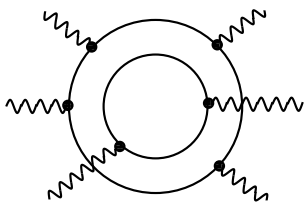}
\caption{Non-planar amplitude}
\label{nonplan}
\end{figure}

If we focus for the moment on gauge anomalies in ten dimensions, the potentially anomalous one-loop string
amplitudes are the planar (fig. (\ref{planar})), non-orientable
(fig. (\ref{nonori})) and non-planar (fig. (\ref{nonplan})) amplitudes, with six vectors 
inserted on the boundaries. 

\noindent
In the first two cases, the vectors are all inserted on one boundary of the annulus or on 
the single boundary of the M\"obius strip,
and thus the Chan-Paton factor is ${\rm tr} F^6$, while in the third case the vectors 
are inserted in both boundaries of the annulus, and the resulting amplitude is proportional
to ${\rm tr} F^2 {\rm tr} F^4$ (all traces here are in the fundamental representation). 
Thus, the first two amplitudes give the irreducible part of the gauge anomaly, that is canceled
if the gauge group is $SO(32)$ in the supersymmetric model \cite{gs} or $USp(32)$ in the Sugimoto model \cite{sugimoto}. 
The third amplitude, that should in principle give the reducible part of the gauge anomaly, 
is in fact non-anomalous, since it can be regulated by the momunetum flowing in the tube \cite{gs}.   
The apparent contraddiction between this result and the field theory analysis, in which the 
reducible part of the hexagon gauge anomaly (fig (\ref{hexagon})) is canceled by the anomalous exchange
of the B field (fig. (\ref{bfield})), is explained because of the open-closed string duality of the 
annulus diagram: the amplitude of fig. \ref{nonplan}, that in the direct channel is a one-loop 
open-string amplitude, in the transverse channel becomes a tree-level closed-string amplitude. 
The two interpretations are valid in two different regions of the integration variable 
$\t$, and the anomalous contributions from these regions exactly cancel.  

Turning to ${\cal N} =(1,0)$ six-dimensional models, we have seen that the condition for the cancellation 
of the irreducible term ${\rm tr} R^4$ in the anomaly polynomial is
\be
n_H - n_V + 29 n_T = 273 \quad ,\label{d6irr}
\ee
where $n_T$, $n_V$ and $n_H$ are the number of tensor, vector and hypermultiplets, respectively.
Perturbative heterotic models, obtained by reduction on $K3$ of the ten-dimensional heterotic
strings, always contain a single tensor multiplet.
The antiself-dual tensor in this tensor multiplet adds to the self-dual tensor of the gravity 
multiplet to form an unconstrained tensor $B$. In these models with $n_T =1$, 
the residual anomaly polynomial 
always factorizes, as in the ten-dimensional case, assuming the form
\be
I_8 = c^z \tilde{c}^{z^\prime} \tr_z F^2 \tr_{z^\prime} F^2  \quad ,\label{polhet}
\ee
where $c^z$'s and $\tilde{c}^z$'s 
are constants and the index $z$ runs over the various factors of the gauge 
group and over the Lorentz group. 
The consistency of the model, that is granted at the string level because of modular invariance, 
from the point of view of the low-energy effective action corresponds to the addition of the
Wess-Zumino term 
\be
B \wedge c^z \tr_z (F \wedge F ) \quad .\label{wzhet}
\ee
As in the ten-dimensional case this term is anomalous, since gauge-invariance of 
the field strength 
\be
H = d B - \tilde{c}^z \w^z
\ee
requires 
\be
\d B = \tilde{c}^z \tr_z (\L d A )\quad .
\ee
Here we denote with $\w$ the Chern-Simons 3-form (for all the notations we remind the reader 
to the next chapter). The anomaly resulting from eq. (\ref{wzhet}) has exactly the form 
of the one originating from the anomaly polynomial (\ref{polhet}), and so can exactly cancel it, 
thus resulting in a Green-Schwarz mechanism exactly as in the ten-dimensional case. The difference, 
here, is that the Wess-Zumino term has two derivatives, and thus is already present in the low-energy
effective action. In the next chapter we shall see a number of peculiar  
consequences of this result.

We now want to perform an anomaly analysis for six-dimensional type-I models. 
We refer in particular to supersymmetric models, even if the same conclusions apply to 
six-dimensional brane supersymmetry breaking vacua, in which supersymmetry is realized 
on the bulk and broken on some branes (we will see in the last chapter that in these models 
supersymmetry is non-linearly realized on these branes). An important feature of these models, 
with respect to the heterotic case, is that one can obtain vacua with several tensor multiplets.
In all cases, the cancellation of R-R tadpoles corresponds to the absence of
irreducible anomalies. 
Analyzing the chiral content of all these six-dimensional vacua, and using the results
of the previous chapter, one can write down the residual anomaly polynomial, 
that in general does not factorize, but can be written in the form \cite{as}
\be
\eta_{rs} c^{r z} c^{s z^\prime} \tr_z F^2 \tr_{z^\prime} F^2 \quad ,\label{anompolsixdim}
\ee
where $\eta$ is the Minkowski metric for $SO(1, n_T )$ $(r,s=0,...,n_T)$, the $c$'s are constant and the index 
$z$ runs over the various simple factors of the gauge group and over the Lorentz group. 
The resulting anomaly is canceled since several tensors take part  in a 
{\it generalized} Green-Schwarz mechanism \cite{as}, corresponding to the inclusion in the low-energy
effective action of the term
\be
c^{rz}  B_r \wedge \tr_z F^2 \quad ,
\ee
that is anomalous since the fields $B^r$ transform as 
\be
\d B^r = c^{rz} \tr_z (\L d A )
\ee
under gauge and Lorentz transformations\footnote{The connection associated to Lorentz transformations is 
of course the spin-connection.}. 

More precisely, only the tensors connected to the characters that are present 
in the transverse annulus amplitude with reflection coefficients not identically vanishing
take part to the anomaly cancellation. Among the various terms in the anomaly polynomial, only one contains
the Riemann curvature, and this factor, always present, is canceled by the antisymmetric tensor 
in the gravity multiplet. For the type-I models analyzed in \cite{as},  
the $c$'s are related to the rows of the matrix $S$.  

\section{Six-dimensional vacua and dualities}
\label{outlook}
\fancyhead[LO]{{\footnotesize 3.5~~{\it Six-dimensional vacua and dualities}}}

At the end of the first chapter we have considered  ten-dimensional ${\cal N}=1$ supergravity 
coupled to vector multiplets, and we have derived the field redefinitions that map the theory in the 
heterotic frame to the same theory in the type-I frame. Since the heterotic $SO(32)$ and the supersymmetric 
type-I ten-dimensional strings have the same massless field content, this field redefinition 
relates their low-energy effective actions. 
This relation is the low-energy realization of the duality between the two complete theories
\cite{pw}, that maps the strong coupling regime of one into the weak coupling regime of the other, 
as can be argued from the relation
\be
\phi_H = -\phi_I \quad ,
\ee
that corresponds to
\be
g_H = g_I^{-1} \quad .
\ee
The fact that this duality is a strong-weak coupling duality 
is in accordance with the fact that, at a perturbative level,
these two string theories have vastly different spectra of massive states. Since the heterotic 2-form 
is mapped to the type-I RR 2-form, the fundamental 
heterotic string, that is charged with respect to this form, is mapped to the D1-string in the type-I model.

Let us now consider these two theories compactified to lower dimensions. 
Denoting with $h_{10-d}$ the metric 
of the internal manifold, the dilaton in $d$ dimensions is
\be
\phi_d = \phi_{10} - \frac{1}{4} {\log}\ {\det} \ h_{10-d} \quad .
\ee
Combining this relation with the duality transformations of the dilaton and the metric in
eq. (\ref{dualhetso32}), one obtains \cite{abpss}
\be
\phi_{I,d} = \frac{6-d}{4} \phi_{H,d} +\frac{2-d}{16} {\log} \ {\det} \ h_{H,10-d} \quad .
\ee
This relation shows that the duality is a strong-weak coupling duality only if
$d > 6$, while it becomes a perturbative duality for $d <6$ 
(a first pair of four-dimensional heterotic and type-I 
vacua that are perturbatively equivalent was also displayed in \cite{abpss}). 
Moreover, in six dimensions the dilaton of one theory is purely geometric if expressed in terms of the 
fields in the dual theory. This is consistent with the fact that in six-dimensional ${\cal N} =(1,0)$ 
vacua the dilaton belongs to a hypermultiplet in the type-I case and to the (unique) tensor multiplet 
in the heterotic case. Perturbative type-I models without tensor multiplets \cite{abpss2} 
correspond, according to this relation, to non-perturbative heterotic vacua without dilaton fluctuations.   

It is then of interest to understand what 
kind of non-perturbative picture corresponds on the heterotic side to 
perturbative type-I models with different numbers of tensor multiplets. Starting form 
the perturbative heterotic $SO(32)$ model compactified on $K3$, with total instanton number $k=24$, one 
develops an extra $SU(2)$ gauge symmetry when an instanton shrinks to zero size \cite{witt2}. In the type-I 
picture, this corresponds to the appearence of a pair of 
D5-branes, with an $USp(2)$ Chan-Paton group, whose world-volume
fills the six uncompactified dimensions. The condition $k=24$ is replaced by $k=24-n$, 
where $n$ is the number of 
instantons shrinking to zero size. When $r$ of them coincide at a given point of $K3$, 
the non-perturbative gauge 
symmetry is enhanced to $USp(2r)$, and this naturally corresponds 
to the appearence of $2r$ coinciding D5-branes, while
the perturbative gauge group $SO(32)$ is broken by a background with a lower instanton number, $k=24-n$.
Vacua with several tensor multiplets can be obtained compactifying the 
ten-dimensional theory on $K3$ manifolds
with singularities: small instantons on these singularities give rise to Coulomb phases parametrized by 
the real scalars in the tensor multiplets \cite{9603167,9702038,9705044}.  

Now we consider six-dimensional vacua of the $E_8 \times E_8$  
heterotic theory on $K3$. In this case, perturbative 
vacua correspond to the condition $k_1 + k_2 = 24$, where $k_1$ and $k_2$ 
are the instanton numbers corresponding
to the two $E_8$ factors. As already anticipated in Section (2.5), the non-perturbative 
ten-dimensional $E_8 \times E_8$ theory corresponds to M-theory compactified on the interval
${S}^1 / {Z}_2$, 
with a length  proportional to the heterotic string coupling \cite{hw}, 
while the gauge fields of the 
two $E_8$ factors are on 
the two different ``end of the world'' 9-branes. Further compatifications 
on $K3$ are obtained embedding $k_1$ 
instantons in one $E_8$ factor and $k_2$ in the other, but another 
possibility is left, the addition of
$n$ M5-branes, located at points in $K3 \times {S}^1 / {Z}_2$, and filling the six non-compact
dimensions. The condition $k_1 + k_2 = 24$ is then replaced by $k_1 + k_2 + n = 24$, and the corresponding 
six-dimensional vacua have $n+1$ tensor multiplets. An instanton 
shrinking to zero size corresponds to the appearence 
of an M5-brane stuck to the end of the world 9-brane. 
When the 5-brane leaves the boundary, a tensor multiplet appears
in the low-energy action. The M5-brane can travel 
to the other boundary and be reabsorbed, so that all possible values of 
$(k_1 , k_2 )$ are connected \cite{sw}. 
The presence of an antiself-dual tensor in the effective field theory of the M5-brane 
means that M2-branes 
can end on the M5-brane, with one direction tangent to it \cite{openpbranes}. 
The corresponding string on the world-volume of the M5-brane is antiself-dual. Since the tension of the 
string is proportional to the distance between the M5-brane and the boundary,  
the singularity associated to a shrinking instanton signals the appearence of tensionless strings. 
This phenomenon actually manifests itself in the low-energy effective action because of the appearence of 
a singularity in the gauge kinetic term, corresponding to an infinite gauge coupling 
constant \cite{as}. As an example, consider the case in which a single tensor 
multiplet is present. 
In this case, the kinetic term in the Einstein frame is proportional to $a 
e^{\phi}+b e^{-\phi}$, where $a$ and $b$ are constants and $\phi$ is 
the scalar in the tensor multiplet. If $a$ and $b$ have opposite sign, the kinetic
term diverges for a particular 
value of the scalar, and singularities of the same type appear if several tensor multiplets are 
present. 
We have now seen in the $E_8 \times E_8$ picture that these singularities 
signal a new kind of phase transition \cite{dmw}, reflecting the  presence 
in the vacuum of string excitations with vanishing tension
\cite{sw,dlp}. The singularity appears for particular values of the scalars in the 
tensor multiplets, since these scalars parametrize the distance between the M5-brane and the 9-brane.

An analogous geometrical picture of this low-energy phenomenon can be given in the type-I 
theory. In this case the starting points are the duality between heterotic theory on ${T}^4$
and type IIA on $K3$, and the T-duality between type IIA and type IIB. Since the former implies 
that type IIA have enhanced gauge symmetry at certain points in $K3$ moduli space, 
one can ask how type IIB behaves at the corresponding 
points. If one further compactifies IIA and IIB theories 
on T-dual circles to five dimensions, T-duality implies
\be
g_{A,6} = \frac{g_{B,6}}{R_B} \quad ,
\ee
and at a distance $\e$ from a point at which IIA gets enhanced gauge symmetry, one has W-bosons of mass
$\e / g_{A,6}$. In five dimensions, in IIB units this is then a W-boson of mass 
\be
M_W = \frac{\e R_B}{g_{B,6}} \quad ,
\ee
that corresponds to a selfdual string in six-dimensions wrapped on a circle of radius $R_B$.
This selfdual string is in fact a selfdual D3-brane of type IIB wrapped on a collapsing 2-sphere in $K3$ \cite{wit}.
After the orientifold projection, these D-strings manifest themselves as singularities 
in the low-energy effective action (see \cite{iu} for a review).

Following \cite{sw}, we want to make a final comment about what kind of physics one should expect
at the singularity. In six dimensions, every conventional gauge theory is free in the infrared.
This is because the gauge coupling has the dimension of a length, and 
consequently for scales larger than this length the theory is usually expected to be 
free. When one approaches the singularity, however, 
this length scale becomes larger and larger, and at the singularity it diverges, 
so that the theory can indeed be non-free in the infrared.

% \newpage ~                  %  per lasciare una pagina bianca
% \thispagestyle{empty}       %  

\chapter{Minimal six-dimensional supergravity}
\label{cap4}
% \newpage
\vspace*{2cm}
\fancyhead[RO,LE]{\thepage}
\fancyhead[RE]{{\footnotesize {\rm Chapter 4.}~~{\it Minimal six-dimensional supergravity}}} 
\fancyhead[LO]{}

\noindent
As we have seen in the previous chapters, one of the most striking 
features of perturbative superstring theory in ten  dimensions
is the absence of anomalies. In type-IIB theory this is  realized by miraculous
cancellations between various contributions \cite{agw}, while in the type-I and
heterotic theories the  Green-Schwarz mechanism \cite{gs} generates anomalous couplings
that exactly cancel the contributions of fermion loops, once one restricts the gauge
group to be $SO(32)$ for the type I theory  and $SO(32)$ or
$E_8 \times E_8$ for the heterotic theory. All these ${\cal N}=1$ theories  are very
interesting, since they can be naturally compactified to $N=1$ theories
in four dimensions. In this context, an interesting intermediate step is the study of
$(1,0)$  vacua in six dimensions, since in these compactifications the absence of 
anomalies is a very strong restriction on the low-energy physics.

The massless representations of $(1,0)$ supersymmetry in six dimensions
are the gravity multiplet $(e_\mu^a,
\psi_{L \mu}, B^+_{\mu\nu})$, the
tensor multiplet $(B^-_{\mu\nu},
\chi_R,\varphi)$, the vector multiplet $(A_\mu,\lambda_L)$ and the hypermultiplet $(4 \phi, \Psi_R )$. 
We have seen in the last chapter that, letting $n_T$, $n_V$ and $n_H$ denote 
the numbers of tensor, vector and hypermultiplets,
the condition that the term $\tr R^4$ be absent  in the anomaly polynomial is \cite{rdsss} 
$$ 
n_H - n_V +29 n_T = 273 \quad ,
$$ 
and compared to the ten dimensional case this 
allows a large number of possible vacua. Perturbative heterotic vacua in six
dimensions can be obtained by orbifold compactifications or by compactifications on
smooth $K3$ manifolds with instanton backgrounds. Anomaly cancellation requires that the
total  instanton number be 24, and these vacua include a single tensor multiplet, as one
can easily see reducing the ten-dimensional  low-energy theory. Repeating what we 
said in the last chapter, we would like to stress that the situation is quite
different in perturbative six-dimensional  type I vacua since, as suggested in
\cite{cargese}, these models are  determined  by a parameter space orbifold
(orientifold) construction, that naturally allows several tensor multiplets \cite{bs,severaltensors}.
In general the residual anomaly  polynomial does not factorize, and several
antisymmetric tensors contribute  to the cancellation via a generalized Green-Schwarz 
mechanism \cite{as}. 

The anomaly polynomial of eq. (\ref{anompolsixdim}) reveals an important difference with 
respect to ten dimensions:
while the ten-dimensional Green-Schwarz coupling, 
$B \wedge (F^4- R^4)$, is a higher derivative term, the gauge portion of the
corresponding six-dimensional coupling, $B \wedge (F^2 - R^2 )$, 
belongs to the low-energy effective action. 
One is thus facing a case of unprecedented complexity in supergravity
constructions, whereby the model is determined by Wess-Zumino  conditions
\cite{wz},  rather than by the usual requirement of local supersymmetry. 
In this chapter, we shall see that one important consequence of this 
is that 
the resulting equations are not unique, since a quartic coupling 
for the gauginos is undetermined, and the construction is consistent for any 
choice of this coupling. Correspondingly, the commutator of two supersymmetry
transformations on the gauginos contains an extension, that plays a 
crucial role in ensuring that the Wess-Zumino consistency conditions close 
on-shell.   
Moreover, in this model 
the divergence of the energy-momentum 
tensor is non-vanishing, as is properly the case for a theory that 
has  gauge anomalies but no gravitational anomalies (gravitational anomalies 
could be accounted for introducing  higher-derivative couplings) \cite{fms,frs,rs1}.
Once more, the low-energy couplings are obtained by consistency once one 
includes the Green-Schwarz term in the low-energy theory, but 
the complete theory, supersymmetric and gauge-invariant,
would also include additional non-local couplings arising from fermion loops.   
Let us stress that this is exactly as in the ten-dimensional case, but in 
these six-dimensional models the anomalous terms belong 
to the low-energy effective action.

One can even consider a slight modification of these couplings, 
resulting from the inclusion in the low-energy Lagrangian of more
general Green-Schwarz couplings to {\it abelian} 
vectors of the form \cite{fr2}
$$
B^r \ c_r^{ab} \ F^a \ F^b \quad , 
$$
where the indices $a,b$ run over the different $U(1)$ gauge groups, while the 
symmetric matrices
$c^{r}$ may not be 
simultaneously diagonalized.
This naturally corresponds to the inclusion of 
non-diagonal Chern-Simons couplings \cite{cjlp},
and in this case the residual 
anomaly polynomial has the more general form 
$$
\ c^r_{ab} \ c^s_{cd} \ \eta_{rs} F^a \wedge F^b \wedge F^c \wedge F^d
\quad .
$$ 
The corresponding kinetic terms of vectors and gauginos are also 
non-diagonal, compatibly with the abelian gauge invariances.

At the end of the last chapter we remarked that 
another interesting feature of these models is the fact that they
exhibit singularities in
the moduli space of tensor multiplets, corresponding to infinite  gauge coupling
constants \cite{as}. As an example, consider the case in which a single tensor 
multiplet is present. 
In this case the kinetic term is proportional to $a 
e^{\phi}+b e^{-\phi}$, where $a$ and $b$ are constants and $\phi$ is 
the scalar in the tensor multiplet, and if $a$ and $b$ have opposite sign, the kinetic
term diverges for a particular 
value of the scalar. The same singularities appear if several tensor multiplets are 
present. 
These singularities 
signal a new kind of phase transition \cite{dmw}, reflecting the  presence 
in the vacuum of string excitations with vanishing tension
\cite{sw}. 

The coupling of $(1,0)$ supergravity
to $n$ tensor multiplets was originally studied
in \cite{romans} to lowest order in the Fermi 
fields, while \cite{ns1} considered the coupling 
to a single tensor multiplet and to vector and hypermultiplets 
to all orders in the Fermi fields.  
As we anticipated, in this case the kinetic term is generally proportional to $a 
e^{\phi}+b e^{-\phi}$ \cite{as}, while \cite{ns1} actually deals with 
the particular case $a=0$, in which the anomaly 
polynomial vanishes and no tensionless string transition occurs. 
The general coupling to non-abelian vectors and self-dual tensors
was worked out to lowest order in the 
Fermi fields in \cite{as}. In this covariant formulation the 
requirement of supersymmetry gives non-integrable equations, while the 
divergence of the vector equation gives the covariant anomaly.
The same model was then reconsidered, again to lowest order in the Fermi fields,
in \cite{fms} in the consistent formulation, requiring the closure of the
Wess-Zumino conditions, that relate the consistent gauge anomaly to the
supersymmetry anomaly.
Additional couplings, as well as the 
inclusion of hyper-multiplets, were then considered in \cite{ns2}.
The complete coupling to non-abelian vector and tensor multiplets 
was then obtained in \cite{frs} in the 
consistent formulation and in \cite{rs2} in the covariant formulation. 
\cite{fr2} contains the most general couplings for the case in which also abelian vectors 
are present, and finally in \cite{fr3} supergravity coupled to vector, tensor and hypermultiplets
was constructed to all orders in the Fermi fields.

In this chapter we construct the complete $(1,0)$ supergravity coupled to tensor, vector
and hypermultiplets. 
Since all the subtleties of the construcion are already present in the absence of 
hypermultiplets, in Section 2 
we construct minimal supergravity coupled to non-abelian vector and $n_T$
tensor multiplets. In Section 3 we will apply to this model the PST construction \cite{rs3}.
Section 4 considers the case in which abelian vectors are present, and finally Section 5
contains the complete coupling of minimal six-dimensional supergravity to tensor, vector 
and hypermultiplets.

\section{Supergravity coupled to tensor and vector multiplets}
\label{yminstantons}
\fancyhead[LO]{{\footnotesize 4.1~~{\it Supergravity coupled to tensor and vector multiplets}}}

In this section we describe minimal $(1,0)$  six-dimensional supergravity coupled to $n$
tensor multiplets and non-abelian vector multiplets \cite{frs}. 
Simple supersymmetry in six dimensions is generated by a $USp(2)$
doublet of chiral spinorial charges $Q^A$ $(A=1,2)$ obeying the symplectic Majorana
condition
\be Q^A = \e^{AB} C \bar{Q}^T_B \quad ,
\ee where $\e^{AB}$ is the $USp(2)$  antisymmetric invariant tensor. Since all  Fermi
fields are $USp(2)$ doublets, in this section we will mostly write $\psi$ to denote a
doublet 
$\psi^A$.

\subsection{Supergravity coupled to tensor multiplets}

We start our analysis from the simpler case in which only tensor multiplets are present. 
Let us begin by reviewing the work of Romans \cite{romans}. The  theory includes the
vielbein
$e_\m{}^m$, a left-handed gravitino $\psi_\m$, $(n+1)$  antisymmetric tensors
$B^r_{\m\n}$ $(r=0,...,n)$  obeying (anti)self-duality conditions, $n$ right-handed ``tensorinos''
$\chi^M$  $(M=1,...,n)$, and $n$ scalars. The scalars parameterize the coset space
$SO(1,n)/SO(n)$, and are  thus associated to the $SO(1,n)$ matrix $(r=0,...n)$
\be 
\left( \begin{array}{c}  v_r \\  x^M_r  \end{array} \right)
\quad ,
\ee 
whose elements satisfy the constraints
\bea 
& & v^r v_r =1 \quad , \nonumber\\ 
& & v_r v_s - x^M_r x^M_s = \eta_{rs}\quad ,
\nonumber \\ 
& & v^r x^M_r =0 \quad . \label{scalars}
\eea 
Defining
\be 
G_{rs} = v_r v_s + x^M_r x^M_s \quad ,
\ee 
the tensor (anti)self-duality conditions can be succinctly written
\be 
G_{rs} H^{s \m\n\r} =\frac{1}{6e} \e^{\m\n\r\a\b\g} H_{r \a\b\g}\quad ,
\label{selfdual}
\ee 
where $H^r_{\m\n\r}= 3 \de_{[\m} B^r_{\n\r ]}$. These relations only hold to lowest
order in the Fermi fields, and imply that  $v_r H^r_{\m\n\r}$ is self dual, while the
$n$ tensors $x^M_r H^r_{\m\n\r}$ are antiself dual, as one can see using eqs.
(\ref{scalars}). The divergence of eq. (\ref{selfdual})  yields the second-order tensor
equation
\be 
\nabla_\m (G_{rs} H^{s\m\n\r} )=0  
\label{tensoreq}
\ee 
while, to lowest order, the fermionic equations are
\be
-i \g^{\m\n\r} D_\n \psi_\r -i v_r H^{r \m\n\r} \g_\n \psi_\r -
\frac{1}{2} x^M_r H^{ r \m\n\r} \g_{\n\r} \chi^M + 
\frac{1}{2} x^M_r \de_\n v^r \g^\n \g^\m \chi^M =0 
\ee 
and
\be
i \g^\m D_\m \chi^M -\frac{i}{12} v_r H^{r \m\n\r} \g_{\m\n\r} \chi^M +\frac{1}{2} x^M_r
H^{r
\m\n\r} \g_{\m\n} \psi_\r+
\frac{1}{2} x^M_r \de_\n v^r \g^\m \g^\n \psi_\m =0\quad. 
\ee 
Varying the Fermi fields in them with the supersymmetry transformations 
\bea 
& & \delta e_\m{}^m = -i ( \bar{\e} \g^m \psi_\m ) \quad , \nonumber\\ 
& & \delta B^r_{\m\n} =i v^r ( \bar{\psi}_{[\m} \g_{\n]} \e )
+\frac{1}{2} x^{Mr} ( \bar{\chi}^M \g_{\m\n} \e ) \quad , \nonumber\\ 
& & \delta v_r = x^M_r ( \bar{\e} \chi^M )\nonumber\\
& & \delta \psi_\m = D_\m \e +\frac{1}{4} v_r H^r_{\m\n\r} \g^{\n\r} \e \quad , 
\nonumber\\ & & \delta \chi^M =\frac{i}{2} x^M_r
\de_\m v^r \g^\m \e +\frac{i}{12} x^M_r H^r_{\m\n\r} \g^{\m\n\r} \e \quad ,  \label{susy}
\eea 
generates the bosonic equations, using also eqs. (\ref{selfdual}) and
(\ref{tensoreq}). Thus, the scalar field equation is
\be 
x^M_r D_\m (\de^\m v^r ) +\frac{2}{3} x^M_r v_s H^r_{\a\b\g}  H^{s \a\b\g} =0 \quad ,
\ee 
while the Einstein equation is
\be 
R_{\m\n} -\frac{1}{2} g_{\m\n} R + \de_\m v^r \de_\n v_r -
\frac{1}{2} g_{\m\n} \de_\a v^r \de^\a v_r - G_{rs} H^r_{\m\a\b} H^s{}_\n{}^{\a\b}
=0\quad.
\ee 
To this order, this amounts to a proof of  supersymmetry,  and it is also possible
to show that the  commutator of two supersymmetry transformations on the bosonic fields
closes on the local symmetries:
\bea 
& & [ \delta_1 , \delta_2 ] = {\delta}_{gct}( \xi^\m = -i ({\bar{\e}}_1 
\g^\m \e_2 )) +
\delta_{tens} (\L^r_\m = -\frac{1}{2} v^r \xi_\m -\xi^\n B^r_{\m\n}) 
\nonumber \\ 
& & +\delta_{SO(n)}(A^{MN} = \xi^\m x^{Mr}(\de_\m x^N_r ) ) \nonumber \\
& & +\delta_{Lorentz} (\W^{mn} =-\xi_\m (\w^{\m mn} - v_r H^{r \m mn})) \quad
.\label{susyalg}
\eea 
To this order, one can not  see the local supersymmetry transformation in the gauge
algebra, since the expected parameter,  $\xi^\m \psi_\m$, is generated  by bosonic
variations. As usual, the spin connection satisfies its equation of motion, that to
lowest order in the Fermi  fields is 
\be 
D_\m e_\n{}^m - D_\n e_\m{}^m =0\quad ,
\ee 
and implies the absence of torsion.

Completing these equations will require terms cubic in the Fermi fields in the fermionic
equations, and  terms quadratic in the Fermi fields in their supersymmetry
transformations.  Supersymmetry will then determine corresponding modifications of the 
bosonic equations, and the (anti)self-duality conditions 
(\ref{selfdual}) will also be
modified by terms quadratic in the Fermi fields. Supercovariance actually   fixes  all
terms containing the gravitino in the first-order equations and in the supersymmetry
variations of Fermi fields.

The supercovariant forms
\be
\hat{\w}_{\m\n\r} = \w^0_{\m\n\r}  -\frac{i}{2} (\bar{\psi}_\m \g_\n \psi_\r 
+\bar{\psi}_\n \g_\r
\psi_\m +\bar{\psi}_\n \g_\m \psi_\r )\quad ,
\ee
\bea 
& & \hat{H}^r_{\m\n\r} = H^r_{\m\n\r}  -\frac{1}{2} x^{Mr} ( \bar{\chi}^m \g_{\m\n}
\psi_\r +
\bar{\chi}^M 
\g_{\n\r}\psi_\m + \bar{\chi}^M \g_{\r\m} \psi_\n )  \nonumber\\ 
& & - \frac{i}{2} v^r
(\bar{\psi}_\m \g_\n \psi_\r +\bar{\psi}_\n \g_\r
\psi_\m +\bar{\psi}_\r \g_\m \psi_\n ) \quad ,
\eea
\be
\hat{\de_\m v^r} = \de_\m v^r -x^{Mr} (\bar{\chi}^M \psi_\m )\quad ,
\ee 
where
\be
\w^0_{\m\n\r} =\frac{1}{2} e_{\r m}(\de_\m e_\n{}^m -\de_\n e_\m{}^m ) -\frac{1}{2} e_{\m
m}(\de_\n e_\r{}^m -\de_\r e_\n{}^m ) +\frac{1}{2} e_{\n m} (\de_\r e_\m{}^m - \de_\m
e_\r{}^m )
\ee 
is the standard spin connection in the absence of  torsion, do not generate
derivatives of the  parameter under supersymmetry. In the same spirit, one can consider
the supercovariant transformations 
\bea 
& & 
\delta \psi_\m = \hat{D}_\m \e +\frac{1}{4} v_r \hat{H}^r_{\m\n\r} 
\g^{\n\r} \e \quad , \nonumber \\ 
& & \delta \chi^M = \frac{i}{2} x^M_r (\hat{\de_\m
v^r} ) \g^\m
\e +\frac{i}{12} x^M_r \hat{H}^r_{\m\n\r} \g^{\m\n\r} \e \quad . \label{newsusy}
\eea 
The tensorino transformation is complete, while the gravitino  transformation could
include additional terms quadratic in the tensorinos. On the other hand, one does not
expect modifications of  the bosonic transformations in the complete theory.

The algebra (\ref{susyalg}) has been obtained varying only the Fermi fields in the
bosonic supersymmetry transformations. The next step is to compute  the commutator
varying the bosonic fields as well.  There is no important novelty in the complete
commutator on $v^r$ and on the vielbein $e_\m{}^m$. However, the local Lorentz parameter
is modified and takes the form
\be
\W^{mn}= -\xi^\m (\hat{\w}_\m{}^{mn} -v_r \hat{H}^r_\m{}^{mn})\quad 
\ee 
while, as anticipated, the supersymmetry parameter is
\be
\zeta = \xi^\m \psi_\m \quad .
\ee 
These results are obtained using the torsion equation for $\hat{\w}$,
\be
\hat{D}_\m e_\n{}^m - \hat{D}_\n e_\m{}^m = 2S^m{}_{\m\n}= -i(\bar{\psi}_\m
\g^m \psi_\n ) \quad .
\ee

One can also compute the commutator on $x^M_r$. Eqs. (\ref{scalars}) determine its
supersymmetry variation 
\be
\delta x^M_r = v_r (\bar{\e} \chi^M ) \quad ,
\ee 
and the resulting commutator includes a  local $SO(n)$ transformation of parameter
\be A^{MN} =\xi^\m x^{Mr} (\de_\m x^N_r ) +(\bar{\chi}^M \e_2 )(\bar{\chi}^N
\e_1 ) -(\bar{\chi}^M \e_1 )(\bar{\chi}^N \e_2 ) \quad .
\ee

New results come from the complete commutator on $B^r_{\m\n}$, where one needs to use the
(anti)self-duality conditions. Supercovariantization is at work here, since these
conditions are first-order equations, that become
\be  
G_{rs} \hat{H}^s_{\m\n\r} =\frac{1}{6e} \e_{\m\n\r\a\b\g} \hat{H}_r^{\a\b\g}
\quad .\label{selfdual2}
\ee 
It is actually possible to alter these conditions demanding that the modified tensor
\be
\hat{\cal H}^{r}_{\m\n\r}=
\hat{H}^r_{\m\n\r} +i\a v^r (\bar{\chi}^M \g_{\m\n\r}\chi^M )\label{newtensor}
\ee 
satisfy (anti)self-duality conditions as in eq. (\ref{selfdual2}).  Using eqs.
(\ref{scalars}), one can see that the new $\chi^2$ terms contribute only to the
self-duality condition, while the tensors
$x^M_r \hat{H}^r_{\m\n\r}$ remain antiself dual without extra $\chi^2$  terms.
Consequently,  since the commutator on $B^r_{\m\n}$  uses only the antiself-duality
conditions, the result does not contain  terms proportional to $\a$. The commutator on
the tensor fields  generates all local symmetries in the proper form, aside from the 
extra terms
\be 
[\delta_1 , \delta_2 ]_{extra} B^r_{\m\n} =\frac{1}{2}v^r (\bar{\e}_1 \chi^M
)(\bar{\chi}^M
\g_{\m\n} \e_2 ) -\frac{1}{2} v^r (\bar{\e}_2 \chi^M )(\bar{\chi}^M \g_{\m\n} \e_1
)\quad ,
\ee 
that may be canceled adding $\chi^2$ terms to the transformation
of the gravitino. The
most general expression one can add is
\be
\delta^\prime \psi_\m = ia \ \g_\m \chi^M (\bar{\e} \chi^M )+  i b \ \g_\n \chi^M (\bar{\e}
\g_\m{}^\n \chi^M )  + i c \ \g_{\m\n\r} \chi^M (\bar{\e} \g^{\n\r} \chi^M ) \quad ,
\ee 
with $a$, $b$ and $c$ real coefficients, and the total commutator on $B^r_{\m\n}$
then leads to the relations
\be 
a + b = -\frac{1}{2} \quad ,\qquad b + 2c = 0 \quad .
\ee 
The commutator on $e_\m{}^m$ now closes with a local Lorentz parameter modified by
the addition of
\be
\Delta{\W}^{mn}  =-\frac{1}{2}[(\bar{\chi}^M \e_1 )(\bar{\e}_2 \g^{mn} \chi^M)
-(\bar{\chi}^M \e_2 )(\bar{\e}_1 \g^{mn} \chi^M )] \quad , 
\ee 
while the commutators on the scalar fields are not modified.

One can now start to compute the commutators on Fermi fields, that as usual close only
on shell. Following \cite{schwarz}, we will actually use this result to  derive the
complete fermionic equations. Let us begin with the  commutator on the tensorinos, using
eq. (\ref{newsusy}). This fixes the free parameter in the gravitino variation and the
parameter $\a$ in eq. (\ref{newtensor}), so that
\be 
a = -\frac{3}{8} \quad , \qquad  b = -\frac{1}{8} \quad , 
\qquad c = \frac{1}{16} \quad , \qquad \a = -\frac{1}{8} \quad .
\ee 
Supercovariance determines the field equation of the tensorinos up to a term
proportional to
$\chi^3$. Closure of  the algebra fixes this additional term, and the end result is
\bea 
& & i \g^\m \hat{D}_\m \chi^M -\frac{i}{12}v_r \hat{H}^r_{\m\n\r} \g^{\m\n\r}
\chi^M + \frac{1}{2} x^M_r \hat{H}^{r \m\n\r} \g_{\m\n} \psi_\r \nonumber \\ 
& &+ \frac{1}{2} x^M_r ( \hat{\de_\n v^r} ) \g^\m \g^\n \psi_\m  + \frac{1}{2} \g^\m \chi^N
(\bar{\chi}^N \g_\m \chi^M ) = 0
\quad .\label{chieq}
\eea 
The complete commutator of two supersymmetry transformations on the tensorinos is
then
\be 
[\delta_1 ,\delta_2 ] \chi^M = \delta_{gct} \ \chi^M +\delta_{Lorentz} \ \chi^M
+\delta_{SO(n)} \ \chi^M +\delta_{susy} \ \chi^M
- \frac{i}{4} \g^\m \xi_\m \ [{\rm eq.} \ \chi^M ] \quad .
\ee 
A similar result can be obtained for the gravitino. In this case the complete
equation,
\bea 
& & -i \g^{\m\n\r} \hat{D}_\n \psi_\r - \frac{i}{4} v_r \hat{H}^r_{\n\s\tau}
\g^{\m\n\r} \g^{\s\tau} \psi_\r -\frac{1}{2}x^M_r \hat{H}^{r \m\n\r} \g_{\n\r}
\chi^M  +\frac{1}{2} x^M_r (\hat{\de_\n v^r}) \g^\n \g^\m \chi^M \nonumber \\ 
& & +\frac{3}{2}
 \g^{\m\n} \chi^M  (\bar{\chi}^M \psi_\n ) - \frac{1}{4} \g^{\m\n} \chi^M (\bar{\chi}^M
\g_{\n\r}
\psi^\r ) + \frac{1}{4} \g_{\n\r} \chi^M (\bar{\chi}^M \g^{\m\n} \psi^\r ) \nonumber\\ 
& & - \frac{1}{2} \chi^M (\bar{\chi}^M \g^{\m\n} \psi_\n ) =0 \quad ,
\label{gravitinoeq}
\eea 
is fixed by  supercovariance, and the commutator closes up to  terms proportional
to a particular combination of eq. (\ref{gravitinoeq}) and its $\g$-trace. Moreover,  a
non-trivial symplectic structure makes its first appearance in  a commutator, so that
the final result is (see the Appendix for the notations)
\bea 
& & [\delta_1 ,\delta_2 ] \psi_\m^A = \delta_{gct} \psi_\m^A
+\delta_{Lorentz}\psi_\m^A +\delta_{susy} \psi_\m^A \nonumber\\ & & 
+\frac{3i}{8} \xi^\m
\g_\m ([eq.\  \psi_\m ] -\frac{1}{4}
\g_\m [\g -{\rm trace} ])^A \nonumber \\ 
& & +\frac{i}{96} \sigma^i_B{}^A \g^{\m\n\r}
\xi^i_{\m\n\r} ([eq.\  \psi_\m ] -\frac{1}{4} \g_\m [\g -{\rm trace} ])^B \quad ,
\label{gravitinoalg}
\eea 
where
\be
\xi^i_{\m\n\r} = -i [\bar{\e}_1 \g_{\m\n\r} \e_2 ]^i \quad . \label{anomalousxi}
\ee

Summarizing, from the algebra we have obtained the complete fermionic equations of
$(1,0)$ six-dimensional supergravity coupled to $n$ tensor multiplets. In addition, the
modified 3-form
\be
\hat{\cal H}^{r}_{\m\n\r}=
\hat{H}^r_{\m\n\r} -\frac{i}{8}v^r (\bar{\chi}^M \g_{\m\n\r}\chi^M )
\ee 
satisfies the (anti)self-duality conditions
\be 
G_{rs} \hat{\cal H}^{s}_{\m\n\r}=\frac{1}{6e}\e_{\m\n\r\s\d\t}
\hat{\cal H}^{\s\d\t}_r \quad. \label{finalselfdual}
\ee
We have also identified the complete supersymmetry transformations, 
that we collect here for convenience:
\bea 
& & \delta e_\m{}^m =-i(\bar{\e} \g^m \psi_\m ) \quad,\nonumber\\ 
& & \delta B^r_{\m\n} =i v^r (\bar{\psi}_{[\m} \g_{\n]} \e )+ \frac{1}{2} x^{Mr} (\bar{\chi}^M
\g_{\m\n} \e ) \quad,
\nonumber\\ 
& & \delta v_r = x^M_r (\bar{\chi}^M \e )\quad,\nonumber\\ 
& & \delta \psi_\m
=\hat{D}_\m \e +\frac{1}{4} v_r \hat{H}^r_{\m\n\r}
\g^{\n\r}\e -\frac{3i}{8} \g_\m \chi^M (\bar{\e} \chi^M ) \nonumber \\
& & \quad \quad -\frac{i}{8} \g^\n \chi^M
(\bar{\e}
\g_{\m\n} \chi^M )+\frac{i}{16}
\g_{\m\n\r} \chi^M (\bar{\e} \g^{\n\r} \chi^M ) \quad ,\nonumber\\ 
& & \delta \chi^M =
\frac{i}{2} x^M_r (\hat{\de_\m v^r} ) \g^\m \e +
\frac{i}{12} x^M_r \hat{H}^r_{\m\n\r} \g^{\m\n\r}\e \quad .
\eea

In order to  obtain the bosonic equations, it is convenient to associate the  fermionic
equations to the Lagrangian 
\bea
 e^{-1} {\cal{L}}_{fer} &=& -\frac{i}{2}\bar{\psi}_\m \g^{\m\n\r} D_\n [\frac{1}{2}(\w
+\hat{\w} )]
\psi_\r -\frac{i}{8}v_r [H+\hat{H}]^{r \m\n\r}(\bar{\psi}_\m \g_\n \psi_\r)
\nonumber \\ 
&+& \frac{i}{48} v_r [H+\hat{H} ]^r_{\r\s\t} (\bar{\psi}_\m
\g^{\m\n\r\s\t} \psi_\n )+\frac{i}{2} \bar{\chi}^M \g^\m D_\m (\hat{\w})
\chi^M \nonumber \\ 
&-& \frac{i}{24}v_r \hat{H}^r_{\m\n\r} (\bar{\chi}^M \g^{\m\n\r}
\chi^M ) +\frac{1}{4} x^M_r [\de_\n v^r +\hat{\de_\n v^r} ](\bar{\psi}_\m \g^\n
\g^\m \chi^M ) \nonumber\\ 
&-& \frac{1}{8} x^M_r [H+\hat{H}]^{r \m\n\r} ( \bar{\psi}_\m
\g_{\n\r}
\chi^M ) +\frac{1}{24} x^M_r [H+\hat{H}]^{r \m\n\r} (\bar{\psi}^\s \g_{\s\m\n\r}
\chi^M ) \nonumber\\ 
&+& \frac{1}{8}(\bar{\chi}^M \g^{\m\n\r} \chi^M )(\bar{\psi}_\m
\g_\n \psi_\r )-\frac{1}{8}(\bar{\chi}^M \g^\m \chi^N )(\bar{\chi}^M \g_\m \chi^N )\quad
,
\label{fermilag}
\eea 
where, in the 1.5 order formalism,  the spin connection
\bea 
\w_{\m\n\r} &=& \w^0_{\m\n\r} -\frac{i}{2}(\bar{\psi}_\m \g_\n \psi_\r
+\bar{\psi}_\n \g_\r
\psi_\m+\bar{\psi}_\n \g_\m \psi_\r ) \nonumber\\ 
&-& \frac{i}{4}(\bar{\psi}^\s 
\g_{\m\n\r\s\t} \psi^\t  )-\frac{i}{4} (\bar{\chi}^M \g_{\m\n\r} \chi^M )
\label{1.5}
\eea 
satisfies its equation of motion, and is thus kept fixed in all variations.

In order to  derive the bosonic equations, one can add to (\ref{fermilag})
\be 
e^{-1}{\cal{L}}_{bose}=-\frac{1}{4}R +\frac{1}{12}G_{rs} H^{r \m\n\r} H^s_{\m\n\r}
-\frac{1}{4} \de_\m v^r \de^\m v_r \quad . \label{boselag}
\ee 
One can then obtain from ${\cal{L}}_{fer}+{\cal{L}}_{bose}$  the equations for  the
vielbein and the scalars, with the prescription that the  
(anti)self-duality conditions be used
only after varying.  Actually, ignoring momentarily  eq. (\ref{finalselfdual}) and
varying
${\cal{L}}_{fer}+{\cal{L}}_{bose}$  with respect to the antisymmetric tensor
$B^r_{\m\n}$ yields  the second-order tensor equation, the divergence of eq.
(\ref{finalselfdual}),
\bea 
& & \nabla_\m (G_{rs} \hat{H}^{s \m\n\r}) =\frac{1}{2} \nabla_\m [ x^M_r (\bar{\chi}^M
\g^{\m\n\r\s}
\psi_\s )] \nonumber\\ 
& & - \frac{i}{4} \nabla_\m [v_r (\bar{\psi}_\s \g^{\s\t\m\n\r}\psi_\t
)]+
\frac{i}{4} \nabla_\m [v_r (\bar{\chi}^M \g^{\m\n\r} \chi^M )] \quad .
\eea 
In a similar fashion, the scalar equation is
\bea 
& & x^M_r [\frac{1}{2}D_\m (\de^\m v^r ) +\frac{1}{3} v_s H^{r \m\n\r} H^s_{\m\n\r}
-\frac{i}{4} H^{r \m\n\r}(\bar{\psi}_\m \g_\n \Psi_\r ) 
 +\frac{i}{24}H^r_{\r\s\t} (\bar{\psi}_\m \g^{\m\n\r\s\t}\psi_\n )
\nonumber\\ 
& & -\frac{i}{24} H^r_{\m\n\r} (\bar{\chi}^N \g^{\m\n\r}\chi^N )  -
\frac{1}{2} \nabla_\n (x^N_r(\bar{\psi}_\m \g^\n \g^\m \chi^N ))]
\nonumber\\  
& & + v_r [ -\frac{1}{4} H^{r \m\n\r} (\bar{\psi}_\m \g_{\n\r} \chi^M ) +
\frac{1}{12} H^{r \m\n\r} (\bar{\psi}^\s \g_{\s\m\n\r} \chi^M ) ] = 0
\label{scalarcomplete}
\quad ,
\eea 
while the Einstein equation is
\bea 
& & \frac{1}{2}e_{\n m} [R^{\m\n} -\frac{1}{2}g^{\m\n} R -G_{rs}H^{r \m\r\s} H^{s
\n}{}_{\r\s}  +\frac{1}{6}g^{\m\n}G_{rs} H^r_{\r\s\t}H^{s \r\s\t} 
\nonumber\\ 
& & +\de^\m v^r
\de^\n v_r -\frac{1}{2}g^{\m\n} \de_\r v^r \de^\r v_r ] 
 -\frac{i}{2}e^\m{}_m (\bar{\psi}_\r \g^{\r\s\t} \hat{D}_\s \psi_\t )+
\frac{i}{2} (\bar{\psi}_m \g^{\m\n\r} \hat{D}_\n \psi_\r )\nonumber\\ 
& & -\frac{i}{2}
(\bar{\psi}_\n \g^{\m\n\r} \hat{D}_m \psi_\r )+
\frac{i}{2} (\bar{\psi}_\n \g^{\m\n\r} \hat{D}_\r \psi_m  )-
\frac{i}{4} e^\m{}_m v_r \hat{H}^r_{\n\r\s} (\bar{\psi}^\n \g^\r \psi^\s ) \nonumber \\
& & +\frac{i}{4}v_r
\hat{H}^r_{\n m \r} (\bar{\psi}^\n \g^\m \psi^\r ) +\frac{i}{2} v_r
\hat{H}^{r \m\n\r} (\bar{\psi}_m \g_\n \psi_\r )+
\frac{i}{2}v_r \hat{H}^r_{m \n\r} (\bar{\psi}^\m \g^\n \psi^\r )\nonumber \\
& &    +\frac{i}{24}
e^\m{}_m v_r
\hat{H}^r_{\n\r\s} (\bar{\psi}_\d 
\g^{\n\r\s\d \t} \psi_\t ) - \frac{i}{12} v_r \hat{H}^r_{\r\s\t}
(\bar{\psi}_m \g^{\m\n\r\s\t} \psi_\n )   -\frac{i}{8}v_r \hat{H}^r_{m \n\r}
(\bar{\psi}_\s
\g^{\m\n\r\s\t} \psi_\t ) \nonumber \\
& & +\frac{i}{2} e^\m{}_m (\bar{\chi}^M \g^\n \hat{D}_\n \chi^M
) -\frac{i}{2}( \bar{\chi}^M \g^\m \hat{D}_m \chi^M )-\frac{i}{24}
e^\m{}_m  v_r
\hat{H}^r_{\n\r\s} (\bar{\chi}^M \g^{\n\r\s} \chi^M ) \nonumber \\  
& & +\frac{i}{8} v_r \hat{H}^r_{m
\n\r} (\bar{\chi}^M \g^{\m\n\r} \chi^M )
+
\frac{1}{2} e^\m{}_m x^M_r (\hat{\de_\n v^r} )(\bar{\psi}_\r \g^\n \g^\r \chi^M ) \nonumber \\
& & -\frac{1}{2} x^M_r (\hat{\de_m v^r })(\bar{\psi}_\n \g^\m \g^\n \chi^M ) 
-\frac{1}{2}x^M_r
(\hat{\de_\n v^r })(\bar{\psi}_m \g^\n \g^\m \chi^M )
\nonumber\\ 
& & -\frac{1}{4}e^\m{}_m x^M_r \hat{H}^r_{\n\r\s} (\bar{\psi}^\n \g^{\r\s} 
\chi^M ) +\frac{1}{2} x^M_r \hat{H}^r_{\n m \r}(\bar{\psi}^\n \g^{\m\r} \chi^M )
 +\frac{1}{4} x^M_r \hat{H}^{r \m}{}_{\n\r}(\bar{\psi}_m \g^{\n\r} \chi^M )
\nonumber\\ 
& & + \frac{1}{4}x^M_r \hat{H}^r_{m \n\r}(\bar{\psi}^\m \g^{\n\r} \chi^M )
 +\frac{1}{12}e^\m{}_m x^M_r \hat{H}^r_{\n\r\s}(\bar{\psi}_\t
\g^{\t \n\r\s}\chi^M ) -\frac{1}{12}x^M_r \hat{H}^r_{\n\r\s} (\bar{\psi}_m
\g^{\m\n\r\s}\chi^M ) \nonumber\\ 
& & -\frac{1}{4}x^M_r \hat{H}^r_{m
\n\r}(\bar{\psi}_\sigma \g^{\sigma \m\n\r}
\chi^M ) + (fermi)^4 =0 \quad . \label{einsteincomplete}
\eea 
For the sake of brevity, a number of quartic fermionic couplings, fully determined
by the lagrangian of eqs. (\ref{fermilag}) and (\ref{boselag}),  are not written
explicitly. It then takes a direct, if somewhat tedious, calculation to prove local
supersymmetry, showing that
\be 
\delta F \frac{\delta {\cal{L}}}{\delta F}+\delta B 
\frac{\delta {\cal{L}}}{\delta B}=0
\quad,\label{susyproof}
\ee 
where $F$ and $B$ denote collectively the Fermi  and  Bose fields 
aside from the
antisymmetric tensors.

\subsection{Inclusion of vector multiplets} 

A $(1,0)$ Yang-Mills  multiplet in six
dimensions comprises  gauge vectors $A_\m$ and pairs of left-handed spinors $\l^A$
satisfying a  symplectic Majorana condition, all in the adjoint  representation of
the gauge group. In this subsection we write the complete field  equations for $N=1$
supergravity coupled to $n$ tensor multiplets and to  vector multiplets.  
This setting plays a crucial role in
six-dimensional perturbative type-I vacua, that naturally include a number of tensor
multiplets \cite{bs}, and more generally in the context of string dualities relating
these to non-perturbative vacua of other strings and to $M$ theory \cite{dmw}.  In all
these cases, the anomaly polynomial comprises in principle an irreducible part, that in
perturbative  type-I vacua is removed by tadpole conditions, and a residual reducible 
part of the form
\be 
I_8 = - \sum_{x,y} \ c^r_x \ c^s_y \ \eta_{rs} \ {\rm tr}_x F^2 \  {\rm tr}_y F^2
\quad ,
\label{residualanomaly}
\ee 
with the $c$'s a collection of constants and $\eta$ the Minkowski metric for
$SO(1,n)$.  In general, this residual anomaly should also include gravitational and mixed
contributions, but we leave them aside, since they would contribute higher-derivative
couplings not part of the low-energy effective supergravity.

The antisymmetric tensors are not inert under vector gauge transformations, as demanded
by the Chern-Simons couplings
\be H^r = dB^r - c^{rz} \w_z\quad , \label{newtensorfieldstrength}
\ee 
where the index $z$ runs over the various factors of the gauge  group.  Gauge
invariance of
$H^r$ indeed requires that $B^r_{\m\n}$ transform under vector gauge transformations
according to
\be
\delta B^r = c^{rz} tr_z (\L dA) \quad .\label{new3form}
\ee 
To lowest order, the (anti)self-duality conditions (\ref{selfdual}) are not
affected, while their divergence becomes 
\be 
\nabla_\m (G_{rs} H^{s \m\n\r} ) =-\frac{1}{4e} \e^{\n\r\a\b\g\delta} c_r^z \tr_z (F_{\a\b}
F_{\g\delta}) \quad.\label{eqtensorgs}
\ee 
In a similar fashion, the fermionic equations become
\bea 
& & -i \g^{\m\n\r} D_\n \psi_\r -i v_r H^{r \m\n\r} \g_\n \psi_\r -
\frac{1}{2} x^M_r H^{ r \m\n\r} \g_{\n\r} \chi^M  
\nonumber\\ 
& & + \frac{1}{2} x^M_r \de_\n v^r \g^\n \g^\m \chi^M +\frac{i}{\sqrt{2}}
v_r c^{rz} \tr_z (F_{\n \r} \g^{\n \r} \g^\m \l )=0 
\eea 
for the gravitino,
\bea 
& & i \g^\m D_\m \chi^M -\frac{i}{12} v_r H^{r \m\n\r} \g_{\m\n\r} \chi^M
+\frac{1}{2} x^M_r H^{r \m\n\r} \g_{\m\n} \psi_\r  \nonumber\\ 
& & + \frac{1}{2} x^M_r
\de_\n v^r \g^\m \g^\n \psi_\m +\frac{1}{\sqrt{2}} x^M_r c^{rz} \tr_z (F_{\m\n}\g^{\m\n}
\l ) =0 
\eea 
for the tensorinos and
\bea 
& & 2i v_r c^{rz}\g^\m D_\m \l + i (\de_\m v_r ) c^{rz} \g^\m \l+
\frac{i}{\sqrt{2}} v_r c^{rz} F_{\n\r} \g^\m \g^{\n\r} \psi_\m \nonumber\\ 
& & - \frac{1}{\sqrt{2}} x^M_r c^{rz} F_{\m\n} \g^{\m\n} \chi^M -
\frac{i}{6} c^{rz} H_{r \m\n\r} \g^{\m\n\r} \l =0 
\eea 
for the gauginos. The  supersymmetry transformations of the vector multiplet are
\bea 
& & \delta A_\m =-\frac{i}{\sqrt{2}} (\bar{\e} \g_\m \l ) \quad ,\nonumber\\ 
& & \delta \l =-\frac{1}{2\sqrt{2}} F_{\m\n} \g^{\m\n} \e \quad ,
\eea 
while the tensor transformation becomes
\be
\delta B^r_{\m\n} =i v^r (\bar{\psi}_{[\m} \g_{\n]} \e )+\frac{1}{2} x^{Mr} (\bar{\chi}^M
\g_{\m\n} \e )-2 c^{rz} \tr_z (A_{[\m} \delta A_{\n]})
\quad . 
\ee 
The other transformations are not modified, 
aside from the change induced  by
(\ref{newtensorfieldstrength}) in the definition of $H^r$. 
Varying the Fermi fields  in the fermionic
equations then gives the bosonic equations
\be 
x^M_r D_\m (\de^\m v^r ) +\frac{2}{3} x^M_r v_s H^r_{\m\n\r}  H^{s \m\n\r} -x^m_r
c^{rz} \tr_z (F_{\m\n} F^{\m\n})=0 
\ee 
for the scalar,
\beq 
& & R_{\m\n} -\frac{1}{2} g_{\m\n} R + \de_\m v^r \de_\n v_r -
\frac{1}{2} g_{\m\n} \de_\r v^r \de^\r v_r - G_{rs} H^r_{\m\r\s} H^s{}_\n{}^{\r\s}
\nonumber\\ & & + 4 v_r c^{rz} \tr_z (F_{\r\m} F^\r{}_\n -\frac{1}{4} g_{\m\n}
F_{\r\s}F^{\r\s})=0
\eeq 
for the metric, and
\be 
D_\m (v_r c^{rz} F^{\m\n} ) -c^{rz} G_{rs} H^{s \n\r \sigma} F_{\r\sigma}=0
\label{vectoreq}
\ee 
for the vectors. The commutator of two supersymmetry transformations now includes a
gauge  transformation of parameter
\be
\L = \xi^\m A_\m \quad.
\ee

The novelty here is the non-vanishing divergence of eq. (\ref{vectoreq}) 
\be 
D_\m J^\m = -\frac{1}{2e} \e^{\m\n\a\b\g\delta} c^{rz}c_r^{z^\prime} F_{\m\n}
\tr_{z^\prime}(F_{\a\b} F_{\g\delta} ) \quad,\label{covanomaly}
\ee 
that reflects the presence of the residual gauge anomaly \cite{as,fms}.  In
particular,  eq. (\ref{covanomaly}) gives the  covariant anomaly.  Leaving aside
momentarily the (anti)self-duality conditions, one  might expect to derive eq.
(\ref{vectoreq}) from
\be 
e^{-1}{\cal{L}} = -\frac{1}{2} v_r c^{rz} \tr_z F_{\m\n} F^{\m\n} 
+ \frac{1}{12}
 G_{rs}H^{r\m\n\r}H^s_{\m\n\r} \quad,
\ee 
but this is actually not the case. In fact, eq. (\ref{vectoreq}) is not integrable,
while the inclusion of a Wess-Zumino term
\bea 
e^{-1}{\cal{L}} &=& -\frac{1}{2} v_r c^{rz} \tr_z F_{\m\n} F^{\m\n} +\frac{1}{12}
 G_{rs}H^{r\m\n\r}H^s_{\m\n\r} \nonumber \\
&-& \frac{1}{8e } \e^{\m\n\a\b\g\delta} c_r^z B^r_{\m\n}
\tr_z (F_{\a\b} F_{\g\delta}) \quad, \label{greenschwarz}
\eea 
turns the vector equation into
\beq 
& & D_\m (v_r c^{rz} F^{\m\n} ) -  G_{rs} H^{s \n\r\sigma} c^{rz} F_{\r\sigma}  -
\frac{1}{8e} \e^{\n\r\a\b\g\delta} c_r^z A_\r c^{rz^\prime}  \tr_{z^\prime}
(F_{\a\b}F_{\g\delta})
\nonumber\\ & & -\frac{1}{12e}\e^{\n\r\a\b\g\delta} c_r^z F_{\r\a} c^{rz^\prime}
\w^{z^\prime}_{\b\g\delta} = 0 \quad, \label{newvectoreq}
\eeq 
and now the divergence of the gauge current is the consistent anomaly \cite{fms}
\be 
{\cal{A}}_\L =- \frac{1}{4} \e^{\m\n\a\b\g\delta} c_r^z c^{rz^\prime} \tr_z (\L
\de_\m A_\n ) \tr_{z^\prime} (F_{\a\b} F_{\g\delta} )\quad .
\label{consanomaly}
\ee 
As an aside, one can observe that, ignoring the (anti)self-duality conditions, eq.
(\ref{greenschwarz}) yields the second-order tensor equations (\ref{eqtensorgs}) when
varied with respect to  the antisymmetric fields.

The Wess-Zumino consistency condition \cite{wz}
\be
\delta_{\L} {\cal{A}}_{\e} =\delta_\e {\cal{A}}_\L
\ee 
now implies the presence of a supersymmetry anomaly of the form
\bea
{\cal{A}}_\e &=& -\frac{1}{4} \e^{\m\n\a\b\g\delta} c_r^z c^{r z^\prime} \tr_z (
\delta_\e A_\m A_\n ) \tr_{z^\prime} (F_{\a\b} F_{\g\delta}) \nonumber \\
&-& \frac{1}{6} \e^{\m\n\a\b\g\delta}
c_r^z c^{r z^\prime} \tr_z (
\delta_\e A_\m F_{\n\a} ) \w^{z^\prime}_{\b\g\delta} \ ,
\label{susyanomaly}
\eea 
and indeed the  supersymmetry variation of the lagrangian is exactly eq. 
(\ref{susyanomaly}). Moreover, the divergence of the gravitino field equation,
proportional to eq. (\ref{susyanomaly}), reflects the presence of the induced
supersymmetry anomaly. We shall now complete this construction to all orders  in the
Fermi fields.

Let us begin by noting that the supercovariant  Yang-Mills field strength is
\be
\hat{F}_{\m\n} = F_{\m\n} +\frac{i}{\sqrt{2}}(\bar{\l} \g_\m \psi_\n ) -
\frac{i}{\sqrt{2}}(\bar{\l} \g_\n \psi_\m ) \quad ,
\ee 
while the other supercovariant fields are not modified.  The supersymmetry
transformations 
\beq 
& & \delta e_\m{}^m =-i(\bar{\e} \g^m \psi_\m ) \quad,\nonumber\\ 
& & \delta B^r_{\m\n} =i v^r (\bar{\psi}_{[\m} \g_{\n]} \e )+ \frac{1}{2} x^{Mr} (\bar{\chi}^M
\g_{\m\n} \e )-2c^{rz} \tr_z (A_{[\m}\delta A_{\n]}) \quad,
\nonumber\\ 
& & \delta v_r = x^M_r (\bar{\chi}^M \e )\quad,\nonumber\\ 
& & \delta A_\m =
-\frac{i}{\sqrt{2}} (\bar{\e} \g_\m \l ) \quad ,\nonumber\\ 
& & \delta \psi_\m
=\hat{D}_\m \e +\frac{1}{4} v_r \hat{H}^r_{\m\n\r}
\g^{\n\r}\e -\frac{3i}{8} \g_\m \chi^M (\bar{\e} \chi^M ) \nonumber \\
& & \quad \quad -\frac{i}{8} \g^\n \chi^M
(\bar{\e}
\g_{\m\n} \chi^M )+\frac{i}{16}
\g_{\m\n\r} \chi^M (\bar{\e} \g^{\n\r} \chi^M ) \quad ,\nonumber\\ 
& & \delta \chi^M =
\frac{i}{2} x^M_r (\hat{\de_\m v^r} ) \g^\m \e +
\frac{i}{12} x^M_r \hat{H}^r_{\m\n\r} \g^{\m\n\r}\e \quad ,\nonumber\\ 
& & \delta \l
=-\frac{1}{2\sqrt{2}}\hat{F}_{\m\n} \g^{\m\n} \e \quad,
\eeq 
could in principle include additional terms proportional to $\l^2$. To be precise,
one could add to $\delta \psi$ a term proportional to
$v_r c^{rz} \tr_z (\l^2 \e)$, and to $\delta \chi$ a term proportional to $x^M_r c^{rz}
\tr_z (\l^2
\e )$. Moreover, the (anti)self-duality conditions could be modified by a self-dual term
of the form $c^{rz} \tr_z (\bar{\l} \g_{\m\n\r} \l )$.

Let us proceed to study the supersymmetry algebra completely. On the  scalar, the
vielbein and the gauge field, the algebra closes with no subtleties, while additional
information  comes from  the algebra on the tensor fields. Using the (anti)self-duality
conditions satisfied by the 3-forms  in eq. (\ref{newtensor}), one can show that the
algebra on $B^r$ closes up to the extra terms
\be 
[\delta_1 , \delta_2 ]_{extra} B^r_{\m\n}=c^{rz} \tr_z [(\bar{\e}_1 \g_\m \l )
(\bar{\e}_2 \g_\n
\l )-(\bar{\e}_1 \g_\n \l )(\bar{\e}_2 \g_\m \l )] \quad .
\ee 
These can be  canceled modifying the transformations of the gravitino and of the
tensorinos according to
\beq 
& & \delta^{\prime} \psi_\m =i v_r c^{rz}\lbrace a \ \tr_z [\l (\bar{\e} \g_\m \l
)]+  b \ \tr_z [\g_{\m\n} \l (\bar{\e} \g^\n \l )]+ c \ \tr_z [\g^{\n\r} \l (\bar{\e}
\g_{\m\n\r} \l )] \rbrace
\quad, 
\nonumber\\ &  & \delta^{\prime}\chi^M = d \ x^M_r c^{rz}\tr_z [ \g_\m \l (\bar{\e}
\g^\m \l )] \quad,
\eeq 
and  requiring that the modified 3-form
\be
\hat{\cal H}^{r}_{\m\n\r} =\hat{H}^{r}_{\m\n\r} -\frac{i}{8} v^r (\bar{\chi}^M
\g_{\m\n\r}\chi^M ) +i\a \ c^{rz} \tr_z (\bar{\l} \g_{\m\n\r} \l )
\ee 
satisfy the (anti)self-duality conditions
\be 
G_{rs} \hat{\cal H}^{s}_{\m\n\r} =\frac{1}{6e} \e_{\m\n\r\a\b\g}
\hat{\cal H}_r^{\a\b\g}\quad.\label{finalselfdual2}
\ee 
It should be appreciated that this change in the definition of the field  strengths
only affects the antiself-duality conditions, since
$(\bar{\l}\g_{\m\n\r}\l )$ is self-dual.

Requiring closure of the algebra on $B^r$ then implies the conditions
\be
\a =\frac{1}{4} \quad, 
\qquad d=\frac{1}{2} \quad ,
\qquad a+ b = -1 \quad ,
\qquad b+2c =0 \quad , \label{parameters}
\ee 
and only one of the parameters is still undetermined. These terms have no effect for
the scalars and the vectors, while the commutator on $e_\m{}^m$ shows that the local
Lorentz parameter is modified by the addition of
\be
\Delta^{\prime} {\W}^{mn} =v_r c^{rz}  \tr_z [(\bar{\e}_1 \g^m \l )(\bar{\e}_2 \g^n \l )-
(\bar{\e}_2 \g^m \l )(\bar{\e}_1 \g^n \l ) ] \quad . \label{locallorentz}
\ee

Turning to the Fermi fields, the commutator on the tensorini $\chi^M$ involves
techniques already met in  the case with tensor multiplets only, and fixes the last free
parameter  in eqs. (\ref{parameters}), so that
\be 
a=-\frac{9}{8}\quad,\qquad b =\frac{1}{8} \quad, \qquad  c =-\frac{1}{16} \quad.
\ee 
It closes on the field equation
\beq 
& & i \g^\m \hat{D}_\m \chi^M -\frac{i}{12}v_r \hat{H}^r_{\m\n\r} \g^{\m\n\r}
\chi^M +\frac{1}{12} x^M_r \hat{H}^{r \m\n\r} \g_\s \g_{\m\n\r} \psi^\s
+ \frac{1}{2} x^M_r (\hat{\de_\n v^r}) \g^\m \g^\n \psi_\m 
\nonumber\\ 
& & + \frac{1}{\sqrt{2}}x^M_r c^{rz} \tr_z (\hat{F}_{\m\n} \g^{\m\n} \l )  
-\frac{i}{2} x^M_r c^{rz} \tr_z [ \g^\m \g^\n  \l (\bar{\psi}_\m \g_\n
\l )]  +\frac{1}{2}\g^\m \chi^N (\bar{\chi}^N \g_\m \chi^M )  
\nonumber\\ 
& & -\frac{3}{8} v_r c^{rz} \tr_z [(\bar{\chi}^M \g_{\m\n} \l )\g^{\m\n}
\l ] - \frac{1}{4} v_r c^{rz} \tr_z [(\bar{\chi}^M \l ) \l ] 
 =0\quad ,\label{tensorinoeq2}
\eeq 
where all terms containing the gravitino are exactly determined by supercovariance. 
Moreover, the field equation appears in the commutator as in  the theory without gauge
fields:
\be  
[\delta_1 , \delta_2 ] \chi^m = \delta_{gct} \chi^m +\delta_{Lorentz}\chi^m
+\delta_{SO(n)}\chi^m +\delta_{susy}\chi^m - \frac{i}{4} \g^\m \xi_\m [eq.\ \chi^m ]
\quad.
\ee

Using similar techniques, one can compute the commutator on the  gauginos $\l$. Here,
however, the transformation 
\be
\delta \l = -\frac{1}{2\sqrt{2}}\hat{F}_{\m\n}\g^{\m\n}\e\label{gauginovar}
\ee 
can not produce the terms proportional to $x^M_r$  already present at the lowest
order, and the only way to generate them is to modify eq. (\ref{gauginovar}) by terms of
the form
\be
\frac{x^M_r c^{rz}}{v_s c^{sz}} \ \chi^M \ \l \ \e \quad .
\nonumber
\ee 
Singular couplings of this type were previously introduced in \cite{ns2}. We
therefore add all possible extra terms, that modulo Fierz identities are
\be
\delta^{\prime} \l =  \frac{x^M_r c^{rz}}{v_s c^{sz}}[a(\bar{\chi}^M \l ) \e+ b (\bar{\chi}^M
\g_{\m\n} \l ) \g^{\m\n}
\e + c (\bar{\chi}^M \e ) \l  +d (\bar{\chi}^M \g_{\m\n} \e ) \g^{\m\n}
\l ]\quad,\label{gauginoextra}
\ee 
and determine their coefficients from the algebra. Eq. (\ref{gauginoextra}) should
not affect the vector (and, a fortiori, the  tensor) commutator, and thus the
coefficients are  to obey the three equations
\be 
a -2 c =0\quad , \qquad b=0 \quad ,\qquad c+2d=0\quad .\label{sys1}
\ee 
The other conditions,
\be 
a+2b = -\frac{1}{2}\quad ,\qquad c+2d+4b=0\quad ,\qquad 2d +\frac{1}{8}a+
\frac{1}{4}b=\frac{3}{16} \quad , \label{sys2}
\ee 
are obtained from the algebra on the gauginos, for instance tracking  the terms
generated by eq. (\ref{gauginoextra}) and proportional to  $\de v $. Combining eqs.
(\ref{sys1}) and (\ref{sys2}),  one finally obtains
\be 
a=-\frac{1}{2} \quad ,\qquad c=-\frac{1}{4} \quad ,\qquad d=\frac{1}{8} \quad .
\ee 
As was the case for the gravitino already without vector multiplets,  here the
algebra generates the field  equation with a non trivial symplectic structure,
\be
- \frac{3i}{16} \g^\m \xi_\m [eq. \l^A ] -\frac{i}{192}\g^{\m\n\r}
\sigma^i_B{}^A \xi^i_{\m\n\r} [eq. \l^B ] \quad ,
\ee 
where $\xi^i_{\m\n\r}$ is defined in eq. (\ref{anomalousxi}).

Eq. (\ref{gauginoextra}) also affects the algebra on  the tensorinos, whose field
equation now includes two additional terms, and becomes 
\beq 
& & i \g^\m \hat{D}_\m \chi^M -\frac{i}{12}v_r \hat{H}^r_{\m\n\r} \g^{\m\n\r}
\chi^M +\frac{1}{12} x^M_r \hat{H}^{r \m\n\r} \g_\s \g_{\m\n\r} \psi^\s
+ \frac{1}{2} x^M_r (\hat{\de_\n v^r}) \g^\m \g^\n \psi_\m 
\nonumber\\ 
& & + \frac{1}{\sqrt{2}}x^M_r c^{rz} \tr_z (\hat{F}_{\m\n} \g^{\m\n} \l ) 
 -\frac{i}{2} x^M_r c^{rz} \tr_z [ \g^\m \g^\n  \l (\bar{\psi}_\m \g_\n
\l )] +\frac{1}{2}\g^\m \chi^N (\bar{\chi}^N \g_\m \chi^M ) \nonumber\\ 
& & - \frac{3}{8}
v_r c^{rz} \tr_z [(\bar{\chi}^M \g_{\m\n} \l )\g^{\m\n}
\l ] -\frac{1}{4} v_r c^{rz} \tr_z [(\bar{\chi}^M \l ) \l ] \nonumber\\ 
& & -
\frac{3}{2} \frac{x^M_r c^{rz} x^N_s c^{sz}}{v_t c^{tz}} \tr_z [(\bar{\chi}^N \l ) \l ]  +
\frac{1}{4}
\frac{x^M_r c^{rz} x^N_s c^{sz}}{v_t c^{tz}} \tr_z [(\bar{\chi}^N \g_{\m\n} \l )
\g^{\m\n} \l ]
 =0\quad .\label{tensorinoeq3}
\eeq

In the commutator of two supersymmetry transformations on the gauginos, these terms
complete the algebra and let it close on the field equation, that now includes 
$\chi^2 \l$  terms corresponding to the $\l^2 \chi$ terms in the equation for  the
tensorinos. In addition, the $\l^3$ terms comprise two groups:  those proportional to 
$v_r v_s$ and those proportional to $\eta_{rs}$ (recall, from eqs.  (\ref{scalars}),
that 
$x^M_r x^M_s = v_r v_s -\eta_{rs}$). The former generate local Lorentz transformations
according to eq. (\ref{locallorentz}) and the term
\be 
iv_r v_s c^{rz} c^{s z^\prime} \tr_{z^\prime} [(\bar{\l} \g_\m \l^\prime )
\g^\m \l^\prime ]
\ee 
in the field equation, while the latter are 
\beq 
[\delta_1 , \delta_2 ]_{extra} \l &=& \frac{c_r^z c^{r z^\prime}}{v_s c^{sz}}
\tr_{z^\prime} [-\frac{1}{4}(\bar{\e}_1 \g_\m \l^\prime )(\bar{\e}_2
\g_\n \l^\prime ) \g^{\m\n} \l  +
\frac{1}{4} (\bar{\l} \g_\m \l^\prime ) (\bar{\e}_1 \g^\m \l^\prime ) \e_2 -(1
\leftrightarrow 2)
\nonumber\\ &+& \frac{1}{16} (\bar{\e}_1 \g^\m \e_2 )(\bar{\l}^\prime
\g_{\m\n\r} \l^\prime  )\g^{\n\r}\l ]\quad . \label{centralcharge1}
\eeq 
In general, one could allow for a modified field equation including the 
$\l^3$ term
\be 2\a c_r^z c^{r z^\prime} \tr_{z^\prime} [(\bar{\l}\g_\m \l^\prime )
\g^\m \l^\prime ] \quad , \label{lambda3}
\ee 
with $\a$ an arbitrary parameter. Although the choice $\a =1$ could seem  the
preferred one on account of the rigid limit, since the  supersymmetric Yang-Mills theory
in six dimensions does not contain such a $\l^3$ term,  the $(1,0)$ supergravity is
actually consistent for an arbitrary value  of $\a$, with the corresponding residual
terms
\beq
\delta_{extra(\a)} \l &\equiv& [\delta_1 , \delta_2 ]_{extra(\a)} \l = \frac{c_r^z c^{r
z^\prime}}{v_s c^{sz}} \tr_{z^\prime} [-\frac{1}{4}(\bar{\e}_1 \g_\m \l^\prime
)(\bar{\e}_2
\g_\n \l^\prime ) \g^{\m\n} \l \nonumber\\  
&-&\frac{\a}{2} (\bar{\l} \g_\m \l^\prime
)(\bar{\e}_1 \g_\n \l^\prime )
\g^{\m\n} \e_2  +\frac{\a}{16}(\bar{\l}\g_{\m\n\r}\l^\prime )(\bar{\e}_1 \g^\r 
\l^\prime ) \g^{\m\n} \e_2 \nonumber\\  &+& \frac{\a}{16} (\bar{\l} \g_\r \l^\prime
)(\bar{\e}_1
\g^{\m\n\r} \l^\prime ) \g_{\m\n} \e_2   + \frac{1-\a}{4} (\bar{\l} \g_\m \l^\prime )
(\bar{\e}_1
\g^\m \l^\prime ) \e_2 -(1 \leftrightarrow 2)  
\nonumber\\ 
&+& \frac{1-\a}{16} (\bar{\e}_1 \g^\m \e_2 )(\bar{\l}^\prime
\g_{\m\n\r} \l^\prime  )\g^{\n\r}\l ] \label{centralcharge}
\eeq 
in the commutator of two supersymmetry transformations on the gauginos.   It should
be appreciated that no choice of $\a$ can eliminate all these terms, that  play the role
of a central charge felt only by the gauginos. The Jacobi  identity for this charge is
properly satisfied for any value of
$\a$, and thus we are effectively discovering a 2-cocycle in our problem. It has long
been known that, in general, anomalies in current conservations are accompanied by
related anomalies in current commutators \cite{anom}, but it is amusing to see how this
``classically anomalous'' model displays all these intricacies.

The complete algebra 
\beq [ \delta_1, \delta_2 ] \l^A &=& \delta_{gct} \l^A + \delta_{Lorentz} \l^A +
\delta_{susy}
\l^A + \delta_{gauge} \l^A + \delta_{extra(\a)} \l^A \nonumber \\ 
&-& \frac{3i}{16}
\g^\m \xi_\m [eq. \l^A ]_{(\a )} -\frac{i}{192}\g^{\m\n\r}
\sigma^i_B{}^A \xi^i_{\m\n\r} [eq. \l^B ]_{(\a )} 
\eeq 
determines the complete field equation of the gauginos
\beq 
& & 2i v_r c^{rz} \g^\m \hat{D}_\m \l +i (\hat{\de_\m v_r})c^{rz}
\g^\m \l + \frac{i}{\sqrt{2}}v_r c^{rz} \hat{F}_{\n\r} \g^\m \g^{\n\r} \psi_\m -
\frac{1}{\sqrt{2}}x^M_r c^{rz} \hat{F}_{\m\n} \g^{\m\n}\chi^M \nonumber\\ 
& &
+\frac{i}{6} x^M_r c^{rz} x^M_s \hat{H}^s_{\m\n\r} \g^{\m\n\r}\l +
i x^M_r c^{rz} (\bar{\chi}^M \l ) \g^\m \psi_\m 
 +\frac{i}{2}x^M_r c^{rz} (\bar{\chi}^M \psi_\m )\g^\m \l \nonumber\\ 
& & - \frac{i}{4}
x^M_r c^{rz} (\bar{\chi}^M \g_{\n\r} \psi_\m ) \g^{\m\n\r} \l 
 - \frac{i}{2}x^M_r c^{rz} (\bar{\chi}^M \g_{\m\n}\psi^\m )\g^\n \l -
\frac{1}{4}v_r c^{rz} (\bar{\l} \chi^M )\chi^M \nonumber\\ 
& & - \frac{3}{8} v_r c^{rz}
(\bar{\l} \g_{\m\n} \chi^M ) \g^{\m\n} \chi^M -\frac{3}{2} \frac{x^M_r c^{rz} x^N_s
c^{sz}}{v_t c^{tz}} (\bar{\l} \chi^M )
\chi^N \nonumber \\
& &  +\frac{1}{4} \frac{x^M_r c^{rz} x^N_s c^{sz}}{v_t c^{tz}} (\bar{\l} 
\g_{\m\n} \chi^M ) \g^{\m\n} \chi^N -2 v_r v_s c^{rz} c^{s z^\prime}
\tr_{z^\prime} [(\bar{\l} \g_\m \l^\prime )
\g^\m \l^\prime ] \nonumber \\
& & +2 \a c_r^z c^{r z^\prime} \tr_{z^\prime} [(\bar{\l}\g_\m \l^\prime )
\g^\m \l^\prime ] = 0  
\label{gauginoeq}
\eeq 
where, again, all terms containing the gravitino are fixed by supercovariance,
while the
$\chi^2 \l$ terms are precisely as demanded by the 
$\l^2 \chi$ terms in the field equations of the tensorinos. At last, one can study the
algebra on the gravitino, thus obtaining the  field equation
\beq 
& & -i \g^{\m\n\r} \hat{D}_\n \psi_\r -\frac{i}{4} v_r \hat{H}^r_{\n\s\t}
\g^{\m\n\r}\g^{\s\t} \psi_\r -\frac{1}{12}x^M_r \hat{H}^{r \n\r\s}\g_{\n\r\s} \g^\m
\chi^M +\frac{1}{2} x^M_r (\hat{\de_\n v^r })\g^\n \g^\m \chi^M \nonumber\\ 
& &
+\frac{3}{2}
\g^{\m\n}\chi^M (\bar{\chi}^M \psi_\n )-\frac{1}{4}
\g^{\m\n} \chi^M (\bar{\chi}^M \g_{\n\r} \psi^\r ) 
 +\frac{1}{4} \g_{\n\r} \chi^M (\bar{\chi}^M \g^{\m\n} \psi^\r )-\frac{1}{2}
\chi^M (\bar{\chi}^M \g^{\m\n} \psi_\n ) \nonumber\\ 
& & -i v_r c^{rz} \tr_z
[-\frac{1}{\sqrt{2}}\g^{\n\r}\g^\m \l \hat{F}_{\n\r}  + \frac{3i}{4} \g^{\m\n\r}\l
(\bar{\psi}_\n
\g_\n \l )  -\frac{i}{2} \g^\m \l (\bar{\psi}_\n \g^\n \l ) +\frac{i}{2}\g^\n \l 
(\bar{\psi}_\n
\g^\m \l ) \nonumber\\ & & + \frac{i}{4}\g_\r \l (\bar{\psi}_\n \g^{\m\n\r}\l )]
 - \frac{i}{2} x^M_r c^{rz} \tr_z [\g_\n \l (\bar{\chi}^M \g^\n \g^\m \l )]=0
\quad,
\eeq 
that enters the supersymmetry algebra as in eq. (\ref{gravitinoalg}). Once more,
all terms containing the gravitino are fixed by supercovariance, while the other $\l^2
\chi$ terms are precisely as demanded by the 
$\l^2 \psi$ terms in the tensorino equation and by the $\l \psi \chi$ terms in the
equations of the gauginos. 

Summarizing, from the algebra we have obtained the complete fermionic equations of
$(1,0)$ six-dimensional supergravity coupled to vector and tensor multiplets. 
In addition, the modified 3-form
\be
\hat{\cal H}^{r}_{\m\n\r}=
\hat{H}^r_{\m\n\r} -\frac{i}{8}v^r (\bar{\chi}^m \g_{\m\n\r}\chi^m )
+ \frac{i}{4} c^{rz} \tr_z( \bar{\l} \g_{\m\n\r} \l )
\ee 
satisfies the (anti)self-duality conditions
\be 
G_{rs} \hat{\cal H}^{s}_{\m\n\r}=\frac{1}{6e}\e_{\m\n\r\a\b\g}
\hat{\cal H}^{\a\b\g}_r \quad. \label{finalselfdual3}
\ee
We have also identified the complete supersymmetry transformations, 
that we collect here for convenience:
\beq
& & \delta e_\m{}^m =-i(\bar{\e} \g^m \psi_\m ) \quad,\nonumber\\ 
& & \delta B^r_{\m\n} =i v^r (\bar{\psi}_{[\m} \g_{\n]} \e )+ \frac{1}{2} x^{Mr} (\bar{\chi}^M
\g_{\m\n} \e )-2c^{rz} \tr_z (A_{[\m}\delta A_{\n]}) \quad,
\nonumber\\ 
& & \delta v_r = x^M_r (\bar{\chi}^M \e )\quad,\nonumber\\ 
& & \delta A_\m =
-\frac{i}{\sqrt{2}} (\bar{\e} \g_\m \l ) \quad ,\nonumber\\ 
& & \delta \psi_\m =\hat{D}_\m \e +\frac{1}{4} v_r \hat{H}^r_{\m\n\r}
\g^{\n\r}\e -\frac{3i}{8} \g_\m \chi^M (\bar{\e} \chi^M ) -\frac{i}{8} \g^\n \chi^M
(\bar{\e} \g_{\m\n} \chi^M ) \nonumber \\
& & \quad \quad +\frac{i}{16} \g_{\m\n\r} \chi^M (\bar{\e} 
\g^{\n\r} \chi^M ) - \frac{9i}{8} v_r c^{rz} \tr_z [\l (\bar{\e} \g_\m \l)] \nonumber \\
& & \quad \quad +  
\frac{i}{8} v_r c^{rz} \tr_z [\g_{\m\n} \l (\bar{\e} \g^\n \l )]
- \frac{i}{16} v_r c^{rz} \tr_z [\g^{\n\r} \l (\bar{\e}
\g_{\m\n\r} \l )] \quad ,\nonumber\\ 
& & \delta \chi^M =
\frac{i}{2} x^M_r (\hat{\de_\m v^r} ) \g^\m \e +
\frac{i}{12} x^M_r \hat{H}^r_{\m\n\r} \g^{\m\n\r}\e +
\frac{1}{2} x^M_r c^{rz} \tr_z [ \g_\m \l (\bar{\e} \g^\m \l ) ] \quad ,\nonumber\\ 
& & \delta \l
=-\frac{1}{2\sqrt{2}}\hat{F}_{\m\n} \g^{\m\n} \e 
- \frac{1}{2} \frac{x^M_r c^{rz}}{v_s c^{sz}} (\bar{\chi}^M \l ) \e 
 - \frac{1}{4} \frac{x^M_r c^{rz}}{v_s c^{sz}} (\bar{\chi}^M \e ) \l  
\nonumber \\
& & \quad \quad + \frac{1}{8} \frac{x^M_r c^{rz}}{v_s c^{sz}} (\bar{\chi}^M \g_{\m\n} \e ) \g^{\m\n}
\l \quad .
\eeq 

Proceeding as in the previous subsection, the bosonic
equations  can be derived from a lagrangian, with the prescription of using the tensor
(anti)self-duality conditions only after varying.  The lagrangian is obtained
supplementing ${\cal{L}}_{fermi}+{\cal{L}}_{bose}$  of eqs. (\ref{fermilag}) and
(\ref{boselag}) with the terms
\beq  
& & -\frac{1}{2} v_r c^{rz} \tr_z (F_{\m\n} F^{\m\n}) -\frac{1}{8e}
\e^{\m\n\a\b\g\delta} c_r^z B^r_{\m\n} \tr_z (F_{\a\b} F_{\g\delta}) \nonumber \\  
& & +
\frac{i}{2\sqrt{2}}  v_r c^{rz} \tr_z [(F+\hat{F})_{\sigma \delta}(\bar{\psi}_\m
\g^{\sigma \delta}\g^\m \l )] 
 +\frac{1}{\sqrt{2}}x^M_r c^{rz} \tr_z [(\bar{\chi}^M \g^{\m\n}
\l )\hat{F}_{\m\n} ] \nonumber\\ 
& & +iv_r c^{rz} \tr_z [(\bar{\l} \g^m \hat{D}_\m \l
)+\frac{i}{12} x^M_r x^M_s \hat{H}^r_{\m\n\r} c^{sz} \tr_z (\bar{\l}\g^{\m\n\r} \l )\nonumber \\
& & +\frac{1}{16}v_r c^{rz} \tr_z (\bar{\l} \g_{\m\n\r} \l )(\bar{\chi}^M
\g^{\m\n\r} \chi^M ) - \frac{i}{8}(\bar{\chi}^M \g_{\m\n}\psi_\r )x^M_r
c^{rz} \tr_z (\bar{\l} \g^{\m\n\r} \l ) \nonumber \\ 
& &  - \frac{i}{2} x^M_r c^{rz} \tr_z [(\bar{\chi}^M \g^\m \g^\n \l ) (\bar{\psi}_\m \g_\n \l
)]
- \frac{3}{16}v_r c^{rz} \tr_z [(\bar{\chi}^M \g_{\m\n} \l ) (\bar{\chi}^M
\g^{\m\n}
\l )] \nonumber \\
& &  -\frac{1}{8} v_r c^{rz} \tr_z [(\bar{\chi}^M \l )(\bar{\chi}^M \l)
-\frac{3}{4}\frac{x^M_r c^{rz}x^N_s c^{sz}}{v_t c^{tz}} \tr_z
[(\bar{\chi}^M \l )(\bar{\chi}^N \l )] \nonumber \\
& &  +\frac{1}{8}\frac{x^M_r c^{rz} x^N_s c^{sz}}{v_t c^{tz}} \tr_z [(\bar{\chi}^M \g_{\m\n}
\l )(\bar{\chi}^N \g^{\m\n}\l )]
+\frac{1}{4} (\bar{\psi}_\m \g_\n \psi_\r )(\bar{\l} \g^{\m\n\r}
\l ) \nonumber \\
& & -\frac{1}{2} v_r v_s c^{rz}c^{s z^\prime} \tr_{z,z^\prime} [(\bar{\l}\g_\m \l^\prime
)(\bar{\l} \g^\m \l^\prime ) ]
+\frac{\a}{2}c^{rz} c_r^{z^\prime} \tr_{z,z^\prime} [(\bar{\l} \g_\m
\l^\prime )(\bar{\l} \g^\m \l^\prime )] \quad ,
\label{completelag}
\eeq 
and the 1.5 order formalism  requires that the spin connection $\w_{\m\n\r}$ now
include the additional term
\be
\w^{(\l )}_{\m\n\r} = -\frac{i}{2} v_r c^{rz} \tr_z (\bar{\l} \g_{\m\n\r} \l )
\quad .
\ee 
With the new definition of $\w$, eqs. (\ref{fermilag}), (\ref{boselag}) and 
(\ref{completelag}) then yield the Fermi  equations. Moreover,  varying with respect to
$B^r_{\m\n}$ yields the second-order tensor equations, the divergence of the
(anti)self-duality conditions.  The  vector equation is covariant, aside from the
anomalous couplings introduced by the Wess-Zumino term in  eq. (\ref{greenschwarz}). The 
complete residual gauge anomaly is thus given in eq. (\ref{consanomaly}). As we shall
see, it solves the Wess-Zumino consistency conditions even in the presence of
supersymmetry. 

The complete vector field equation is
\beq 
& & c^{rz} D_\n (v_r F^{\n\m} ) - G_{rs}\hat{H}^{r \m\n\r} c^{sz} F_{\n\r}
-\frac{1}{12e}\e^{\m\n\r\a\b\g} c_r^z c^{r z^\prime} F_{\n\r} \w^{z^\prime}_{\a\b\g}
\nonumber\\ & & -\frac{1}{8e}\e^{\m\n\r\a\b\g} c_r^z c^{r z^\prime} A_\n \tr_{z^\prime}
(F_{\r\a} F_{\b\g} )
 -\frac{i}{4}v_r c^{rz} F_{\n\r} (\bar{\psi}_\s \g^{\m\n\r\s\t} 
\psi_\t ) \nonumber\\ 
& & + \frac{i}{4} v_r c^{rz} F_{\n\r} (\bar{\chi}^M \g^{\m\n\r}
\chi^M )
 -\frac{x^M_r c^{rz}}{2} F_{\n\r} (\bar{\psi}_\s \g^{\s\m\n\r} \chi^M ) 
- \frac{i}{2} x^M_r x^M_s c^{rz} c^{s
z^\prime} F_{\n\r} \tr_{z^\prime} (
\bar{\l} \g^{\m\n\r} \l ) \nonumber\\ 
& & +\frac{i}{\sqrt{2}}  c^{rz} D_\n [v_r (\bar{\psi}_\r
\g^{\m\n}
\g^\r \l )] +\frac{1}{\sqrt{2}} c^{rz}D_\n  [x^M_r (\bar{\chi}^M \g^{\m\n} \l )] = 0 \quad ,
\label{completevectoreq}
\eeq 
the complete scalar equation is obtained adding to eq. (\ref{scalarcomplete})  the
terms
\beq  
& & x^M_r \lbrace \frac{1}{32} c^{rz} \tr_z (\bar{\l} \g_{\n\s\t}\l )(\bar{\psi}_\m
\g^{\m\n\r}\g^{\s\t} \psi_\r) +\frac{i}{2\sqrt{2}}c^{rz} \tr_z[(\bar{\psi}_\m \g^{\n\r} \g^\m
\l )(F +\hat{F} )_{\n\r} )]  \nonumber\\ 
&& + i c^{rz}\tr_z (\bar{\l} \g^\m \hat{D}_\m \l
) -\frac{3}{16} c^{rz} \tr_z  [(\bar{\chi}^N \g^{\m\n} \l )(\bar{\chi}^N \g_{\m\n} \l )]
-\frac{1}{8} c^{rz} \tr_z  [(\bar{\chi}^N \l )(\bar{\chi}^N \l )]\nonumber\\ 
&&
+\frac{1}{4} c^{rz} \tr_z (\bar{\l} \g^{\m\n\r} 
\l ) (\bar{\psi}_\m \g_\n \psi_\r ) - v_s c^{rz} c^{s z^\prime} \tr_{z, z^\prime}
[(\bar{\l} \g^\m
\l^\prime )(\bar{\l} \g_\m \l^\prime )]\nonumber\\ && - \frac{1}{8}\frac{x^N_s c^{s z}
x^P_t c^{t z}}{(v\cdot c^z )^2 }c^{rz} \tr_z [(\bar{\chi}^N \g^{\m\n} \l )(\bar{\chi}^P
\g_{\m\n} \l )] +\frac{3}{4} \frac{x^N_s c^{s z} x^P_t c^{t z}}{(v\cdot c^z )^2 }c^{rz}
\tr_z [(\bar{\chi}^N \l )(\bar{\chi}^P \l )] \rbrace \nonumber\\ && +v_r \lbrace
\frac{1}{\sqrt{2}}c^{rz} \tr_z [(\bar{\chi}^M \g^{\m\n} \l )
\hat{F}_{\m\n} ] +\frac{i}{12} x^M_s \hat{H}^{r \m\n\r} c^{sz} \tr_z (\bar{\l}
\g_{\m\n\r }  \l )\nonumber\\ 
&&+ \frac{i}{12} x^M_s \hat{H}^{s \m\n\r} c^{rz} \tr_z
(\bar{\l}
\g_{\m\n\r }  \l ) -\frac{i}{8} c^{rz} \tr_z (\bar{\l} \g_{\m\n\r} \l ) (\bar{\chi}^M
\g^{\m\n}
\psi^\r )\nonumber\\ 
&& -\frac{i}{2} c^{rz} \tr_z [(\bar{\chi}^M \g^\m \g^\n \l
)(\bar{\psi}_\m
\g_\n \l )] +\frac{1}{4}\frac{x^N_s c^{rz}c^{sz}}{v_t c^{tz}} \tr_z  [(\bar{\chi}^M
\g^{\m\n} \l )(\bar{\chi}^N \g_{\m\n} \l )]\nonumber\\ 
&&-\frac{3}{2}\frac{x^N_s
c^{rz}c^{sz}}{v_t c^{tz}} \tr_z [(\bar{\chi}^M \l ) (\bar{\chi}^N \l )] \rbrace \quad ,
\eeq 
and the Einstein equation is obtained adding to eq. (\ref{einsteincomplete})  the
terms
\beq 
&& c^{rz} \tr_z \lbrace 2e_{\n m} v_r ( F_{\r}{}^{\m} F^{\r\n} -\frac{1}{2}g^{\m\n} 
F_{\r\s} F^{\r\s} ) + 
\frac{i}{\sqrt{2}} e^\m{}_m v_r (\bar{\psi}_\n \g^{\r\s} \g^\n \l )F_{\r\s}\nonumber\\ 
&&
-\frac{2i}{\sqrt{2}} v_r (\bar{\psi}_\n \g^{\m\r} \g^\n \l ) F_{m \r}-\frac{i}{\sqrt{2}}
v_r (\bar{\psi}_m \g^{\n\r} \g^\m \l ) F_{\n\r}  +\frac{1}{\sqrt{2}} x^M_r e^\m{}_m
(\bar{\chi}^M
\g^{\n\r} \l ) F_{\n\r}\nonumber\\ 
&& - \sqrt{2} x^M_r (\bar{\chi}^M \g^{\m\n}
\l ) F_{m
\n} +i e^\m{}_m v_r (\bar{\l} \g^\n D_\n \l ) -iv_r (\bar{\l} \g^\m D_n \l )\nonumber\\
&& +\frac{i}{12} e^\m{}_n x^M_r x^M_s H^s_{\n\r\s} (\bar{\l} \g^{\n\r\s} \l )
-\frac{i}{4} x^M_r x^M_s H^s_{m \n\r} (\bar{\l} \g^{\m\n\r} \l )\nonumber\\ &&
-\frac{i}{4} e_{\n m}D_\r [v_r (\bar{\l}\g^{\m\n\r} \l )]+(fermi)^4 \rbrace
\quad .
\eeq 
We would like to stress that this result is expressed  in terms of the previous
definition of
$\w$, not corrected by bilinears in the gauginos.  Moreover, for the sake of brevity, a
number of quartic fermionic couplings, fully determined by the lagrangian of eqs.
(\ref{fermilag}), (\ref{boselag}) and  (\ref{completelag}), are not written explicitly.
Letting $F$ and $B$ denote all the Fermi and Bose fields aside from  the antisymmetric 
tensors, the supersymmetry variation of the lagrangian, after using the
(anti)self-duality conditions of eq. (\ref{finalselfdual2}), is
\be
\delta B \frac{\delta {\cal{L}}}{\delta B} +\delta F 
\frac{\delta {\cal{L}}}{\delta F }={\cal{A}}_\e \quad,
\ee 
where ${\cal{A}}_\e$ is the complete supersymmetry anomaly. Neglecting the last term
in eq. (\ref{completelag}), {\it i.e.} setting $\a$ to zero, 
\beq 
{\cal{A}}_\e &=& c_r^z c^{r z^\prime} \tr_{z, z^\prime} \lbrace -\frac{1}{4}
\e^{\m\n\a\b\g\delta}\delta_\e A_\m A_\n F^\prime_{\a\b} F^\prime_{\g\delta} -\frac{1}{6}
\e^{\m\n\a\b\g\delta} \delta_\e A_\m F_{\n\a} 
\w^\prime_{\b\g\delta} \nonumber\\ &+& \frac{i e}{2} \delta_\e A_\m F_{\n\r}
(\bar{\l}^\prime
\g^{\m\n\r} 
\l^\prime )+\frac{i e}{2} \delta_\e A_\m (\bar{\l} \g^{\m\n\r} \l^\prime )
F^\prime_{\n\r}  + ie\delta_\e A_\m (\bar{\l}\g_\n \l^\prime ) F^{\prime \m\n}
\nonumber\\ &+&
\frac{e}{32} \delta_\e e_\m{}^m (\bar{\l} \g^{\m\n\r} \l )(\bar{\l}^\prime
\g_{m \n\r} \l^\prime )  -\frac{e}{2\sqrt{2}} \delta_\e A_\m (\bar{\l} \g^\m \g^\n \g^\r 
\l^\prime )(\bar{\l}^\prime \g_\n \psi_\r )  \nonumber\\ 
&+& \frac{e x^M_s c^{s
z^\prime}}{v_t c^{t z^\prime}} [-\frac{3 i}{2\sqrt{2}}
 \delta_\e A_\m (\bar{\l} \g^\m \l^\prime )(\bar{\l}^\prime \chi^M )
 -\frac{i}{4 \sqrt{2}} \delta_\e A_\m (\bar{\l} \g^{\m\n\r} \l^\prime )(\bar{\l}^\prime
\g_{\n\r}\chi^M ) \nonumber\\ 
&-& \frac{i}{2\sqrt{2}} \delta_\e A_\m  (\bar{\l} \g_\n
\l^\prime )(\bar{\l}^\prime \g^{\m\n} \chi^M )] \rbrace \quad ,
\label{theanomaly}
\eeq 
while including the last term in eq. (\ref{completelag}) would give the additional
contribution
\be
\Delta {\cal{A}}_\e = \delta_\e {\cal{L}}_{\l^4} \quad ,\label{extraanomaly}
\ee where
\be {\cal{L}}_{\l^4} =  \frac{e \a}{2} \ c_r^z c^{r z^\prime} tr_{z, z^\prime}
[(\bar{\l} \g^\a
\l^\prime )(\bar{\l}\g_\a \l^\prime ) ] \quad.
\ee 

In verifying the supersymmetry anomaly, the equations for the fermi
fields and for the vector field are presented here must be rescaled by suitable
overall factors that may be simply identified.

We now turn to show that ${\cal{A}}_\e$ satisfies the complete Wess-Zumino
consistency conditions.

\subsection{Wess-Zumino Consistency Conditions} 

In general, the Wess-Zumino consistency
conditions follow from the requirement that the symmetry algebra be realized on the
effective action. For locally supersymmetric theories this implies
\beq & & \delta_{\L_1} {\cal{A}}_{\L_2} -\delta_{\L_2} {\cal{A}}_{\L_1}=
{\cal{A}}_{[\L_1 ,\L_2 ]}
\quad ,\nonumber \\ & & \delta_\e {\cal{A}}_\L = \delta_\L {\cal{A}}_\e \quad ,\nonumber
\\ & &
\delta_{\e_1} {\cal{A}}_{\e_2} -\delta_{\e_2} {\cal{A}}_{\e_1}= {\cal{A}}_{\tilde{\e}}+
{\cal{A}}_{\tilde{\L}} \quad ,\label{wesszumino}
\eeq where only gauge and supersymmetry anomalies are considered, and where 
$\tilde{\e}$ and $\tilde{\L}$ are the parameters of supersymmetry and gauge
transformations determined by the supersymmetry algebra.

In global supersymmetry the analysis is somewhat simpler, since the r.h.s. of the  last
of eqs. (\ref{wesszumino}) does not contain the (global) supersymmetry anomaly.  Let us
therefore begin by reviewing the case of supersymmetric  Yang-Mills theory in four
dimensions \cite{wzsusy}.  From the 6-form anomaly polynomial
\be 
I_6 =\tr F^3 \quad ,
\ee 
in the language of forms, one obtains the four-dimensional gauge anomaly
\be {\cal{A}}^{(4)}_\L =\tr [\L (dA)^2 +\frac{ig}{2} \ d\L A^3 ] \quad,
\ee and from eqs. (\ref{wesszumino}) one can determine the form of the global
supersymmetry anomaly.  With the classical lagrangian
\be  {\cal{L}}_{SYM} ={\rm tr} \left[ -\frac{1}{2} F_{\m\n} F^{\m\n} + 2 i 
\bar{\l} \g^\m D_\m \l \right]  \quad ,
\ee and $\l$ a right-handed Weyl spinor, the supersymmetry transformations are
\beq & & \delta A_\m =\frac{i}{\sqrt{2}}(\bar{\e} \g_\m \l -\bar{\l}
\g_\m \e )\quad ,\nonumber\\ & & \delta \l =\frac{1}{2\sqrt{2}} F_{\m\n} \g^{\m\n} \e
\quad .
\eeq 
The second of eqs. (\ref{wesszumino}) (with ${\cal A}_{\tilde{\e}}$ absent in this
global case), then determines the supersymmetry anomaly up to terms cubic in $\l$,
\be 
{\cal{A}}^{(4)}_\e =\tr [\delta_\e A A (dA) +\delta_\e A(dA) A-\frac{3ig}{2}\delta_\e A
A^3 ]\quad ,
\ee 
and indeed
\be
\delta_{\e_2}{\cal{A}}^{(4)}_{\e_1} -\delta_{\e_1} {\cal{A}}^{(4)}_{\e_2}
={\cal{A}}^{(4)}_{\tilde{\L}}+3 \ \tr [\delta_{\e_1} A \delta_{\e_2} A F -
\delta_{\e_2} A \delta_{\e_1} A F ] \quad . \label{susy4dpartial}
\ee In order to compensate the second term in eq. (\ref{susy4dpartial}), one is to add to
${\cal{A}}^{(4)}_\e$ the gauge-invariant term
\be
\Delta {\cal{A}}^{(4)}_\e = -\frac{i}{2} \tr [\delta_\e A \bar{\l} \g^{(3)} \l + \bar{\l}
\delta_\e A
\g^{(3)} \l ] \quad ,
\ee so that ${\cal{A}}^{(4)}_\e  +\Delta {\cal{A}}^{(4)}_\e$ is the proper global
supersymmetry anomaly. Although the supersymmetry algebra closes only on the  field
equation of $\l$, in four dimensions a simple dimensional counting shows that eqs.
(\ref{wesszumino}) can not generate a term proportional to $\g^\m D_\m \l$. Therefore,
in this case the Wess-Zumino consistency conditions close  accidentally even off-shell,
as pointed out in \cite{wzsusy}.

The situation is quite different in six dimensions. In this case, in the spirit of the 
previous section, let us restrict our attention to the 8-form residual anomaly
polynomial 
\be I_8 = - \  c^{rz} c_r^{z^\prime} \tr_z (F^2 ) \tr_{z^\prime} (F^2 ) \quad ,
\ee where the sums are left implicit, so that the gauge anomaly is
\be {\cal{A}}^{(6)}_{\L} =-c^{rz} c_r^{z^\prime} \tr_z (\L dA ) \tr_{z^\prime} (F^2 ) \quad .
\ee Then, from the second of eqs. (\ref{wesszumino}),
\be {\cal{A}}^{(6)}_\e =-c^{rz} c_r^{z^\prime } [ \tr_z (\delta_\e A A ) \tr_{z^\prime} (F^2 )
+2 \tr_z (\delta_\e AF )
\w^{z^\prime}_3 ]\quad , \label{d=6globalanomaly}
\ee 
but there are residual terms, so that
\bea 
(\delta_{\e_1} {\cal{A}}^{(6)}_{\e_2} -\delta_{\e_2}
{\cal{A}}^{(6)}_{\e_1} )_{extra} &=& -4 c^{rz} c_r^{z^\prime } [ \tr_z
(\delta_{\e_2}A \delta_{\e_1}A )\tr_{z^\prime}(F^2) \nonumber \\
&+&  2 \tr_z (\delta_{\e_2} A
F)\tr_{z^\prime}(\delta_{\e_1} AF) ] \quad .
\eea 
Consequently, eq. (\ref{d=6globalanomaly})  is to be modified by terms cubic in the 
gauginos, and the complete result, written in component notation, is finally
\beq 
{\cal{A}}^{(6)}_\e &=& -\frac{1}{4}\e^{\m\n\a\b\g\delta} c_r^z c^{r z^\prime} \tr_z (
\delta_\e A_\m A_\n ) \tr_{z^\prime} ( F^\prime_{\a\b}F^\prime_{\g\delta} ) \nonumber\\ &-& 
\frac{1}{6}\e^{\m\n\a\b\g\delta} c_r^z c^{rz^\prime} \tr_z ( \delta_\e A_\m F_{\n\a}
)\w^{z^\prime}_{\b\g\delta} \nonumber\\ &+& A c_r^z c^{r z^\prime} \tr_z (\delta_\e A_\m
F_{\n\r} ) \tr_{z^\prime} ( \bar{\l}^\prime \g^{\m\n\r} \l^\prime ) \nonumber\\ &+& B
c_r^z c^{r z^\prime} \tr_z ( \delta_\e A_\m \bar{\l})
\g^{\m\n\r} \tr_{z^\prime} (\l^\prime F^\prime_{\n\r} ) \nonumber\\ &+& C c_r^z c^{r
z^\prime } \tr_z (\delta_\e A_\m \bar{\l}  )
\g_\n  \tr_{z^\prime} (\l^\prime F^{\prime \m\n } )\quad ,
\label{completesusyanomaly}
\eeq 
where the coefficients $A$, $B$ and $C$ satisfy the relations
\beq & & A+B =i \quad, \nonumber\\ & & C=4A - 2B \quad . \label{relations}
\eeq These leave one undetermined parameter, in agreement with the well-known fact that
anomalies are defined up to the variation of local functionals. Indeed, adding to the 
supersymmetry anomaly the term
\be
\delta_\e [(\bar{\l}\g^\a \l^\prime )(\bar{\l} \g_\a \l^\prime )]
\ee 
corresponds to adding terms like the last three in eq.  (\ref{completesusyanomaly})
with coefficients satisfying  the relations $A+B=0$ and $C=4A-2B$, that thus preserve
eqs. (\ref{relations}).  One can then show that the last of eqs.  (\ref{wesszumino})
generates terms containing one derivative and   four gauginos, that cancel using the
Dirac equation $\g^\m D_\m \l =0 $. Naturally, something similar also happens in
six-dimensional supergravity,  as we are about to verify.

Returning to the supersymmetry anomaly of eq. (\ref{theanomaly}), one can observe  that
the coefficients of the third, fourth  and fifth terms are consistent with eqs. 
(\ref{relations}). Moreover, demanding that the last of  eqs. (\ref{wesszumino}) be
satisfied  fixes the other gauge-invariant terms to give exactly the anomaly in  eq.
(\ref{theanomaly}). Finally, the Wess-Zumino condition is satisfied only  on-shell, and
one obtains
\beq & & (\delta_{\e_1}{\cal{A}}_{\e_2} -\delta_{\e_2}{\cal{A}}_{\e_1}) = {\cal
A}_{\tilde{\e}} + {\cal A}_{\tilde{\L}}  + c_r^z c^{r z^\prime} \tr_{z,z^{\prime}}
 \lbrace
\frac{ie}{16}\xi_\sigma ( \bar{\l} \g_\m \l^\prime ) (\bar{\l}\g^\m
\g^\sigma [eq.\l^\prime ]_{(\a = 0) }) \nonumber\\ & & -\frac{ie}{32} 
\xi_{i \sigma \delta \tau} \lbrace [\bar{\l}\g^\tau
\l^\prime ]_i (\bar{\l}\g^{\sigma\delta} [eq.\l^\prime ]_{(\a = 0) })+ [\bar{\l}\g^\tau
\l ]_i (\bar{\l}^\prime \g^{\sigma\delta} [ eq. \l^\prime ]_{(\a = 0) }) \rbrace
\rbrace  \quad ,\label{openwz}
\eeq where we have stressed that the corresponding field equation for the gaugini  is
determined by eq. (\ref{completelag}) with $\a = 0$.  To reiterate, the anomaly obtained
for $\a=0$ naturally closes on the corresponding field equation for $\l$. Still, the
identity
\beq 
&& c_r^z c^{r z^\prime} \tr_{z,z^{\prime}} \lbrace 
\frac{ie}{16} \xi_\sigma ( \bar{\l} \g_\m \l^\prime ) (\bar{\l}\g^\m
\g^\sigma [eq.\l^\prime ]_{(\a) }) -\frac{ie}{32} \xi_{i \sigma \delta \tau}
 \lbrace [\bar{\l}\g^\tau
\l^\prime ]_i (\bar{\l}\g^{\sigma\delta} [eq.\l^\prime ]_{(\a) })\nonumber\\ &&+
[\bar{\l}\g^\tau
\l ]_i (\bar{\l}^\prime \g^{\sigma\delta} [ eq. \l^\prime ]_{(\a) }) \rbrace  \rbrace =
c_r^z c^{r z^\prime} \tr_{z,z^{\prime}} \lbrace 
\frac{ie}{16} \xi_\sigma ( \bar{\l} \g_\m \l^\prime ) (\bar{\l}\g^\m
\g^\sigma [eq.\l^\prime ]_{(\a = 0) }) \nonumber\\ & & -\frac{ie}{32} 
\xi_{i \sigma \delta \tau} \lbrace [\bar{\l}\g^\tau
\l^\prime ]_i (\bar{\l}\g^{\sigma\delta} [eq.\l^\prime ]_{(\a = 0) })+ [\bar{\l}\g^\tau
\l ]_i (\bar{\l}^\prime \g^{\sigma\delta} [ eq. \l^\prime ]_{(\a = 0) }) \rbrace \rbrace
\nonumber\\ && +
\frac{ie \a}{8} 
\frac{c_r^z c^{r z^\prime}c_s^{z^\prime}c^{s z^{\prime\prime}}}{v_t c^{t z^\prime}}
\xi_i^{\m\n\r}
\tr_{z,z^\prime ,z^{\prime\prime} }[\bar{\l}\g_\m \l ]_i 
[\bar{\l}^\prime \g_\n
\l^\prime ]_j [\bar{\l}^{\prime\prime}\g_\r
\l^{\prime\prime} ]_j 
\eeq implies that the last term should somehow be generated in the anomaly,  if the
Wess-Zumino condition is to close for any value of $\a$.  In the presence of ${\cal
L}_{\l^4}$, however, the anomaly is modified by eq. (\ref{extraanomaly}), and applying
the last of eqs. (\ref{wesszumino}) to  this term gives
\beq 
[\delta_{\e_1},\delta_{\e_2}]{\cal{L}}_{\l^4} &=& ([ \delta_{\e_1},\delta_{\e_2} 
] e) {\a
\over 2} c_r^z c^{r z^\prime } \tr_{z, z^\prime} [(\bar{\l} \g_\a 
\l^\prime )(\bar{\l} \g^\a \l^\prime )] \nonumber\\ 
&+& 2e \a c_r^z c^{r
z^\prime} \tr_{z,z^\prime}[(\bar{\l}\g_\a \l^\prime )(\bar{\l}\g^\a 
[\delta_{\e_1},\delta_{\e_2}]\l^\prime )] \quad . \label{extrawz}
\eeq 
The commutator in eq. (\ref{extrawz}) is fully known: in particular, 
the coordinate transformation
in the second term  combines with the commutator on $e$ to give a total  divergence,
while gauge and local Lorentz transformations give a vanishing  result. Moreover, the
field equation is obtained from eq.  (\ref{completelag}). The charge in eq. 
(\ref{centralcharge}) thus plays a crucial role: it generates in eq. (\ref{extrawz}) 
precisely
\be
\frac{i e \a}{8}
\frac{c_r^z c^{r z^\prime}c_s^{z^\prime}c^{s z^{\prime\prime}}}{v_t c^{t z^\prime}}
\xi_i^{\m\n\r} \tr_{z,z^\prime ,z^{\prime\prime} } [\bar{\l}\g_\m \l ]_i 
[\bar{\l}^\prime \g_\n
\l^\prime ]_j [\bar{\l}^{\prime\prime}\g_\r
\l^{\prime\prime} ]_j \quad ,
\ee as needed for consistency. Thus, one can understand the rationale behind the
occurrence of the extension  in the  algebra on the gauginos: it lets the Wess-Zumino
conditions close precisely on  the field equations determined by the algebra. Since  the
Wess-Zumino conditions close on the equation of the gauginos,  only these fields perceive
the additional transformation.

\subsection{The energy-momentum tensor} 

The gauge anomaly ${\cal A}_\L = \delta_\L {\cal L}$ naturally satisfies the
condition
\be {\cal A}_\L= -tr (\L D_\m J^\m )\quad, \label{div1}
\ee where $J^\m =0$ is the complete field equation of the vector field. One can
similarly show that the supersymmetry anomaly is related to the field equation
of the gravitino, that we write succinctly
${\cal J}^\m =0$, according to
\be {\cal A}_\e = -(\bar{\e} D_\m {\cal J}^\m ) \quad .\label{div2}
\ee

We would like to stress that the Noether identities (\ref{div1}) and
(\ref{div2}) relate the anomalies to the equations of the fields whose
transformations contain derivatives. This observation has 
a natural application to
gravitational anomalies, that we would now like to elucidate. In fact, 
in analogy with the previous cases one would expect that
\be {\cal A}_{\xi} = \delta_{\xi}{\cal L}= 2 \xi_\m D_\n T^{\m\n} \quad ,
\ee 
where the variation of the metric under general 
coordinate transformations is
\be
\delta g_{\m\n} = -\xi^\a \de_\a g_{\m\n} -g_{\a\n} \de_\m \xi^\a -g_{\m\a} 
\de_\n \xi^\a \quad .
\ee 
Thus, for models {\it without} gravitational anomalies one
would expect that the divergence of the energy-momentum tensor  vanish. Actually,
this is no longer true if other anomalies are present, since  {\it all}
fields, not only the metric, have derivative variations under  coordinate
transformations. For instance, in a theory with gauge and
supersymmetry anomalies, the gravitational anomaly is actually
\be 
{\cal A}_{\xi} = \delta_{\xi}{\cal L}= 2 \xi_\n D_\m T^{\m\n} +
\xi_\n \tr (A^\n D_\m J^\m ) + \xi_\n  (\bar{\psi}^\n D_\m
{\cal J}^\m )\quad. \label{emtnoether}
\ee 
In particular, in our case we are not accounting for gravitational
anomalies, that would  result in higher-derivative couplings, and indeed one can
verify that the divergence of the energy-momentum tensor does not vanish, but
satisfies the relation
\be
D_\m T^{\m\n} =-\frac{1}{2} \tr (A^\n D_\m J^\m )- \frac{1}{2} 
(\bar{\psi}^\n D_\m {\cal J}^\m )\quad. \label{emtdiv}
\ee

\section{Covariant field equations and covariant anomalies} 
\label{comments}
\fancyhead[LO]{{\footnotesize 4.2~~{\it Covariant field equations and anomalies}}}

It is well known
that consistent and covariant gauge anomalies are related  by the divergence of
a local functional \cite{bz}. In six dimensions the residual covariant gauge
anomaly is \cite{as}
\be 
{\cal A}_\L^{cov}=\frac{1}{2} \e^{\m\n\a\b\g\delta} c^{rz}c_r^{z^\prime} \tr_z
(\L F_{\m\n}) \tr_{z^\prime} (F^\prime_{\a\b} F^\prime_{\g\delta} ) 
\quad , \label{covanomaly}
\ee 
and is related to the consistent anomaly by a local counterterm,
\be {\cal A}_\L^{cons} +\tr [\L D_\m f^\m ] ={\cal A}_\L^{cov}\quad
,\label{covcons}
\ee 
where
\be f^\m =c_r^z c^{r z^\prime}  \lbrace -\frac{1}{4}
\e^{\m\n\a\b\g\delta} A_\n \tr_{z^\prime}( F^\prime_{\a\b} F^\prime_{\g\delta} )
-\frac{1}{6} \e^{\m\n\a\b\g\delta}  F_{\n\a} \ 
\w^\prime_{\b\g\delta} \rbrace \quad .\label{fmu}
\ee  
Comparing eq. (\ref{fmu}) with eq. (\ref{theanomaly}) one can see that, to
lowest order in the Fermi fields,
\be {\cal A}_\e = \tr(\delta_\e A_\m f^\m )  \quad,
\ee 
and this implies that the transition from consistent  to  covariant anomalies
turns a model with a supersymmetry anomaly into another without any
\cite{as,fms}. Indeed, six-dimensional supergravity coupled to vector and tensor
multiplets was originally formulated in this fashion in \cite{as} to lowest
order in the Fermi fields, extending the results of Romans 
\cite{romans}. The
resulting vector equation is not integrable. Moreover, the corresponding gauge
anomaly is not the gauge variation of a local functional and does not satisfy
Wess-Zumino consistency conditions.

This result can be generalized naturally, if somewhat tediously, to include terms
of all
orders in the Fermi fields \cite{fr2}. The complete supersymmetry anomaly of eq.
(\ref{theanomaly}) has the form
\be {\cal A}_\e = \tr(\delta_\e A_\m f^\m ) +\delta_\e e_\m{}^a g^\m{}_a \quad ,
\ee 
where  to lowest order $f^\m$ is defined in eq. (\ref{fmu}).  Modifying the
vector equation so that
\be 
(eq. \ A^\m )_{(cov)} \equiv J^\m_{(cov)}
=\frac{\delta {\cal L}}{\delta A_\m } -f^\m \quad ,
\label{modifiedvector}
\ee 
and similarly for the Einstein equation, the resulting theory is
supersymmetric but no longer integrable. The covariant vector field equation
is
\beq 
& & 2 D_\n (v_r F^{\m\n} )-2 G_{rs}\hat{H}^{s\m\n\r}F_{\n\r}
-\frac{i}{2}v_r (\bar{\psi}_\a \g^{\a\b\m\n\r}\psi_\b )F_{\n\r}\nonumber\\ & &
+\frac{i}{2}v_r (\bar{\chi}^m \g^{\m\n\r} \chi^m ) F_{\n\r} -x^M_r
(\bar{\psi}_\a \g^{\a\m\n\r} \chi^M ) F_{\n\r}-i x^M_r x^M_s c^{s z^\prime} 
tr_{z^\prime}
(\bar{\l}^\prime \g^{\m\n\r} \l^\prime )  
F_{\n\r}\nonumber\\ & & +i \sqrt{2} D_\n [v_r
(\bar{\psi}_\r \g^{\m\n} \g^\r \l )]+
\sqrt{2} D_\n [x^M_r (\bar{\chi}^M \g^{\m\n} \l )]\nonumber\\ & &
-\frac{i}{2} F_{\n\r}c_r^{z^\prime} \tr_{z^\prime} (\bar{\l}^\prime
\g^{\m\n\r} 
\l^\prime ) -\frac{i}{2} 
c_r^{z^\prime}\tr_{z^\prime}[ (\bar{\l} \g^{\m\n\r} \l^\prime )
F^\prime_{\n\r}]  - i c_r^{z^\prime}[ (\bar{\l}\g_\n \l^\prime )  F^{\prime
\m\n}]
\nonumber\\  & &  +\frac{1}{2\sqrt{2}} c_r^{z^\prime}\tr_{z^\prime} [(\bar{\l}
\g^\m \g^\n \g^\r 
\l^\prime )(\bar{\l}^\prime \g_\n \psi_\r )]  \nonumber\\  & & + \frac{x^M_s
c^{s z^\prime}}{v_t c^{t z^\prime}} c_r^{z^\prime} \tr_{z^\prime} [\frac{3
i}{2\sqrt{2}} (\bar{\l} \g^\m \l^\prime )(\bar{\l}^\prime \chi^M )
 +\frac{i}{4 \sqrt{2}} (\bar{\l} \g^{\m\n\r} \l^\prime )(\bar{\l}^\prime
\g_{\n\r}\chi^M ) \nonumber\\ & & +\frac{i}{2\sqrt{2}} (\bar{\l} \g_\n
\l^\prime )(\bar{\l}^\prime \g^{\m\n} \chi^M )] \nonumber \\ & & +
c_r^{z^\prime} \tr_{z^\prime} [i\a \hat{F}_{\n\r}(\bar{\l}^\prime
\g^{\m\n\r}\l^\prime )-  i\a(\bar{\l} \g^{\m\n\r}\l^\prime ) 
\hat{F}^\prime_{\n\r} + 6i\a(\bar{\l}\g^\n 
\l^\prime ) \hat{F}^\prime_{\m\n}]\nonumber\\ & & + c_{r}^{z^\prime} \frac{x^M_s
c^{s z^\prime}}{v_t c^{t z^\prime}} \tr_{z^\prime}[i \a \sqrt{2} (\bar{\l}\g^\m
\l^\prime ) (\bar{\l}^\prime \chi^M )
-\frac{i\a}{2\sqrt{2}}(\bar{\l}\g_{\n\r}\chi^M )(\bar{\l}^\prime 
\g^{\m\n\r}\l^\prime )]\nonumber\\ & & + c_{r}^{z^\prime} \frac{x^M_s c^{s
z}}{v_t c^{t z}} \tr_{z^\prime} [-\frac{i \a}{\sqrt{2}}(\bar{\l}\g^\m \l^\prime
)(\bar{\l}^\prime
\chi^M ) +\frac{i \a}{2\sqrt{2}}(\bar{\l} \g^{\m\n\r} \l^\prime )(\bar{\l}^\prime
\g_{\n\r} \chi^M )\nonumber\\ & & -\frac{i \a}{\sqrt{2}}(\bar{\l}\g_\n \l^\prime
)(\bar{\l}^\prime \g^{\m\n}
\chi^M ) ] = 0 \quad ,\label{covvectoralpha}
\eeq
and completes the results in \cite{as} to all orders in the Fermi fields.  Its
divergence satisfies
\be \tr (\L D_\m J^\m_{(cov)} )=-{\cal A}_\L^{cov}\quad ,
\ee where ${\cal A}_\L^{cov}$ contains higher-order Fermi terms:
\beq  
{\cal{A}}_\L^{cov} &=& c^{rz}c_r^{z^\prime} \tr_{z,z^\prime}\lbrace
\frac{1}{2} \e^{\m\n\a\b\g\delta} (\L F_{\m\n})  (F^\prime_{\a\b}
F^\prime_{\g\delta} ) \nonumber\\  
&+& i e \L F_{\n\r} (\bar{\l}^\prime
\g^{\m\n\r}D_\m 
\l^\prime )+\frac{i e}{2} \L D_\m (\bar{\l} \g^{\m\n\r} \l^\prime )
F^\prime_{\n\r}  + ie \L D_\m [ (\bar{\l}\g_\n \l^\prime ) F^{\prime \m\n}]
\nonumber\\  
&-& \frac{e}{2\sqrt{2}} \L D_\m [ (\bar{\l} \g^\m \g^\n
\g^\r 
\l^\prime )(\bar{\l}^\prime \g_\n \psi_\r )  ]\nonumber\\  
&+&  e \L D_\m \lbrace
\frac{ x^M_s c^{s z^\prime}}{v_t c^{t z^\prime}} [-\frac{3 i}{2\sqrt{2}}
(\bar{\l} \g^\m \l^\prime )(\bar{\l}^\prime \chi^M )
 -\frac{i}{4 \sqrt{2}} (\bar{\l} \g^{\m\n\r} \l^\prime )(\bar{\l}^\prime
\g_{\n\r}\chi^M ) \nonumber\\ &-& \frac{i}{2\sqrt{2}}  (\bar{\l} \g_\n
\l^\prime )(\bar{\l}^\prime \g^{\m\n} \chi^M )] \rbrace \nonumber\\
&+& e \L D_\m [-i\a
\hat{F}_{\n\r}(\bar{\l}^\prime \g^{\m\n\r}\l^\prime )+   i\a(\bar{\l}
\g^{\m\n\r}\l^\prime ) 
\hat{F}^\prime_{\n\r} - 6i\a(\bar{\l}\g^\n 
\l^\prime ) \hat{F}^\prime_{\m\n}]\nonumber\\ 
&+&  e \L D_\m \lbrace
\frac{x^M_s c^{s z^\prime}}{v_t c^{t z^\prime}} [-i \a \sqrt{2}
(\bar{\l}\g^\m \l^\prime ) (\bar{\l}^\prime \chi^M )
+\frac{i\a}{2\sqrt{2}}(\bar{\l}\g_{\n\r}\chi^M )(\bar{\l}^\prime 
\g^{\m\n\r}\l^\prime )]\rbrace \nonumber\\ 
&+& e \L D_\m \lbrace \frac{x^M_s c^{s
z}}{v_t c^{t z}}  [\frac{i \a}{\sqrt{2}}(\bar{\l}\g^\m \l^\prime
)(\bar{\l}^\prime
\chi^M ) -\frac{i \a}{2\sqrt{2}}(\bar{\l} \g^{\m\n\r} \l^\prime )(\bar{\l}^\prime
\g_{\n\r} \chi^M )\nonumber\\ &+& \frac{i \a}{\sqrt{2}}(\bar{\l}\g_\n \l^\prime
)(\bar{\l}^\prime \g^{\m\n}
\chi^M ) ]\rbrace
\rbrace \quad .
\label{completecovanomaly}
\eeq 

Finally, one can study the divergence of the Rarita-Schwinger and Einstein 
equations in the covariant model. To this end, let us begin by stating that
the derivation of Noether identities for a system of non-integrable 
equations does
not present difficulties of principle, since these involve only first 
variations. 
Indeed, the only difference with respect to
the standard case of integrable equations is that now 
$\delta {\cal L}$ is not an exact differential in field space.  
Still, all invariance principles reflect themselves
in linear dependencies of the field equations.  Thus, for instance, 
with the covariant
equations obtained from the consistent ones by the redefinition of eq. 
(\ref{modifiedvector}) and by
\be 
(eq. \ e^\m{}_m )_{(cov)}=\frac{\delta {\cal L}}{\delta e_\m{}^m } -g^\m{}_m 
\quad ,\ee
the total $\delta_\e {\cal L}$ vanishes by construction. 
The usual procedure then proves that the 
divergence of the Rarita-Schwinger
equation vanishes for any value of the parameter $\a$. 
On the other hand, the divergence
of the energy-momentum tensor presents some subtleties, already anticipated 
in the previous section, that we 
would now like to 
describe.  In particular, it vanishes to lowest order in the Fermi couplings, 
while it gives a covariant non-vanishing result if all fermion 
couplings are taken into
account.  The subtlety has to do with the behavior of the vector 
under general coordinate transformations,
\be
\delta _\xi A_\m = - \xi^\a \de_\a A_\m - \de_\m \xi^\a A_\a \quad ,
\ee
and with the corresponding full (off-shell) form of the identity of 
eq. (\ref{emtnoether}).
Starting again from the consistent equations, one finds
\be {\cal A}_{\xi} = \delta_{\xi}{\cal L}= 2 \xi_\n D_\m T^{\m\n} +
\xi_\n \tr (A^\n D_\m J^\m ) + \xi_\n \tr (F^{\m\n} J_\m ) 
+ \xi_\n  (\bar{\psi}^\n D_\m
{\cal J}^\m )\quad. \label{emtnoetheroff}
\ee 
Reverting to the covariant form eliminates the divergence of the Rarita-Schwinger equation
and alters the vector equation, so that the third
term has to be retained. The final result is then
\be
D_\m T^{\m\n}_{(cov)} = -\frac{1}{2}  \tr(A^\n D_\m J^\m_{(cov)} )-
\frac{1}{2} \tr(f_\m  F^{\m\n} ) -\frac{1}{2} \tr( A^\n D_\m f^\m )
-\frac{1}{2} e^{\n m} D_\m g^\m{}_m \quad ,
\ee
and is nicely verified by our equations.  In particular, this implies that, to lowest
order in the Fermi couplings, the divergence of $T^{\m\n}_{(cov)}$ vanishes.

\section{PST construction}
\label{comments}
\fancyhead[LO]{{\footnotesize 4.3~~{\it PST construction}}}

In the previous section we have reviewed a number of properties of six-dimensional
$(1,0)$ supergravity coupled to vector and tensor multiplets~\cite{as,fms,frs,rs1,rs2}.  We have 
always confined our
attention to the field equations, thus evading the traditional difficulties met
with the action principles for (anti)self-dual tensor fields.
In this section we would like to complete our discussion, presenting an action
principle for the consistent field equations. What follows is an
application \cite{rs3} of a general method introduced by
Pasti, Sorokin and Tonin (PST), that have shown how to obtain
Lorentz-covariant Lagrangians for (anti)self-dual tensors 
with a single auxiliary field~\cite{pst}. Alternative constructions~\cite{infinite},
some of which preceded the work of PST, need an infinite number of
auxiliary fields, and bear a closer relationship to the BRST formulation
of closed-string spectra~\cite{berkovits}. 
This method has already been applied to a number of systems, including 
$(1,0)$ six-dimensional supergravity 
coupled to tensor multiplets~\cite{6bpst} and type IIB 
ten-dimensional supergravity~\cite{pst2b}, whose (local) gravitational anomaly 
has been shown to reproduce~\cite{anomalypst} the 
well-known results~\cite{agw} of Alvarez-Gaum\'e and Witten.
Still, an action principle for the consistent equations reviewed in the
previous section is of some interest since, as we have seen, these 
six-dimensional models have a number of unfamiliar properties. 

Let us begin by considering a single 2-form with a self-dual  
field strength in six-dimensional Minkowski space. 
The PST lagrangian~\cite{pst}
\be
{\cal L} =\frac{1}{12}H_{\m\n\r} H^{\m\n\r} 
-\frac{1}{4(\de \Xi)^2}
\de^\m \Xi H^-_{\m\n\r} H^{- \s\n\r} \de_\s \Xi \quad ,
\ee
where $H = d B$ and $H^- = H - *H$, is invariant under 
the gauge transformation $\delta B = d \L $,
as well as under the additional gauge transformations 
\be
\delta B = (d \Xi ) \L^\prime \label{1} 
\ee 
and
\ba
& & \delta \Xi = \L^{\prime \prime} \quad , \nonumber \\
& & \delta B_{\m\n} = \frac{\L^{\prime \prime}}{(\de \Xi )^2 } H^-_{\m\n\r} \de^\r \Xi \quad .
\label{2}
\ea
The last two types of gauge transformations can be used to recover
the usual field equation of a self-dual 2-form.
Indeed, the scalar equation results from the 
tensor equation contracted with
\be
\frac{H^-_{\m\n\r} \de^\r \Xi }{(\de \Xi )^2 } \quad ,
\ee
and consequently does not introduce any additional degrees of freedom.
The invariance of eq. (\ref{2}) can then be used to eliminate the scalar field. This
field, however, can not be set to zero, since this choice would clearly make the 
Lagrangian of eq. (\ref{1}) inconsistent. With this proviso, using eq. (\ref{1}) 
one can see that the only solution of the tensor equation 
is precisely the self-duality condition for its field strength.

We now want to apply this construction to six-dimensional supergravity coupled
to vector and tensor multiplets. The theory describes a single self-dual 2-form
\be
\hat{\cal{H}}_{\m\n\r} =v_r \hat{H}^r_{\m\n\r} -\frac{i}{8}(\bar{\chi}^m 
\g_{\m\n\r} \chi^m )
\ee
and $n$ antiself-dual 2-forms
\be
\hat{\cal{H}}^M_{\m\n\r} =x^M_r \hat{H}^r_{\m\n\r} +\frac{i}{4}x^M_r c^{rz}
\tr_z (\bar{\l} \g_{\m\n\r} \l ) \quad .
\ee
The complete Lagrangian is obtained adding the term
\be
-\frac{1}{4}\frac{\de^\m \Xi \de^\s \Xi }{(\de \Xi )^2 }
[ \hat{\cal H}^-_{\m\n\r} \hat{\cal H}^-_{\s}{}^{\n\r} +
\hat{\cal H}^{M+}_{\m\n\r} \hat{\cal H}^{M+}{}_\s{}^{\n\r} ] 
\ee
to the lagrangian derived in Section (4.2).
It can be shown \cite{6bpst} that the 3-form
\be
\hat{K}_{\m\n\r} =\hat{\cal{H}}_{\m\n\r}-
3\frac{\de_{[\m} \Xi \de^\s \Xi }{(\de \Xi )^2 }\hat{\cal{H}}^-_{\n\r ]\s}
\ee
is {\it identically} self-dual, while the 3-forms
\be
\hat{K}^M_{\m\n\r} =\hat{\cal{H}}^M_{\m\n\r}-
3\frac{\de_{[\m} \Xi \de^\s \Xi }{(\de \Xi )^2 }\hat{\cal{H}}^{M +}_{\n\r ]\s}
\ee
are  {\it identically} antiself-dual.
With these definitions, we can display rather simply the complete 
supersymmetry transformations of the fields. Actually, only the transformations of the 
gravitino and of the tensorinos are affected, and become
\ba
& & \delta \psi_\m =\hat{D}_\m \e +\frac{1}{4} \hat{K}_{\m\n\r}
\g^{\n\r}\e +\frac{i}{32}( \bar{\chi}^M \g_{\m\n\r} \chi^M )\g^{\n\r} \e -
\frac{3i}{8} (\bar{\e}\chi^M ) \g_\m \chi^M \nonumber\\
& & \qquad -\frac{i}{8}(\bar{\e}\g_{\m\n}
\chi^M)\g^\n \chi^M 
+\frac{i}{16} (\bar{\e}\g^{\n\r} \chi^M )\g_{\m\n\r} \chi^M -\frac{9i}{8}
v_r c^{rz} \tr_z [(\bar{\e}\g_\m \l ) \l ] \nonumber\\
& & \qquad +\frac{i}{8}v_r c^{rz} \tr_z [(\bar{\e}\g^\n \l ) 
\g_{\m\n} \l ]  -\frac{i}{16}v_r c^{rz} \tr_z [(\bar{\e}\g_{\m\n\r} \l ) \g^{\n\r}\l ]
\quad  ,\nonumber\\
& & \delta \chi^M = \frac{i}{2} x^M_r \hat{\de_\m v^r } \g^\m \e +
\frac{i}{12} \hat{K}^M_{\m\n\r} \g^{\m\n\r}\e  +\frac{1}{2} x^M_r c^{rz} \tr_z 
[(\bar{\e} \g_\m \l) \g^\m \l ]
\quad ,
\ea
while the scalar field $\Xi$
is invariant under supersymmetry~\cite{6bpst,pst2b}.
It can be shown that the complete lagrangian transforms under supersymmetry as
dictated by the Wess-Zumino consistency conditions.

We now turn to describe the corresponding modifications of the supersymmetry algebra. 
In addition to general coordinate, gauge and supersymmetry 
transformations, the commutator of two supersymmetry 
transformations on $B^r_{\mu\nu}$ now generates 
two local PST transformations with parameters
\be
\L^\prime_{r \m} =\frac{\de^\s \Xi}{(\de \Xi )^2 }
(v_r \hat{\cal H}^-_{\s\m\r }-x^M_r \hat{\cal H}^{M+}_{\s\m\r} )\xi^\r 
\qquad , \quad
\L = \xi^\m \de_\m \Xi \quad .\label{pst2}
\ee
The transformation of eq. (\ref{pst2}) on the scalar field $\phi$ 
is opposite to its coordinate transformation, and this gives an interpretation
of the corresponding commutator~\cite{6bpst,pst2b}
\be
[ \delta_1 , \delta_2 ] \Xi =\delta_{gct}\Xi +\delta_{PST}\Xi =0\quad ,
\ee
that vanishes
consistently with the invariance of $\Xi$ under supersymmetry.
Finally, the commutator on the vielbein determines
the parameter of the local Lorentz transformation, that is now
\ba
& & \Omega_{mn} = -\xi^\n (\w_{\n mn} -\hat{K}_{\n mn}
-\frac{i}{8}(\bar{\chi}^M \g_{\n mn }\chi^M ) )\nonumber\\
& & +\frac{1}{2}
(\bar{\chi}^M \e_1 )(\bar{\chi}^M \g_{mn}\e_2 )
-\frac{1}{2}(\bar{\chi}^M \e_2 )(\bar{\chi}^M \g_{mn}\e_1 )\nonumber\\
& & +v_r c^{rz} \tr_z [(\bar{\e}_1 \g_m \l )(\bar{\e}_2 \g_n \l )
-(\bar{\e}_2 \g_m \l )(\bar{\e}_1 \g_n \l ) ]\quad .
\ea
All other parameters remain unchanged while, aside from the extension~\cite{frs}, 
the algebra closes on-shell on the modified field equations of the Fermi fields.
Of course, the resulting field equations reduce to those of Section 1 once
one fixes the PST gauge invariances in order to recover 
the conventional equations for (anti)self-dual tensor fields.

For completeness, we conclude by displaying
the lagrangian of six-dimensional supergravity
coupled to vector and tensor multiplets with the inclusion of the PST term,
\ba
e^{-1}{\cal{L}} & = & -\frac{1}{4}R +\frac{1}{12}G_{rs} H^{r \m\n\r}
H^s_{\m\n\r} -\frac{1}{4} \de_\m v^r \de^\m v_r-\frac{1}{2} v_r c^{rz} \tr_z
(F_{\m\n} F^{\m\n}) 
\nonumber\\ 
&- &  \frac{1}{8e}
\e^{\m\n\a\b\g\delta} c_r^z B^r_{\m\n} \tr_z (F_{\a\b} F_{\g\delta}) 
 -\frac{i}{2} ( \bar{\psi}_\m \g^{\m\n\r} D_\n [\frac{1}{2}(\w +\hat{\w} )]
\psi_\r ) \nonumber\\
& - &\frac{i}{8}v_r [H+\hat{H}]^{r \m\n\r}(\bar{\psi}_\m \g_\n \psi_\r)
+\frac{i}{48} v_r [H+\hat{H} ]^r_{\a\b\g} (\bar{\psi}_\m
\g^{\m\n\a\b\g}\psi_\n ) \nonumber\\
& + & \frac{i}{2} ( \bar{\chi}^M \g^\m D_\m (\hat{\w})
\chi^M )  -\frac{i}{24}v_r \hat{H}^r_{\m\n\r} (\bar{\chi}^M
\g^{\m\n\r}
\chi^M ) \nonumber\\
& +& \frac{1}{4}x^M_r [\de_\n v^r +\hat{\de_\n v^r} ](\bar{\psi}_\m \g^\n
\g^\m \chi^M )  -\frac{1}{8} x^M_r [H+\hat{H}]^{r \m\n\r} (
\bar{\psi}_\m \g_{\n\r} \chi^M )\nonumber\\
& +& \frac{1}{24} x^M_r [H+\hat{H}]^{r \m\n\r} (\bar{\psi}^\a \g_{\a\m\n\r}
\chi^M )+iv_r c^{rz} \tr_z (\bar{\l} \g^\m
\hat{D}_\m \l )\nonumber\\
& +&
\frac{i}{2\sqrt{2}}  v_r c^{rz} \tr_z [(F+\hat{F})_{\n\r}(\bar{\psi}_\m
\g^{\n\r}\g^\m \l )] 
 \nonumber\\  & +& \frac{i}{12} x^M_r x^M_s \hat{H}^r_{\m\n\r} c^{sz} \tr_z
(\bar{\l}\g^{\m\n\r} \l )- \frac{i}{2} x^M_r c^{rz} \tr_z [(\bar{\chi}^M \g^\m \g^\n \l ) 
(\bar{\psi}_\m \g_\n \l )]
\nonumber\\ 
&+&  \frac{1}{16}v_r c^{rz} \tr_z (\bar{\l} \g_{\m\n\r} \l
)(\bar{\chi}^M
\g^{\m\n\r} \chi^M ) 
+\frac{1}{\sqrt{2}}x^M_r c^{rz} \tr_z [(\bar{\chi}^M \g^{\m\n}
\l )\hat{F}_{\m\n} ]
\nonumber\\  
& -& \frac{i}{8}(\bar{\chi}^M
\g_{\m\n}\psi_\r ) x^M_r c^{rz} \tr_z (\bar{\l} \g^{\m\n\r} \l ) 
\nonumber\\ 
& - &  \frac{3}{16}v_r c^{rz}\tr_z [(\bar{\chi}^M \g_{\m\n} \l )
(\bar{\chi}^M \g^{\m\n} \l )] 
 -\frac{1}{8} v_r c^{rz} \tr_z [(\bar{\chi}^M \l )(\bar{\chi}^M \l)]
\nonumber\\ 
& - & \frac{3}{4}\frac{x^M_r c^{rz}x^N_s c^{sz}}{v_t c^{tz}} \tr_z
[(\bar{\chi}^M \l )(\bar{\chi}^N \l )] 
+\frac{1}{8}(\bar{\chi}^M \g^{\m\n\r} \chi^M )(\bar{\psi}_\m \g_\n \psi_\r )
\nonumber\\  
& + & \frac{1}{8}\frac{x^M_r c^{rz} x^N_s c^{sz}}{v_t c^{tz}} \tr_z [(\bar{\chi}^M
\g_{\m\n}
\l )(\bar{\chi}^N \g^{\m\n}\l )]
-\frac{1}{8}(\bar{\chi}^M \g^\m \chi^N )(\bar{\chi}^M \g_\m
\chi^N )
\nonumber\\ & +& \frac{1}{4} (\bar{\psi}_\m \g_\n \psi_\r )
v_r c^{rz} \tr_z(\bar{\l} \g^{\m\n\r}
\l )  -\frac{1}{2} v_r v_s c^{rz}c^{s z^\prime} \tr_{z,z^\prime} [(\bar{\l}\g_\m
\l^\prime )(\bar{\l} \g^\m \l^\prime ) ]\nonumber\\
& + &  \frac{\a}{2}c^{rz} c_r^{z^\prime} \tr_{z,z^\prime} [(\bar{\l}
\g_\m
\l^\prime )(\bar{\l} \g^\m \l^\prime )] \nonumber\\
& -& \frac{\de^\m \Xi \de^\s \Xi }{4(\de \Xi )^2 }
[ \hat{\cal H}^-_{\m\n\r} \hat{\cal H}^-_{\s}{}^{\n\r} +
\hat{\cal H}^{M+}_{\m\n\r} \hat{\cal H}^{M+}{}_\s{}^{\n\r} ]
\quad,
\label{completelagwithpst}
\ea
where $\a$ is the (undetermined) coefficient of the quartic
coupling for the gauginos, 
and the corresponding supersymmetry transformations
\ba
& & \delta e_\m{}^m =-i(\bar{\e} \g^m \psi_\m ) \quad,\nonumber\\  & &
\delta B^r_{\m\n} =i v^r (\bar{\psi}_{[\m} \g_{\n]} \e )+ \frac{1}{2} x^{Mr}
(\bar{\chi}^M
\g_{\m\n} \e )-2c^{rz} \tr_z (A_{[\m}\delta A_{\n]}) \quad,
\nonumber\\  & & \delta v_r = x^M_r (\bar{\chi}^M \e )\quad,\nonumber\\  & &
\delta A_\m = -\frac{i}{\sqrt{2}} (\bar{\e} \g_\m \l ) \quad ,\nonumber\\  & &
\delta \psi_\m =\hat{D}_\m \e +\frac{1}{4} \hat{K}_{\m\n\r}
\g^{\n\r}\e +\frac{i}{32}( \bar{\chi}^M \g_{\m\n\r} \chi^M )\g^{\n\r} \e -
\frac{3i}{8} (\bar{\e}\chi^M ) \g_\m \chi^M \nonumber\\
& & \qquad -\frac{i}{8}(\bar{\e}\g_{\m\n}
\chi^M )\g^\n \chi^M 
+\frac{i}{16} (\bar{\e}\g^{\n\r} \chi^M )\g_{\m\n\r} \chi^M -\frac{9i}{8}
v_r c^{rz} \tr_z [(\bar{\e}\g_\m \l ) \l ] \nonumber\\
& & \qquad +\frac{i}{8}v_r c^{rz} \tr_z [(\bar{\e}\g^\n \l ) 
\g_{\m\n} \l ]  -\frac{i}{16}v_r c^{rz} \tr_z [(\bar{\e}\g_{\m\n\r} \l ) \g^{\n\r}\l ]
\quad  ,\nonumber\\
& & \delta \chi^M = \frac{i}{2} x^M_r \hat{\de_\m v^r } \g^\m \e +
\frac{i}{12} \hat{K}^M_{\m\n\r} \g^{\m\n\r}\e  +\frac{1}{2} x^M_r c^{rz} 
\tr_z [(\bar{\e} \g_\m \l) \g^\m \l ]
\quad
,\nonumber\\  
& & \delta \l =-\frac{1}{2\sqrt{2}}\hat{F}_{\m\n} \g^{\m\n} \e  -
\frac{1}{2} \frac{x^M_r c^{rz}}{v_s c^{sz}} (\bar{\chi}^M \l ) \e 
 - \frac{1}{4} \frac{x^M_r c^{rz}}{v_s c^{sz}} (\bar{\chi}^M \e ) \l  
\nonumber \\ & & \quad \quad + \frac{1}{8} \frac{x^M_r c^{rz}}{v_s c^{sz}}
(\bar{\chi}^M \g_{\m\n} \e ) \g^{\m\n}
\l \quad .
\ea
One further comment is in order.
Kavalov 
and Mkrtchyan \cite{kavmkrt}
obtained long ago a complete action for pure 
d=6 (1,0) supergravity in terms of a single tensor auxiliary field.
Their work may be connected to this special case of our result 
via an ansatz relating their tensor
to the PST scalar. Still, the PST formulation has the virtue of simplicity 
and makes it manifest that the extra degrees of freedom may be locally
eliminated via additional gauge transformations.

\section{Inclusion of abelian vector multiplets}
\label{comments}
\fancyhead[LO]{{\footnotesize 4.4~~{\it Inclusion of abelian vector multiplets}}}

In this section we 
construct the general coupling of $(1,0)$ six-dimensional supergravity 
to $n$ tensor multiplets and {\it abelian} vector multiplets \cite{fr2}. We will see the 
the inclusion of abelian vectors allows the presence of more general couplings, with
respect to the ones we have derived so far.

In this case, indeed, the field strengths of the 2-forms include 
generalized
Chern-Simons 3-forms of the vector fields \cite{cjlp} according to
\be
H^r_{\m\n\r}  = 3 \partial_{[\m} B^r_{\n\r ]} - 3 c^{rab} A_{[\m}^{a}
\partial_{\n}A_{\r ] }^{b}\quad ,
\ee
where the $c^{rab}$ are the constants that determine the gauge part of the
residual anomaly polynomial.
In the complete theory, the anomaly induced by this term would 
cancel against the contribution of fermion loops, 
while the irreducible part of the anomaly polynomial is directly 
absent in consistent models \cite{rdsss,as}. 

The model can be constructed using the same method as before:
the completion to all orders in the Fermi fields of the equations of motion
is obtained requiring the closure of the commutator of two supersymmetry 
transformations on the fermionic field equations. 
All the resulting equations may be conveniently derived from the Lagrangian
\beq  
e^{-1}{\cal{L}} &=& -\frac{1}{4}R +\frac{1}{12}G_{rs} H^{r \m\n\r}
H^s_{\m\n\r} -\frac{1}{4} \de_\m v^r \de^\m v_r-\frac{1}{4} v_r c^{rab} 
F^{a}_{\m\n} F^{b\m\n} 
\nonumber\\ 
&-& \frac{1}{16e}
\e^{\m\n\a\b\g\delta} c_r^{ab} B^r_{\m\n} F^{a}_{\a\b} F^{b}_{\g\delta}  
-\frac{i}{2}\bar{\psi}_\m \g^{\m\n\r} D_\n [\frac{1}{2}(\w +\hat{\w} )]
\psi_\r \nonumber \\
&-& \frac{i}{8}v_r [H+\hat{H}]^{r \m\n\r}(\bar{\psi}_\m \g_\n \psi_\r)
+\frac{i}{48} v_r [H+\hat{H} ]^r_{\a\b\g} (\bar{\psi}_\m
\g^{\m\n\a\b\g}\psi_\n ) \nonumber \\ 
&+&\frac{i}{2} \bar{\chi}^M \g^\m D_\m (\hat{\w})
\chi^M  -\frac{i}{24}v_r \hat{H}^r_{\m\n\r} (\bar{\chi}^M
\g^{\m\n\r}
\chi^M ) \nonumber \\ 
&+&\frac{1}{4}x^M_r [\de_\n v^r +\hat{\de_\n v^r} ](\bar{\psi}_\m \g^\n
\g^\m \chi^M ) -\frac{1}{8} x^M_r [H+\hat{H}]^{r \m\n\r} (
\bar{\psi}_\m
\g_{\n\r}
\chi^M )\nonumber \\ 
&+& \frac{1}{24}x^M_r [H+\hat{H}]^{r \m\n\r} (\bar{\psi}^\a \g_{\a\m\n\r}
\chi^M ) 
+\frac{1}{8}(\bar{\chi}^M \g^{\m\n\r} \chi^M )(\bar{\psi}_\m
\g_\n \psi_\r )\nonumber \\ 
&-& \frac{1}{8}(\bar{\chi}^M \g^\m \chi^N )(\bar{\chi}^M \g_\m
\chi^N ) +
\frac{i}{4\sqrt{2}}  v_r c^{rab} (F+\hat{F})^{a}_{\n\r}(\bar{\psi}_\m
\g^{\n\r}\g^\m \l^{b} ) \nonumber \\ 
&+&\frac{1}{2\sqrt{2}}x^m_r c^{rab}(\bar{\chi}^m \g^{\m\n}
\l^{a} )\hat{F}^{b}_{\m\n}  +\frac{i}{2} v_r c^{rab}  (\bar{\l}^{a} \g^\m
\hat{D}_\m \l^{b} )\nonumber \\ 
&+& \frac{i}{24} x^M_r x^M_s \hat{H}^r_{\m\n\r} c^{sab} 
(\bar{\l}^{a}\g^{\m\n\r} \l^{b} )
+\frac{1}{32}v_r c^{rab} (\bar{\l}^{a} \g_{\m\n\r} 
\l^{b}
)(\bar{\chi}^M
\g^{\m\n\r} \chi^M ) \nonumber\\  
&-& \frac{i}{16}(\bar{\chi}^M
\g_{\m\n}\psi_\r )x^M_r c^{rab}  (\bar{\l}^{a} \g^{\m\n\r} \l^{b} ) 
 - \frac{i}{4} x^M_r c^{rab} (\bar{\chi}^M \g^\m \g^\n \l^{a} ) (\bar{\psi}_\m
\g_\n \l^{b})
\nonumber\\ 
&-& \frac{1}{16} v_r c^{rab}(\bar{\chi}^M \l^{a} )
(\bar{\chi}^M \l^{b})- \frac{3}{32}v_r c^{rab}(\bar{\chi}^M \g_{\m\n} 
\l^{a} )
(\bar{\chi}^M
\g^{\m\n}
\l^{b} ) 
\nonumber\\ 
&+& [(x^M \cdot c )(v \cdot c )^{-1} (x^N \cdot c )]^{ab} \lbrace
-\frac{1}{4}(\bar{\chi}^M \l^a )(\bar{\chi}^N \l^b )+
\frac{1}{16} (\bar{\chi}^N \g_{\m\n} \l^a )(\bar{\chi}^M \g^{\m\n} \l^b )
\nonumber\\
&-& \frac{1}{8} (\bar{\chi}^N \l^a )(\bar{\chi}^M \l^b )\rbrace
+\frac{1}{8} v_{r}c^{rab}(\bar{\psi}_\m \g_\n \psi_\r 
)(\bar{\l}^{a} \g^{\m\n\r}
\l^{b} ) 
\nonumber \\
&-& \frac{1}{8} v_r v_s c^{rab}c^{s cd} (\bar{\l}^{a}\g_\m
\l^c )(\bar{\l}^{b} \g^\m \l^d ) 
+ \frac{\a}{8}c^{rab} c_r^{cd} (\bar{\l}^{a}
\g_\m
\l^c )(\bar{\l}^{b} \g^\m \l^d )
\quad,
\eeq 
after imposing the (anti)self duality conditions. 
The last term, proportional to the arbitrary parameter $\a$, 
vanishes identically in the case of a single 
abelian vector multiplet. Since the kinetic terms of the vector fields 
are non-diagonal, this generalization is only possible 
in the abelian case. 

The variation of this Lagrangian with respect to gauge transformations 
gives the gauge anomaly
\be
{\cal{A}}_{\L}=-\frac{1}{32}\e^{\m\n\a\b\g\d}c_{r}^{ab}c^{rcd}\L^{a}F^{b}_{\m\n}
F^{c}_{\a\b}  F^{d}_{\g\d}\quad ,   
\ee
while the variation with respect to the supersymmetry transformations
\beq 
& & \delta e_\m{}^m =-i(\bar{\e} \g^m \psi_\m ) \quad,\nonumber\\  
& & \delta B^r_{\m\n} =i v^r (\bar{\psi}_{[\m} \g_{\n]} \e )+ \frac{1}{2} x^{Mr}
(\bar{\chi}^M
\g_{\m\n} \e )- c^{rab}  (A^{a}_{[\m}\delta A^{b}_{\n]}) \quad,
\nonumber\\  
& & \delta v_r = x^M_r (\bar{\chi}^M \e )\quad,\nonumber\\  & &
\delta A^{a}_\m = -\frac{i}{\sqrt{2}} (\bar{\e} \g_\m \l^{a} ) \quad ,
\nonumber\\  & &
\delta \psi_\m =\hat{D}_\m \e +\frac{1}{4} v_r \hat{H}^r_{\m\n\r}
\g^{\n\r}\e -\frac{3i}{8} \g_\m \chi^M (\bar{\e} \chi^M ) -\frac{i}{8} \g^\n
\chi^M (\bar{\e} \g_{\m\n} \chi^M ) \nonumber \\
& & \quad \quad +\frac{i}{16} \g_{\m\n\r} \chi^M (\bar{\e} 
\g^{\n\r} \chi^M ) - \frac{9i}{16} v_r 
c^{rab} 
\l^{a} (\bar{\e} \g_\m \l^{b}) \nonumber \\
& & \quad \quad +  
\frac{i}{16} v_r c^{rab} \g_{\m\n} \l^{a} (\bar{\e} \g^\n \l^{b} ) - 
\frac{i}{32}
v_r c^{rab}  \g^{\n\r} \l^{a} (\bar{\e}
\g_{\m\n\r} \l^{b} ) \quad ,\nonumber\\  
& & \delta \chi^M =
\frac{i}{2} x^M_r (\hat{\de_\m v^r} ) \g^\m \e +
\frac{i}{12} x^M_r \hat{H}^r_{\m\n\r} \g^{\m\n\r}\e +
\frac{1}{4} x^M_r c^{rab} \g_\m \l^{a} (\bar{\e} \g^\m \l^{b} )  \quad
,\nonumber\\  
& & \delta \l^{a} =-\frac{1}{2\sqrt{2}}\hat{F}^{a}_{\m\n} 
\g^{\m\n} \e  +[(v \cdot c )^{-1} (x^M \cdot c )]^{ab} \lbrace -
\frac{1}{2} (\bar{\chi}^M \l^{b} ) \e 
- \frac{1}{4} (\bar{\chi}^M \e ) \l^{b}  
\nonumber \\ 
& & \quad \quad + \frac{1}{8} 
(\bar{\chi}^M \g_{\m\n} \e ) \g^{\m\n}
\l^{b} \rbrace 
\eeq 
gives the supersymmetry anomaly
\beq  
{\cal{A}}_\e & &=c_r^{ab} c^{r cd}  \lbrace -\frac{1}{16}
\e^{\m\n\a\b\g\delta}\delta_\e A^{a}_\m A^{b}_\n F^c_{\a\b} F^d_{\g\delta}
-\frac{1}{8}
\e^{\m\n\a\b\g\delta} \delta_\e A^{a}_\m F^{b}_{\n\a} 
A_{\b}^{c}F_{\g\d}^{d} \nonumber\\ & & +\frac{i e}{8} \delta_\e A^{a}_\m 
F^{b}_{\n\r}
(\bar{\l}^c
\g^{\m\n\r} 
\l^d )+\frac{i e}{8} \delta_\e A^{a}_\m (\bar{\l}^{b} \g^{\m\n\r} \l^c )
F^d_{\n\r}  + \frac{ie}{4}\delta_\e A^{a}_\m (\bar{\l}^{b}\g_\n \l^c ) 
F^{d \m\n}
\nonumber\\ & & -
\frac{ie}{128} (\bar{\e}\g^\a \psi_\m ) (\bar{\l}^{a} \g^{\m\n\r} \l^{b} )
(\bar{\l}^c
\g_{\a \n\r} \l^d )  -\frac{e}{8\sqrt{2}} \delta_\e A^{a}_\m 
(\bar{\l}^{b} \g^\m
\g^\n \g^\r 
\l^c )(\bar{\l}^d \g_\n \psi_\r ) \rbrace  \nonumber\\  
& & + c_r^{ab} [c^r (v \cdot c )^{-1} (x^M \cdot c )]^{cd}
\lbrace -\frac{ i}{4\sqrt{2}}
\delta_\e A^{a}_\m (\bar{\l}^{b} \g^\m \l^c )(\bar{\l}^d \chi^M )
\nonumber\\
& & +\frac{i}{16 \sqrt{2}}\delta_\e A^{a}_\m (\bar{\l}^b \g^\m \g^{\n\r}
\l^d )(\bar{\chi}^M \g_{\n\r}\l^c ) 
-\frac{i}{8\sqrt{2}} \delta_\e A^a_\m (\bar{\l}^b \g^\m \l^d )(\bar{\chi}^M
\l^c ) \rbrace \nonumber\\
& & +\frac{\a}{8} c_r^{ab}c^{rcd}\delta_\e \lbrace
e (\bar{\l}^a \g_\m \l^c )(\bar{\l}^b \g^\m \l^d )\rbrace \quad . 
\eeq 
Once again, the complete theory would contain additional non-local 
couplings induced by fermion loops, whose variation would cancel the 
anomalous contribution of the contact terms. Thus,  the low-energy 
couplings that we are displaying are properly 
neither gauge-invariant nor supersymmetric. 
However, gauge and supersymmetry anomalies are related by Wess-Zumino 
consistency conditions, and this grants the coherence of the construction.
The presence of the arbitrary parameter $\a$ reflects the
freedom of adding to the anomaly the variation of a local functional, 
consistently with all Wess-Zumino conditions. 
We have already seen that this anomalous behavior of the low-energy Lagrangian is related to
another remarkable property of these models \cite{frs}:
aside from local symmetry transformations and the equation of motion, 
the commutator of two supersymmetry transformations on the  gauginos generates
the two-cocycle
\beq
 \delta_{(\a)} \l^a &=&
[(v \cdot c )^{-1} c_r ]^{ab} c^{rcd}[-\frac{1}{8}(\bar{\e}_1
\g_\m \l^c )(\bar{\e}_2
\g_\n \l^d ) \g^{\m\n} \l^b -\frac{\a}{4} (\bar{\l}^b \g_\m
\l^c )(\bar{\e}_1 \g_\n \l^d )
\g^{\m\n} \e_2 \nonumber\\  
&+& \frac{\a}{32}(\bar{\l}^b\g_{\m\n\r}\l^c )(\bar{\e}_1 \g^\r 
\l^d ) \g^{\m\n} \e_2 +\frac{\a}{32} (\bar{\l}^b \g_\r
\l^c )(\bar{\e}_1
\g^{\m\n\r} \l^d ) \g_{\m\n} \e_2  \nonumber\\  
&+&  \frac{1-\a}{8} (\bar{\l}^b \g_\m
\l^c ) (\bar{\e}_1
\g^\m \l^d ) \e_2 -(1 \leftrightarrow 2)  \nonumber \\
&+& \frac{1-\a}{32} (\bar{\e}_1 \g^\m \e_2 )(\bar{\l}^c
\g_{\m\n\r} \l^d  )\g^{\n\r}\l^b ] \quad ,
\eeq 
different from zero for any value of $\a$.
In six dimensions the Wess-Zumino conditions
close only on the field equations of the gauginos, and
this two-cocycle actually makes these conditions close for any value 
of $\a$.
In the case of a 
single vector multiplet, in which the term of the Lagrangian 
proportional to $\l^{4}$
disappears, the two-cocycle is still present, 
although it is properly independent of $\a$.

Of course, it is possible to apply the construction of Pasti, Sorokin and 
Tonin (PST) \cite{pst} also in this case, following the results of the 
previous section.
Since the theory describes a single self-dual 2-form
\be
\hat{\cal{H}}_{\m\n\r} =v_r \hat{H}^r_{\m\n\r} -\frac{i}{8}(\bar{\chi}^M 
\g_{\m\n\r} \chi^M )
\ee
and $n$ antiself-dual 2-forms
\be
\hat{\cal{H}}^M_{\m\n\r} =x^M_r \hat{H}^r_{\m\n\r} +\frac{i}{8}x^M_r c^{rab}
(\bar{\l}^a \g_{\m\n\r} \l^b ) \quad ,
\ee
the complete Lagrangian is obtained by the addition of the term
\be
-\frac{1}{4}\frac{\de^\m \Xi \de^\s \Xi }{(\de \Xi )^2 }
[ \hat{\cal H}^-_{\m\n\r} \hat{\cal H}^-_{\s}{}^{\n\r} +
\hat{\cal H}^{M+}_{\m\n\r} \hat{\cal H}^{M +}{}_\s{}^{\n\r} ] \quad ,
\ee
where $\Xi$ is an auxiliary scalar field \cite{pst,6bpst} and 
$H^{\pm}=H \pm *H $. This lagrangian has PST gauge 
invariances needed to cancel the additional degrees of 
freedom.
As before, once the transformations of the gravitino and of the tensorinos
are properly modified,
the supersymmetry algebra generates also the PST 
gauge transformations
\cite{6bpst}.
The field equations obtained from the complete Lagrangian reduce to those
obtained from the Lagrangian without the PST term, once these gauge
invariances are fixed. 

As supersymmetry does not constrain the values of the 
coefficients $c^{r}$, we have obtained a class of models whose anomaly 
polynomials can contain odd powers of the individual 
field strengths $F^{a}$. It is interesting to compare these results with 
\cite{cjlp}. Although for generic values of the $c$'s
the $SO(1,n)$ global symmetry is broken, the authors of \cite{cjlp} 
consider the amusing case of $n=2$ with two abelian vector 
multiplets transforming in the spinorial representation of $SO(1,2)$.
Identifying this group with the one that transforms the tensor fields, 
one obtains an $SO(1,2)$-invariant
Lagrangian if $c^{r}=\g^{0}\g^{r}$. In particular the results of 
\cite{cjlp} correspond to the Majorana representation of $SO(1,2)$:
\be
\g^{0}=\s_{2}\quad , \qquad \g^{1}=i \s_{1} \quad , \qquad \g^{2}=i \s_{3}
\quad ,
\ee
and for this choice the anomaly polynomial vanishes identically.

The transition to tensionless strings corresponds to values of the 
scalar fields for which the gauge coupling vanishes \cite{sw,dlp}. In our 
Lagrangian, this would correspond to the vanishing of some eigenvalues of the 
matrix $v_r c^{r ab}$. In the case of \cite{cjlp} the moduli space
is a two-dimensional hyperboloid, described by the equation $v_0^2 -v_1^2 -v_2^2
=1 $, and one can show that 
the eigenvalues of the matrix $v_r c^{r ab}$ are both positive for
$v_0 \geq 1$ and both negative for $v_0 \leq -1$, so that the transition 
is not reached. 

\section{Inclusion of hypermultiplets}
\label{comments}
\fancyhead[LO]{{\footnotesize 4.5~~{\it Inclusion of hypermultiplets}}}

In this section we describe the full coupling of six-dimensional supergravity 
to vector, tensor and hypermultiplets \cite{fr3}.
In the description of the coupling to hypermultiplets we will follow 
the notation of 
\cite{ns1}. Some details about our conventions are contained in the Appendix.

The spinors in the theory are the
left-handed gravitino $\psi_\m^A$, $n_T$ right-handed tensorinos $\chi^{MA}$,
the left-handed gauginos $\l^A$  
the right-handed hyperinos $\Psi^a$, where $a=1,..., 2n_H$. 
The index $A=1,2$
is in the fundamental representation of $USp(2)$, and the gravitino, the 
tensorinos and the gauginos are $USp(2)$ doublets satisfying 
the symplectic-Majorana condition
\be
\psi^A =\e^{AB} C \bar{\psi}^T_B \quad. 
\ee
The index $a$, instead, is a $USp(2n_H)$ index, and the 
hyperinos satisfy the symplectic-Majorana condition
\be
\Psi^a =\W^{ab} C \bar{\Psi}^T_b \quad,
\ee
where $\W^{ab}$ is the 
antisymmetric invariant tensor of $USp(2n_H)$ (see the appendix for more details). 
The hyper-scalars $\phi^\a$, $\a=1,...,4n_H$, are coordinates of a quaternionic
manifold, that is a manifold whose holonomy group is contained in 
$USp(2) \times USp(2n_H)$.

If the quaternionic manifold parametrized 
by the hyper-scalars has isometries, 
these correspond to global symmetries of the supergravity theory. 
Then the global symmetry group, or a subgroup thereof, can be gauged. 
We recall the notations used to describe 
the scalars in the hypermultiplets.
We denote by $V_\a^{aA}(\phi )$ the vielbein of the quaternionic manifold, 
where the index structure corresponds to the requirement that the 
holomony be contained
in $USp(2)\times USp(2n_H )$. The internal $USp(2)$ and
$USp(2n_H)$ connections are then denoted, respectively, by 
${\cal{A}}_\a^A{}_B$  and ${\cal{A}}_\a^a{}_b$, that in our conventions 
are anti-hermitian matrices. The index $\a =1,...,4 n_H$ is a curved 
index on the quaternionic manifold. 
The field-strengths of the connections are
\beq
& & {\cal{F}}_{\a\b}{}^A{}_B = \de_{\a} {\cal{A}}_\b^A{}_B -\de_{\b} 
{\cal{A}}_\a^A{}_B
+[ {\cal{A}}_\a , {\cal{A}}_\b ]^A{}_B \quad ,\nonumber\\
& & {\cal{F}}_{\a\b}{}^a{}_b = \de_{\a} {\cal{A}}_\b^a{}_b -\de_{\b} 
{\cal{A}}_\a^a{}_b
+[ {\cal{A}}_\a , {\cal{A}}_\b ]^a{}_b \quad ,
\eeq
where $\de_\a = \de / \de \phi^\a $. 
The request that the vielbein $V_\a^{aA}(\phi )$ 
be covariantly constant gives the following relations \cite{baggerwitten}:
\beq
& & V^{\a}_{aA} V^{\b}_{bB}g_{\a\b}=\W_{ab} \e_{AB} \quad ,\nonumber\\
& & V^{\a}_{aA} V^{\b bA} + V^{\b}_{aA} V^{\a bA} =\frac{1}{n_{H}}
g^{\a\b}\delta^a_b 
\quad , \nonumber\\
& & V^{\a}_{aA} V^{\b aB} +V^{\b}_{aA} V^{\a aB}=g^{\a\b}\delta^A_B \quad ,
\eeq 
where $\W_{ab}$ is the antisymmetric invariant tensor of $USp(2n_H)$.
The raising and lowering conventions are collected in the appendix. 
The field-strength of the $USp(2)$ connection ${\cal{A}}_\a^A{}_B$  
is naturally constructed in terms of $V_{\a}^{aA}$ by the relation: 
\be
{\cal{F}}_{\a\b AB}= V_{\a aA}V_\b^a{}_B +V_{\a aB}V_\b^a{}_A \quad ,
\ee
and then the cyclic identity for the internal curvature tensor implies 
that the field-strength of
the $USp(2n_H)$ connection ${\cal{A}}_\a^a{}_b$ has the form
\be
{\cal{F}}_{\a\b ab}= V_{\a aA}V_{\b b}{}^A +V_{\a bA}V_{\b a}{}^A  + \W_{abcd}
V_\a^{dA}V_\b^c{}_A \quad , 
\ee 
where $\W_{abcd}$ is totally symmetric in its indices \cite{baggerwitten}.

In order to describe the gauging of a subgroup of the isometry group, 
we denote the gauge fields 
of this group by $A_\m^i$,
where $i$ takes values in the adjoint representation, 
and the corresponding field-strengths are
\be
F_{\m\n}^i =\de_\m A_\n^i -\de_\n A_\m^i + f^{ijk} A_\m^j A_\n^k \quad,
\ee
where $f^{ijk}$ are the structure constants of the gauge group.
Under the gauge transformation 
\be
\delta A_\m^i = D_\m \L^i 
\ee
the scalars transform as
\be
\delta \phi^\a = \L^i \xi^{\a i} \quad ,
\ee
where $\xi^{\a i}$ are the Killing vectors corresponding to
the isometries that we are gauging.
The covariant derivative for the scalars is then
\be
D_\m \phi^\a =\de_\m \phi^\a - A_\m^i \xi^{\a i} \quad .
\ee

One can correspondingly define the covariant derivatives for the spinors in 
a natural way, adding the composite connections
$D_\m \phi^\a {\cal{A}}_\a$. 
For instance, the covariant derivative for the hyperinos 
$\Psi^a$ will contain the connections $D_\m \phi^\a {\cal{A}}_\a^a{}_b$, 
while the covariant derivative for the gravitino and the tensorinos 
will contain the connections
$D_\m \phi^\a {\cal{A}}_\a^A{}_B$. The covariant derivatives for the 
gauginos $\l^{iA}$ are 
\be
D_\m \l^{iA}=\de_\m \l^{iA} +\frac{1}{4}\w_{\m mn}\g^{mn}\l^{iA}
+ D_\m \phi^\a {\cal{A}}_\a^A{}_B \l^{i B}+ f^{ijk} A_\m^j \l^{kA}
\quad .
\ee
Notice that the gravitino, the tensorinos and the hyperinos 
are not coupled to the gauge vectors through terms
that do not contain the hyper-scalars. 

We now proceed to the construction of the model.
We assume that the gauge group has the form 
$G= \prod_z G_z$, with $G_z$ semi-simple. The scalars in the hypermultiplets
are charged with respect to   
$G_1$.
To lowest order in the Fermi fields, we reproduce the construction of Section 1, 
adding the hypermultiplet couplings. The equations for all 
fields, with the exception of the 2-forms, can be obtained from the lagrangian
\beq
e^{-1}{\cal{L}} &=& -\frac{1}{4}R +\frac{1}{12}G_{rs} H^{r \m\n\r} H^s_{\m\n\r}
-\frac{1}{4} \de_\m v^r \de^\m v_r 
- \frac{1}{2} v_r c^{rz} \tr_z (F_{\m\n}F^{\m\n}) \nonumber\\
& & - \frac{1}{8e}\e^{\m\n\r\sigma\delta\tau}B^r_{\m\n}c_r^z \tr_{z}(F_{\r\sigma}
F_{\delta\tau})
+\frac{1}{2} g_{\a\b}(\phi ) D_\m \phi^\a D^\m \phi^\b 
+\frac{1}{4v_r c^{r1}} {\cal{A}}_\a^A{}_B 
{\cal{A}}_\b^B{}_A \xi^{\a i} \xi^{\b i} 
\nonumber \\
& & -\frac{i}{2}(\bar{\psi}_\m \g^{\m\n\r} D_\n 
\psi_\r )-\frac{i}{2}v_r H^{r \m\n\r}(\bar{\psi}_\m \g_\n \psi_\r)
+\frac{i}{2} (\bar{\chi}^M \g^\m D_\m \chi^M )\nonumber \\ 
& & -\frac{i}{24}v_r H^r_{\m\n\r} (\bar{\chi}^M \g^{\m\n\r}
\chi^M ) +\frac{1}{2}x^M_r \de_\n v^r (\bar{\psi}_\m \g^\n
\g^\m \chi^M) -\frac{1}{2} x^M_r H^{r \m\n\r} ( \bar{\psi}_\m
\g_{\n\r} \chi^M )\nonumber\\
& & +\frac{i}{2}(\bar{\Psi}_a \g^\m D_\m \Psi^a ) +\frac{i}{24}
v_r H^r_{\m\n\r} (\bar{\Psi}_a \g^{\m\n\r}\Psi^a ) - V_\a^{aA}D_\n \phi^\a
(\bar{\psi}_{\m A} \g^\n \g^\m \Psi_a )\nonumber \\
& & +i v_r c^{rz} \tr_z(\bar{\l} \g^\m D_\m \l )
+\frac{i}{\sqrt{2}} v_r c^{rz} \tr_z[F_{\n\r} (\bar{\psi}_\m \g^{\n\r}\g^\m \l )]
\nonumber\\
&+& \frac{1}{\sqrt{2}} x^M_r c^{rz} \tr_z [F_{\m\n} (\bar{\chi}^M \g^{\m\n}\l )]
- \frac{i}{12}c_r^z H^r_{\m\n\r} tr_z (\bar{\l} \g^{\m\n\r} \l)
\nonumber\\
&-& \sqrt{2} V_\a^{aA} \xi^{\a i}(\bar{\l}^i_A \Psi_a ) +\frac{1}{\sqrt{2}}{\cal{A}}_\a^A{}_B 
\left[ i \xi^{\a i }(\bar{\l}^i_A \g^\m \psi_\m^B )
+ \frac{x^M_r c^{r1}}{v_s c^{s1}}\xi^{\a i}
(\bar{\l}^i_A \chi^{MB} ) \right],
\label{lag}
\eeq
after imposing the (anti)self-duality conditions. 
With this prescription, its variation under the supersymmetry transformations
\beq 
& & \delta e_\m{}^m = -i ( \bar{\e} \g^m \psi_\m ) \quad , \nonumber\\ 
& & \delta B^r_{\m\n} =i v^r ( \bar{\psi}_{[\m} \g_{\n]} \e )
+\frac{1}{2} x^{Mr} ( \bar{\chi}^M \g_{\m\n} \e ) 
- 2c^{rz} \tr_z( A_{[\m} \delta A_{\n]} )
\quad , \nonumber\\ 
& & \delta v_r = x^M_r ( \bar{\e} \chi^M ) \quad , \qquad \delta x^M_r = v_r
(\bar{\e} \chi^M )\quad , \nonumber\\
& & \delta \phi^\a = V^\a_{aA} ({\bar{\e}}^A \Psi^a ) \quad ,\nonumber\\
& & \delta A_\m = -\frac{i}{\sqrt{2}} (\bar{\e} \g_\m \l )\quad, 
\nonumber\\
& & \delta \psi_\m^A = D_\m \e^A +\frac{1}{4} v_r H^r_{\m\n\r} \g^{\n\r} \e^A 
\quad , \nonumber\\
& & \delta \chi^{MA} =\frac{i}{2} x^M_r
\de_\m v^r \g^\m \e^A +\frac{i}{12} x^M_r H^r_{\m\n\r} \g^{\m\n\r} 
\e^A \quad ,\nonumber\\
& & \delta \Psi^a = i \g^\m \e_A V_\a^{aA} D_\m \phi^\a \quad , \nonumber\\
& & \delta \l^A = -\frac{1}{2\sqrt{2}} F_{\m\n} \g^{\m\n} \e^A \qquad 
\quad (z \neq 1 ) \quad , \nonumber \\
& & \delta \l^{iA}= -\frac{1}{2\sqrt{2}} F^{i}_{\m\n} \g^{\m\n} \e^A 
-\frac{1}{\sqrt{2}v_r c^{r1}} {\cal{A}}_\a^A{}_B \xi^{\a i} \e^B 
\eeq
gives the supersymmetry anomaly
\bea 
{\cal{A}}_\e &=& -\frac{1}{4} \e^{\m\n\r\sigma\delta\tau} c_r^z c^{r z^\prime} \tr_z (
\delta_\e A_\m A_\n ) \tr_{z^\prime} (F_{\r\sigma} F_{\delta\tau})  \nonumber \\
&-& \frac{1}{6} \e^{\m\n\r\sigma\delta\tau}
c_r^z c^{r z^\prime} \tr_z (
\delta_\e A_\m F_{\n\r} ) \w^{z^\prime}_{\sigma\delta\tau} \quad  ,
\label{susyanomalywithhypers}
\eea
related by the Wess-Zumino conditions to the consistent gauge anomaly
\be 
{\cal{A}}_\L =- \frac{1}{4} \e^{\m\n\r\sigma\delta\tau} c_r^z c^{rz^\prime} \tr_z (\L
\de_\m A_\n ) \tr_{z^\prime} (F_{\r\sigma} F_{\delta\tau} )\quad .
\label{consanomalywithhypers}
\ee
Notice the presence in the lagrangian of the scalar potential
\be
V(\phi )=-\frac{1}{4v_r c^{r1}} {\cal{A}}_\a^A{}_B 
{\cal{A}}_\b^B{}_A \xi^{\a i} \xi^{\b i} \quad .
\label{pot}
\ee
As in rather more conventional gauged models, the potential contains 
interesting informations, and it may be very instructive to study its extrema 
in special cases.  

We now want to extend the results to all orders in the Fermi fields. 
First of all, we define the supercovariant quantities
\beq
& & \hat{\w}_{\m\n\r} = \w^0_{\m\n\r}  -\frac{i}{2} (\bar{\psi}_\m \g_\n \psi_\r 
+\bar{\psi}_\n \g_\r
\psi_\m +\bar{\psi}_\n \g_\m \psi_\r )\quad ,\nonumber\\
& & \hat{H}^r_{\m\n\r} = H^r_{\m\n\r}  -\frac{1}{2} x^{Mr} ( \bar{\chi}^M \g_{\m\n}
\psi_\r +
\bar{\chi}^M 
\g_{\n\r}\psi_\m + \bar{\chi}^M \g_{\r\m} \psi_\n )  \nonumber\\ 
& & \quad \quad - \frac{i}{2} 
v^r (\bar{\psi}_\m \g_\n \psi_\r +\bar{\psi}_\n \g_\r
\psi_\m +\bar{\psi}_\r \g_\m \psi_\n ) \quad ,\nonumber\\
& & \hat{\de_\m v^r} = \de_\m v^r -x^{Mr} (\bar{\chi}^M \psi_\m )\quad ,\nonumber\\
& & \hat{D_\m \phi^\a } =D_\m \phi^a - V^\a_{aA} (\bar{\psi}_\m^A \Psi^a )\quad,
\nonumber\\
& & \hat{F}_{\m\n}= F_{\m\n}+\frac{i}{\sqrt{2}}(\bar{\l} \g_\m 
\psi_\n )-\frac{i}{\sqrt{2}}(\bar{\l} \g_\n \psi_\m )\quad ,
\eeq
and require that the transformation rules for the Fermi fields be supercovariant.
All fermionic terms in the supersymmetry transformations of the Fermi fields 
that are not determined by supercovariance are then obtained 
requiring the closure of the supersymmetry algebra 
on Bose and Fermi fields. 
Moreover, since the supersymmetry algebra on the Fermi 
fields closes only on-shell, in this way one can determine the complete fermionic 
field equations, and from these the complete lagrangian, up to some subtleties
related to the (anti)self-dual forms, that will be described in section 4. 

The complete supersymmetry transformations of the Fermi fields are
\beq
& & \delta \psi_\m^A =D_\m (\hat{\w})\e^A +\frac{1}{4} v_r \hat{H}^r_{\m\n\r}
\g^{\n\r}\e^{A} -\frac{3i}{8} \g_\m \chi^{MA} (\bar{\e} \chi^M ) -\frac{i}{8} 
\g^\n \chi^{MA}
(\bar{\e} \g_{\m\n} \chi^M )\nonumber\\
& & \quad \quad +\frac{i}{16} \g_{\m\n\r} \chi^{MA} (\bar{\e} 
\g^{\n\r} \chi^M ) 
- \frac{9i}{8} v_r c^{rz} \tr_z [\l^A (\bar{\e} \g_\m \l)] +  
\frac{i}{8} v_r c^{rz} \tr_z [\g_{\m\n} \l^A (\bar{\e} \g^\n \l )]\nonumber\\
& & \quad \quad - \frac{i}{16} v_r c^{rz} \tr_z [\g^{\n\r} \l^A (\bar{\e}
\g_{\m\n\r} \l )] -\delta \phi^\a {\cal{A}}_\a^A{}_B \psi_\m^B \quad ,\nonumber\\ 
& & \delta \chi^{MA} =
\frac{i}{2} x^M_r (\hat{\de_\m v^r} ) \g^\m \e^A +
\frac{i}{12} x^M_r \hat{H}^r_{\m\n\r} \g^{\m\n\r}\e^A \nonumber\\
& & \quad \quad + \frac{1}{2} x^M_r c^{rz} \tr_z [ \g_\m \l^A (\bar{\e} \g^\m \l ) ] 
-\delta \phi^\a {\cal{A}}_\a^A{}_B \chi^{MB} \quad ,\nonumber\\ 
& & \delta \Psi^a = i \g^\m \e_A V_\a^{aA} \hat{D_\m \phi^\a} -\delta \phi^\a
{\cal{A}}_\a^a{}_b \Psi^b \quad , \nonumber \\
& & \delta \l^{A}
=-\frac{1}{2\sqrt{2}}\hat{F}_{\m\n} \g^{\m\n} \e^A 
- \frac{x^M_r c^{rz}}{2v_s c^{sz}} (\bar{\chi}^M \l ) \e^A 
- \frac{x^M_r c^{rz}}{4v_s c^{sz}} (\bar{\chi}^M \e ) \l^{A}  
\nonumber \\
& & \quad \quad + 
\frac{x^M_r c^{rz}}{8v_s c^{sz}} (\bar{\chi}^M \g_{\m\n} \e ) \g^{\m\n}
\l^{A} -\delta \phi^\a {\cal{A}}_\a^A{}_B \l^{B}
\qquad \qquad \qquad \quad (z \neq 1 )\quad , \nonumber\\
& & \delta \l^{iA}
=-\frac{1}{2\sqrt{2}}\hat{F}^i_{\m\n} \g^{\m\n} \e^A 
- \frac{x^M_r c^{r1}}{2v_s c^{s1}} (\bar{\chi}^M \l^i ) \e^A 
- \frac{x^M_r c^{r1}}{4v_s c^{s1}} (\bar{\chi}^M \e ) \l^{iA}  
\nonumber \\
& & \quad \quad + 
\frac{x^M_r c^{r1}}{8v_s c^{s1}} (\bar{\chi}^M \g_{\m\n} \e ) \g^{\m\n}
\l^{iA} -\delta \phi^\a {\cal{A}}_\a^A{}_B \l^{iB}
-\frac{1}{\sqrt{2}v_r
c^{r1}} {\cal{A}}_\a^A{}_B \xi^{\a i} \e^B  \quad . 
\eeq
One can compute the commutators of two supersymmetry transformations 
on the Bose fields using these relations, and show that they generate 
the local symmetries:
\be
[\delta_1 , \delta_2 ] = \delta_{gct} + \delta_{Lorentz}+\delta_{susy}+
\delta_{tens}+\delta_{gauge} +\delta_{SO(n)}\quad ,
\ee
where the parameters of generic coordinate, local Lorentz, 
supersymmetry,
tensor gauge, vector gauge and composite $SO(n)$ transformations are respectively
\beq
& & \xi_\m = -i (\bar{\e}_1 \g_\m \e_2 ) \quad ,\nonumber \\
& & \W^{mn} = - i \xi^\m (\hat{\w}_\m{}^{mn} - v_r \hat{H}^r_\m{}^{mn} )
-\frac{1}{2}[(\bar{\chi}^M \e_1 )(\bar{\e}_2 \g^{mn} \chi^M )-(\bar{\chi}^M \e_2 )
(\bar{\e}_1 \g^{mn} \chi^M )] \nonumber \\
& & \quad \quad -v_r c^{rz} tr_z [ (\bar{\e}_1 \g^m \l )(\bar{\e}_2 \g^n \l )
- (\bar{\e}_2 \g^m \l )(\bar{\e}_1 \g^n \l )] \quad,\nonumber\\
& & \zeta^A = \xi^\m \psi^A_\m + V^\a_{aC} 
{\cal{A}}_\a^A{}_B \e_2^B (\bar{\e}_1^C \Psi^a ) 
-V^\a_{aC} {\cal{A}}_\a^A{}_B \e_1^B 
(\bar{\e}_2^C \Psi^a )\quad ,\nonumber\\
& & \L^r_\m = -\frac{1}{2}v^r \xi_\m -\xi^\n B^r_{\m\n} \quad , \nonumber\\
& & \L = \xi^\m A_\m\quad, \nonumber\\
& & A^{MN} = \xi^\m x^{Mr} (\de_\m x^N_r )+(\bar{\chi}^M \e_2 )(\bar{\chi}^N \e_1 )
-(\bar{\chi}^M \e_1 )(\bar{\chi}^N \e_2 )\quad .
\eeq
In order to prove this result, one has to use the (anti)self-duality 
condition for the tensor fields, that to all orders in the Fermi fields 
is
\be
G_{rs} {\hat{\cal{H}}}^s_{\m\n\r} =\frac{1}{6e}\e_{\m\n\r\sigma\delta\tau}
\hat{\cal{H}}_r^{\sigma\delta\tau} \label{asd}
\ee
in terms of the 3-forms \cite{ns1}
\be
\hat{\cal{H}}^r_{\m\n\r}=\hat{H}^r_{\m\n\r} -\frac{i}{8}v^r (\bar{\chi}^M 
\g_{\m\n\r} \chi^M ) +\frac{i}{8} v^r (\bar{\Psi}_a \g_{\m\n\r} \Psi^a )
+\frac{i}{4} c^{rz} \tr_z (\bar{\l} \g_{\m\n\r} \l ) \quad.\label{fs}
\ee
Requiring that the commutator of two supersymmetry transformations 
on the Fermi fields close on-shell then 
determines the complete Fermi field equations.
The equations  obtained in this way are
\beq & & -i\g^{\m\n\r} D_\n (\hat{\w})\psi_\r^A -\frac{i}{4} v_r 
\hat{H}^r_{\n\sigma\delta}
\g^{\m\n\r}\g^{\sigma\delta} \psi_\r^A -\frac{1}{12}x^M_r \hat{H}^{r 
\n\r\sigma}
\g_{\n\r\sigma} \g^\m
\chi^{MA} \nonumber\\ & & +\frac{1}{2} x^M_r (\hat{\de_\n v^r })\g^\n \g^\m 
\chi^{MA} 
+\frac{3}{2}
\g^{\m\n}\chi^{MA} (\bar{\chi}^M \psi_\n )-\frac{1}{4}
\g^{\m\n} \chi^{MA} (\bar{\chi}^M \g_{\n\r} \psi^\r )\nonumber \\
& &  
+\frac{1}{4} \g_{\n\r} \chi^{MA} (\bar{\chi}^M \g^{\m\n} \psi^\r )
-\frac{1}{2}
\chi^{MA} (\bar{\chi}^M \g^{\m\n} \psi_\n ) - i v_r c^{rz} \tr_z
[-\frac{1}{\sqrt{2}}\g^{\n\r}\g^\m \l^A \hat{F}_{\n\r} \nonumber \\ 
& &  + \frac{3i}{4} \g^{\m\n\r}
\l^A (\bar{\psi}_\n
\g_\r \l )-\frac{i}{2} \g^\m \l^A (\bar{\psi}_\n \g^\n \l ) 
+\frac{i}{2}\g^\n 
\l^A (\bar{\psi}_\n
\g^\m \l ) + \frac{i}{4}\g_\r \l^A (\bar{\psi}_\n \g^{\m\n\r}\l )]
\nonumber \\ 
& &  - \frac{i}{2} x^M_r c^{rz} \tr_z [\g_\n \l^A (\bar{\chi}^M 
\g^\n \g^\m \l )]
-V_\a^{aA} \hat{D_\n \phi^\a} \g^\n \g^\m \Psi_a \nonumber \\ & & 
+\frac{i}{\sqrt{2}}{\cal{A}}_\a^a{}_B \xi^{\a i}\g^\m \l^{iB} =0
\eeq
for the gravitino,
\beq 
& & i\g^\m D_\m (\hat{\w})\chi^{M A}-\frac{i}{12}v_r \hat{H}^r_{\m\n\r} \g^{\m\n\r}
\chi^{MA} +\frac{1}{12} x^M_r \hat{H}^r_{ \m\n\r} \g^\sigma \g^{\m\n\r} 
\psi_\sigma^A
+ \frac{1}{2} x^M_r (\hat{\de_\n v^r}) \g^\m \g^\n \psi_\m^A 
\nonumber\\ 
& & + \frac{1}{\sqrt{2}}x^M_r c^{rz} \tr_z (\hat{F}_{\m\n} 
\g^{\m\n} \l^A ) 
- \frac{i}{2} x^M_r c^{rz} \tr_z [ \g^\m \g^\n  \l^A (\bar{\psi}_\m \g_\n
\l )] +\frac{1}{2}\g^\m \chi^{NA} (\bar{\chi}^N \g_\m \chi^M ) \nonumber\\ 
& & -
\frac{3}{8}
v_r c^{rz} \tr_z [(\bar{\chi}^M \g_{\m\n} \l )\g^{\m\n}
\l^A ] - \frac{1}{4} v_r c^{rz} \tr_z [(\bar{\chi}^M \l ) \l^A ] \nonumber\\ 
& & -
\frac{3}{2} \frac{x^M_r c^{rz} x^N_s c^{sz}}{v_t c^{tz}} \tr_z 
[(\bar{\chi}^N \l ) \l^A ]  +
\frac{1}{4}
\frac{x^M_r c^{rz} x^N_s c^{sz}}{v_t c^{tz}} \tr_z [(\bar{\chi}^N \g_{\m\n} \l )
\g^{\m\n} \l^A ]
 \nonumber\\
& & -\frac{x^M_r c^{r1}}{\sqrt{2}v_s c^{s1}}
{\cal{A}}_\a^A{}_B \xi^{\a i} \l^{iB}
=0
\eeq
for the tensorinos,and 
\beq
& & i \g^\m D_\m (\hat{\w})\Psi^a +\frac{i}{12}v_r \hat{H}^r_{\m\n\r}\g^{\m\n\r}
\Psi^a +\g^\m \g^\n \psi_{\m A} V_\a^{aA} \hat{D_\n \phi^\a} - \frac{1}{48} v_r 
c^{rz} \tr_z (\bar{\l} \g_{\m\n\r} \l ) \g^{\m\n\r} \Psi^a \nonumber\\
& &  +\frac{1}{12}
\W^{abcd} \g^\m \Psi_b (\bar{\Psi}_c \g_\m \Psi_d )+\sqrt{2}V_\a^{aA}
\xi^{\a i} \l^i_A  =0
\eeq
for the hyperinos. As usual, more care is needed in order to derive the equations for the 
gauginos, since the $c_r^z c^{r z^\prime}$ terms in the commutator 
of two supersymmetry transformations are
\beq
& & \frac{c_r^z c^{r z^\prime}}{v_s c^{sz}}
\tr_{z^\prime} [-\frac{1}{4}(\bar{\e}_1 \g_\m \l^\prime )(\bar{\e}_2
\g_\n \l^\prime ) \g^{\m\n} \l^A  +
\frac{1}{4} (\bar{\l} \g_\m \l^\prime ) (\bar{\e}_1 \g^\m \l^\prime ) \e_2^A -(1
\leftrightarrow 2)
\nonumber\\ & &  + \frac{1}{16} (\bar{\e}_1 \g^\m \e_2 )(\bar{\l}^\prime
\g_{\m\n\r} \l^\prime  )\g^{\n\r}\l^A ]\quad . \label{extrawithhypers}
\eeq 
If one allows for the term
\be
\a c_r^z c^{r z^\prime} \tr_{z^\prime} [(\bar{\l} \g_\m \l^\prime )
\g^\m \l^{\prime A}]
\ee
in the gaugino field equation, then what remains of eq. (\ref{extrawithhypers}) is
\beq
\delta_{extra(\a)} \l^A &=& \frac{c_r^z c^{r
z^\prime}}{v_s c^{sz}} \tr_{z^\prime} [-\frac{1}{4}(\bar{\e}_1 \g_\m \l^\prime
)(\bar{\e}_2
\g_\n \l^\prime ) \g^{\m\n} \l^A \nonumber\\  
&+& \frac{\a}{2} (\bar{\l} \g_\m \l^\prime
)(\bar{\e}_1 \g_\n \l^\prime )
\g^{\m\n} \e_2^A  + \frac{\a}{16}(\bar{\l}\g_{\m\n\r}\l^\prime )(\bar{\e}_1 \g^\r 
\l^\prime ) \g^{\m\n} \e_2^A \nonumber\\  
&+& \frac{\a}{16} (\bar{\l} \g_\r \l^\prime
)(\bar{\e}_1
\g^{\m\n\r} \l^\prime ) \g_{\m\n} \e_2^A   + \frac{1-\a}{4} (\bar{\l} 
\g_\m \l^\prime )
(\bar{\e}_1
\g^\m \l^\prime ) \e_2^A -(1 \leftrightarrow 2)  
\nonumber\\ 
&+& \frac{1-\a}{16} (\bar{\e}_1 \g^\r \e_2 )(\bar{\l}^\prime
\g_{\m\n\r} \l^\prime  )\g^{\m\n}\l^A ] \quad .\label{central}
\eeq 
As explained in Section (4.1), no choice of $\a$ can eliminate all these terms, 
that play the role of a central charge felt only by the gauginos. This is the 
``classical'' realization of a general feature: anomalies in current 
conservations are accompanied by related anomalies in current 
commutators \cite{anom}. 
When this is properly taken into account, the field
equations for the gauginos are 
\beq 
& & i v_r c^{rz} \g^\m D_\m (\hat{\w})\l^A +\frac{i}{2} (\hat{\de_\m v_r})c^{rz}
\g^\m \l^A + \frac{i}{2\sqrt{2}}v_r c^{rz} \hat{F}_{\n\r} \g^\m \g^{\n\r} 
\psi_\m^A \nonumber \\
& & -
\frac{1}{2\sqrt{2}}x^M_r c^{rz} \hat{F}_{\m\n} \g^{\m\n}\chi^{MA} 
+\frac{i}{12} x^M_r c^{rz} x^M_s \hat{H}^s_{\m\n\r} \g^{\m\n\r}\l^A +
\frac{i}{2} x^M_r c^{rz} (\bar{\chi}^M \l ) \g^\m \psi_\m^A \nonumber \\
& &  +\frac{i}{4}x^M_r c^{rz} (\bar{\chi}^M \psi_\m )\g^\m \l^A - 
\frac{i}{8} x^M_r c^{rz} (\bar{\chi}^M \g_{\n\r} \psi_\m ) \g^{\m\n\r} \l^A 
 - \frac{i}{4}x^M_r c^{rz} (\bar{\chi}^M \g_{\m\n}\psi^\m )\g^\n \l^A \nonumber \\
& & -
\frac{1}{8}v_r c^{rz} (\bar{\l} \chi^M )\chi^{MA} -\frac{3}{16} v_r c^{rz}
(\bar{\l} \g_{\m\n} \chi^M ) \g^{\m\n} \chi^{MA} -\frac{3}{4} 
\frac{x^M_r c^{rz} x^N_s
c^{sz}}{v_t c^{tz}} (\bar{\l} \chi^M )
\chi^{NA} \nonumber\\ & &  
+\frac{1}{8} \frac{x^M_r c^{rz} x^N_s c^{sz}}{v_t c^{tz}} (\bar{\l} 
\g_{\m\n} \chi^M ) \g^{\m\n} \chi^{NA} 
-\frac{1}{96} (\bar{\Psi}_a \g_{\m\n\r} \Psi^a ) \g^{\m\n\r} \l^A
\nonumber \\ & & 
- v_r v_s c^{rz} c^{s z^\prime}
\tr_{z^\prime} [(\bar{\l} \g_\m \l^\prime )
\g^\m \l^{\prime A} ] + \a c_r^z c^{r z^\prime} \tr_{z^\prime} 
[(\bar{\l}\g_\m \l^\prime )
\g^\m \l^{\prime A} ] = 0  \quad .
\label{gauginoeq}
\eeq
Actually to the left-hand side of this equation, valid for the case 
$z\neq 1$, one has to add the terms 
\be
- \sqrt{2} V_\a^{aA} \xi^{\a i} \Psi_a 
+\frac{i}{\sqrt{2}} {\cal{A}}_\a^A{}_B \xi^{\a i } \g^\m \psi_\m^B +
\frac{x^M_r c^{r1}}{\sqrt{2} v_s c^{s1}} 
{\cal{A}}_\a^A{}_B \xi^{\a i} \chi^{MB}
\ee
in the remaining case, i.e. for $\l^{i}$.

Having obtained the complete fermionic field equations, one can add to 
eq. (\ref{lag}) all the terms quartic in the Fermi fields, thus obtaining 
the complete lagrangian
\beq 
e^{-1}{\cal{L}} &=& -\frac{1}{4}R +\frac{1}{12}G_{rs} H^{r \m\n\r} H^s_{\m\n\r}
-\frac{1}{4} \de_\m v^r \de^\m v_r + \frac{1}{2} g_{\a\b}(\phi ) 
D_\m \phi^\a D^\m \phi^\b \nonumber\\
&-&\frac{1}{2} v_r c^{rz} \tr_z (F_{\m\n} F^{\m\n}) -\frac{1}{8e}
\e^{\m\n\r\sigma\delta\tau} c_r^z B^r_{\m\n} \tr_z (F_{\r\sigma} 
F_{\delta\tau}) 
+\frac{1}{4v_r c^{r1}} {\cal{A}}_\a^A{}_B 
{\cal{A}}_\b^B{}_A \xi^{\a i} \xi^{\b i} 
\nonumber\\
&-& \frac{i}{2}(\bar{\psi}_\m \g^{\m\n\r} D_\n [\frac{1}{2}(\w
+\hat{\w} )]
\psi_\r )-\frac{i}{8}v_r [H+\hat{H}]^{r \m\n\r}(\bar{\psi}_\m \g_\n \psi_\r)
\nonumber \\ 
&+& \frac{i}{48} v_r [H+\hat{H} ]^r_{\r\sigma\delta} (\bar{\psi}_\m
\g^{\m\n\r\sigma\delta}\psi_\n )+\frac{i}{2} (\bar{\chi}^M \g^\m D_\m (\hat{\w})
\chi^M ) \nonumber \\ 
&-& \frac{i}{24}v_r \hat{H}^r_{\m\n\r} (\bar{\chi}^M \g^{\m\n\r}
\chi^M ) +\frac{1}{4}x^M_r [\de_\n v^r +\hat{\de_\n v^r} ](\bar{\psi}_\m \g^\n
\g^\m \chi^M ) \nonumber\\ 
&-& \frac{1}{8} x^M_r [H+\hat{H}]^{r \m\n\r} ( \bar{\psi}_\m
\g_{\n\r}
\chi^M )+\frac{1}{24}x^M_r [H+\hat{H}]^{r \m\n\r} (\bar{\psi}^\sigma 
\g_{\sigma\m\n\r} \chi^M ) \nonumber\\ 
&+& \frac{i}{2} (\bar{\Psi}_a \g^\m D_\m (\hat{\w}) \Psi^a ) + \frac{i}{24}
v_r \hat{H}^r_{\m\n\r} (\bar{\Psi}_a \g^{\m\n\r} \Psi^a ) \nonumber\\
&-&  \frac{1}{2} V_\a^{aA}
[ D_\n \phi^\a + \hat{D_\n \phi^\a} ] ( \bar{\psi}_{\m A}\g^{\n} 
\g^{\m} \Psi_{a} )
\nonumber \\
&+& iv_r c^{rz} \tr_z (\bar{\l} \g^\m D_\m (\hat{\w}) \l
)+\frac{i}{12} x^M_r x^M_s \hat{H}^r_{\m\n\r} c^{sz} \tr_z (\bar{\l}\g^{\m\n\r} \l )
\nonumber \\ 
&+&  \frac{i}{2\sqrt{2}}  v_r c^{rz} \tr_z [(F+\hat{F})_{\n\r}(\bar{\psi}_\m
\g^{\n\r} \g^\m \l )] 
 +\frac{1}{\sqrt{2}}x^M_r c^{rz} \tr_z [(\bar{\chi}^M \g^{\m\n}
\l )\hat{F}_{\m\n} ] \nonumber \\
&-& \sqrt{2} V_\a^{aA} \xi^{\a i}(\bar{\l}^i_A \Psi_a ) +\frac{1}{\sqrt{2}}
{\cal{A}}_\a^A{}_B [i \xi^{\a i }(\bar{\l}^i_A \g^\m \psi_\m^B )
+\frac{x^M_r c^{r1}}{v_s c^{s1}}\xi^{\a i}
(\bar{\l}^i_A \chi^{MB} )] \nonumber \\
&+& \frac{1}{8}(\bar{\chi}^M \g^{\m\n\r} \chi^M )(\bar{\psi}_\m
\g_\n \psi_\r )-\frac{1}{8}(\bar{\chi}^M \g^\m \chi^N )(\bar{\chi}^M \g_\m \chi^N )
\nonumber \\  
&+& \frac{1}{8} (\bar{\Psi}_a \g^{\m\n\r} \Psi^a )(\psi_\m \g_\n \psi_\r )
+\frac{1}{48} \W^{abcd} (\bar{\Psi}_a \g_\m \Psi_b )(\bar{\Psi}_c \g^\m \Psi_d )
\nonumber \\
&+& \frac{1}{16}v_r c^{rz}\tr_z (\bar{\l} \g_{\m\n\r} \l )(\bar{\chi}^M
\g^{\m\n\r} \chi^M )  - \frac{i}{8}(\bar{\chi}^M \g_{\m\n}\psi_\r )x^M_r
c^{rz} \tr_z (\bar{\l} \g^{\m\n\r} \l )\nonumber \\
&-&  \frac{i}{2} x^M_r c^{rz} \tr_z 
[(\bar{\chi}^M \g^\m \g^\n \l ) (\bar{\psi}_\m \g_\n \l )]  +\frac{1}{4} 
(\bar{\psi}_\m \g_\n \psi_\r ) v_r c^{rz} \tr_z (\bar{\l} \g^{\m\n\r}
\l )\nonumber\\ 
&-& \frac{1}{8} v_r c^{rz} \tr_z [(\bar{\chi}^M \l )(\bar{\chi}^M \l) ]
- \frac{3}{16}v_r c^{rz}\tr_z [(\bar{\chi}^M \g_{\m\n} \l ) (\bar{\chi}^M
\g^{\m\n}
\l )] 
\nonumber\\ 
&-& \frac{3 x^M_r c^{rz}x^N_s c^{sz}}{4 v_t c^{tz}} \tr_z
[(\bar{\chi}^M \l )(\bar{\chi}^N \l )] 
+\frac{x^M_r c^{rz} x^N_s c^{sz}}{8 v_t c^{tz}} \tr_z [(\bar{\chi}^M 
\g_{\m\n} \l )(\bar{\chi}^N \g^{\m\n}\l )] \nonumber\\ 
&+& \frac{5}{96} v_r c^{rz} \tr_z (\bar{\l} \g_{\m\n\r}\l )(\bar{\Psi}_a 
\g^{\m\n\r}\Psi^a )
 -\frac{1}{2} v_r v_s c^{rz}c^{s z^\prime} tr_{z,z^\prime} [(\bar{\l}\g_\m \l^\prime
)(\bar{\l} \g^\m \l^\prime ) ] \nonumber\\ 
&+& \frac{\a}{2}c^{rz} c_r^{z^\prime} tr_{z,z^\prime} [(\bar{\l} \g_\m
\l^\prime )(\bar{\l} \g^\m \l^\prime )] \quad .
\label{completelagwithhyper}
\eeq
From this lagrangian, in the 1.5 order formalism and using the (anti)self-duality 
conditions of eqs. (\ref{asd}) and (\ref{fs}), one can obtain the remaining
complete bosonic field equations. 
Once more, it is important to notice that this lagrangian in neither gauge 
invariant nor supersymmetric: its variation under gauge transformations produces 
the gauge anomaly of eq. (\ref{consanomalywithhypers}), while its variation under 
the complete supersymmetry transformations produces the complete 
supersymmetry anomaly
\beq {\cal{A}}_\e &=& c_r^z c^{r z^\prime} tr_{z, z^\prime} \lbrace -\frac{1}{4}
\e^{\m\n\r\sigma\delta\tau}\delta_\e A_\m A_\n F^\prime_{\r\sigma} 
F^\prime_{\delta\tau} -\frac{1}{6}
\e^{\m\n\r\sigma\delta\tau} \delta_\e A_\m F_{\n\r} 
\w^\prime_{\sigma\delta\tau} \nonumber\\ 
&+&\frac{i e}{2} \delta_\e A_\m F_{\n\r}
(\bar{\l}^\prime
\g^{\m\n\r} 
\l^\prime )+\frac{i e}{2} \delta_\e A_\m (\bar{\l} \g^{\m\n\r} \l^\prime )
F^\prime_{\n\r}  + ie\delta_\e A_\m (\bar{\l}\g_\n \l^\prime ) F^{\prime \m\n}
\nonumber\\ 
&+& 
\frac{e}{32} \delta_\e e_\m{}^m (\bar{\l} \g^{\m\n\r} \l )(\bar{\l}^\prime
\g_{m \n\r} \l^\prime )  -\frac{e}{2\sqrt{2}} \delta_\e 
A_\m (\bar{\l} \g^\m \g^\n \g^\r 
\l^\prime )(\bar{\l}^\prime \g_\n \psi_\r )  \nonumber\\ 
&+& \frac{e x^M_s c^{s
z^\prime}}{v_t c^{t z^\prime}} [-\frac{3 i}{2\sqrt{2}}
 \delta_\e A_\m (\bar{\l} \g^\m \l^\prime )(\bar{\l}^\prime \chi^M )
 -\frac{i}{4 \sqrt{2}} \delta_\e A_\m (\bar{\l} \g^{\m\n\r} \l^\prime )
(\bar{\l}^\prime \g_{\n\r}\chi^M ) \nonumber\\ 
&-&  \frac{i}{2\sqrt{2}} \delta_\e A_\m  (\bar{\l} \g_\n
\l^\prime )(\bar{\l}^\prime \g^{\m\n} \chi^M )] +\frac{\a}{2} \delta_\e
[ e (\bar{\l} \g_\m \l^\prime )(\bar{\l} \g^\m \l^\prime ) ] \rbrace 
\nonumber \\
&-&\frac{i e c_r^1 c^{rz}}{2 v_s c^{s 1}}
{\cal{A}}_\a^A{}_B \xi^{\a i}  \tr_z [\delta_\e A_\m
(\bar{\l}^i_A \g^\m \l^B )] 
\quad .
\label{completesusyanomalywithhyper}
\eeq

The presence of a term proportional to the parameter $\a$ in eq. 
(\ref{completelagwithhyper}) reflects the general fact that anomalies are defined up to 
the variation of a local functional. 
Gauge and supersymmetry anomalies are in general related by the Wess-Zumino 
consistency conditions \cite{wz}
\beq
& & \delta_\e {\cal{A}}_\L = \delta_\L {\cal{A}}_\e \quad , \nonumber \\
& & \delta_{\e_1} {\cal{A}}_{\e_2} -  \delta_{\e_2} {\cal{A}}_{\e_1}=
{\cal{A}}_\L + {\cal{A}}_\zeta \quad .
\eeq
So the inclusion of hypermultiplets does not alter the peculiarity 
of these six dimensional models: the second 
condition closes only on-shell, and precisely on the gaugino field equations 
\cite{frs}. 
Since the inclusion of the term proportional to $\a$ in the lagrangian 
modifies both these 
equations and the supersymmetry anomaly, there must be some extra terms
that permit the Wess-Zumino conditions to close on-shell 
for every value of $\a$.    
This is precisely the role of the terms in eq. (\ref{central}) in the 
commutator of two supersymmetry transformations on the gauginos, that thus can 
be seen as a transformation needed in order to close the Wess-Zumino conditions
precisely on the field equations determined by the algebra. Since the Wess-Zumino 
conditions need only the equation of the gauginos, only these fields sense the 
additional transformation.

For completion, we observe that the PST method can be naturally applied here 
following the results of Section (4.3):
our theory describes a single self-dual 3-form
\be
\hat{\cal{H}}_{\m\n\r}=v_r \hat{H}^r_{\m\n\r} -\frac{i}{8}(\bar{\chi}^M \g_{\m\n\r}
\chi^M )+ \frac{i}{8}(\bar{\Psi}_a \g_{\m\n\r}\Psi^a ) 
\ee
and $n_T$ antiself-dual 3-forms
\be
\hat{\cal{H}}^M_{\m\n\r}= x^M_r \hat{H}^r_{\m\n\r} +\frac{i}{4}x^M_r 
c^{rz} \tr_z (\bar{\l} \g_{\m\n\r} \l ) \quad .
\ee
The complete Lagrangian is obtained adding to eq. (\ref{completelagwithhyper}) the term
\be
-\frac{\de^\m \Xi \de^\s \Xi}{4 (\de \Xi )^2}[\hat{\cal{H}}^-_{\m\n\r}
\hat{\cal{H}}^-_\s{}^{\n\r} +\hat{\cal{H}}^{M+}_{\m\n\r} 
\hat{\cal{H}}^{M+}{}_\s{}^{\n\r} ]
\quad ,
\ee
where $\Xi$ is an auxiliary field and $H^{\pm} = H \pm * H$. 
The resulting lagrangian is invariant under the additional gauge transformations
\cite{pst}
\be
\delta B^r_{\m\n}= (\de_\m \Xi )\L^r_\n - (\de_\n \Xi )\L^r_\m
\ee
and
\be
\delta \Xi =\L \quad , \qquad \delta B^r_{\m\n} =\frac{\L}{(\de \Xi )^2}
[v^r \hat{\cal{H}}^-_{\m\n\r} -x^{Mr}\hat{\cal{H}}^{M+}_{\m\n\r}]\de^\r \Xi \quad ,
\ee
used to recover the usual field equations for (anti)self-dual forms.
The 3-form
\be
\hat{K}_{\m\n\r} =\hat{\cal{H}}_{\m\n\r}-
3\frac{\de_{[\m} \Xi \de^\s \Xi }{(\de \Xi )^2 }\hat{\cal{H}}^-_{\n\r ]\s}
\ee
is {\it identically} self-dual, while the 3-forms
\be
\hat{K}^M_{\m\n\r} =\hat{\cal{H}}^M_{\m\n\r}-
3\frac{\de_{[\m} \Xi \de^\s \Xi }{(\de \Xi )^2 }\hat{\cal{H}}^{M +}_{\n\r ]\s}
\ee
are  {\it identically} antiself-dual \cite{6bpst}. In order to obtain the complete 
supersymmetry transformations, we have to substitute $\hat{\cal{H}}$ with
$\hat{K}$ in the transformation  of the gravitino and $\hat{\cal{H}}^M$ with
$\hat{K}^M$ in the transformations of the tensorinos. Moreover, the auxiliary scalar
is invariant under supersymmetry \cite{6bpst,pst2b}. 
It can be shown that the complete lagrangian transforms under supersymmetry 
as dictated by the Wess-Zumino consistency conditions.
The commutator of two 
supersymmetry transformations on $B^r_{\m\n}$ 
now generates the local PST transformations
with parameters
\be
\L_{r \m} =\frac{\de^\s \Xi}{(\de \Xi )^2 }
(v_r \hat{\cal H}^-_{\s\m\n }-x^M_r \hat{\cal H}^{M+}_{\s\m\n} )\xi^\n 
\quad , \qquad
\L = \xi^\m \de_\m \Xi \quad ,
\ee
while in the parameter of the local Lorentz transformation the term 
$\hat{\cal{H}}$ is replaced by $\hat{K}$. All other parameters 
remain unchanged. 

Following the results of Section (4.4), one can also generalize these results 
to the case in which abelian vectors are present. We will then consider the gauging 
with respect to abelian subgroups of the isometry group. There are no 
subtleties when the symmetric matrices $c^r_{IJ}$ are diagonal (or 
simultaneously diagonalizable), since in this situation the previous results 
can be straightforwardly applied. We are thus interested in 
the case in which the $c^r_{IJ}$ can not be simultaneously diagonalized.  
To this end, we will consider a model in which only 
these abelian gauge groups are present. 
The most general situation can be obtained  
combining the following results with those obtained previously.

We denote with $A_\m^I$, $I=1,...,m$, the set of abelian vectors, and
the gauginos are correspondingly denoted by $\l^{IA}$. 
We collect here only the final results, since the construction 
follows the same lines 
as in the non-abelian case. All the field equations may then be derived from the 
lagrangian
\beq 
e^{-1}{\cal{L}} & & =-\frac{1}{4}R +\frac{1}{12}G_{rs} H^{r \m\n\r} H^s_{\m\n\r}
-\frac{1}{4} \de_\m v^r \de^\m v_r \nonumber\\
& &  -\frac{1}{4} v_r c^{rIJ} F_{\m\n}^I F^{J \m\n} - \frac{1}{16e}
\e^{\m\n\r\sigma\delta\tau} c_r^{IJ} B^r_{\m\n} F_{\r\sigma}^I 
F_{\delta\tau}^J \nonumber\\
& &  + \frac{1}{2} g_{\a\b}(\phi ) D_\m \phi^\a D^\m \phi^\b 
+\frac{1}{4}[(v \cdot c )^{-1}]^{IJ} 
{\cal{A}}_\a^A{}_B {\cal{A}}_\b^B{}_A \xi^{\a I} \xi^{\b J}
\nonumber\\
& & -\frac{i}{2}(\bar{\psi}_\m \g^{\m\n\r} D_\n [\frac{1}{2}(\w
+\hat{\w} )]
\psi_\r )-\frac{i}{8}v_r [H+\hat{H}]^{r \m\n\r}(\bar{\psi}_\m \g_\n \psi_\r)
\nonumber \\ 
& & +\frac{i}{48} v_r [H+\hat{H} ]^r_{\r\sigma\delta} (\bar{\psi}_\m
\g^{\m\n\r\sigma\delta}\psi_\n )+\frac{i}{2} (\bar{\chi}^M \g^\m D_\m (\hat{\w})
\chi^M ) \nonumber \\ 
& & -\frac{i}{24}v_r \hat{H}^r_{\m\n\r} (\bar{\chi}^M \g^{\m\n\r}
\chi^M ) +\frac{1}{4}x^M_r [\de_\n v^r +\hat{\de_\n v^r} ](\bar{\psi}_\m \g^\n
\g^\m \chi^M ) \nonumber\\ 
& & -\frac{1}{8} x^M_r [H+\hat{H}]^{r \m\n\r} ( \bar{\psi}_\m
\g_{\n\r}
\chi^M )+\frac{1}{24}x^M_r [H+\hat{H}]^{r \m\n\r} (\bar{\psi}^\sigma 
\g_{\sigma\m\n\r} \chi^M ) \nonumber\\ 
& & +\frac{i}{2} (\bar{\Psi}_a \g^\m D_\m (\hat{\w}) \Psi^a ) + \frac{i}{24}
v_r \hat{H}^r_{\m\n\r} (\bar{\Psi}_a \g^{\m\n\r} \Psi^a ) \nonumber\\
& & - \frac{1}{2} V_\a^{aA}
[ D_\n \phi^\a + \hat{D_\n \phi^\a} ] (\bar{\psi}_{\m A} \g^{\n} \g^{\m}
\Psi_{a} )\nonumber \\
& &  +\frac{i}{2}v_r c^{rIJ} (\bar{\l}^I \g^\m D_\m (\hat{\w}) \l^J
)+\frac{i}{24} x^M_r x^M_s \hat{H}^r_{\m\n\r} c^{sIJ} (\bar{\l}^I\g^{\m\n\r} \l^J )
\nonumber \\ 
& & + \frac{i}{4\sqrt{2}}  v_r c^{rIJ} (F+\hat{F})_{\n\r}^I (\bar{\psi}_\m
\g^{\n\r} \g^\m \l^J ) 
 +\frac{1}{2\sqrt{2}}x^M_r c^{rIJ} (\bar{\chi}^M \g^{\m\n}
\l^I )\hat{F}_{\m\n}^J  \nonumber\\ 
& & -\sqrt{2} V_\a^{aA} \xi^{\a I} (\bar{\l}^I_A
\Psi_a ) +\frac{i}{\sqrt{2}} {\cal{A}}_\a^A{}_B 
\xi^{\a I} (\bar{\l}^I_A \g^\m \psi_\m^B )\nonumber \\
& &   +\frac{1}{\sqrt{2}}[(v \cdot c )^{-1} (x^M \cdot c )]^{IJ}
{\cal{A}}_\a^A{}_B \xi^{\a I} ( \bar{\l}^J_A \chi^{MB} ) \nonumber \\
& & +\frac{1}{8}(\bar{\chi}^M \g^{\m\n\r} \chi^M )(\bar{\psi}_\m
\g_\n \psi_\r )-\frac{1}{8}(\bar{\chi}^M \g^\m \chi^N )(\bar{\chi}^M \g_\m \chi^N )
\nonumber \\  
& & +\frac{1}{8} (\bar{\Psi}_a \g^{\m\n\r} \Psi^a )(\psi_\m \g_\n \psi_\r )
+\frac{1}{48} \W^{abcd} (\bar{\Psi}_a \g_\m \Psi_{b} )(\bar{\Psi}_c 
\g^\m \Psi_{d} )
\nonumber \\
& &
 +\frac{1}{32}v_r c^{rIJ} (\bar{\l}^I \g_{\m\n\r} \l^J )(\bar{\chi}^M
\g^{\m\n\r} \chi^M )  - \frac{i}{16}(\bar{\chi}^M \g_{\m\n}\psi_\r )x^M_r
c^{rIJ} (\bar{\l}^I \g^{\m\n\r} \l^J )\nonumber \\
& &  -\frac{i}{4} x^M_r c^{rIJ} 
(\bar{\chi}^M \g^\m \g^\n \l^I ) (\bar{\psi}_\m \g_\n \l^J )  
+\frac{1}{8} 
(\bar{\psi}_\m \g_\n \psi_\r ) v_r c^{rIJ}(\bar{\l}^I \g^{\m\n\r}
\l^J )\nonumber\\ 
& &  -\frac{1}{16} v_r c^{rIJ}(\bar{\chi}^M \l^I )(\bar{\chi}^M \l^J ) 
- \frac{3}{32}v_r c^{rIJ} (\bar{\chi}^M \g_{\m\n} \l^I ) (\bar{\chi}^M
\g^{\m\n} \l^J ) \nonumber\\ 
& & +[ (x^M \cdot c )(v \cdot c )^{-1} (x^N \cdot c )]^{IJ}
[-\frac{1}{4}(\bar{\chi}^M \l^I )(\bar{\chi}^N \l^J ) \nonumber \\
& & +\frac{1}{16} (\bar{\chi}^N
\g^{\m\n} \l^I )(\bar{\chi}^M \g_{\m\n} \l^J ) -\frac{1}{8}(\bar{\chi}^N \l^I )
(\bar{\chi}^M \l^J )]
\nonumber\\ 
& & 
+\frac{5}{192} v_r c^{rIJ} (\bar{\l}^I \g_{\m\n\r}\l^J )(\bar{\Psi}_a 
\g^{\m\n\r}\Psi^a )
 -\frac{1}{8} v_r v_s c^{rIJ}c^{s KL} (\bar{\l}^I \g_\m \l^K
)(\bar{\l}^J \g^\m \l^L )  \nonumber\\ 
& & +\frac{\a}{8}c^{rIJ} c_r^{KL} (\bar{\l}^I \g_\m
\l^K )(\bar{\l}^J \g^\m \l^L )] \quad .
\eeq
The variation of this lagrangian with respect to gauge 
transformations gives the
abelian gauge anomaly
\be
{\cal{A}}_{\L}=-\frac{1}{32}\e^{\m\n\r\sigma\d\tau}c_{r}^{IJ}c^{rKL}\L^{I}F^{J}_{\m\n}
F^{K}_{\r\sigma}  F^{L}_{\d\tau}\quad , \label{abeliananomaly}  
\ee
while its variation with respect to the supersymmetry transformations 
\beq 
& & \delta e_\m{}^m = -i ( \bar{\e} \g^m \psi_\m ) \quad , \nonumber\\ 
& & \delta B^r_{\m\n} =i v^r ( \bar{\psi}_{[\m} \g_{\n]} \e )
+\frac{1}{2} x^{Mr} ( \bar{\chi}^M \g_{\m\n} \e ) 
+ 2c^{rIJ}  A_{[\m}^I \delta A_{\n]}^J 
\quad , \nonumber\\ 
& & \delta v_r = x^M_r ( \bar{\e} \chi^M ) \quad , \qquad \delta x^M_r = v_r
(\bar{\e} \chi^M )\quad , \nonumber\\
& & \delta \phi^\a = V^\a_{aA} ({\bar{\e}}^A \Psi^a ) \quad ,\nonumber\\
& & \delta A_\m^I = -\frac{i}{\sqrt{2}} (\bar{\e} \g_\m \l^I )\quad, 
\nonumber\\
& & \delta \psi_\m^A =D_\m (\hat{\w})\e^A +\frac{1}{4} v_r \hat{H}^r_{\m\n\r}
\g^{\n\r}\e^{A} -\frac{3i}{8} \g_\m \chi^{MA} (\bar{\e} \chi^M ) -\frac{i}{8} 
\g^\n \chi^{MA}
(\bar{\e} \g_{\m\n} \chi^M )\nonumber\\
& & \quad \quad +\frac{i}{16} \g_{\m\n\r} \chi^{MA} (\bar{\e} 
\g^{\n\r} \chi^M ) 
- \frac{9i}{16} v_r c^{rIJ} \l^{IA} (\bar{\e} \g_\m \l^J) + 
\frac{i}{16} v_r c^{rIJ} \g_{\m\n} \l^{IA} (\bar{\e} \g^\n \l^J )\nonumber\\
& & \quad \quad - \frac{i}{32} v_r c^{rIJ} \g^{\n\r} \l^{IA} (\bar{\e}
\g_{\m\n\r} \l^J ) -\delta \phi^\a 
{\cal{A}}_\a^A{}_B \psi_\m^B \quad ,\nonumber\\ 
& & \delta \chi^{MA} =
\frac{i}{2} x^M_r (\hat{\de_\m v^r} ) \g^\m \e^A +
\frac{i}{12} x^M_r \hat{H}^r_{\m\n\r} \g^{\m\n\r}\e^A \nonumber\\
& & \quad \quad +\frac{1}{4} x^M_r c^{rIJ}  \g_\m \l^{IA} (\bar{\e} \g^\m \l^J )  
-\delta \phi^\a {\cal{A}}_\a^A{}_B \chi^{MB} \quad ,\nonumber\\ 
& & \delta \Psi^a = i \g^\m \e_A V_\a^{aA} \hat{D_\m \phi^\a} -\delta \phi^\a
{\cal{A}}_\a^a{}_b \Psi^b \quad , \nonumber \\
& & \delta \l^{IA}
=-\frac{1}{2\sqrt{2}}\hat{F}_{\m\n}^I \g^{\m\n} \e^A +[(v \cdot c)^{-1} (x^M \cdot
c )]^{IJ}[
- \frac{1}{2} (\bar{\chi}^M \l^J ) \e^A 
- \frac{1}{4} (\bar{\chi}^M \e ) \l^{JA}  
\nonumber \\
& & \quad \quad + 
\frac{1}{8} (\bar{\chi}^M \g_{\m\n} \e ) \g^{\m\n}
\l^{JA}] -\delta \phi^\a 
{\cal{A}}_\a^A{}_B \l^{IB} -\frac{1}{\sqrt{2}}[(v \cdot c)^{-1}
]^{IJ} {\cal{A}}_\a^A{}_B \xi^{\a J } \e^B
\eeq
gives the supersymmetry anomaly
\beq {\cal{A}}_\e & &=c_r^{IJ} c^{r KL}  \lbrace -\frac{1}{16}
\e^{\m\n\r\sigma\delta\tau}\delta_\e A_\m^I A_\n^J F^K_{\r\sigma} 
F^L_{\delta\tau} -\frac{1}{8}
\e^{\m\n\r\sigma\delta\tau} \delta_\e A_\m^I F_{\n\r}^J A_\sigma^K F_{\delta\tau}^L
\nonumber\\ 
& & +\frac{i e}{8} \delta_\e A_\m^I F_{\n\r}^J
(\bar{\l}^K
\g^{\m\n\r} 
\l^L )+\frac{i e}{8} \delta_\e A_\m^I (\bar{\l}^J \g^{\m\n\r} \l^K )
F^L_{\n\r}  + \frac{ie}{4}\delta_\e A_\m^I (\bar{\l}^J
\g_\n \l^K ) F^{L \m\n}
\nonumber\\ 
& & 
+\frac{e}{128} \delta_\e e_\m{}^m (\bar{\l}^I \g^{\m\n\r} \l^J )(\bar{\l}^K
\g_{m \n\r} \l^L )  -\frac{e}{8\sqrt{2}} \delta_\e A_\m^I (\bar{\l}^J \g^\m \g^\n 
\g^\r 
\l^K )(\bar{\l}^L \g_\n \psi_\r ) \rbrace \nonumber\\ 
& & + e c_r^{IJ}[ c^r (v \cdot c )^{-1} (x^M \cdot c)]^{KL}
\lbrace -\frac{i}{4\sqrt{2}}\delta_\e A_\m^I (\bar{\l}^J \g^\m \l^K )(
\bar{\chi}^M \l^L )\nonumber \\
& & +\frac{i}{16\sqrt{2}} \delta_\e A_\m^I (\bar{\l}^J
\g^\m \g^{\n\r} \l^L )(\bar{\chi}^M \g_{\n\r} \l^K ) -\frac{i}{8 \sqrt{2}}
\delta_\e A_\m^I (\bar{\l}^J \g^\m \l^L )(\bar{\chi}^M \l^K ) \rbrace\nonumber\\
& & -\frac{ie}{4}c_r^{IJ} [(v \cdot c )^{-1} c^r ]^{KL}
\delta_\e A_\m^I {\cal{A}}_\a^A{}_B \xi^{\a K} (\bar{\l}^L_A \g^\m \l^{JB} )
\nonumber\\
& & +\frac{\a}{8} c_r^{IJ} c^{rKL}\delta_\e
[ e (\bar{\l}^I \g_\m \l^K )(\bar{\l}^J \g^\m \l^L ) ]  \quad .
\eeq
Once again, in the case of the gauginos, aside from 
local symmetry transformations and field equations,
the commutator of two supersymmetry transformations generates the additional 
two-cocycle 
\beq
 \delta_{(\a)} \l^I & &=
[(v\cdot c )^{-1}c_r ]^{IJ}c^{rKL}  [-\frac{1}{8}(\bar{\e}_1
\g_\m \l^K )(\bar{\e}_2
\g_\n \l^L ) \g^{\m\n} \l^J -\frac{\a}{4} (\bar{\l}^J \g_\m
\l^K )(\bar{\e}_1 \g_\n \l^L )
\g^{\m\n} \e_2 \nonumber\\  
& & +\frac{\a}{32}(\bar{\l}^J \g_{\m\n\r}\l^K )(\bar{\e}_1 \g^\r 
\l^L ) \g^{\m\n} \e_2 +\frac{\a}{32} (\bar{\l}^J \g_\r
\l^K )(\bar{\e}_1
\g^{\m\n\r} \l^L ) \g_{\m\n} \e_2  \nonumber\\  
& & + \frac{1-\a}{8} (\bar{\l}^J \g_\m
\l^K ) (\bar{\e}_1
\g^\m \l^L ) \e_2 -(1 \leftrightarrow 2)  \nonumber\\
& & 
+ \frac{1-\a}{32} (\bar{\e}_1 \g^\m \e_2 )(\bar{\l}^K
\g_{\m\n\r} \l^L  )\g^{\n\r}\l^J ] \quad .\label{twococycle}
\eeq 
All the observations made for the non-abelian case are naturally valid also 
here: the theory is obtained by the requirement that the Wess-Zumino
conditions close on-shell, and, as we have already shown, it is determined 
up to an arbitrary quartic coupling for the gauginos. 
In the case of a single vector 
multiplet, in which this quartic coupling
vanishes, the two-cocycle of eq. (\ref{twococycle}) is still present, 
although it is properly independent of $\a$. The tensionless string 
phase transition point in the moduli space of the scalars in the tensor multiplets
now would correspond to the vanishing of some of the 
eigenvalues of the matrix $(v \cdot c )^{IJ}$
\cite{fr2}. 

\section{Discussion}
\label{comments}
\fancyhead[LO]{{\footnotesize 4.6~~{\it Discussion}}}

In the previous Sections we have completed the coupling  of $(1,0)$  six-dimensional
supergravity to tensor, vector and hypermultiplets.  The coupling to tensor multiplets only,
initiated by Romans
\cite{romans},  is of a more conventional nature, and parallels similar constructions in
other supergravity models.   On the other hand, the
coupling  to vector multiplets \cite{as}, originally suggested by perturbative type-I
vacua \cite{bs}, is of a more unconventional nature, since it is induced by the residual
anomaly polynomial left over after tadpole conditions are imposed,
\be 
I_8 = \ - \ \sum_{x,y} \ c^r_x \ c^s_y \ \eta_{rs} \ {\rm tr}_x F^2 \ {\rm tr}_y F^2
\quad .
\ee 
The corresponding Chern-Simons couplings of the two-forms, 
\be 
H^r = dB^r - c^{rz} \w_z \quad ,
\ee 
involve the constants $c^r_z$ and determine related couplings of the other fields. In
particular, the Yang-Mills currents are not conserved, and the  consistent residual
gauge anomaly is accompanied by a corresponding anomaly  in the supersymmetry current
\cite{fms}.  In completing these results to all orders in the Fermi fields, we have come
to terms with another peculiar feature of anomalies, neatly displayed by these
``classical'' field equations: anomalous divergences of gauge currents are typically
accompanied by corresponding anomalies in current commutators \cite{anom}. Indeed, we
have discovered an amusing extension of the supersymmetry algebra on  the gaugini, and
we have linked its presence to an ambiguity in the definition of the supergravity model
via Wess-Zumino consistency conditions.   Whereas typical supergravity constructions
yield a unique result, here one is  free to add to the theory a quartic coupling for the
gaugini
\be 
{\cal{L}}_{\l^4} =  \frac{e \a}{2} \ c_r^z c^{r z^\prime} \tr_{z, z^\prime}
[(\bar{\l} \g^\a
\l^\prime )(\bar{\l}\g_\a \l^\prime ) ] \quad ,
\ee 
whose presence affects only the supersymmetry anomaly.  The Wess-Zumino conditions
for six-dimensional supergravity close only on the field equation of the gaugini, and
are consistent for any choice of $\a$ only thanks to the presence of the extension, as
discussed in Section 3.3. Finally, we should mention that the singular gauge couplings
$v^r c_r^z$ of ref. \cite{as} are accompanied by corresponding divergent fermionic
couplings, while the inclusion of charged hypermultiplets gives an additional contribution to the
supersymmetry anomaly.
 
It would be interesting to study in some detail the vacua of the 
lagrangian (\ref{completelagwithhyper}), analyzing the extrema of the potential 
(\ref{pot}). 
As a simple example, consider the model without
hypermultiplets, in which one can gauge the global 
R-symmetry group $USp(2)$ of the theory. Formally, the gauged theory 
without hypermultiplets is obtained from the theory described previously
putting $n_H=0$ and making the identification
\be
{\cal{A}}_\a^A{}_B \xi^{\a i} \rightarrow - i T^{iA}{}_B \quad,
\ee
where $T^i$ are the hermitian generators of $USp(2)$. This 
corresponds to the replacement of the previous 
couplings between gauge fields and spinors, 
dressed by the scalars in the case $n_H \neq 0$, with ordinary minimal couplings:
\be
D_\m \phi^\a {\cal{A}}_\a^A{}_B \e^B \rightarrow i A_\m^A{}_B \e^B \quad .
\ee
Implementing this identification gives in this case the positive-definite
potential  
\be
V =\frac{3}{8 v_r c^{r1}} 
\ee
for the scalars in the tensor multiplets. 
One would thus expect that in these models supersymmetry be spontaneously broken. 
Notice that this potential diverges at the tensionless string phase transition 
point. 
Similarly, one could  try to study explicitly the behavior of the potential 
in simple models containing charged hypermultiplets. 
Their dimensional reduction  gives N=2 supergravity
coupled to vector and hypermultiplets in five dimensions, and 
in the context of the AdS/CFT
correspondence  and its generalizations \cite{m}
there is a renewed interest in studying the explicit
gauging  of these five-dimensional models (see, for instance, 
\cite{bc} and references therein). Notice that in five dimensions
the anomaly that results from
the dimensional reduction of our model  can be canceled by a 
local counterterm, and thus the low-energy effective action does not 
present the subtleties of the six-dimensional case 
\cite{fms}.

The couplings we have derived here are the most general couplings of 
$(1,0)$ six-dimensional supergravity to vector, tensor and hypermultiplets. 
One may wonder if one had the option to gauge a subgroup of $SO(1,n_T)$, 
the isometry group of the scalars in the tensor multiplets. Of course, 
we do not know 
how to write a gauge covariant field-strength for interacting antisymmetric tensor fields, 
but there is a more direct reason why this gauging is not expected  
to work, namely
the fact that once we couple vector and tensor multiplets, the $SO(1,n_T)$ 
transformations are no longer  global symmetries of the theory, because of 
the presence of the matrices $c^r$.

\newpage ~                  %  per lasciare una pagina bianca
\thispagestyle{empty}       %  

\chapter{Low-energy actions for brane supersymmetry breaking}
\label{cap5}
% \newpage
\vspace*{2cm}
\fancyhead[RO,LE]{\thepage}
\fancyhead[RE]{{\footnotesize {\rm Chapter 5.}~~{\it Low-energy actions for brane supersymmetry breaking}}} 
\fancyhead[LO]{}

\noindent

In Chapter 3 we have shown that 
orientifold vacua \cite{cargese,dudrev} allow the simultaneous presence of 
supersymmetric bulks, with one or more gravitinos, and
non-supersymmetric combinations of BPS branes. The resulting 
``brane supersymmetry breaking'' can be realized in stable
configurations, in ten dimensions with 
only anti-$D9$ (${\bar{D}}9$) branes \cite{sugimoto}, 
and in six and four dimensions, up to 
T-dualities, 
with tachyon-free combinations of $D9$ branes and anti-$D5$ (${\bar{D}}5$) 
branes \cite{bsb}. Since this is only one of the options offered
by this class of models for the breaking of supersymmetry, a 
fundamental issue in attempting to relate string
theory to low-energy physics, it is instructive to briefly review our 
current knowledge in this respect.
 
In perturbative string vacua, one has actually 
four options for supersymmetry breaking. The first is to break
supersymmetry from the start, 
so that no gravitinos are present, and the resulting
models, descendants of the type-0 models of \cite{0a0b}, have in general
tachyonic modes \cite{bs}, although a special Klein-bottle projection, 
suggested by
the WZW constructions of \cite{pss}, leads to the $0'$B
model \cite{augusto0b}, that is free of tachyons, a property shared by its
compactifications \cite{carlo} and neatly rooted in 
its brane content \cite{dm1}.
The second option is the Scherk-Schwarz 
mechanism \cite{ss}, in which the breaking, induced by
deformed harmonic expansions in the internal space, is at the 
compactification scale.  In this setting, widely studied in the context
of models of oriented closed strings \cite{ssclosed}, the
presence of branes allows the new 
option of correlating the Scherk-Schwarz deformations
to the brane geometry, giving rise, in particular, to the phenomenon of
``brane supersymmetry'', whereby one or more residual 
global supersymmetries are left, to lowest order, for the brane modes
\cite{ssopen}.
The third option, magnetic deformations \cite{ft}, resorts to the
different magnetic moments of the various fields to induce supersymmetry
breaking  \cite{bachas}, again at the compactification scale, 
but the resulting vacua, that have also T-dual descriptions in terms
of branes at angles \cite{douglas}, generally contain tachyons \cite{berlin}, 
aside from some special instanton-like stable configurations that
recover supersymmetry, albeit with gauge groups of reduced rank \cite{aads}.
Finally, one has the option of brane supersymmetry breaking 
\cite{sugimoto,bsb}, made possible by the presence
of two types of $O$-planes. Together with the conventional $O_-$,
with negative tension and negative R-R charge, there are indeed additional
BPS objects, the  $O_+$ planes,
with positive tension and positive R-R charge, and while the two can
coexist in supersymmetric Klein-bottle projections, the saturation
of the $O_+$ charge requires the presence of anti-branes, with the
result that supersymmetry is broken on the latter {\it at the string
scale}. It is the rigidity of the breaking scale, together with some
special features of the resulting low-energy effective field theories, 
that typically do not allow a gravitino mass term,
that makes the explicit construction of the goldstino couplings
quite interesting in this case. 

Dudas and Mourad \cite{dm2} have shown that, in 
the simplest model with brane supersymmetry breaking, the 
$USp(32)$ model of \cite{sugimoto}, the low-energy gravitino
couplings reflect a non-linear realization of local supersymmetry 
{\it \`a la} Volkov-Akulov, 
along the lines of \cite{samwess}, and their work
is the starting point for our considerations. Let us stress that, while
all branes, including the supersymmetric ones, result in the non-linear
realization {\it \`a la} Born-Infeld of the supersymmetries broken by their 
presence, here one arrives at a complete breaking, and the peculiarity
with respect to lower-dimensional settings for the super-Higgs mechanism
is the absence of a gravitino mass term. This feature is common to the
case analyzed in \cite{dm2} and to the lower-dimensional models of \cite{bsb},
that we shall also discuss in this Chapter \cite{pr}. Actually, all these
configurations, even the supersymmetric ones, can accommodate 
additional brane-antibrane pairs of identical
dimensions, that are to be
spatially separated in order to lift the resulting tachyons. These 
additional pairs provide in their own right additional ways 
to realize brane supersymmetry breaking, but have clearly potential 
tachyon instabilities \cite{abg} for their geometric moduli, in view of 
the mutual attraction of identical branes
and antibranes. We shall thus confine our attention to 
the ``minimal'' configurations of \cite{sugimoto,bsb} demanded by tadpole 
cancellation,
although the other pairs could be described along similar lines. 
Still, we should mention that non-minimal brane-antibrane configurations 
are also quite interesting, and are currently the object of a considerable 
activity as a string setting for brane-world extensions of the Standard Model \cite{berlin,ibanez}.

All models with brane supersymmetry breaking contain a candidate goldstino
among their brane modes, and in \cite{dm2} Dudas and Mourad indeed
constructed the low-energy couplings of the goldstino for 
the ten-dimensional $USp(32)$ Sugimoto model of \cite{sugimoto} up to quartic 
fermionic terms. These were all shown to be of a geometric nature,
being induced by the dressing
of bulk fields with additional terms depending on the goldstino in all
their couplings to the non-supersymmetric brane matter,
aside from some Wess-Zumino-like terms resulting from the supersymmetrization 
of the Chern-Simons couplings. The geometric nature of the
dressing implies that
non-linear supersymmetries of the matter sector take the form of
gaugino-dependent general coordinate transformations.
In this chapter, following the results of \cite{pr}, 
we extend the work of \cite{dm2}, showing
that, up to quartic fermionic terms, the {\it{whole}} low energy effective
Lagrangian of the Sugimoto model, including the Chern-Simons couplings, 
has a geometric nature when expressed 
in terms of the dual $6$-form gauge field, rather than of the more familiar 
$2$-form.  The starting point in this case is thus the
low-energy supergravity built long ago by Chamseddine \cite{chamseddine},
rather than the model of \cite{bdrdwvn,chapmant}. The ten-dimensional Chern-Simons 
terms become in this way higher derivative couplings that, as such, 
do not appear in the low-energy effective action, while a
Wess-Zumino term must be added, and this can be simply ``geometrized'' 
dressing the six-form along the lines of \cite{dm2}.  We also 
extend the analysis of the low-energy effective action to 
six dimensional models with brane supersymmetry breaking.
The starting point in this case is provided by the low-energy (1,0)
effective actions of \cite{as,fms,frs,rs1,rs2,fr3}. These, however, include both
Wess-Zumino and Chern-Simons couplings for the gauge fields, and as
a result have the subtle feature of embodying reducible gauge anomalies to 
be canceled by fermion loops. This peculiar feature,
not present in the earlier constructions of \cite{ns1} motivated by
perturbative heterotic strings, links these constructions to the
Wess-Zumino conditions for the anomalies, with the end result that
many familiar properties of current algebra find in this case an
explicit local realization. In order to 
write a covariant action for the resulting (anti)self-dual 3-forms,
we shall resort to the method of Pasti-Sorokin-Tonin \cite{pst}. The remaining 
couplings are determined requiring that supersymmetry be
non-linearly realized as in the ten-dimensional case, but the
simultaneous presence of Chern-Simons and Wess-Zumino terms produces
a novel effect.  Indeed, while the action is still determined by the 
underlying geometrical structure, as is often the case with Wess-Zumino 
terms, only the field equations are geometrical in this case, aside from 
anomalous terms that arise in the presence of vectors from both 
supersymmetric and non-supersymmetric sectors.

The chapter is organized as follows.  In Section 1 we review the 
low-energy effective couplings built by Dudas and Mourad \cite{dm2}
for the ten-dimensional model and exhibit their geometric nature 
in terms of the 6-form potential. Section  2 is devoted to the
six-dimensional non-linear realizations, and  
finally, Section 3 contains a discussion of the results.

\section{Low-energy couplings for the Sugimoto model}
\label{comments}
\fancyhead[LO]{{\footnotesize 5.1~~{\it Low-energy couplings for the Sugimoto model}}}

This section builds on \cite{dm2}, where 
the low-energy effective action for the $USp(32)$ model was 
constructed, to lowest order in the Fermi fields, requiring that supersymmetry 
be non-linearly realized on the ${\bar{D}}9$ branes, and thus obtaining 
consistent couplings for the gravitino. Our aim is to show  
how {\it all} the couplings of \cite{dm2} can be written in a geometric form \cite{pr}. 

Let us start by reviewing the work of \cite{dm2}. 
As we have seen in Chapter 3, the Sugimoto model 
results from a different IIB orientifold projection
with respect to the one leading to the type-I $SO(32)$ theory. The closed 
spectrum is identical in the two cases, and comprises at the massless
level the (1,0) supergravity multiplet, with the vielbein
$e_\m{}^m$, a 2-form $B_{\m\n}$, the dilaton $\phi$, 
a left-handed gravitino $\psi_\m$ and a right-handed dilatino $\chi$, while 
the open sector describes a $USp(32)$ gauge group, whose gauge
boson $A_\m$ is accompanied by a massless Majorana-Weyl spinor 
in the {\it reducible} antisymmetric representation, that contains a 
spinor $\l$ in the {\bf 495} and a
singlet $\th$.   
Dudas and Mourad identified in \cite{dm2} this singlet as the goldstino of 
supersymmetry, that is non-linearly realized on the brane modes, 
consistently with its breaking at the string scale. 
Starting from this, they constructed the low-energy effective action for 
the Sugimoto model up to quartic terms in the spinors.

Let us briefly review how a single spinor can be treated {\it \`a la} 
Volkov-Akulov \cite{samwess} as a goldstino of global supersymmetry. Let us 
restrict our attention to the ten dimensional case, considering a 
Majorana-Weyl fermion $\th$ with the 
supersymmetry transformation
\be
\d \th =\e  - \frac{i}{2}(\bar{\e} \g^\m \th ) \de_\m \th \label{va}.
\ee
The commutator of two such transformations is a translation,
\be
[\d_1 , \d_2 ] \th = -i (\bar{\e}_2 \g^\m \e_1 )  \de_\m \th \quad ,
\ee
and thus eq. (\ref{va}) provides a realization of supersymmetry.
In order to write a 
Lagrangian for $\th$ invariant under eq. (\ref{va}), let us define
the 1-form
\be
e_\m{}^m= \d_\m^m -\frac{i}{2}(\bar{\th} \g^m \de_\m \th )\quad ,
\ee
whose supersymmetry transformation is 
\be
\d e^m = -L_\xi e^m \quad ,
\ee
with $L_\xi$ the Lie derivative with respect to
\be
\xi_\m= -\frac{i}{2}(\bar{\th}\g_\m \e ) \quad . \label{xi}
\ee
The action of supersymmetry on $e$ is thus a general coordinate 
transformation, with a parameter depending on $\th$, and therefore
\be
{\cal L} = -\det e 
\ee
is clearly an invariant Lagrangian. 
Expanding the determinant, one can see that the energy has a positive 
vacuum expectation value, and supersymmetry is thus spontaneously broken.
Using the same technique, for a generic field $A$ that transforms 
under supersymmetry as 
\be
\d A = - L_\xi A \quad , \label{susygct}
\ee
defining the induced metric as $g_{\m\n}=e_\m{}^m e_{\n m}$, a 
supersymmetric Lagrangian in flat space is determined by
the substitution
\be
{\cal L}(\eta,A) \rightarrow e{\cal L}(g,A) \quad .
\ee

We can now review the results of \cite{dm2}, and to this end
we begin by considering the Lagrangian for the closed sector
\bea
e^{-1} {\cal{L}}_{closed}  =
&-& \frac{1}{4} R +\frac{1}{2}\de_\m \phi \de^\m \phi +\frac{1}{6}
e^{-2\phi}H_{\m\n\r} H^{\m\n\r} \nonumber\\
&-&\frac{i}{2} (\bar{\psi}_\m \g^{\m\n\r} D_\n \psi_\r )+\frac{i}{2}(\bar{\chi}
\g^\m D_\m \chi )+\frac{1}{\sqrt{2}}(\bar{\psi}_\m \g^\n \g^\m \chi )\de_\n \phi 
\nonumber\\
&-& \frac{i}{12\sqrt{2}} e^{-\phi} H_{\m\n\r} (\bar{\psi}_\s \g^{\s\d\m\n\r}\psi_\d
)+\frac{i}{2\sqrt{2}}e^{-\phi} H_{\m\n\r} (\bar{\psi}^\m \g^\n \psi^\r)\nonumber \\
&+& \frac{1}{12}
e^{-\phi}H_{\m\n\r} (\bar{\psi}_\s \g^{\m\n\r} \g^\s \chi) \quad ,
\label{sugra}
\eea
that provides a linear realization of the minimal (1,0) ten-dimensional
supersymmetry, and is thus invariant under the local 
supersymmetry transformations
\bea
& & \d e_\m{}^m = -i (\bar{\e} \g^m \psi_\m )\quad , \nonumber\\
& & \d B_{\m\n}= -\frac{i}{\sqrt{2}}e^\phi (\bar{\e}\g_{[\m}\psi_{\n]} )-\frac{1}{4}
e^\phi (\bar{\e}\g_{\m\n} \chi )\quad, \nonumber \\
& & \delta \phi =-\frac{1}{\sqrt{2}} (\bar{\e} \chi )\quad ,\nonumber\\
& & \d \psi_\m =D_\m \e +\frac{1}{24\sqrt{2}}e^{-\phi}H^{\n\r\s}\g_{\m\n\r\s} \e
-\frac{3}{8\sqrt{2}}e^{-\phi} H_{\m\n\r} \g^{\n\r}\e \quad , \nonumber\\
& & \d \chi =-\frac{i}{\sqrt{2}}\de_\m \phi \g^\m \e-\frac{i}{12}e^{-\phi}
H^{\m\n\r} \g_{\m\n\r} \e \quad . 
\eea

In the supersymmetric case,
when these bulk modes are coupled to a gauge multiplet supported on the 9-branes and
containing a vector $A_\m$ and a left-handed gaugino $\l$ both in the adjoint representation 
of $SO(32)$,
supersymmetry requires that the 3-form 
$H_{\m\n\r}$ include a Chern-Simons coupling, so that 
\be
H_{\m\n\r}= 3 \de_{[\m} B_{\n\r ]} +\sqrt{2} \w_{\m\n\r} \quad ,\label{3form}
\ee
where $\w_{\m\n\r}$ is the Chern-Simons 3-form defined as 
\be
\w = A dA -\frac{2i}{3} A^3 \quad ,
\ee
and this leads to the modified Bianchi identity
\be
\de_{[\m }H_{\n\r\s ]} =\frac{3}{\sqrt{2}} \tr (F_{[\m\n} F_{\r\s ]} )\quad .
\label{bianchi}
\ee
The Lagrangian for supergravity coupled to vector multiplets is then \cite{bdrdwvn,chapmant}
(see Section (1.4))
\bea
e^{-1} {\cal{L}} = 
&-& \frac{1}{4} R +\frac{1}{2}\de_\m \phi \de^\m \phi +\frac{1}{6}
e^{-2\phi}H_{\m\n\r} H^{\m\n\r} -\frac{1}{2}e^{-\phi} \tr (F_{\m\n}F^{\m\n} )
\nonumber\\
&-& \frac{i}{2} (\bar{\psi}_\m \g^{\m\n\r} D_\n \psi_\r )+\frac{i}{2}(\bar{\chi}
\g^\m D_\m \chi )+\frac{1}{\sqrt{2}}(\bar{\psi}_\m \g^\n \g^\m \chi )\de_\n \phi 
\nonumber\\
&-& \frac{i}{12\sqrt{2}} e^{-\phi} H_{\m\n\r} (\bar{\psi}_\s \g^{\s\d\m\n\r}\psi_\d
)+\frac{i}{2\sqrt{2}}e^{-\phi} H_{\m\n\r} (\bar{\psi}^\m \g^\n \psi^\r)\nonumber \\
&+& \frac{1}{12}
e^{-\phi}H_{\m\n\r} (\bar{\psi}_\s \g^{\m\n\r} \g^\s \chi) +i \tr (\bar{\l}\g^\m
D_\m \l )\nonumber \\
&-& \frac{1}{2} e^{-\frac{1}{2}\phi } \tr [F^{\m\n} (\bar{\l} \g_{\m\n} \chi)]
+\frac{i}{\sqrt{2}}e^{-\frac{1}{2}\phi }\tr [F^{\m\n} (\bar{\l} \g_\r \g_{\m\n}
\psi^\r )]\nonumber\\
&-& \frac{i}{6\sqrt{2}}e^{-\phi }H^{\m\n\r}\tr (\bar{\l} \g_{\m\n\r} \l ) \ , 
\label{lag1}
\eea
up to quartic terms in the fermions.
The supersymmetry transformations of the bulk fields
$e_\m{}^m$, $\phi$, $\psi_\m$ and $\chi$ are as before, while 
for the gauge multiplet
\bea
& & \d A_\m =-\frac{i}{\sqrt{2}} e^{\frac{1}{2}\phi}(\bar{\e} \g_\m \l ) \quad ,
\nonumber\\
& & \d \l =-\frac{1}{2\sqrt{2}}  e^{-\frac{1}{2}\phi}F^{\m\n}\g_{\m\n}\e \quad . 
\label{susy1}
\eea
Gauge invariance of $H$ requires that under vector gauge
transformations $B$ transform as 
\be
\d B = -\sqrt{2} \tr (\L dA ) \ ,
\ee
and in order that gauge and 
supersymmetry transformations commute, up to a tensor gauge transformation,
one has to add a term to the supersymmetry variation of $B_{\m\n}$, obtaining
\be
\d B_{\m\n}= -\frac{i}{\sqrt{2}}e^\phi (\bar{\e}\g_{[\m}\psi_{\n]} )-\frac{1}{4}
e^\phi (\bar{\e}\g_{\m\n} \chi ) +2\sqrt{2} \tr (A_{[\m } \d A_{\n ]})
\quad. \label{B2}
\ee

In order to couple the Lagrangian (\ref{sugra})
to non-supersymmetric matter, one must construct from the fields of the 
supergravity multiplet quantities whose supersymmetry variations are
general coordinate transformations with the parameter $\xi_\m$ of
eq. (\ref{xi}). 
We thus define
\be
\hat{\phi}=\phi +\frac{1}{\sqrt{2}}(\bar{\th}\chi) +\frac{i}{24\sqrt{2}}e^{-\phi}
(\bar{\th} \g_{\m\n\r} \th ) H^{\m\n\r} \label{phihat}
\ee
so that
\be
\d \hat{\phi} = -\xi^{\m} \de_\m \hat{\phi} =\d_{gct}\hat{\phi}
\ee
and
\bea
\hat{e}_\m{}^m &=& e_\m{}^m +i (\bar{\th}\g^m \psi_\m )-\frac{i}{2}(\bar{\th}\g^m 
D_\m \th )-\frac{i}{48\sqrt{2}}e_\m{}^m e^{-\phi} (\bar{\th}\g_{\n\r\s}\th )
H^{\n\r\s}
\nonumber\\
&+& \frac{i}{16\sqrt{2}}e^{-\phi} (\bar{\th} \g_{\m\n\r} \th )H^{m \n\r}+
\frac{3i}{16\sqrt{2}} e^{-\phi} (\bar{\th} \g^{m \n\r} \th )H_{\m\n\r} 
\label{ehat}
\eea
so that
\be
\d \hat{e}_\m{}^m = \d_{gct} \hat{e}_\m{}^m +\L^m{}_n \hat{e}_\m{}^n \quad ,
\ee
where the parameter of the local Lorentz transformation is
\be
\L^{mn} =\frac{i}{2}(\bar{\th} \g^\r \e )\w_\r{}^{mn} 
+\frac{i}{24\sqrt{2}} e^{-\phi}
(\bar{\th} \g^{mn\n\r\s}\e )H_{\n\r\s} +\frac{3i}{4\sqrt{2}} e^{-\phi}
(\bar{\th} \g_\r \e )H^{\r mn} \quad .
\ee 
In constructing a Lagrangian invariant under non-linear 
supersymmetry that couples supergravity to non-supersymmetric matter,
it is important to notice that 
eq. (\ref{3form}) still holds, because of anomaly cancellation. For the same 
reason, the variation of $B$ is still given by eq. (\ref{B2}), once one 
uses the new transformation for $A_\m$, 
\be
\d A_\m = F_{\m\n} \xi^\n \quad .
\ee
Observe that this covariant expression for $\d A_\m$ contains the proper 
coordinate transformation, together with
an additional gauge transformation of parameter
\be
\L = \xi^\m A_\m \quad .
\ee
The supersymmetry transformation of the spinor $\l$ in the {\bf 495} of $USp(32)$ 
will not be taken 
into account in this discussion, since it contains higher-order Fermi terms.
One can now include \cite{dm2} the kinetic term for $A_\m$ and the 
dilaton tadpole in a Lagrangian that is supersymmetric up to terms  
quartic in the fermions, considering 
\bea
{\cal{L}}  &=& {\cal{L}}_{closed} - \frac{1}{2} \hat{e} e^{-\hat{\phi}}
\hat{g}^{\m\r} \hat{g}^{\n\s} \tr( F_{\m\n} F_{\r\s} ) -\L \hat{e}
e^{\frac{3}{2}\hat{\phi}} \nonumber \\
&+& ie \, \tr (\bar{\l} \g^\m D_\m \l ) 
-\frac{ie}{6\sqrt{2}}e^{-\phi}H^{\m\n\r} 
\, \tr(\bar{\l} \g_{\m\n\r} \l ) \quad ,
\label{geom}
\eea
where in the Sugimoto model $\L = 64 T_9$, with $T_9$ the anti-brane tension. 
By string considerations, one can show that the coefficient of the coupling 
of $H$ to $\l^2$, not constrained by supersymmetry at this level, is the 
same as in the supersymmetric case \cite{dm2}. 
Actually, the Lagrangian of eq. (\ref{geom}) is still 
not invariant under supersymmetry, since the 
inclusion of the Chern-Simons term and the consequent
modification of the Bianchi identity for $H$ generate contributions 
proportional to $F \wedge F$ in the variation of ${\cal{L}}_{closed}$. 
Up to higher order fermionic terms, however, these are exactly canceled by 
the variation of the additional terms
\bea
& & \frac{1}{6 !\sqrt{2}}\e^{\m_1 ...\m_6 \m\n\r\s}
[\frac{3i}{\sqrt{2}} e^{-\phi} (\bar{\th} \g_{\m_1 ...\m_5} \psi_{\m_6})
-\frac{1}{4}  e^{-\phi}(\bar{\th} \g_{\m_1 ...\m_6} \chi )\nonumber\\ 
& & +\frac{i}{8\sqrt{2}}
e^{-2\phi} \de_\t \phi (\bar{\th} \g_{\m_1 ...\m_6}{}^\t \th ) 
-\frac{3i}{2\sqrt{2}} e^{-\phi} (\bar{\th} \g_{\m_1 ...\m_5}D_{\m_6}\th )
\nonumber \\
& & -\frac{3i}{8} e^{-2\phi} (\bar{\th} \g_{\m_1 ...\m_5 \t\d} \th )
H_{\m_6}{}^{\t\d} -\frac{5i}{2}e^{-2\phi} 
(\bar{\th} \g_{\m_1 \m_2 \m_3} \th )
H_{\m_4 \m_5 \m_6} ] \tr(F_{\m\n} F_{\r\s} ) \quad .
\label{nongeom}
\eea
To summarize,  up to quartic fermionic terms the Lagrangian is 
\bea
{\cal{L}}  &=& {\cal{L}}_{closed} - \frac{1}{2} \hat{e} e^{-\hat{\phi}}
\hat{g}^{\m\r} \hat{g}^{\n\s}\tr( F_{\m\n} F_{\r\s} ) -\L \hat{e}
e^{\frac{3}{2}\hat{\phi}} \nonumber \\
&+& ie \, 
\tr (\bar{\l} \g^\m D_\m \l ) -\frac{ie}{6\sqrt{2}}e^{-\phi}H^{\m\n\r}
\, \tr (\bar{\l} \g_{\m\n\r} \l ) \nonumber\\
&+&  \, \frac{1}{6 !\sqrt{2}}\e^{\m_1 ...\m_6 \m\n\r\s}
[\frac{3i}{\sqrt{2}} e^{-\phi} (\bar{\th} \g_{\m_1 ...\m_5} \psi_{\m_6})
-\frac{1}{4}  e^{-\phi}(\bar{\th} \g_{\m_1 ...\m_6} \chi )\nonumber\\ 
&+& \frac{i}{8\sqrt{2}}
e^{-2\phi} \de_\t \phi (\bar{\th} \g_{\m_1 ...\m_6}{}^\t \th ) 
-\frac{3i}{2\sqrt{2}} e^{-\phi} (\bar{\th} \g_{\m_1 ...\m_5}D_{\m_6}\th )
\nonumber \\
&-& \frac{3i}{8} e^{-2\phi} (\bar{\th} \g_{\m_1 ...\m_5 \t\d} \th )
H_{\m_6}{}^{\t\d} -\frac{5i}{2}e^{-2\phi} 
(\bar{\th} \g_{\m_1 \m_2 \m_3} \th )
H_{\m_4 \m_5 \m_6} ] \tr(F_{\m\n} F_{\r\s} ) \quad .\label{lagB2}
\eea
As noticed in \cite{dm2}, in this formulation 
a geometric description for the terms in eq. (\ref{nongeom}) is  
not possible, 
{\it i.e.} it is not possible to rewrite them in terms of properly
dressed bulk 
fields adding fermionic bilinears containing the goldstino. 
We can now explain why this is the case,
and moreover we can also show how a geometric description is possible, 
after performing a duality transformation 
to a 6-form gauge field. 

Let us again begin with standard results: performing a duality transformation 
on eq. (\ref{lag1}), one  obtains a new Lagrangian,
with a 6-form rather than a 2-form, coupled
to vector multiplets \cite{chamseddine}.
Technically, this is performed starting from the first-order Lagrangian
\bea
{\cal L} &=& 
\frac{1}{6} e^{-2\phi} H_{\m\n\r}H^{\m\n\r} +\frac{1}{3 \cdot 6!}
\e^{\m_1 ...\m_7 \m\n\r } \de_{\m_1} \tilde{B}_{\m_2 ...\m_7} H_{\m\n\r} 
\nonumber\\
&+& \frac{1}{6! \sqrt{2}}\e^{\m_1 ...\m_6 \m\n\r\s }\tilde{B}_{\m_1 ...\m_6}
\tr (F_{\m\n}
F_{\r\s})
\label{lduality}
\eea
that contains both the 2-form and the 6-form.
The field equation for $\tilde{B}_6$ is then exactly the Bianchi 
identity of eq. (\ref{bianchi})
for $H_3$, while the field equation for $H_3$ is
\be
e^{-\phi} H_3 = e^\phi * \tilde{H}_7 \quad ,
\label{dual}
\ee
where $\tilde{H}_7 =d \tilde{B}_6$.
The Lagrangian obtained substituting this relation in (\ref{lduality}) and
redefining $\tilde{H}_7 \rightarrow H_7$,
\be
{\cal L}=\frac{1}{7 !}e^{2\phi}H_{\m_1 ...\m_7 }H^{\m_1 ...\m_7 }+
\frac{1}{6! \sqrt{2}}\e^{\m_1 ...\m_6 \m\n\r\s }B_{\m_1 ...\m_6} \tr (F_{\m\n}
F_{\r\s}) \quad ,
\ee
shows how the Chern-Simons term in $H_3$ is replaced by the 
Wess-Zumino term $B \wedge F \wedge F$.

If one performs this duality transformation in (\ref{lag1}),
one ends up with the Lagrangian originally 
obtained by Chamseddine \cite{chamseddine}:
\bea
e^{-1} {\cal{L}} =  
&-&\frac{1}{4} R +\frac{1}{2}\de_\m \phi \de^\m \phi +\frac{1}{7!}
e^{2\phi}H_{\m_1 ...\m_7} H^{\m_1 ...\m_7} 
-\frac{1}{2}e^{-\phi} \tr (F_{\m\n}F^{\m\n} )
\nonumber\\
&+& \frac{1}{6! \sqrt{2}}\e^{\m_1 ...\m_6 \m\n\r\s } B_{\m_1 ...\m_6}
\tr (F_{\m\n}
F_{\r\s}) -\frac{i}{2} (\bar{\psi}_\m \g^{\m\n\r} D_\n \psi_\r ) \nonumber \\
&+&\frac{i}{2}(\bar{\chi}
\g^\m D_\m \chi )+\frac{1}{\sqrt{2}}(\bar{\psi}_\m \g^\n \g^\m \chi )\de_\n \phi 
+\frac{i}{240\sqrt{2}} e^{\phi} H^{\m_1 ...\m_7} 
(\bar{\psi}_{\m_1} \g_{\m_2 ...\m_6} \psi_{\m_7} ) \nonumber \\
&+& \frac{i}{2 \cdot 7!\sqrt{2}}e^{\phi} 
H^{\m_1 ...\m_7} (\bar{\psi}^\m \g_{\m\m_1 ... \m_7 \n} \psi^\n) -\frac{1}{2 \cdot 7!}
e^{\phi}H^{\m_1 ...\m_7} (\bar{\psi}^\s \g_{\m_1 ...\m_7} \g_\s \chi) \nonumber \\
&+& i \tr (\bar{\l}\g^\m
D_\m \l )-\frac{1}{2} e^{-\frac{1}{2}\phi } \tr [F^{\m\n} (\bar{\l} \g_{\m\n} \chi)] 
\nonumber \\
&+& \frac{i}{\sqrt{2}}e^{-\frac{1}{2}\phi }\tr [F^{\m\n} (\bar{\l} \g_\r \g_{\m\n}
\psi^\r )]+\frac{i}{7!\sqrt{2}}e^{\phi }H^{\m_1 ...\m_7} \tr(\bar{\l} \g_{\m_1 ...\m_7} \l ) 
\quad .
\label{lag2}
\eea
The corresponding supersymmetry transformations are obtained 
from eq. (\ref{susy1}) performing the redefinition of eq. (\ref{dual}) 
on the variations of $\psi_\m$ and $\chi$, leaving the variations of 
$e_\m{}^m$, $\phi$, $A_\m$ and $\l$ unaffected
and replacing the variation of the 2-form with 
\be
\d B_{\m_1 ...\m_6}= -\frac{3i}{\sqrt{2}}e^{-\phi} (\bar{\e}\g_{[\m_1 ...\m_5}
\psi_{\m_6 ]})+\frac{1}{4} e^{-\phi} (\bar{\e}\g_{\m_1 ...\m_6 }\chi )\quad .
\label{B6}
\ee
Notice that the supersymmetry variation of the 6-form does not include a term 
depending on the vector field. This reflects the  
fact that the 6-form is inert under gauge transformations, 
since its field-strength does not contain a Chern-Simons form, that
in this case would enter higher-derivative couplings not present in the
effective supergravity.

One can now couple the supergravity multiplet expressed in terms of the
6-form to non-supersymmetric matter. In order to do this, 
together with $\hat{\phi}$ and $\hat{e}_\m{}^m$ of eqs. (\ref{phihat}) and 
(\ref{ehat}), one must define the dressed 6-form
\bea
\hat{B}_{\m_1 ...\m_6}  &=& B_{\m_1 ...\m_6} + 
\frac{3i}{\sqrt{2}} e^{-\phi} (\bar{\th} \g_{[\m_1 ...\m_5} \psi_{\m_6 ]})
-\frac{1}{4}  e^{-\phi}(\bar{\th} \g_{\m_1 ...\m_6} \chi )\nonumber\\ 
&+& \frac{i}{8\sqrt{2}}
e^{-2\phi} \de_\t \phi (\bar{\th} \g_{\m_1 ...\m_6}{}^\t \th ) 
-\frac{3i}{2\sqrt{2}} e^{-\phi} (\bar{\th} \g_{[ \m_1 ...\m_5}D_{\m_6 ]}\th )
\nonumber \\
&-& \frac{3i}{8} e^{-2\phi} (\bar{\th} \g_{[ \m_1 ...\m_5 \t\d} \th )
H_{\m_6 ]}{}^{\t\d} -\frac{5i}{2}e^{-2\phi} 
(\bar{\th} \g_{[ \m_1 \m_2 \m_3} \th )
H_{\m_4 \m_5 \m_6 ]}\quad , \label{B6hat}
\eea
whose supersymmetry transformation is a coordinate 
transformation, up to
an additional tensor gauge transformation of parameter
\be
\L_{\m_1 ... \m_5 } =-\frac{i}{4 \sqrt{2}}e^{-\phi} (\bar{\th}\g_{\m_1 ... \m_5 }\e ) \quad.
\ee
We have intentionally written the last line of eq. (\ref{B6hat}) in terms of the dual
3-form, using eq. (\ref{dual}), so that the similarity 
with eq. (\ref{nongeom}) be more transparent. 
The Lagrangian for the closed sector,
\bea
e^{-1} \tilde{\cal{L}}_{closed}  =
&-& \frac{1}{4} R +\frac{1}{2}\de_\m \phi \de^\m \phi +\frac{1}{6}
e^{2\phi}H_{\m_1 ...\m_7} H^{\m_1 ...\m_7} \nonumber\\
&-& \frac{i}{2} (\bar{\psi}_\m \g^{\m\n\r} D_\n \psi_\r )+\frac{i}{2}(\bar{\chi}
\g^\m D_\m \chi )+\frac{1}{\sqrt{2}}(\bar{\psi}_\m \g^\n \g^\m \chi )\de_\n \phi 
\nonumber\\
&+& \frac{i}{240\sqrt{2}} e^{\phi} H^{\m_1 ...\m_7} 
(\bar{\psi}_{\m_1} \g_{\m_2 ...\m_6}\psi_{\m_7}
)+\frac{i}{2\sqrt{2}7!}e^{\phi} 
H^{\m_1 ...\m_7} (\bar{\psi}^\m \g_{\m \m_1 ...\m_7 \n} \psi^\n)\nonumber \\
&+& \frac{1}{2\cdot 7!}
e^{\phi}H^{\m_1 ...\m_7} (\bar{\psi}^\s \g_{\m_1 ...\m_7} \g_\s \chi) 
\label{sugra2}
\eea
is simply obtained performing the duality 
transformation in eq. (\ref{sugra}), while
the same duality in eq. 
(\ref{lagB2}) gives 
\bea
{\cal{L}}  &=& \tilde{\cal{L}}_{closed} 
 - \frac{1}{2} \hat{e} e^{-\hat{\phi}}
\hat{g}^{\m\r} \hat{g}^{\n\s} \tr( F_{\m\n} F_{\r\s} ) -\L \hat{e}
e^{\frac{3}{2}\hat{\phi}} \nonumber \\
&+& ie \tr (\bar{\l} \g^\m D_\m \l ) 
+\frac{i e}{7!\sqrt{2}}
e^{\phi }H^{\m_1 ...\m_7} \tr(\bar{\l} \g_{\m_1 ...\m_7} \l ) \nonumber\\
&+& \frac{1}{6!\sqrt{2}}\e^{\m_1 ...\m_6 \m\n\r\s} \hat{B}_{\m_1 ... \m_6 }
\tr (F_{\m\n} F_{\r\s} ) \quad .\label{lagB6}
\eea
Note that in this Lagrangian {\it all} terms containing the goldstino are
grouped in redefinitions of the bulk fields, and therefore 
all couplings are written in a geometric form.
This result concludes this section: for the
ten-dimensional $USp(32)$ model a fully geometric description is possible if
one formulates it in terms of the 6-form, since in this case the 
Chern-Simons term is higher derivative, and thus is not in the 
low-energy effective action. More precisely, as we have seen, 
duality maps the Chern-Simons term 
into the Wess-Zumino term, and this falls simply into a geometric
form.  The result is still valid in presence of additional
brane-antibrane pairs, since
the introduction of supersymmetric vectors does not modify the 
field strength relative to the 6-form potential in the low-energy effective action.
In the dual theory, although the field strength of the 2-form is modified, no additional
terms containing the goldstino have to be added to the low-energy lagangian. 
As we shall see, this feature is common to the six-dimensional case. 

\section{Geometric couplings in six-dimensional models}
\label{comments}
\fancyhead[LO]{{\footnotesize 5.2~~{\it Geometric couplings in six dimensions}}}

In this section we construct the low-energy couplings for 
six-dimensional type-I models with brane supersymmetry breaking 
\cite{bsb}. 
All the features 
of brane supersymmetry breaking are present in the 
$T^4 / Z_2$ orientifold of \cite{bsb}, where a change of the 
orientifold projection leads to $D9$ branes and $\bar{D}5$ branes. 
The spectrum has $(1,0)$ supersymmetry in the closed 
and 9-9 sectors, while supersymmetry is broken in the 9-$\bar{5}$ and
$\bar{5}$-$\bar{5}$ sectors. The gauge group is $SO(16) \times SO(16)$ on 
the $D9$ branes and $USp(16) \times USp(16)$ on the ${\bar{D}}5$ branes, 
if all the ${\bar{D}}5$ branes are at a fixed point.

One of the peculiar features of low-energy effective actions for six-dimensional
type-I models with minimal supersymmetry is the fact that they embody reducible gauge and 
supersymmetry anomalies, to be canceled by fermion loops. Consequently, the 
Lagrangian is determined imposing the closure of the Wess-Zumino consistency 
conditions, rather than
by the requirement of supersymmetry.
We use the notations of the previous chapter, and we denote
the  
vector multiplet from the 9-9 sector as $A^{(9)}_\m , \l^{(9)A}$
Denoting with  $\Phi^{\bar{\a}}$ $(\bar{\a}=1,...,n_T )$ the scalars in the tensor multiplets, parametrizing 
the coset $SO(1, n_T ) / SO( n_T )$, 
the vielbein $V_{\bar{\a}}^M$ of the internal manifold 
is related to $v^r$ and $x^{Mr}$ of eq. (\ref{scalars}) by 
\be
V_{\bar{\a}}^M =v^r \de_{\bar{\a}} x^M_r \quad ,
\ee
where $\de_{\bar{\a}} = \de / \de \Phi^{\bar{\a}}$.
The metric of the internal manifold is $g_{\bar{\a}\bar{\b}} = V_{\bar{\a}}^M
V_{\bar{\b}}^M$.

Denoting with  $A_\m^{(9)i}$ the gauge fields under which the hypermultiplets are
charged (the index $i$ runs in the adjoint of the gauge group), 
under the gauge transformations 
\be
\delta A_\m^{(9)i} = D_\m \L^{(9)i}  
\ee
the scalars transform as
\be
\delta \phi^\a = \L^{(9)i} \xi^{\a i} \quad ,
\ee
where $\xi^{\a i}$ are the Killing vectors corresponding to
the isometry that we are gauging.
The covariant derivative for the scalars is then
\be
D_\m \phi^\a =\de_\m \phi^\a - A_\m^{(9)i} \xi^{\a i} \quad .
\ee
The covariant derivatives for the 
gauginos $\l^{(9)iA}$ are 
\be
D_\m \l^{(9)iA}=\de_\m \l^{(9)iA} +\frac{1}{4}\w_{\m mn}\g^{mn}\l^{(9)iA}
+ D_\m \phi^\a {\cal{A}}_\a^A{}_B \l^{(9)i B}+ f^{ijk} A_\m^{(9)j} \l^{(9)kA}
\quad ,
\ee
where $f^{ijk}$ are the structure constants of the group.

We use the method of Pasti, Sorokin and Tonin (PST) \cite{pst} 
in order to write a covariant action for 
fields that satisfy self-duality conditions. 
For a self-dual 3-form in six dimensions the PST action 
\be
{\cal L}_{PST} = \frac{1}{12}H_{\m\n\r} H^{\m\n\r} -
\frac{1}{4}\frac{\de^\m \Xi \de^\s \Xi}{(\de \Xi )^2 } H^-_{\m\n\r} 
H^-_\s{}^{\n\r}\quad ,
\ee
where $H^- = H- *H$ and $\Xi$ is a scalar auxiliary field,  
is invariant under the standard gauge transformations
for a 2-form,
\be
\d B= d \L,
\ee
and under the additional PST gauge transformations 
\be
\d B_{\m\n} =(\de_\m \Xi )\L_\n - (\de_\n \Xi )\L_\m
\ee
and 
\be
\d \Xi = \L \quad , \qquad \d B_{\m\n}=\frac{\L}{(\de \Xi )^2 } H^-_{\m\n\r}
\de^\r \Xi \quad .
\ee
We have a single self-dual 3-form and $n_T$ antiself-dual 
3-forms, where
$n_T$ is equal to 17 in the $T^4/Z_2$ model of \cite{bsb}. These forms
are obtained dressing with the scalars in the tensor multiplets the
3-forms
\be
H^r = d B^r - c^{rz} \w^{(9)z} \quad ,\label{cs6}
\ee
where the index $z$ runs over the various semi-simple factors of the 
gauge group in the 9-9 sector, $\w$ is the Chern-Simons 3-form and the $c$'s are constants
(we denote with $z=1$ the group under which the hypermultiplets are
charged). 
More precisely, the combinations
\be
H_{\m\n\r}=v_r H^r_{\m\n\r}
\ee
and
\be
H^M_{\m\n\r}= x^M_r  H^r_{\m\n\r}
\ee
are respectively self-dual and antiself-dual \cite{romans}, 
to lowest order in the Fermi fields, although in the complete
lagrangian these conditions 
are modified by the inclusion of fermionic bilinears.
As in ten dimensions, the gauge invariance of $H^r$ in eq. (\ref{cs6}) implies
that $B^r$ vary as 
\be
\d B^r = c^{rz} \tr_z ( \L^{(9)} d A^{(9)} ) 
\label{gaugeB}
\ee
under gauge transformations. 

To lowest order in the Fermi fields, 
the Lagrangian describing the coupling of the supergravity  
multiplet to $n_T$ tensor multiplets, vector multiplets and $n_H$ hypermultiplets is
\bea
e^{-1}{\cal{L}}_{susy} = &-& \frac{1}{4}R +\frac{1}{12}G_{rs} H^{r \m\n\r} H^s_{\m\n\r}
+\frac{1}{4} g_{\bar{\a}\bar{\b}} \de_\m \Phi^{\bar{\a}}\de^\m \Phi^{\bar{\b}} 
-\frac{1}{2} v_r c^{rz} \tr_z (F^{(9)}_{\m\n}F^{(9)\m\n}) \nonumber\\
&-& \frac{\de^\m \Xi \de^\s \Xi }{4(\de \Xi )^2 }
[{\cal H}^-_{\m\n\r} {\cal H}^-_{\s}{}^{\n\r} +
{\cal H}^{M+}_{\m\n\r} {\cal H}^{M+}{}_\s{}^{\n\r} ]
-\frac{1}{8e}\e^{\m\n\r\sigma\delta\tau}B^r_{\m\n}c_r^z \tr_{z}
(F^{(9)}_{\r\sigma}
F^{(9)}_{\delta\tau}) \nonumber \\
&+& \frac{1}{2} g_{\a\b}(\phi ) D_\m \phi^\a D^\m \phi^\b 
+\frac{1}{4v_r c^{r1}} {\cal{A}}_\a^A{}_B 
{\cal{A}}_\b^B{}_A \xi^{\a i} \xi^{\b i} 
\nonumber \\
&-& \frac{i}{2}(\bar{\psi}_\m \g^{\m\n\r} D_\n 
\psi_\r )-\frac{i}{2}v_r H^{r \m\n\r}(\bar{\psi}_\m \g_\n \psi_\r)
+\frac{i}{2} (\bar{\chi}^M \g^\m D_\m \chi^M )\nonumber \\ 
&-&\frac{i}{24}v_r H^r_{\m\n\r} (\bar{\chi}^M \g^{\m\n\r}
\chi^M ) +\frac{1}{2}x^M_r \de_\n v^r (\bar{\psi}_\m \g^\n
\g^\m \chi^M) \nonumber \\
&-& \frac{1}{2} x^M_r H^{r \m\n\r} ( \bar{\psi}_\m
\g_{\n\r} \chi^M )
+ \frac{i}{2}(\bar{\Psi}_a \g^\m D_\m \Psi^a ) +\frac{i}{24}
v_r H^r_{\m\n\r} (\bar{\Psi}_a \g^{\m\n\r}\Psi^a ) \nonumber \\
&-&  V_\a^{aA}D_\n \phi^\a
(\bar{\psi}_{\m A} \g^\n \g^\m \Psi_a ) + i v_r c^{rz} \tr_z(\bar{\l}^{(9)} \g^\m D_\m \l^{(9)} )
\nonumber \\
&+& \frac{i}{\sqrt{2}} v_r c^{rz} \tr_z[F^{(9)}_{\n\r} (\bar{\psi}_\m \g^{\n\r}
\g^\m \l^{(9)} )]
+ \frac{1}{\sqrt{2}} x^M_r c^{rz} \tr_z [F^{(9)}_{\m\n} 
(\bar{\chi}^M \g^{\m\n}\l^{(9)} )] \nonumber \\
&-& \frac{i}{12}c_r^z H^r_{\m\n\r} \tr_z (\bar{\l}^{(9)} \g^{\m\n\r} \l^{(9)})
- \sqrt{2} V_\a^{aA} \xi^{\a i}(\bar{\l}^{(9)i}_A \Psi_a ) \nonumber \\
&+& \frac{i}{\sqrt{2}}{\cal{A}}_\a^A{}_B 
\xi^{\a i }(\bar{\l}^{(9)i}_A \g^\m \psi_\m^B ) 
+ \frac{1}{\sqrt{2}}{\cal{A}}_\a^A{}_B 
\frac{x^M_r c^{r1}}{v_s c^{s1}}\xi^{\a i}
(\bar{\l}^{(9)i}_A \chi^{MB} ) \quad,
\label{pstlag99}
\eea
where $G_{rs} = v_r v_s + x^M_r x^M_s$, while
\be
{\cal H}_{\m\n\r} = v_r {H}^{r}_{\m\n\r}
-\frac{3i}{2} (\bar{\psi}_{[\m} \g_\n \psi_{\r ]})
-\frac{i}{8} 
(\bar{\chi}^M \g_{\m\n\r} \chi^M )+\frac{i}{8} 
(\bar{\Psi}_a \g_{\m\n\r} \Psi^a)
\ee
and
\be
{\cal H}^M_{\m\n\r} = x^M_r {H}^{r}_{\m\n\r} 
+ \frac{3}{2} (\bar{\chi}^M\g_{[\m\n}
\psi_{\r ]}) +\frac{i}{4} x^M_r c^{rz} \tr_z (\bar{\l}^{(9)} \g_{\m\n\r}
\l^{(9)} )
\ee
satisfy on-shell self-duality and antiself-duality conditions,
respectively.
Finally, $\Xi$ is the PST auxiliary field.

Due to eq. (\ref{gaugeB}), the Wess-Zumino term 
$B \wedge F \wedge F$ is not gauge invariant, and thus the variation 
of eq. (\ref{pstlag99}) under gauge transformations produces the consistent
gauge anomaly
\be 
{\cal{A}}_\L =- \frac{1}{4} \e^{\m\n\r\sigma\delta\tau} c_r^z c^{rz^\prime} \tr_z ({\L}^{(9)} 
\de_\m A^{(9)}_\n ) \tr_{z^\prime} (F^{(9)}_{\r\sigma} F^{(9)}_{\delta\tau} )\quad ,
\label{consanomaly99}
\ee 
related by the Wess-Zumino conditions to the supersymmetry anomaly
\be 
{\cal{A}}_\e = \e^{\m\n\r\sigma\delta\tau} c_r^z c^{r z^\prime}
[-\frac{1}{4} \tr_z (
\delta_\e A^{(9)}_\m A^{(9)}_\n ) \tr_{z^\prime} (F^{(9)}_{\r\sigma} 
F^{(9)}_{\delta\tau})  
-\frac{1}{6} \tr_z (
\delta_\e A^{(9)}_\m F^{(9)}_{\n\r} ) \w^{(9)z^\prime}_{\sigma\delta\tau} ]\ ,
\label{susyanomaly99}
\ee
that one can recover varying the Lagrangian of eq. (\ref{pstlag99}) under
the supersymmetry transformations 
\bea
& & \delta e_\m{}^m = -i ( \bar{\e} \g^m \psi_\m ) \quad , \nonumber\\ 
& & \delta B^r_{\m\n} =i v^r ( \bar{\psi}_{[\m} \g_{\n]} \e )
+\frac{1}{2} x^{Mr} ( \bar{\chi}^M \g_{\m\n} \e ) 
- 2c^{rz}\tr_z( A^{(9)}_{[\m} \delta A^{(9)}_{\n]} )
\quad , \nonumber\\ 
& & \d \Phi^{\bar{\a}} = V^{\bar{\a}M} (\bar{\chi}^M \e ) \quad , \nonumber\\
& & \delta \phi^\a = V^\a_{aA} ({\bar{\e}}^A \Psi^a ) \quad ,\nonumber\\
& & \d \Xi = 0 \quad , \nonumber \\
& & \delta A^{(9)}_\m = -\frac{i}{\sqrt{2}} (\bar{\e} \g_\m \l^{(9)} )\quad, 
\nonumber\\
& & \d \psi_\m^A =D_\m \e^A +
\frac{1}{4} K_{\m\n\r} \g^{\n\r} \e^A \quad , \nonumber \\
& & \d \chi^{MA} = -\frac{i}{2} V_{\bar{\a}}^M \de_\m \Phi^{\bar{\a}}\g^\m 
\e^A +\frac{i}{12}
K^M_{\m\n\r} \g^{\m\n\r} \e^A  \quad ,\nonumber \\
& & \delta \Psi^a = i \g^\m \e_A V_\a^{aA} D_\m \phi^\a \quad , \nonumber\\
& & \delta \l^{(9)A} = -\frac{1}{2\sqrt{2}} F^{(9)}_{\m\n} \g^{\m\n} \e^A \qquad 
\quad (z \neq 1 ) \quad , \nonumber \\
& & \delta \l^{(9)iA}= -\frac{1}{2\sqrt{2}} F^{(9)i}_{\m\n} \g^{\m\n} \e^A 
-\frac{1}{\sqrt{2}v_r c^{r1}} {\cal{A}}_\a^A{}_B \xi^{\a i} \e^B 
\label{susy6}
\eea
where
\be
K_{\m\n\r} ={\cal{H}}_{\m\n\r} -3 \frac{\de_{[\m} \Xi \de^\s  \Xi}{(\de \Xi )^2}
{\cal H}^-_{\s\n\r ]}
\ee
and
\be
K^M_{\m\n\r} ={\cal{H}}^M_{\m\n\r} -3 
\frac{\de_{[\m} \Xi \de^\s \Xi }{(\de \Xi)^2}
{\cal H}^{M+}_{\s\n\r ]}
\ee
are {\it identically} self-dual and antiself-dual, respectively.
In the complete theory, the anomalous terms would be exactly canceled by the 
anomalous contributions of fermion loops. 

Following the same reasoning as for 
the ten dimensional case of \cite{dm2}, we can describe the couplings to 
non-supersymmetric matter requiring that local supersymmetry be non-linearly 
realized on the ${\bar D}5$-branes, and requiring that the supersymmetry
variation of the non-supersymmetric fields be as in eq. (\ref{susygct}). 
As explained in \cite{dm2} and 
reviewed in the previous section, to lowest order in the fermions the
coupling between the supersymmetric sector and the non-supersymmetric
one is obtained dressing the bosonic fields in the supersymmetric sector with 
fermionic bilinears containing the goldstino, whose supersymmetry 
transformation is $\d \th =\e$ to lowest order in the fermionic fields. 
As a result, the supersymmetry variation of the dressed scalars in the tensor multiplets
\be
\hat{\Phi}^{\bar{\a}} = \Phi^{\bar{\a}} -V^{\bar{\a}M}(\bar{\th}
\chi^M ) +\frac{i}{24} V^{\bar{\a}M} x^M_r H^r_{\m\n\r}
(\bar{\th} \g^{\m\n\r} \th )
\ee
is a general coordinate transformation of parameter
\be
\xi_\m =-\frac{i}{2} (\bar{\th} \g_\m \e ) \quad .
\label{xi6}
\ee
This definition of $\hat{\Phi}$ then induces the corresponding dressing
\be
\hat{v}^r = v^r - x^{Mr} (\bar{\th} \chi^M ) -\frac{i}{24} H^r_{\m\n\r}
(\bar{\th} \g^{\m\n\r} \th ) \quad ,
\ee
and, in a similar fashion, the supersymmetry transformation of
\be
\hat{\phi}^\a = \phi^\a - V^\a_{aA} (\bar{\th}^A \Psi^a ) -\frac{i}{2}
V_{\b aA} V^{\a a B} (\bar{\th}^A \g^\m \th_B ) D_\m \phi^\b
\ee
is again a coordinate transformation with the same parameter, 
together with an additional gauge transformation of parameter
\be
\L^{(9)} = \xi^\m A^{(9)}_\m \quad .
\label{gauge9}
\ee 
Similarly, the supersymmetry variation of
\be
\hat{e}_\m{}^m  =  e_\m{}^m +i(\bar{\th}\g^m \psi_\m )-\frac{i}{2}
(\bar{\th} \g^m D_\m \th ) -\frac{i}{8} v_r H^r_{\m\n\r} 
(\bar{\th} \g^{m\n\r} \th ) 
\ee
contains also an additional
local Lorentz transformation of parameter
\be
\L^{mn} = -\xi^\m[ \w_\m{}^{mn} -v_r H^r_\m{}^{mn} ]
\label{lolo}
\ee
where $\w$ denotes the spin connection. 
Since the scalars in the non-supersymmetric 9-$\bar{5}$ sector are 
charged with respect to the vectors in the 9-9 sector, we define also
\bea
& & \hat{A}_\m^{(9)} = A^{(9)}_\m + \frac{i}{\sqrt{2}}(\bar{\th} \g_\m 
\l^{(9)} ) + \frac{i}{8} F^{(9)\n\r} (\bar{\th} \g_{\m\n\r} \th )\quad ( 
z \neq 1)\quad ,\nonumber \\
& & \hat{A}_\m^{(9)i} = A^{(9)i}_\m + \frac{i}{\sqrt{2}}(\bar{\th} \g_\m 
\l^{(9)i} ) + \frac{i}{8} F^{(9)i\n\r} (\bar{\th} \g_{\m\n\r} \th )\nonumber \\
& & \quad \quad +\frac{i}{4 v_r c^{r1}}{\cal{A}}_\a^A{}_B \xi^{\a i} (\bar{\th}_A \g_\m 
\th^B ) \ ,
\eea
whose supersymmetry transformation is a general coordinate transformation of 
parameter as in eq. (\ref{xi6}), aside from a gauge transformation of parameter
as in (\ref{gauge9}).
If one requires that the supersymmetry variation of the vector $A^{(5)}_\m$ from the
non-supersymmetric $\bar{5}$-$\bar{5}$ sector be
\be
\d A^{(5)}_\m = F^{(5)}_{\m\n}\xi^\n \quad ,
\label{efxi}
\ee
namely a general coordinate transformation together with a gauge transformation of 
parameter
\be
\L^{(5)} = \xi^\m A^{(5)}_\m \quad ,\label{gauge5}
\ee
one obtains a supersymmetrization of the kinetic term for $A^{(5)}_\m$ writing
\be
 -\frac{1}{2} \hat{e} \hat{v}^r c_r^w \tr_w (F^{(5)}_{\m\n} F^{(5)}_{\r\s} )
\hat{g}^{\m\r} \hat{g}^{\n\s} \quad ,\label{kin}
\ee
where
\be
\hat{g}_{\m\n} = \hat{e}_\m{}^m \hat{e}_{\n m} \quad ,
\ee
and the index $w$ runs over the various semi-simple factors of the gauge
group in the $\bar{5}$-$\bar{5}$ sector.  
In analogy with the ten-dimensional case, the uncanceled
NS-NS tadpole translates, in the 
low-energy theory, in the presence of a term 
\be
- \L \hat{e} f(\hat{\Phi}^{\bar{\a}}, {\hat\phi}^\a ) \quad ,\label{tadpole}
\ee
that depends on the scalars of the closed sector and contains the 
dilaton, that belongs to a hypermultiplet in type-I vacua. 
Thus, supersymmetry breaking 
naturally corresponds in this case also to a breaking of 
the isometries of the scalar manifolds.

Denoting with $S$ the scalars in the 9-$\bar{5}$ sector, charged with 
respect to the gauge fields in both the 9-9 and $\bar{5}$-$\bar{5}$ sectors, 
we define their covariant derivative as  
\be
\hat{D}_\m S = \de_\m S -i \hat{A}^{(9)}_\m  S  - i  {A}^{(5)}_\m  S 
\quad,
\ee  
so that the term
\be
\frac{1}{2} \hat{e}(\hat{D}_\m S )^\dagger (\hat{D}_\n S) \hat{g}^{\m\n}
\ee
is supersymmetric, if again the supersymmetry 
transformation of $S$ is a general
coordinate transformation, together with 
a gauge transformation of the right parameters.
As in the ten-dimensional case, if one considers terms up to quartic couplings
in the fermionic fields, one does not have to supersymmetrize terms that are 
quadratic in the additional fermions from the non-supersymmetric 
$\bar{5}$-$\bar{5}$ and $9$-$\bar{5}$ sectors. 
Denoting with $\l^{(5)}$ these fermions, the coupling of $\l^{(5)2}$
to the 3-forms is not determined by supersymmetry, and can only be determined 
by string considerations, as in \cite{dm2}. 

The inclusion of additional non-supersymmetric 
vectors modifies $H^r$, that now includes the Chern-Simons 3-forms 
corresponding to these fields, so that eq. (\ref{cs6}) becomes
\be
H^r = d B^r - c^{rz} \w^{(9)z} - c^{rw} \w^{(5)w} \quad .
\ee
The gauge invariance of $H^r$ then
requires that
\be
\d B^r = c^{rw} \tr_w (\L^{(5)} d A^{(5)} ) \label{cs6bis}
\ee
under gauge transformations in the $\bar{5}$-$\bar{5}$ sector.
Consequently, the supersymmetry transformation of $B^r$ is 
also modified, and becomes
\bea
\delta B^r_{\m\n} &=& i v^r ( \bar{\psi}_{[\m} \g_{\n]} \e )
+\frac{1}{2} x^{Mr} ( \bar{\chi}^M \g_{\m\n} \e ) \nonumber\\ 
&-&2c^{rz} \tr_z( A^{(9)}_{[\m} \delta A^{(9)}_{\n]} )
- 2c^{rw} \tr_w( A^{(5)}_{[\m} \delta A^{(5)}_{\n]} )
\quad .
\eea
The complete reducible gauge anomaly
\bea 
{\cal{A}}_\L = & - & \frac{1}{4} \e^{\m\n\r\sigma\delta\tau}\{ c_r^z c^{rz^\prime} \tr_z ({\L}^{(9)} 
\de_\m A^{(9)}_\n ) \tr_{z^\prime} (F^{(9)}_{\r\sigma} F^{(9)}_{\delta\tau} )\nonumber\\
& + & c_r^z c^{rw} \tr_z ({\L}^{(9)} 
\de_\m A^{(9)}_\n ) \tr_{w} (F^{(5)}_{\r\sigma} F^{(5)}_{\delta\tau} )\nonumber\\
& + & c_r^w c^{rz} \tr_w ({\L}^{(5)} 
\de_\m A^{(5)}_\n ) \tr_{z} (F^{(9)}_{\r\sigma} F^{(9)}_{\delta\tau} )\nonumber\\
& + & c_r^w c^{rw^\prime} \tr_w ({\L}^{(5)} 
\de_\m A^{(5)}_\n ) \tr_{w^\prime} (F^{(5)}_{\r\sigma} F^{(5)}_{\delta\tau} ) \} \quad ,
\label{complconsanomaly95}
\eea 
related by the Wess-Zumino conditions to the supersymmetry anomaly
\bea 
{\cal{A}}_\e & = & \e^{\m\n\r\sigma\delta\tau} \{ c_r^z c^{r z^\prime}
[-\frac{1}{4} \tr_z (
\delta_\e A^{(9)}_\m A^{(9)}_\n ) \tr_{z^\prime} (F^{(9)}_{\r\sigma} 
F^{(9)}_{\delta\tau})  
-\frac{1}{6} \tr_z (
\delta_\e A^{(9)}_\m F^{(9)}_{\n\r} ) \w^{(9)z^\prime}_{\sigma\delta\tau} ]\nonumber \\
& + & c_r^z c^{r w}
[-\frac{1}{4} \tr_z (
\delta_\e A^{(9)}_\m A^{(9)}_\n ) \tr_{w} (F^{(5)}_{\r\sigma} 
F^{(5)}_{\delta\tau})  
-\frac{1}{6} \tr_z (
\delta_\e A^{(9)}_\m F^{(9)}_{\n\r} ) \w^{(5)w}_{\sigma\delta\tau} ]\nonumber \\
& +& c_r^w c^{r z}
[-\frac{1}{4} \tr_w (
\delta_\e A^{(5)}_\m A^{(5)}_\n ) \tr_{z} (F^{(9)}_{\r\sigma} 
F^{(9)}_{\delta\tau})  
-\frac{1}{6} \tr_w (
\delta_\e A^{(5)}_\m F^{(5)}_{\n\r} ) \w^{(9)z}_{\sigma\delta\tau} ]\nonumber \\
& +& c_r^w c^{r w^\prime}
[-\frac{1}{4} \tr_w (
\delta_\e A^{(5)}_\m A^{(5)}_\n ) \tr_{w^\prime} (F^{(5)}_{\r\sigma} 
F^{(5)}_{\delta\tau})  
-\frac{1}{6} \tr_w (
\delta_\e A^{(5)}_\m F^{(5)}_{\n\r} ) \w^{(5)w^\prime}_{\sigma\delta\tau} ] \} \ ,
\label{complsusyanomaly95}
\eea
is induced by the Wess-Zumino term
\be
-\frac{1}{8} \e^{\m\n\r\s\d\t}B^r_{\m\n} c_{r}^w \tr_w (F_{\r\s}^{(5)} F_{\d\t}^{(5)})\quad .
\label{gs6}
\ee
It should be noticed that, as in the case of linearly realized supersymmetry, 
eqs. (\ref{complconsanomaly95}) and (\ref{complsusyanomaly95}) satisfy
the Wess-Zumino condition
\be
\d_\L {\cal{A}}_\e = \d_\e {\cal{A}}_\L \quad ,
\ee
since the explicit form of the gauge field supersymmetry variation 
plays no role in its proof. We expect that, to higher order in the fermions, 
the supersymmetry anomaly will be modified by gauge-invariant terms 
as in \cite{frs,fr3}.
From the definition of $H^r$, one can deduce the Bianchi identities  
\be
\de_{[\m} H^r_{\n\r\s ]} =-\frac{3}{2} c^r_z \tr_z (F^{(9)}_{[\m\n} F^{(9)}_{\r\s ]} ) 
-\frac{3}{2} c^r_w \tr_w (F^{(5)}_{[\m\n} F^{(5)}_{\r\s ]} )\quad .
\ee

We now want to determine the terms proportional
to $F \wedge F$ containing the goldstino 
that one has to add for the consistency of the model.
Unlike the ten dimensional case, where duality maps the 2-form theory 
with Chern-Simons couplings
to the 6-form theory with Wess-Zumino couplings, in this case 
the low-energy effective action contains both Chern-Simons and Wess-Zumino  
couplings.
First of all, we observe that for the quantity
\bea
\hat{B}^r_{\m\n} &=& B^r_{\m\n} -iv^r (\bar{\psi}_{[\m}\g_{\n ]}\th )
-\frac{1}{2} x^{Mr} (\bar{\chi}^M \g_{\m\n} \th ) -\frac{2i}{\sqrt{2}}c^{rz}\tr_z 
[A_{[\m}^{(9)} (\bar{\th} \g_{\n ]}\l^{(9)} ) ]\nonumber \\
&+& \frac{i}{8} (\de_\r v^r ) (\bar{\th}\g_{\m\n}{}^\r \th )
+\frac{i}{8} x^{Mr} H^M_{[\m}{}^{\r\s} (\bar{\th} \g_{\n ] \r\s} \th )+
\frac{i}{2} v^r (\bar{\th} \g_{[\m }D_{\n ]}\th )\nonumber \\
&-& \frac{i}{4}c^{rz}\tr_z [A_{[\m}^{(9)} F^{(9)\r\s} ](\bar{\th} \g_{\n ] \r\s} \th )
-\frac{i c^{r1}}{4 v_s c^{s1}} {\cal A}_\a^A{}_B \xi^{\a i} A_{[ \m}^{(9)i}
(\bar{\th}^A \g_{\n ]} \th^B ) 
\label{bextra}
\eea
the supersymmetry variation is a general coordinate 
transformation of the correct  
parameter, together with an additional tensor gauge transformation 
of parameter
\be
\L^r_\m = -\frac{1}{2} v^r \xi_\m -\xi^\n B^r_{\m\n} \quad ,
\label{tega}
\ee
as well as PST gauge transformations of parameters 
\be
\L^{(PST)r}_\m =\frac{\de^\s \Xi}{(\de \Xi )^2 }[v^r v_s H^{s-}_{\s\m\r}
- x^{Mr} x^M_s H^{s+}_{\s\m\r}] \xi^\r 
\label{pst1}
\ee
and 
\be
\L^{(PST)} = \xi^\m \de_\m \Xi 
\label{pst2}
\ee
and gauge transformations of the form  (\ref{gaugeB}) and 
(\ref{cs6bis}) whose parameters are
as in eq. (\ref{gauge9}) and (\ref{gauge5}). 
We should now consider all the terms proportional to $F \wedge F$
that arise, those directly introduced by 
the inclusion of the Chern-Simons 3-form for the fields in the $\bar{5}$-$\bar{5}$ sector, those originating from 
the consequent modification of the Bianchi identities,
and finally those introduced by the variation of the Wess-Zumino term.

The end result is that the variation of all these contributions gives
\bea
\d {\cal L} &=& \e^{\m\n\r\s\d\t} \{ -\frac{i}{4} v_r (\bar{\e} \g_\m \psi_\n )
 + \frac{1}{8} x^{M}_{r}(\bar{\e}\g_{\m\n} \chi^M ) \} c^{rw} \tr_w (F^{(5)}_{\r\s} F^{(5)}_{\d\t})
\nonumber\\
&-& 2v_r c^{rw} \tr_w (\d A^{(5)}_\m F^{(5)}_{\n\r} )K^{\m\n\r}-2 x^M_r c^{rw} \tr_w
(\d A^{(5)}_\m F^{(5)}_{\n\r} )K^{M \m\n\r}  \quad .\label{deltalag}
\eea
The first two terms are canceled by the goldstino variation in the
additional couplings
\be
{\cal L}^\prime = \e^{\m\n\r\s\d\t} 
\{ \frac{i}{4}v_r(\bar{\th} \g_\m \psi_\n )
- \frac{1}{8} x^{M}_{r}(\bar{\th}\g_{\m\n} \chi^M ) \} c^{rw} \tr_w (F^{(5)}_{\r\s} F^{(5)}_{\d\t})
\quad ,
\ee
where, however, the variations of the gravitino and the tensorinos 
produce additional terms. Some of these 
cancel the last two terms in eq. (\ref{deltalag}), while the remaining ones
are canceled by the goldstino variation in
\bea
{\cal L}^{\prime \prime} &=& \e^{\m\n\r\s\d\t} \{ -\frac{i}{32}\de^\r v_r
(\bar{\th}\g_{\m\n\r} \th )  -\frac{i}{8} v_r (\bar{\th} \g_\m
D_\n \th )  \nonumber \\
&-& \frac{i}{32}x^M_r K^M_\m{}^{\a\b} (\bar{\th}\g_{\n\a\b} \th ) 
\} c^{rw} \tr_w (F^{(5)}_{\r\s} F^{(5)}_{\d\t} ) \quad .
\eea
If one restricts the attention to terms up to quartic fermion couplings, 
no further contributions are produced.
We can thus conclude that the non-linear realization of supersymmetry is 
granted by the inclusion of ${\cal L}^\prime$ and ${\cal L}^{\prime\prime}$ 
in the low-energy effective action. From eq. (\ref{bextra}) we also
see that these two contributions can be written in the compact form
\be
{\cal L}^\prime + {\cal L}^{\prime\prime}= -\frac{1}{4}\e^{\m\n\r\s\d\t}
B^r_{\m\n}{}^{extra} c_r^{w} \tr_w (F^{(5)}_{\r\s} F^{(5)}_{\d\t} )\quad ,\label{2gs}
\ee
where
\bea
B^r_{\m\n}{}^{extra} = &-& iv^r (\bar{\psi}_{[\m}\g_{\n ]}\th )
-\frac{1}{2} x^{Mr} (\bar{\chi}^M \g_{\m\n} \th ) -\frac{2i}{\sqrt{2}}c^{rz} \tr_z 
[A_{[\m}^{(9)} (\bar{\th} \g_{\n ]}\l^{(9)} ) ]\nonumber \\
&+&  \frac{i}{8} (\de_\r v^r ) (\bar{\th}\g_{\m\n}{}^\r \th )
+\frac{i}{8} x^{Mr} K^M_{[\m}{}^{\r\s} (\bar{\th} \g_{\n ] \r\s} \th )+
\frac{i}{2} v^r (\bar{\th} \g_{[\m }D_{\n ]}\th )
\eea
coincides with the counterterm of $B^r$ only if no 9-9 vectors are present.

We now want to interpret these non-geometric terms along the lines
of Section (5.1). To this end, observe that, if no 9-9 vectors are 
present, eq. (\ref{2gs}) is exactly twice the 
term that one should add to eq. (\ref{gs6}) in order to geometrize the 
Wess-Zumino term, substituting $B$ with $\hat{B}$. 
This means, roughly speaking, that half of the contribution in eq. (\ref{2gs})
comes from the Green-Schwarz term, and half from the Chern-Simons couplings. 
This interpretation is in perfect agreement with self-duality, and thus in six 
dimensions there is no duality transformation that can give a fully geometric 
Lagrangian. If also 9-9 vectors are in the spectrum, no additional terms are
produced in the lagrangian, in agreement with the fact that the additional terms of $\hat{B}^r$ in eq. 
(\ref{bextra}) are not gauge invariant.

To resume, the Lagrangian for supergravity coupled to tensor multiplets, 
hypermultiplets and non-supersymmetric vectors is
\bea
{\cal L} &=& {\cal L}_{susy} -\frac{1}{2} \hat{e}\hat{v}^r c_r^w \tr_w (F^{(5)}_{\m\n} F^{(5)}_{\r\s} )
\hat{g}^{\m\r} \hat{g}^{\n\s} - \L \hat{e}f(\hat{\Phi}^{\bar{\a}}, \hat{\phi}^\a )\nonumber \\
&+&  \frac{1}{2} \hat{e}(\hat{D}_\m S )^\dagger ( \hat{D}_\n S ) \hat{g}^{\m\n}
-\frac{1}{8} \e^{\m\n\r\s\d\t} B^r_{\m\n} c_r^w \tr_w (F^{(5)}_{\r\s}F^{(5)}_{\d\t}) \nonumber \\
&-& \frac{1}{4}\e^{\m\n\r\s\d\t}
B^r_{\m\n}{}^{extra} c_r^w \tr_w (F^{(5)}_{\r\s} F^{(5)}_{\d\t} ) \quad . 
\label{complag6}
\eea
Since the supersymmetry transformation of other non-supersymmetric
fermions is of higher order in the Fermi fields, at this level we can always
add them in the construction, while 
the couplings that can not be determined by supersymmetry could in 
principle be determined by string inputs, as in \cite{dm2}.

Finally, it is important to observe that without 9-9 vectors, 
although the Lagrangian 
(\ref{complag6}) is not completely geometric, the corresponding 
field equations are. 
Indeed, if one fixes the PST gauge in such a way 
that the 3-forms satisfy the standard (anti)self-duality conditions, 
the equation for the vector fields, up to terms quartic in the fermions, 
is 
\bea
& & \hat{e} D_\n [ \hat{v}^r c_{rw} F^{(5)}_{\r\s} \hat{g}^{\m\r}
\hat{g}^{\n\s}
]+\frac{1}{6} \e^{\m\n\r\s\d\t} 
c_{rw} F^{(5)}_{\n\r}
\hat{H}^r_{\s\d\t} \nonumber\\
& & + \frac{1}{12}\e^{\m\n\r\s\d\t} c_{rw} F^{(5)}_{\n\r} c^{rw^\prime}
\w^{(5)w^\prime}_{\s\d\t} 
+ \frac{1}{8}\e^{\m\n\r\s\d\t} c_{rw} A^{(5)}_{\n} c^{rw^\prime}
\tr_{w^\prime}(F^{(5)}_{\r\s} F^{(5)}_{\d\t} ) =0 \ ,
\eea
where 
\be
\hat{H}^r_{\m\n\r} = 3 \de_{[\m} \hat{B}^r_{\n\r ]} - c^r_w \w^{(5)w}_{\m\n\r} 
\quad ,
\ee
and this is nicely of geometric form.
It should be noticed that no additional counterterms 
containing the goldstino have to be added if also 9-9 vectors are present.
In fact, all the terms in $\hat{B}^r$ induced by $A^{(9)}$ are 
not gauge invariant,
and their inclusion in the lagrangian is forbidden because it would modify
the gauge anomaly.  The resulting equation for $A^{(5)}$ is then
\bea
& & \hat{e} D_\n [ \hat{v}^r c_{rw} F^{(5)}_{\r\s} \hat{g}^{\m\r}
\hat{g}^{\n\s}
]+\frac{1}{6} \e^{\m\n\r\s\d\t} 
c_{rw} F^{(5)}_{\n\r}
\tilde{H}^r_{\s\d\t} \nonumber\\
& & + \frac{1}{12}\e^{\m\n\r\s\d\t} c_{rw} F^{(5)}_{\n\r} c^{rz}
\w^{(9)z}_{\s\d\t} 
+ \frac{1}{8}\e^{\m\n\r\s\d\t} c_{rw} A^{(5)}_{\n} c^{rz}
\tr_{z}(F^{(9)}_{\r\s} F^{(9)}_{\d\t}) \nonumber \\
& & + \frac{1}{12}\e^{\m\n\r\s\d\t} c_{rw} F^{(5)}_{\n\r} c^{rw^\prime}
\w^{(5)w^\prime}_{\s\d\t} 
+ \frac{1}{8}\e^{\m\n\r\s\d\t} c_{rw} A^{(5)}_{\n} c^{rw^\prime}
\tr_{w^\prime}(F^{(5)}_{\r\s} F^{(5)}_{\d\t} ) =0 ,
\label{eqcons}
\eea
where
\be
\tilde{H}^r_{\m\n\r} = 3 \de_{[\m} B^r_{\n\r ]} + 3 \de_{[\m} B^r_{\n\r ]}{}^{extra}
- c^{rz} \w^{(9)z}_{\m\n\r}- c^{rw} \w^{(5)w}_{\m\n\r}
\ee
is geometric up to gauge-invariant terms proportional to $c^{rz}$.
The result is thus in agreement with what expected
by anomaly considerations.  If gauge and supersymmetry anomalies are 
absent, the $A^{(5)}$ equation is mapped into itself by supersymmetry:
this is the very reason why this equation is geometric.  In the presence of
gauge and supersymmetry anomalies, as long as 9-9 vectors are absent, 
the equation
for $A^{(5)}$ is still geometric, albeit not gauge invariant.  
The supersymmetry anomaly, in this case, results from the gauge 
transformation contained in eq. (\ref{efxi}).
When also 9-9 vectors are present,  these arguments do not apply, 
and thus in eq. (\ref{eqcons}) 
the geometric structure is violated 
by terms proportional to $c^{rz} c_{r}^{w}$.  

The consistent formulation described above 
can be reverted to a supersymmetric formulation in terms of 
covariant non-integrable field equations \cite{as,rs2}, 
that embody the corresponding covariant gauge anomaly
\bea
{\cal A}_\L^{cov} & = & \frac{1}{2} \e^{\m\n\r\s\d\t} [c^{rz} c_r^{z^\prime} \tr_z (\L^{(9)} F^{(9)}_{\m\n} )
\tr_{z^\prime} (F^{(9)}_{\r\s} F^{(9)}_{\d\t} )\nonumber \\
& + & c^{rz} c_r^{w} \tr_z (\L^{(9)} F^{(9)}_{\m\n} )
\tr_{w} (F^{(5)}_{\r\s} F^{(5)}_{\d\t} )\nonumber \\
& + & c^{rw} c_r^{z} \tr_w (\L^{(5)} F^{(5)}_{\m\n} )
\tr_{z} (F^{(9)}_{\r\s} F^{(9)}_{\d\t} )\nonumber \\
& + & c^{rw} c_r^{w^\prime} \tr_w (\L^{(5)} F^{(5)}_{\m\n} )
\tr_{w^\prime} (F^{(5)}_{\r\s} F^{(5)}_{\d\t} ) ]\quad ,
\label{covan}
\eea
given by the divergence of the covariant equation for $A_\m^{(5)}$,
\be
\hat{e} D_\n [ \hat{v}^r c_{rw} F^{(5)}_{\r\s}
\hat{g}^{\m\r} \hat{g}^{\n\s} ]+\frac{1}{6} \e^{\m\n\r\a\b\g} c_{rw} F^{(5)}_{\n\r}
\tilde{H}^r_{\a\b\g} =0 \quad ,
\label{coveq}
\ee
and the divergence of the covariant equation for $A_\m^{(9)}$ \cite{rs2}.
Without 9-9 vectors, eq. (\ref{coveq}) is both geometric and 
gauge-covariant, while, if 9-9 vectors are present, 
the geometric structure is violated 
by gauge-covariant terms proportional again to $c^{rz} c_{r}^{w}$.

\section{Discussion}
\label{comments}
\fancyhead[LO]{{\footnotesize 5.3~~{\it Discussion}}}

In this chapter we have reviewed the results of \cite{pr}, extending the work of \cite{dm2} on the
low-energy effective action for models with brane supersymmetry
breaking. In this class of models a supersymmetric bulk is
coupled to a non-supersymmetric open sector, and as a result
local supersymmetry is non-linearly realized {\it \`a la}
Volkov-Akulov. In particular, we have shown that, up to quartic order
in the fermions, the low-energy
couplings between the supersymmetric bulk and the non-supersymmetric open
sector in the ten-dimensional $USp(32)$ model of \cite{sugimoto} are {\it all}
of geometric origin, being induced by the dressing of the bulk fields
in terms of the goldstino, provided one turns to the dual 6-form
\cite{chamseddine} formulation. Thus, in retrospect, the non-geometric
terms in \cite{dm2} are precisely what is needed to geometrize the dual
form of the theory, where the (high-derivative)
Chern-Simons couplings are absent. We have completed a similar construction
for six-dimensional models with brane supersymmetry breaking. Since in
this case both Chern-Simons and Wess-Zumino terms are simultaneously
present, not all couplings in the Lagrangian can be related to 
goldstino-dependent dressings of bulk fields. 
However, in the absence of supersymmetric vectors, the 
field equations exhibit this geometric structure, that is
naturally violated in the general case by anomalous terms.  

It would be interesting to apply the same construction to the four-dimensional 
brane supersymmetry breaking models of \cite{bsb}, and in general to brane-world scenarios (see 
\cite{dudrev} and references therein) in which 
supersymmetry is linearly realized in the gravitational sector and non-linearly
realized in the brane universe.  However,  
in four dimensions the gravitino can acquire a Majorana
mass through a super-Higgs mechanism \cite{suhig}, 
and it is this difference with respect to minimal 
supergravity in ten and six dimensions that makes the models studied 
in this paper rather peculiar.

A general property of all this class of models is the presence of 
a dilaton tadpole of positive sign, required in order 
to have a correct kinetic term for the goldstino \cite{dm2}, and
guaranteed by the residual tension of anti-branes and orientifolds. 
In ten dimensions the tadpole signals the impossibility of having
maximally symmetric vacuum configurations \cite{dm3}, and one should try 
to analyze the same effects in the six-dimensional models discussed in 
this paper.

Finally, it should be observed that it is always possible, in any 
supersymmetric theory coupled to a goldstino, to dress the fields in 
the linear sector by terms containing the goldstino itself.
This is a property of the commutator of two supersymmetries:
by construction, a supersymmetry transformation on the dressed fields 
exactly corresponds to the commutator of two supersymmetries on
the linear fields, producing general coordinate transformations together 
with all the other local symmetry transformations. 
In the six dimensional case discussed in Section $4$, this can be 
explicitly verified:
the parameters of eqs. (\ref{xi6}), (\ref{gauge9}), (\ref{lolo}), 
(\ref{tega}), (\ref{pst1}) and (\ref{pst2}) coincide with those
coming from the supersymmetry algebra, provided one
substitutes  $ -\frac{i}{2}(\bar{\th} \g_\m \e )$ with 
$- i (\bar{\e}_1 \g_\m \e_2 )$ \cite{frs,rs1,fr3}.
Following this way of reasoning, one could try to generalize the results 
obtained here to all orders in the Fermi fields.

\newpage ~                  %  per lasciare una pagina bianca
\thispagestyle{empty}       %  

\chapter*{Conclusions}
\label{concl}
\addcontentsline{toc}{chapter}{Conclusions}
% \newpage
\vspace*{2cm}
\fancyhead[RO,LE]{\thepage}
\fancyhead[RE,LO]{{\footnotesize {\rm Conclusions}}} 

\noindent
The main subject of this dissertation has been the analysis of minimal six-dimensional supergravity. 
As we have seen in Chapter 4, these models are ``classically'' anomalous, and are thus determined
requiring the closure of  the Wess-Zumino consistency conditions \cite{wz} rather than requiring supersymmetry invariance. 
As a result, the low-energy effective action is not unique, and a quartic gaugino coupling remains undetermined,
since the model is consistent for any value of this coupling. 
An interesting open problem is then to analyze the same coupling at the string level, that would result from 
an annulus amplitude with the insertion of two gauginos at each boundary. Once one extracts the low-energy limit
from this amplitude (see for instance \cite{dm4} and references therein) one should be able to understand whether this 
coupling is determined by string theory. This result would be quite important, since it would state that string theory
 is more powerful than supersymmetry in determining the low-energy effective action, even for vacua with 8 
supercharges, that a priori should receive string corrections only in the hypermultiplet sector. 

Another very interesting open problem related to these six-dimensional vacua is to understand the physics corresponding 
to the tensionless string phase transition \cite{sw,dlp}. There is an analogous phenomenon in ${\cal N} =(2,0)$ 
six-dimensional models \cite{wit,openpbranes}, and in \cite{henningson} some recent attempts have been made in order to
give a proper definition to interacting superconformal quantum theories in six dimensions. These theories (see \eg 
\cite{9911147,9705117} for reviews) seem also to give the correct framework for a better understanding of ${\cal N} =4$
SYM theory in four dimensions. In particular, the fact that ${\cal N} =4$ four-dimensional theories are obtained by 
compactification of six-dimensional superconformal quantum theories on $T^2$ could provide an explanation for $SL(2,Z)$ S-duality 
(in this respect, a better understanding of these theories could also correspond to a deeper comprehension of F-theory 
\cite{ftheory}).

In the last part of the thesis we have analyzed the low-energy effective action for type-I models with brane 
supersymmetry breaking, both in ten and six dimensions. 
We emphasize once more that these models are very interesting from a phenomenological point of view in the context 
of brane-world scenarios. In would be relevant then to study them in more detail, trying to determine higher order
Fermi terms, and also obtaining in specific cases the scalar potential.

% \newpage ~                  %  per lasciare una pagina bianca
% \thispagestyle{empty}       %  

% APPENDICES

\appendix

\chapter{Notations and spinor algebra}
\label{appa}
% \newpage
\vspace*{2cm}
\fancyhead[RO,LE]{\thepage}
\fancyhead[RE]{{\footnotesize {\rm Appendix A.}~~{\it 
Notations and spinor algebra}}} 
\fancyhead[LO]{{\footnotesize {\it Notations and spinor algebra}}}

\section{Reality properties}
\label{realprop}
\fancyhead[LO]{{\footnotesize A.1~~{\it Reality properties}}}

In our conventions, the Clifford algebra has the form
\be
\{ \g_m ,\g_n \} = 2 \eta_{mn} {\bf 1} \quad ,
\ee
and we use the mostly minus convention for the signature of space-time. 
The matrix $\g^0$ is hermitian, while the 
$\g^i$'s are anti-hermitian, with $(\g^0 )^2 = 1$ and $\g^i \g^i = -1$.

We want to resume here the reality properties of the spinors in various dimensions (see for instance 
\cite{sfspinors} and references therein for details).
We are intersted only in spinors of the Lorentz group $SO(D-1,1)$. In general 
it can be shown that for the group $SO(t,s)$, the reality properties of the 
spinor only depend on $ \vert s - t \vert \ {\rm mod} \ 8 $. 
One can define a real spinor if it is possible to construct consistently a
charge conjugation matrix $C$ such that
\be
C \g_\m C =\pm \g_\m^T \quad .\label{cdef}
\ee
The charge-conjugated spinor is
\be
\psi_C = C \bar{\psi}^T \quad ,
\ee
where $\bar{\psi} = \psi^\dagger \g_0 $, and this definition is consistent if one can consistently 
define a Majorana spinor, satisying
\be
\psi = \psi_C \quad .\label{realspinor}
\ee
This can be obtained it two ways, with either $C$ symmetric satisfying eq. (\ref{cdef}) with the 
plus sign, or $C$ antisymmetric satisfying eq. (\ref{cdef}) with the minus sign. 
In the former case, if $\psi$ and $\chi$ satisfy eq. (\ref{realspinor}), 
the bilinear $(\bar{\psi} \chi )$ is odd under Majorana flip and thus is an imaginary number, while
in the latter case it is even under Majorana flip and thus is a real 
number.\footnote{We define the charge conjugation operation on Grassmann variables as $(ab)^* = b^* a^* $, 
and so if $a$ and $b$ are real, $ab$ is imaginary.}
As we will see, the situation is reversed in the case in which the spinors satisfy symplectic Majorana
conditions.
The properties under flip of the bilinears obtained contracting the spinors with $\g$ matrices can be
straightforwardly obtained: the result is that if the number of $\g$ matrices is odd, the two $C$'s give
the same Majorana flip properties, while  if the number of $\g$ matrices is even the 
flip properties are opposite. 
The fact that the two definitions give the same flip properties 
when an odd number of $\g$ matrices is inserted is fundamental in order to close the supersymmetry
algebra. 
Whenever possible, we always choose $C$  
such that $(\bar{\psi} \chi )$ is even under Majorana flip. 

Let us begin by discussing  the two-dimensional case, in which the $\g$ matrices are two-dimensional. 
One possible choice is
\bea
\g^0 &=& \s_1 \quad ,\nonumber \\
\g^1 &=& i \s_2 \quad .\label{clifford2}
\eea
The chirality matrix is $\g_3 =\s_3$.
The symmetric matric $C_S =\s_1$ satisfies (\ref{cdef}) with the 
plus sign, while the antisymmetric matrix $C_A =\s_2$ satisfies eq. (\ref{cdef}) with the minus sign, so
they are both good definitions for a real spinor. Observe that in our base, 
the choice $C_S$ corresponds in eq. (\ref{realspinor}) to the condition $\psi =\psi^\star$. 
Both these matrices anticommute with $\g_3$, and so both the conditions are compatible with 
the chirality condition: in two dimensions one can define a Majorana-Weyl spinor. 

As usual, one goes from $D=2n$ to $D=2n+1$ adding $\g^D = i \hat{\g}$. So in three dimensions we add 
\be
\g^2 = i\s_3 
\ee
to eq. (\ref{clifford2}). The only possible choice is $C= C_A = \s_2$, satisfying eq. (\ref{cdef}) 
with the minus sign, and so in three dimensions one can define a Majorana spinor. 

In four dimensions the $\g$ matrices are $4 \times 4$, and can be written as the tensor
product of two Pauli matrices:
\bea
\g^0 &=& \s_1 \otimes {\bf 1}\quad ,\nonumber \\
\g^1 &=& i \s_2 \otimes \s_1 \quad , \nonumber \\
\g^2 &=& i \s_2 \otimes \s_2 \quad , \nonumber \\
\g^3 &=& i \s_2 \otimes \s_3 \quad .\label{clifford4}
\eea
The chirality matrix is $\g_5 = \s_3 \otimes {\bf 1}$. The condition (\ref{cdef}) is satisfied by
$C_1 = {\bf 1} \otimes \s_2$ with the plus sign and by $C_2 =\s_3 \otimes \s_2$ with the minus sign. Since 
both these matrices are antisymmetric, only the second gives a consistent reality condition. The fact that 
this matrix commutes with $\g_5$ means that the Majorana and Weyl conditions are not compatible. 

In five dimensions we add 
\be
\g^4 = i \s_3 \otimes {\bf 1} 
\ee
to eq. (\ref{clifford4}). Then we are left with only the matrix $C_1 = {\bf 1} \otimes \s_2$, and so 
it is not possible to impose a Majorana condition on a single spinor. 
This is analogous to the six-dimensional case, in which the $8 \times 8$ $\g$ matrices can be written in the
form 
\bea
\g^0 &=& \s_1 \otimes {\bf 1} \otimes \s_1 \quad ,\nonumber \\
\g^1 &=& i \s_2 \otimes \s_1 \otimes {\bf 1}\quad , \nonumber \\
\g^2 &=& i \s_2 \otimes \s_2 \otimes {\bf 1}   \quad , \nonumber \\
\g^3 &=& i \s_2 \otimes \s_3 \otimes {\bf 1} \quad , \nonumber \\
\g^4 &=& i \s_1 \otimes {\bf 1} \otimes \s_2 \quad , \nonumber \\
\g^5 &=& i \s_1 \otimes {\bf 1} \otimes \s_3 \quad .\label{clifford6}
\eea
The matrix $C_1 =  \s_2 \otimes \s_2 \otimes \s_2$ is antisymmetric and satisfies eq. (\ref{cdef}) with
the plus sign, while the matrix  $C_2 =  \s_1 \otimes \s_2 \otimes \s_2$ is symmetric 
and satisfies eq. (\ref{cdef}) with the minus sign, and neither give a consistent Majorana condition for
a single spinor.
the seven-dimensional case is analogous to five dimensions.

In five, six and seven dimensions one can define for a $USp(2)$ doublet of 
spinors the symplectic Majorana condition
\be
\psi^A =\e^{AB} C \bar{\psi}^T_B \quad,\label{majoranacond}
\ee 
where
\be
\bar{\psi}_A =(\psi^A )^\dagger \g_0
\ee 
and $\e^{12} =\e_{12}=1$. 
Any bilinear
$\bar{\psi}_A \chi^B$ carries a pair of $USp(2)$ indices, and can be decomposed in terms of the
identity and of the three Pauli matrices. Indeed, one can form the bilinears
\be 
(\bar{\psi} \chi )=\bar{\psi}_A \chi^A \quad ,\qquad [\bar{\psi} \chi ]_i =
\sigma_{i A}{}^B \bar{\psi}_B \chi^A \quad,
\ee 
and standard properties imply that
\be
\bar{\psi}_A \chi^B =\frac{1}{2}\delta_A^B (\bar{\psi} \chi ) +\frac{1}{2}\sigma_{i
A}{}^B [\bar{\psi} \chi ]_i\quad.\label{sp2exp}
\ee 
Using eq. (\ref{majoranacond}), and choosing $C$ to be symmetric, 
one can then see that the Fermi bilinear
$(\bar{\psi} \chi )$ has standard behavior under Majorana-flip, namely
\be 
(\bar{\psi} \chi ) =(\bar{\chi} \psi ) \quad ,
\ee 
while all three bilinears $[\bar{\psi} \chi ]_i $ have the anomalous behavior
\be 
[\bar{\psi} \chi ]_i = -[\bar{\chi} \psi ]_i \quad .
\ee 
Corresponding relations hold for all Fermi bilinears, that naturally display pairs
of opposite behaviors under Majorana flip.  In particular, these properties imply that
\be 
[ \bar{\psi} \g_{\m\n\r} \psi ]_i =0\quad ,
\ee 
a relation often used in deriving the results of Chapter 4. 
In six dimensions the symplectic Majorana condition and the Weyl condition can be imposed together, 
since $C$ anticommutes with the chirality matrix $\g_7 = \s_3 \otimes {\bf 1} \otimes {\bf 1}$.

Following the same arguments as before, one can show that the eight-dimensional case is similar 
to four dimensions, while the nine-dimensional case is similar to three dimensions. 
Let us analyze in more detail the ten-dimensional case, that is completely analogous 
to the two-dimensional case by general arguments. Making the choice
\bea
\g^0 &=&   \s_1 \otimes {\bf 1} \otimes \s_1    \otimes \s_1   \otimes {\bf 1}\quad ,\nonumber \\
\g^1 &=& i \s_2 \otimes \s_1    \otimes {\bf 1} \otimes {\bf 1}\otimes \s_1   \quad , \nonumber \\
\g^2 &=& i \s_2 \otimes \s_2    \otimes {\bf 1} \otimes {\bf 1}\otimes \s_1   \quad , \nonumber \\
\g^3 &=& i \s_2 \otimes \s_3    \otimes {\bf 1} \otimes {\bf 1}\otimes \s_1   \quad , \nonumber \\
\g^4 &=& i \s_1 \otimes {\bf 1} \otimes \s_2    \otimes \s_1   \otimes {\bf 1}\quad , \nonumber \\
\g^5 &=& i \s_1 \otimes {\bf 1} \otimes \s_3    \otimes \s_1   \otimes {\bf 1}\quad ,\nonumber\\
\g^6 &=& i \s_1 \otimes {\bf 1} \otimes {\bf 1} \otimes \s_2   \otimes {\bf 1}\quad , \nonumber \\
\g^7 &=& i \s_1 \otimes {\bf 1} \otimes {\bf 1} \otimes \s_3   \otimes {\bf 1}\quad ,\nonumber\\
\g^8 &=& i \s_2 \otimes {\bf 1} \otimes {\bf 1} \otimes {\bf 1}\otimes \s_2   \quad , \nonumber \\
\g^9 &=& i \s_2 \otimes {\bf 1} \otimes {\bf 1} \otimes {\bf 1}\otimes \s_3   \quad , \label{gammad=10}
\eea
one obtains that  eq. (\ref{cdef}) is satisfied 
by the symmetric matrix $C_S = \s_1 \otimes \s_2 \otimes \s_2    \otimes \s_3   
\otimes \s_3$ with the plus sign, and by the antisymmetric matrix 
$C_A = \s_2 \otimes \s_2 \otimes \s_2    \otimes \s_3  \otimes \s_3$ with the minus sign, and both
these choices are consistent with eq. (\ref{realspinor}). Moreover, both these matrices anticommute
with the chirality matrix 
$\g_{11} = \s_3 \otimes {\bf 1} \otimes {\bf 1}\otimes {\bf 1} \otimes {\bf 1}$, so that 
the Majorana condition can be imposed on a Weyl spinor, as in two dimensions. 
Finally the eleven-dimensional case is completely analogous to three dimensions: the complete set 
of $\g$ matrices can be obtained adding 
\be
\g^{10} = i \g_{11} = i \s_3 \otimes {\bf 1} \otimes {\bf 1}\otimes {\bf 1} \otimes {\bf 1}
\ee
to eq. (\ref{gammad=10}), and from this one obtains the antisymmetric charge-conjugation matrix
\be
C_A = \s_2 \otimes \s_2 \otimes \s_2    \otimes \s_3  \otimes \s_3 \quad, 
\ee
satisfying eq. (\ref{cdef}) with the minus sign.

\section{Fierz identities}
\label{fierz}
\fancyhead[LO]{{\footnotesize A.2~~{\it Fierz identities}}}

In this section we collect the Fierz relations that are used in this dissertation.
We begin with the four dimensional case, in which the product of two spinors has 16 components,
and can be expanded in terms of the bilinears obtained contracting the spinors with the identity,
$\g_5$, $\g_\m$,  $\g_\m \g_5$ and $\g_{\m\n}$. All the other bilinears are related
to these by the relation
\be
\g^{\m_1 ...\m_n}=\frac{i(-1)^{[n/2]}}{e(4-n)!}\e^{\m_1 ...\m_n \n_1
...\n_{4-n}}\g_{\n_1 ...\n_{4-n}} \g_5 \quad ,
\ee
where $\g^{\m_1 ...\m_n} ={\bf 1}$ for $n=0$, and $\e^{0123}=1$. 
The result is
\bea
\psi \bar{\chi} &=& -\frac{1}{4} {\bf 1} (\bar{\chi}\psi ) -\frac{1}{4} \g_5 (\bar{\chi} \g_5 \psi )
-\frac{1}{4} \g_\m (\bar{\chi} \g^\m \psi ) \nonumber\\
&+& \frac{1}{4} \g_\m \g_5 (\bar{\chi} \g^\m \g_5 \psi )
+ \frac{1}{8} \g_{\m\n} (\bar{\chi} \g^{\m\n} \psi ) \quad .
\eea
For Majorana spinors, this implies
\be
\psi \bar{\chi} - \chi \bar{\psi}= -\frac{1}{2} \g_\m (\bar{\chi} \g^\m \psi )
+ \frac{1}{4} \g_{\m\n} (\bar{\chi} \g^{\m\n} \psi ) \quad .
\ee

In six dimensions, using $\e^{012345}=+1 $, one obtains
\be
\g^{\m_1 ... \m_n}=- \frac{(-1)^{[n/2]}}{e(6-n)!}\e^{\m_1 ...\m_n \n_1
...\n_{6-n}}\g_{\n_1 ...\n_{6-n}} \g_7 \quad .\label{gammamatrices6}
\ee 
In particular, eq. (\ref{gammamatrices6}) shows that $\g_{\m\n\r} \psi$ is self-dual
if $\psi$ is  left-handed, \ie $\g_7 \psi =\psi$, and antiself-dual if $\psi$ is right-handed. 
One can study Fierz relations between spinor bilinears  using eq. (\ref{sp2exp}). If
$\psi$ and
$\chi$ have the same chirality
\be
\psi \bar{\chi} = -\frac{1}{4}\bar{\chi} \g^\m \psi \g_\m +\frac{1}{48}
\bar{\chi}
\g^{\m\n\r} \psi \g_{\m\n\r}\quad ,\label{fierz1}
\ee 
while if they have opposite chirality
\be
\psi \bar{\chi} = -\frac{1}{4} \bar{\chi} \psi  +\frac{1}{8} 
\bar{\chi} \g^{\m\\n} \psi \g_{\m\n} \quad .
\ee 
In the case of spinors satisfying the symplectic Majorana condition (\ref{majoranacond}),
interesting results are obtained (anti)symmetrizing these relations. In
particular, eq. (\ref{fierz1}) implies
\be
\psi^A \bar{\chi}_B -\chi^A \bar{\psi}_B = -\frac{1}{4} (\bar{\chi} \g^\m \psi
)\delta_B^A \g_\m +\frac{1}{48}  [\bar{\chi} \g^{\m\n\r} \psi ]_i \sigma_{i B}{}^A
\g_{\m\n\r}\quad .
\ee 

Now we consider the ten-dimensional case. From the definition of $\g_{11}$ one obtains
\be
\g^{\m_1 ... \m_n}= \frac{(-1)^{[n/2]}}{e(10-n)!}\e^{\m_1 ...\m_n \n_1
...\n_{10-n}}\g_{\n_1 ...\n_{10-n}} \g_{11} \quad ,\label{gammamatrices10}
\ee 
that again implies that $\g_{\m_1 ...\m_5} \psi$ is self-dual
if $\psi$ is  left-handed, and antiself-dual if $\psi$ is right-handed. 
The Fierz identity is
\be
\psi \bar{\chi} = 
-\frac{1}{32} \g_\m (\bar{\chi} \g^\m \psi ) +\frac{1}{192} \g_{\m\n\r} 
(\bar{\chi} \g^{\m\n\r} \psi )
- \frac{1}{32 \cdot 120} \g_{\m\n\r\s\t} (\bar{\chi} \g^{\m\n\r\s\t} \psi ) 
\ee
for spinors of the same chirality and
\be
\psi \bar{\chi} = 
-\frac{1}{32} (\bar{\chi} \psi ) +\frac{1}{64} \g_{\m\n} 
(\bar{\chi} \g^{\m\n} \psi )
- \frac{1}{32 \cdot 24} \g_{\m\n\r\s} (\bar{\chi} \g^{\m\n\r\s} \psi ) 
\ee
for spinors of opposite chirality.

Finally, we write the Fierz identities for $D=11$:
\bea
\psi \bar{\chi} &=& 
-\frac{1}{32} (\bar{\chi} \psi ) -\frac{1}{32} (\bar{\chi} \psi )+\frac{1}{64} \g_{\m\n} 
(\bar{\chi} \g^{\m\n} \psi ) +\frac{1}{192} \g_{\m\n\r} 
(\bar{\chi} \g^{\m\n\r} \psi )\nonumber\\
&-& \frac{1}{32 \cdot 24} \g_{\m\n\r\s} (\bar{\chi} \g^{\m\n\r\s} \psi ) 
- \frac{1}{32 \cdot 120} \g_{\m\n\r\s\t} (\bar{\chi} \g^{\m\n\r\s\t} \psi ) \quad .
\eea

\section{Six-dimensional conventions}
\label{six}
\fancyhead[LO]{{\footnotesize A.3~~{\it Six-dimensional conventions}}}

In six dimensions, a 3-form $X_{\m\n\r}$ is (anti)self-dual if  
\be 
X_{\m\n\r}=\ (-) \ \frac{1}{6e}\e_{\m\n\r\a\b\g}X^{\a\b\g}\quad .
\ee 
If $X_{\m\n\r}$ and $Y_{\m\n\r}$ are both (anti)self-dual, 
\be 
X_{\m\n\r}Y^{\m\n\r}=0
\ee 
and
\be X_{\m\n\a}Y^{\m\n}{}_\b - X_{\m\n\b}Y^{\m\n}{}_\a =0\quad ,
\ee 
while if they have opposite duality properties 
\be 
X_{\m\n\a}Y^{\m\n}{}_\b +X_{\m\n\b}Y^{\m\n}{}_\a =\frac{1}{3}g_{\a\b}
X_{\m\n\r}Y^{\m\n\r}\quad .
\ee 
Moreover, an (anti)self-dual antisymmetric tensor $X_{\m\n\r}$ satisfies
\be 
X^{\m\n\r}X_{\a\b\r} = \frac{1}{4} [ -\delta^\m_\b X_{\a\g\delta} X^{\n\g\delta}
+\delta^\n_\b X_{\a\g\delta}X^{\m\g\delta} +
\delta^\m_\a X_{\b\g\delta}X^{\n\g\delta} -\delta^\n_\a X_{\b\g\delta} X^{\m\g\delta } ]
\quad .
\ee

The indices of $USp(2)$ and $USp(2 n_H)$ are raised and lowered by the 
antisymmetric symplectic invariant tensors $\e^{AB}$ and $\W^{ab}$ with 
the following conventions:
\bea
& & V^A = \e^{AB} V_B \quad , 
\qquad V_A = \e_{BA} V^B \quad  \qquad (\e^{AB} \e_{AC} =\d^B_C )\quad ,\nonumber \\
& & 
W^a = \W^{ab} W_b \quad , \qquad W_a = \W_{ba} W^b \quad \qquad
( \W^{ab} \W_{ac}  = \d^b_c )\quad .
\eea
Spinors with $USp(2 n_H)$ indices satisfy the symplectic-Majorana condition
\be
\Psi^a = \W^{ab} C \bar{\Psi}^T_b \quad,
\ee
where
\be
\bar{\Psi}_{a} = (\Psi^{a} )^{\dag} \g_0 \quad .
\ee
From these relations one can deduce the properties of spinor bilinears under 
Majorana flip. For instance:
\be
(\bar{\chi}_A \Psi^a ) = \e_{AB} \W^{ab} (\bar{\Psi}_b \chi^B ) \quad ,
\ee
and similar relations when $\g$-matrices are included.
In our notations a spinor bilinear with two $USp(2)$ indices contracted is 
written without explicit indices, {\it i.e.}
\be
(\bar{\chi}_A \Psi^A ) \equiv (\bar{\chi} \Psi ) \quad,
\ee
while in all the other bilinears the symplectic indices are explicit.

The connections ${\cal{A}}_\a^A{}_B$ and ${\cal{A}}_\a^a{}_b$ are anti-hermitian.
Belonging to the adjoint representation of a symplectic group, 
they are symmetric if considered 
with both upper or both lower indices. 

The hermitian gauge-group generators $T^i$ satisfy 
the commutation relations
\be
[ T^i , T^j ] = i f^{ijk} T^k \quad , 
\ee
as well as the trace conditions
\be
\tr ( T^i T^j ) = \frac{1}{2} \d^{ij} \quad .
\ee

% \newpage ~                  %  per lasciare una pagina bianca
% \thispagestyle{empty}       %  

% \chapter{PST actions for self-dual fields}
% \label{appb}
% \input{appd2.txt}

% \newpage ~                  %  per lasciare una pagina bianca
% \thispagestyle{empty}       %  

% \newpage ~                  %  per lasciare una pagina bianca
% \thispagestyle{empty}       %  

% BIBLIOGRAPHY

\end{document}